\documentclass{article}
\usepackage{amsfonts}
\usepackage{amssymb}
\usepackage{graphicx}
\usepackage{amsmath}
\usepackage{geometry}
\usepackage{appendix}
\usepackage{authordate1-4}

\input{tcilatex}
\begin{document}

\title{Financial Markets and the Real Economy: \\
A Statistical Field Perspective on Capital Allocation and Accumulation}
\author{Pierre Gosselin\thanks{%
Pierre Gosselin : Institut Fourier, UMR 5582 CNRS-UGA, Universit\'{e}
Grenoble Alpes, BP 74, 38402 St Martin d'H\`{e}res, France.\ E-Mail:
Pierre.Gosselin@univ-grenoble-alpes.fr} \and A\"{\i}leen Lotz\thanks{%
A\"{\i}leen Lotz: Cerca Trova, BP 114, 38001 Grenoble Cedex 1, France.\
E-mail: a.lotz@cercatrova.eu} \and Marc Wambst\thanks{%
Marc Wambst : IRMA, UMR 7501 CNRS, Universit\'{e} de Strasbourg, 7 rue Ren%
\'{e} Descartes, 67084, Strasbourg Cedex, France.\ E-Mail:
wambst@math.unistra.fr}}
\date{May 2022}
\maketitle

\begin{abstract}
This paper provides a general method to directly translate a classical
economic framework with a large number of agents into a field-formalism
model. This type of formalism allows the analytical treatment of economic
models with an arbitrary number of agents, while preserving the system's
interactions and microeconomic features of the individual level.

We apply this methodology to model the interactions between financial
markets and the real economy, described in a classical framework of a large
number of heterogeneous agents, investors and firms. Firms are spread among
sectors but may shift between sectors to improve their returns. They compete
by producing differentiated goods and reward their investors by paying
dividends and through their stocks' valuation. Investors invest in firms and
move along sectors based on firms' expected long-run returns.

The field-formalism model derived from this framework allows for collective
states to emerge. We show that the number of firms in each sector depends on
the aggregate financial capital invested in the sector and its firms'
expected long-term returns. Capital accumulation in each sector depends\
both on short-term returns and expected long-term returns relative to
neighbouring sectors.

For each sector, three patterns of accumulation emerge. In the first
pattern, the dividend component of short-term returns is determinant for
sectors with small number of firms and low capital. \ In the second pattern,
both short and long-term returns in the sector drive intermediate-to-high
capital. In the third pattern, higher expectations of long-term returns
drive massive inputs of capital.

Instability in capital accumulation may arise among and within sectors.\ We
therefore widen our approach and study the dynamics of the collective
configurations, in particular interactions between average capital and
expected long-term returns, and show that overall stability crucially
depends on the expectations' formation process.

Expectations that are highly reactive to capital variations stabilize high
capital configurations, and drive low-to-moderate capital sectors towards
zero or a higher level of capital, depending on their initial capital.
Inversely, low-to moderate capital configurations are stabilized by
expectations moderately reactive to capital variations, and drive high
capital sectors towards more moderate level of capital equilibria.

Eventually, the combination of expectations both highly sensitive to
exogenous conditions and highly reactive to variations in capital imply that
large fluctuations of capital in the system, at the possible expense of the
real economy.


\end{abstract}

\section{Introduction}

This paper studies the interactions between financial and physical capital.
A large number of heterogeneous agents is divided into two groups, investors
and firms. The specificity of our work is to model these interactions
between agents as a field-theoretic model.

The field-formalism used in this paper is rooted in a probabilistic
description of economic systems with large number of agents. Classically,
each agent's dynamics is described by an optimal path\textbf{\ }for\textbf{\ 
}some vector variable, say $A_{i}\left( t\right) $, from an initial to a
final point, up to some fluctuations. The same system of agents can however
be seen as a probabilistic: indeed an agent can be described by a \emph{%
probability density }that is, due to idiosyncratic uncertainties, centred
around the classical optimal path\footnote{%
Due to the infinite number of possible paths, each individual path has a
null probability to exist.\ We therefore use the word probability density
rather than probability.} (see Gosselin, Lotz and Wambst 2017, 2020, 2021).
In this probabilistic approach, each possible dynamics for the set of $N$
agents must be taken into account and weighted by its probability. The
system is then described by a \emph{statistical weight}, the probability
density for any configuration of $N$\ arbitrary individual paths. Once this
statistical weight found, we can compute the transition probabilities of the
system, i.e. the probabilities for any number of agents to evolve from an
initial to a final state in $A_{i}$, $B_{i}$ in a given time.

Because this probabilistic approach implies to keep track of the $N$\
agents' probability transitions, it is practically untractable for a large
number of agents. It remains however a necessary step, since it can
conveniently be translated into a more compact \emph{field formalism}\textbf{%
\ }(see Gosselin, Lotz and Wambst 2017, 2020, 2021). This field formalism
preserves the essential information encoded in the model but implements a
change in perspective.\ It does not keep track of the $N$-indexed agents,
but describes their dynamics and interactions as a collective thread of all
possible anonymous paths. This collective thread can be seen as an
environment that itself conditions the dynamics of individual agents from
one state to another. The field formalism eases the computation of
transition functions. More importantly, it detects the collective states or
phases encompassed in the field, that would otherwise remain indetectable
using the probabilistic formulation.

To translate the probabilistic approach into a field model, the $N$\ agents'
trajectories $\mathbf{A}_{i}\left( t\right) $ are replaced by a field $\Psi $%
, which is a complex valued function that solely depends on a single set of
variables, $\mathbf{A}$. The statistical weight of the probabilistic
approach is translated into a probability density on the space of
complex-valued functions of the variables $\mathbf{A}$.\textbf{\ }For the
configuration $\Psi \left( \mathbf{A}\right) $, this probability density has
the form $\exp \left( -S\left( \Psi \right) \right) $. The functional $%
S\left( \Psi \right) $\ is called the \emph{field action functional}. It
encodes the microscopic features of individual agents dynamics and
interactions. The idea is that of a dictionnary that would translate the
various terms of the classical description in terms of their field
equivalent.\ The integral of $\exp \left( -S\left( \Psi \right) \right) $
over the configurations $\Psi $ is the \emph{partition function} of the
system. The fields that minimize the action functional are the \emph{%
classical background fields, }or more simply the\emph{\ background fields}.\
They encapsulate the collective states of the system.\ 

For several types of agents, the generalisation is straightforward. Each
type $\alpha $ is described by a field $\Psi _{\alpha }\left( \mathbf{A}%
_{\alpha }\right) $. The field action depends on the whole set of fields $%
\left\{ \Psi _{\alpha }\right\} $. It accounts for all types of agents and
their interactions, and writes $S\left( \left\{ \Psi _{\alpha }\right\}
\right) $. The form of $S\left( \left\{ \Psi _{\alpha }\right\} \right) $\
is obtained directly from the classical description of our model.

We have detailed in previous papers\textbf{\ }the transitions from the
classical to the probabilistic frameworks, and then from probabilistic to
field models, and studied individual behaviors in given collective states.

\ In this paper, we present a shortcut to directly translate a classical
framework into a field-formalism model and\textbf{\ }find the backgound
fields that describe the collective states and the global features of the
system.

Two groups of agents, producers and investors, each represent the real
economy and the financial markets, respectively.\ The first group is
composed of a large number of firms in different sectors that collectively
own the entire physical capital. The second group, investors, holds the
entire financial capital and allocate it between firms across sectors
according to investment preferences, expected returns, and stock prices
variations on financial markets. In return, firms pay their investors
dividends. Thus financial capital is a function of dividends and stocks'
valuations, whereas physical capital is a function of the overall capital
allocated by the financial sector. They are described by two different
interacting fields and by the system action functional. The solutions to the
minimization equations of the action functional are the background fields of
the system.\ They characterize the collective states of the system,
structure the interactions between the two types of agents, and condition
individual dynamics. From a sector perspective, the collective states
determine the capital average distribution and the firms' concentration
within sectors, given external parameters, such as changes in expected
returns, technological advances, as well as their dynamics when these
external conditions evolve.

We first show that the number of firms per sector depends on the average
level of capital invested in this sector and on its expected long-term
return, i.e. the long-term return relative to those of neighbouring sectors.
Sectoral capital accumulation itself depends on short-term returns, both
absolute and relative, and on relative expected long-term returns.

Equilibrium capital at each point of the sector space characteristizes the
collective configuration of the system. This equilibrium may however be
unstable\textbf{: }due to the limited number of agents, changes in
parameters or expectations may induce changes in portfolio allocation that
may leave some sectors deserted. Both the sectoral equilibrium capital and
number of agents must therefore be interpreted as potential, not actual,
equilibria: they are thresholds. At a macro-timescale, any deviation from an
equilibrium drives the sector towards the next stable equilibrium, zero
included, and if there is none, towards infinity. Note that this looming
potential instability in sectors depends on the position of the sector
relative to its neighbouring sectors. This notion of instability is thus
relative and context-dependent: variations of parameters in some sectors may
propagate to other sectors.

To account for this systemic instability, we adopt a wider approach to our
model:\ we consider a dynamic system involving average capital and
endogenized long-term expected returns, that is the most volatile parameter
of our model.\ In such a dynamic system, average capital per sector interact
with one another, but also with long-term expected returns.\textbf{\ }

This dynamic system differs from those in standard economic: whereas in
economics the dynamics is usually studied around a static equilibria, we
consider the dynamic interactions between potential equilibria and expected
long-term returns.

Some solutions of this dynamic system are oscillatory: changes in one or
several sectors may propagate over the whole sectors' space. We find, for
each sector, the conditions of stable or unstable oscillations for the
system. Depending on the sector's specific characteristics, oscillations in
average capital and expected long-term returns may dampen or increase. Some
characteristic of the system discriminate stable and unstable oscillations:
some formation of expectations favour overall stability in equilibria,
others deter it.

Eventually, fluctuations in financial expectations impose their pace to the
real economy.\ The combination of expectations both highly sensitive to
exogenous conditions and highly reactive to variations in capital imply that
large fluctuations of capital in the system, at the possible expense of the
real economy.

The paper is organized as follows. The second section is a literature
review. Section three presents the microeconomic framework on which our
field model is based. Section four presents the general method of
translation of a model with a large number of agents into a field theoretic
model. This method is applied to our microeconomic framework in section five
to derive the field-theoretic representation of the system. Section 6
exposes the use of the field model in our context, and the various averages%
\textbf{\ }it allows to compute. In section 7, we present the resolution of
the model. We derive the background field for the real economy and the
density of firms par sector. We then compute the background field for the
financial agents and find the density of investors per sector and the
defining equation for average capital per firm per sector. Section 8
investigates the solutions of this equation. It studies its differential
form, expands it around some particular solutions, and finds directly the
solutions for some particular forms of the parameter functions defining the
system. In section 9, the model is extended to a dynamic system at the
macro-time scale by endogenizing the expected long-term revenue. This
dynamic system presents some oscillatory solutions whose stability depend on
the various patterns of accumulation. Section 10 gathers the interpretations
of the model. The results are presented in section 11 and discussed in
Section 12. Section 13 concludes.

\section{Literature review}

Several branches of the economic literature seek to replace the
representative agent by a collection of heterogeneous ones. Among other
things, they differ in the way they model this collection of agents.

A first branch of the literature represent this collection of agents by
probability densities. This is the approach followed by mean field theory,
heterogeneous agents new Keynesian (HANK) models and the information
theoretic approach to economics.

Mean field theory studies the evolution of agents' density in the state
space of economic variables. It includes the interactions between agents and
the population as a whole, but does not consider the direct interactions
between agents. This approach is thus at an intermediate scale between the
macro and micro scale: it does not aggregate agents but replace them by an
overall probability distribution. Mean field theory has been applied to game
theory (Bensoussan et al. 2018, Lasry et al. 2010a, b) and economics (Gomes
et al. 2015). However, these mean fields are actually probability
distributions. In our formalism, the notion of fields refers to some
abstract complex functions defined on the state space and similar to the
second-quantized-wave functions of quantum theory. Besides, our formalism
directly includes the interactions between agents at the individual level.

Heterogeneous agents new Keynesian (HANK) models uses a probabilistic
treatment similar to mean fields theory.\ An equilibrium probability
distribution is derived from a set of optimizing heterogeneous agents\ in a
new Keynesian context (see Kaplan and Violante 2018 for an account). Our
approach, on the contrary, focuses on the direct interactions between agents
at the microeconomic level. We do not look for an equilibrium probability
distribution for each agent, but rather directly build a probability density
for the system of $N$ agents seen as a whole, that includes interactions,
and then translate this probability density in terms of fields. The states'
space we consider is thus much larger than those considered in the above
approaches. Because it is the space of all paths for a large number of
agents, it allows to study the agents' economic structural relations and the
emergence of the particular phases or collective states induced by these
specific micro-relations, that will in turn impact each agent's stochastic
dynamics at the microeconomic level. Other differences are worth mentioning.
While HANK models\ stress the role of an infinite number of
heterogeneously-behaved consumers, our formalism dwells on the relations
between physical and financial capital\footnote{%
Note that our formalism could also include heterogeneous consumers (see
Gosselin, Lotz, Wambst 2020).}. Besides, our formalism does not rely on
agents' rationality assumptions, since for a large number of agents,
behaviours, be they fully or partly rational, can be modelled as random.

The information theoretic approach to economics (see Yang 2018) considers
probabilistic states around the equilibrium.\ It is close to our
methodological stance: it replaces the Walrasian equilibrium by a
statistical equilibrium derived from a entropy maximisation program.\ Our
statistical weight is similar to the one they use, but is directly built
from microeconomic dynamic equations. The same difference stands for the
rational inattention theory (Sims 2006) in which non-gaussian density laws
are derived from limited information and constraints: our setting directly
includes constraints in the random description of an agent (Gosselin, Lotz,
Wambst 2020).

A second branch of the literature is closest to our approach, since it
considers the interacting system of agents in itself. It is the multi-agent
systems literature, notably agent based models (see Gaffard Napoletano 2012,
Mandel et al. 2010 2012) and economic networks (Jackson 2010).

Agent-based models use general macroeconomics models, whereas network models
lower-scale models such as contract theory, behaviour diffusion, information
sharing or learning. In both settings, agents are typically defined by and
follow various sets of rules, leading to the emergence of equilibria and
dynamics otherwise inaccessible to the representative agent set-up. Both
approaches are however highly numerical and model-dependent and relies on
microeconomic relations - such as ad-hoc reaction functions - that may be
too simplistic. Statistical fields theory on the contrary accounts for
transitions between scales. Macroeconomic patterns do not emerge from the
sole dynamics of a large set of agents: they are grounded in behaviours and
interactions structures. Describing these structures in terms of field
theory allows for the emergence of phases at the macro scale, and the study
of their impact at the individual level.

A third branch of the literature, Econophysics, is also related to ours,
since it often considers the set of agents as a statistical system (for a
review, see Abergel et al. 2011a,b and references therein; or Lux 2008,
2016).\ But it tends to focus on empirical laws, rather than apply the full
potential of field theory to economic systems. In the same vein, Kleinert
(2009) uses path integrals to model stock prices' dynamics. Our approach in
contrast keeps track of usual microeconomic concepts, such as utility
functions, expectations and forward-looking behaviours and includes these
behaviours into the analytical treatment of multi-agent systems by
translating the main characteristics of optimizing agents in terms of
statistical systems.

The literature on interactions between finance and real economy or capital
accumulation takes place mainly in the context of DGSE models. (for a review
of the literature, see Cochrane 2006; for further developments see Grassetti
et al. 2022, Grosshans and Zeisberger 2018, B\"{o}hm et al. 2008, Caggese
and Orive, Bernanke e al. 1999, Campello et al. 2010, Holmstrom and Tirole
1997, Jermann, and Quadrini 2012, Khan Thomas 2013, Monacelli et al. 2011).
Theoretical models include several types of agents at the aggregated level.\
They describe the interactions between a few representative agents such as
producers for possibly several sectors, consumers, financial intermediaries,
etc. to determine interest rates, levels of production, asset pricing, in a
context of ad-hoc anticipations.

Our formalism differs from this literature in three ways. First, we consider
several groups of large number of agents to describe the emergence of
collective states and study the continuous space of sectors. Second, we
consider expected returns and the longer-term horizon as somewhat exogeneous
or structural. Expected returns are a combination of elements, such as
technology, returns, productivity, sectoral capital stock, expectations and
beliefs. These returns are also a function\ defined over the sectors' space:
the system's background fields are functionals of these expected returns.\
Taken together, the background fields of a field model describe an economic
environment for a given configuration of expected returns.\ As such,
expected returns are at first seen as exogenous. It is only in a second
step, when we consider the dynamics between capital accumulation and
expectations, that expectations may themselves be seen as endogenous.\textbf{%
\ }Even then, the form of relations between actual and expected variables
specified are general enough to derive some types of possible dynamics.

Last but not least, we do not seek individual or even aggregated dynamics,
but rather background fields that describe potential long-term equilibria
and may evolve with the structural parameters. For such a background,
agents' individual typical dynamics may nevertheless be retrieved through
Green functions (see GLW). These functions compute the transition
probabilities from one capital-sector point to another. But backgrounds
themselves may be considered as dynamical quantities. Structural or
long-term variations in the returns' landscape may modify the background,
and in turn the individual dynamics. Expected returns themselves depend on,
and interact with, capital accumulation.

\section{The Microeconomic Framework}

This section develops a microeconomic framework that will be turned into a
field model. Since our goal is to picture the interactions between the real
and the financial economy, we consider two groups of agents, producers and
investors. In the following, we will refer to producers or firms $i$
indistinctively, and use the upper script $\symbol{94}$ for variables
describing investors.

\subsection{Producers}

Producers are modelled as firms that belong to sectors. Here, both the
notions of firm and sector are versatile: a single firm with subsidiaries in
different countries and/or offering differentiated products can be modelled
as a set of independent firms. Similarly, a sector refers to a group of
firms with similar activities, but this criteria is loose: sectors can be
decomposed in sector per country, to account for local specificities, or in
several sectors for that matter.

Producers move across sectors described by a vector space of arbitrary
dimension. The position of producer $i$ in this space is denoted $X_{i}$ and
his physical capital, $K_{i}$. Producers are defined by these two variables,
which are both subject to dynamic changes. Producers may change their
capital stocks over time, or altogether shift sector.

Each firm produces a single differentiated good.\ However in the following
we will merely consider the return each producer may provide to its
investors.

The return of producer $i$ at time $t$, denoted $r_{i}$, depends on $K_{i}$, 
$X_{i}$ and on the level of competition in the sector.\ It is written: 
\begin{equation}
r_{i}=r\left( K_{i},X_{i}\right) -\gamma \sum_{j}\delta \left(
X_{i}-X_{j}\right) \frac{K_{j}}{K_{i}}  \label{dvd}
\end{equation}%
The first term is an arbitrary function that depends on the sector and the
level of capital per firm in this sector.\ It represents the return of
capital in a specific sector $X_{i}$ under no competition. We deliberately
keep the form of $r\left( K_{i},X_{i}\right) $\ unspecified, since most of
the results of the model rely on general properties of the functions
involved. When needed, we will give a standard Cobb-Douglas form to \ the
returns $r\left( K_{i},X_{i}\right) $.\ The second term in (\ref{dvd}) is
the decreasing return of capital. In any given sector, it is proportional to
both the number of competitors and the specific level of capital per firm
used.

We also assume that, for all $i$, firm $i$ has a market valuation defined by
both its price $P_{i}$ and the variation $\dot{P}_{i}$ of this price on
financial markets.\ This variation is assumed to be a function of an
expected long-term return denoted $R\left( K_{i},X_{i}\right) $, such that: 
\begin{equation}
\frac{\dot{P}_{i}}{P_{i}}=F_{1}\left( \frac{R\left( K_{i},X_{i}\right) }{%
\sum_{l}R\left( K_{l},X_{l}\right) }\right)  \label{pr}
\end{equation}%
where the quantity $\frac{R\left( K_{i},X_{i}\right) }{\sum_{l}R\left(
K_{l},X_{l}\right) }$ is the relative return of firm $i$ relative to the
whole set of firms in its sector. The function $F_{1}$ is arbitrary and
reflects the preferences of the market relatively to the firm's relative
returns.

We assume that firms shift their production in the sector space according to
returns, in the direction of the gradient of the expected long-term return $%
R\left( K_{i},X_{i}\right) $.\ Yet, the accumulation of agents at any point
of the space creates a repulsive force, so that the evolution of $X_{i}$
minimizes, up to some shocks, the following quantity:%
\begin{equation}
L_{i}\left( X_{i},\frac{dX_{i}}{dt}\right) =\left( \frac{dX_{i}}{dt}-\nabla
_{X}R\left( K_{i},X_{i}\right) H\left( K_{i}\right) \right) ^{2}+\tau
\sum_{j}\delta \left( X_{i}-X_{j}\right)  \label{dnp}
\end{equation}%
When $\tau =0$, there are no repulsive forces and the move towards the
gradient of $R$ is given by the expression:

\begin{equation*}
\frac{dX_{i}}{dt}=\nabla _{X}R\left( K_{i},X_{i}\right) H\left( K_{i}\right)
\end{equation*}%
When $\tau \neq 0$, repulsive forces deviate the trajectory.\ The dynamic
equation associated to the minimization of (\ref{dnp}) is given by the
general formula of the dynamic optimization:%
\begin{equation}
\frac{d}{dt}\frac{\partial }{\partial \frac{dX_{i}}{dt}}L_{i}\left( X_{i},%
\frac{dX_{i}}{dt}\right) =\frac{\partial }{\partial X_{i}}L_{i}\left( X_{i},%
\frac{dX_{i}}{dt}\right)  \label{lgd}
\end{equation}%
This last equation does not need to be developed further, since formula (\ref%
{dnp}) is sufficient to switch to the field description of the system. Note
for later purpose that the expression $\frac{dX_{i}}{dt}$\ stands for the
continuous version of a discrete variation, $X_{i}\left( t+1\right)
-X_{i}\left( t\right) $.

\subsection{Investors}

Each investor $j$ is defined by his level of capital $\hat{K}_{j}$ and his
position $\hat{X}_{j}$ in the sector space. Investors can invest in the
entire sector space, but tend to invest in sectors close to their position.

Besides, investors tend to diversify their capital: each investor $j$ chose
to allocate parts of his entire capital $\hat{K}_{j}$ between various firms $%
i$. The capital allocated by investor $j$ to firm $i$ is denoted $\hat{K}%
_{j}^{\left( i\right) }$, and given by:

\begin{equation}
\hat{K}_{j}^{\left( i\right) }\left( t\right) =\left( \frac{F_{2}\left(
R\left( K_{i},X_{i}\right) \right) G\left( X_{i}-\hat{X}_{j}\right) }{%
\sum_{l}F_{2}\left( R\left( K_{l},X_{l}\right) \right) G\left( X_{l}-\hat{X}%
_{j}\right) }\hat{K}_{j}\right) \left( t\right)  \label{grandf2}
\end{equation}%
where $F_{2}$ is an arbitrary function that depends on the expected return
of firm $i$ and the distance between sector $X_{i}$ et $\hat{X}_{j}$.

We define $\varepsilon $ the time scale for capital accumulation.\ The
variation of capital\ of investor $j$ between $t$\ and $t+\varepsilon $\ is
the sum of two terms: the short-term returns $r_{i}$\ of the firms\ in which 
$j$\ invested, and the stock price variations of these same firms:

\begin{equation}
\hat{K}_{j}\left( t+\varepsilon \right) -\hat{K}_{j}\left( t\right)
=\sum_{i}\left( r_{i}+\frac{\dot{P}_{i}}{P_{i}}\right) \hat{K}_{j}^{\left(
i\right) }=\sum_{i}\left( r_{i}+F_{1}\left( \frac{R\left( K_{i},X_{i}\right) 
}{\sum_{l}R\left( K_{l},X_{l}\right) },\frac{\dot{K}_{i}\left( t\right) }{%
K_{i}\left( t\right) }\right) \right) \hat{K}_{j}^{\left( i\right) }
\label{fsn}
\end{equation}%
Incidentally, note that in equation (\ref{dnp}), the time scale of motions
within the sectors space was normalized to one. Here, on the contrary, we
define this motion time scale as $\varepsilon $, and assume $\varepsilon <<1$%
: the mobility in the sector space is lower than capital dynamics. To
rewrite (\ref{fsn}) on the same time-span as $\frac{dX_{i}}{dt}$, we write:

\begin{eqnarray*}
\hat{K}_{j}\left( t+1\right) -\hat{K}_{j}\left( t\right) &=&\sum_{k=1}^{%
\frac{1}{\varepsilon }}\hat{K}_{j}\left( t+k\varepsilon \right) -\hat{K}%
_{j}\left( t\right) \\
&=&\sum_{k=1}^{\frac{1}{\varepsilon }}\sum_{i}\left( r_{i}+\frac{\dot{P}_{i}%
}{P_{i}}\right) \hat{K}_{j}^{\left( i\right) }\left( t+k\varepsilon \right)
\simeq \frac{1}{\varepsilon }\sum_{i}\left( r_{i}+F_{1}\left( \frac{R\left(
K_{i},X_{i}\right) }{\sum_{l}R\left( K_{l},X_{l}\right) },\frac{\dot{K}%
_{i}\left( t\right) }{K_{i}\left( t\right) }\right) \right) \hat{K}%
_{j}^{\left( i\right) }
\end{eqnarray*}%
where the quantities in the sum have to be understood as averages over the
time span $\left[ t,t+1\right] $. Using equation (\ref{pr}), equation (\ref%
{fsn}) becomes in the continuous approximation: 
\begin{equation}
\frac{d}{dt}\hat{K}_{j}\left( t\right) =\frac{1}{\varepsilon }\sum_{i}\left(
r_{i}+F_{1}\left( \frac{R\left( K_{i},X_{i}\right) }{\sum_{l}\delta \left(
X_{l}-X_{i}\right) R\left( K_{l},X_{l}\right) },\frac{\dot{K}_{i}\left(
t\right) }{K_{i}\left( t\right) }\right) \right) \frac{F_{2}\left( R\left(
K_{i},X_{i}\right) \right) G\left( X_{i}-\hat{X}_{j}\right) }{%
\sum_{l}F_{2}\left( R\left( K_{l},X_{l}\right) \right) G\left( X_{l}-\hat{X}%
_{j}\right) }\hat{K}_{j}  \label{nfs}
\end{equation}%
where $\frac{d}{dt}\hat{K}_{j}\left( t\right) =\hat{K}_{j}\left( t+1\right) -%
\hat{K}_{j}\left( t\right) $\ is now normalized to the time scale of $\frac{%
dX_{i}}{dt}$, i.e. 1.

\subsection{Link between financial and physical capital}

The entire financial capital is, at any time, completely allocated by
investors between firms. There is no alternative source of financing for
producers: self-financing is discarded, since it essentially amounts to
consider two agents, a producer and an investor, as one. The physical
capital of a any given firm is thus the sum of all capital allocated to this
firm by all its investors. Physical capital entirely depends on the
arbitrage and allocations of the financial sector. Firms do not own their
capital: they fully return it at the end of each period with a dividend,
though possibly negative. Investors then chose to reallocate their funds in
their entirety between firms at the beginning of the next period.

This set up is a generalisation of the dividend irrelevance theory. It may
not be fully accurate in the short-run but, since physical capital cannot
subsist without investment, it holds in the long-run.\ When investors choose
not to finance a firm, this firm is bound to disappear in the long run.
Under these assumptions, the following identity holds\textbf{:}%
\begin{equation}
K_{i}\left( t+\varepsilon \right) =\sum_{j}\hat{K}_{j}^{\left( i\right)
}=\sum_{j}\frac{F_{2}\left( R\left( K_{i}\left( t\right) ,X_{i}\left(
t\right) \right) \right) G\left( X_{i}\left( t\right) -\hat{X}_{j}\right) }{%
\sum_{l}F_{2}\left( R\left( K_{l}\left( t\right) ,X_{l}\left( t\right)
\right) \right) G\left( X_{l}\left( t\right) -\hat{X}_{j}\right) }\hat{K}%
_{j}\left( t\right)  \label{phs}
\end{equation}%
where $K_{i}$ stands for the physical capital of firm $i$ at time $t$, and $%
\sum_{j}\hat{K}_{j}^{\left( i\right) }$ for the sum of capital invested in
firm $i$ by investors $j$. Recall that the parameter $\varepsilon $ accounts
for the specific time scale of capital accumulation.\ It differs from that
of mobility within the sector space (\ref{dnp}), which is normalized to one.

The dynamics (\ref{phs}) rewrites:%
\begin{equation}
\frac{K_{i}\left( t+\varepsilon \right) -K_{i}\left( t\right) }{\varepsilon }%
=\frac{1}{\varepsilon }\left( \sum_{j}\frac{F_{2}\left( R\left( K_{i}\left(
t\right) ,X_{i}\left( t\right) \right) \right) G\left( X_{i}\left( t\right) -%
\hat{X}_{j}\right) }{\sum_{l}F_{2}\left( R\left( K_{l}\left( t\right)
,X_{l}\left( t\right) \right) \right) G\left( X_{l}\left( t\right) -\hat{X}%
_{j}\right) }\hat{K}_{j}\left( t\right) -K_{i}\left( t\right) \right)
\label{dnK}
\end{equation}%
Using the same token as in the derivation of (\ref{nfs}), we obtain in the
continuous approximation:%
\begin{equation}
\frac{d}{dt}K_{i}\left( t\right) +\frac{1}{\varepsilon }\left( K_{i}\left(
t\right) -\sum_{j}\frac{F_{2}\left( R\left( K_{i}\left( t\right)
,X_{i}\left( t\right) \right) \right) G\left( X_{i}\left( t\right) -\hat{X}%
_{j}\right) }{\sum_{l}F_{2}\left( R\left( K_{l}\left( t\right) ,X_{l}\left(
t\right) \right) \right) G\left( X_{l}\left( t\right) -\hat{X}_{j}\right) }%
\hat{K}_{j}\left( t\right) \right) =0  \label{dnk}
\end{equation}%
where $\frac{d}{dt}K_{i}\left( t\right) $\ stands for $K_{i}\left(
t+1\right) -K_{i}\left( t\right) $.

\subsection{Capital allocation dynamics}

Investors choose to allocate their capital within sectors, and may modify
their portfolio according to the returns of the sector or firms they invest
in. This is modelled by a move along the sectors' space in the direction of
the gradient of $R\left( K_{i},X_{i}\right) $. The move of $\hat{X}_{j}$ is
described by the dynamic equation:

\begin{equation}
\frac{d}{dt}\hat{X}_{j}-\frac{1}{\sum_{i}\delta \left( X_{i}-\hat{X}%
_{j}\right) }\sum_{i}\left( \nabla _{\hat{X}}F_{0}\left( R\left( K_{i},\hat{X%
}_{j}\right) \right) +\nu \nabla _{\hat{X}}F_{1}\left( \frac{R\left( K_{i},%
\hat{X}_{j}\right) }{\sum_{l}R\left( K_{l},X_{l}\right) }\right) \right) =0
\label{prd}
\end{equation}%
where the factor $\sum_{i}\delta \left( X_{i}-\hat{X}_{j}\right) $ is the
agents' density in the sector $\hat{X}_{j}$, so that the more competitors in
a sector, the slower the move.

In equation (\ref{rpd}), the term $\nabla _{\hat{X}}F_{0}\left( R\left(
K_{i},\hat{X}_{j}\right) \right) $ is the tendency of investors to invest in
sectors with the highest returns, which can be described by a move in the
direction defined by a function $F_{0}$\ of long-term returns.

The term $\nu \nabla _{\hat{X}}F_{1}\left( \frac{R\left( K_{i},\hat{X}%
_{j}\right) }{\sum_{l}R\left( K_{l},X_{l}\right) }\right) $ describes the
investors' preference for stocks with the highest price-dividend ratio.

Ultimately, note that unlike $K$\ and $\hat{K}$,\ $X$ and $\hat{X}$\ are not
a strictly standard economic variables, and that their dynamics should thus
be ascribed an ad-hoc form.

\section{Fields formalism}

In the above, we have detailed a standard, classical microeconomic
framework. We will now present a general method to translate such a
framework into a field model.

\subsection{General Method}

To transform our framework into a field model, we must first consider the
types of agents in the model, and rewrite their dynamics as the minimization
equations of some initial functions, in the same way as, for instance,
consumption dynamics could be derived from an utility function.

Since each type of agent in our framework is described by two dynamic
equations, there are four minimization functions to find. These minimization
functions will be translated into four functionals of two independent fields%
\footnote{%
The term functional refers to a function of a function, i.e. a function
whose argument is itself a function.}, one for producers, the real economy,
and one for investors, financial markets. The sum of the four functionals is
the "field action functional" that describes the whole system in terms of
fields\footnote{%
Details about the probabilistic step will be given as a reminder along the
text and in appendix 1.}.

\subsubsection{Minimization functions}

In standard economic frameworks, each type of agent is characterized by one
or more dynamic equations. Some of these dynamic equations can result from a
minimization, while others do not. The translation into fields is built upon
the functions which, once minimized, yields the system's dynamics. Two cases
arise.

When the dynamics are constructed from minimizations, it suffices to use the
function which, minimized, gave the equation of the dynamics. These
functions, related to the probabilistic interpretation, represent the
deviation between the trajectory of an agent and an average or optimal
trajectory. They are therefore directly linked to the number of agents in
the system.

However, dynamics may not always result from a minimization. In this second
case, we must ad-hoc reconstruct functions whose minimization would restore
the dynamic equations. Such functions refer to the probabilistic
interpretation of the system.\ They are not unique. When modelling
heterogeneous agents, quadratic functions allow to translate the quadratic
deviation from the mean trajectory of agents subject to idiosyncratic
shocks. This quadratic deviation represents the variance of these shocks and
is directly related to the probability of deviating from an average
trajectory (see GLW). The construction of these quadratic functions is
straightforward.

In general, we assume agents are described by vectors $\mathbf{A}_{i}\left(
t\right) $\ of arbitrary dimension, where $\mathbf{A}_{i}\left( t\right) $\
satisfies a dynamic equation characterizing agent $i$:%
\begin{equation}
\frac{d\mathbf{A}_{i}\left( t\right) }{dt}-\sum_{j,k,l...}f\left( \mathbf{A}%
_{i}\left( t\right) ,\mathbf{A}_{j}\left( t\right) ,\mathbf{A}_{k}\left(
t\right) ,\mathbf{\hat{A}}_{l}\left( t\right) ,\mathbf{\hat{A}}_{m}\left(
t\right) ...\right) =0  \label{gauche}
\end{equation}%
This type of equation, which involves the whole set of other agents, is
characteristic of models with a large number of interacting agents. Its
associated minimization function is obtained directly, by first squaring the
lhs of (\ref{gauche}), and then summing over the whole set of agents to
describe the full system:

\begin{equation}
\sum_{i}\left( \frac{d\mathbf{A}_{i}\left( t\right) }{dt}-\sum_{j,k,l...}f%
\left( \mathbf{A}_{i}\left( t\right) ,\mathbf{A}_{j}\left( t\right) ,\mathbf{%
A}_{k}\left( t\right) ,\mathbf{\hat{A}}_{l}\left( t\right) ,\mathbf{\hat{A}}%
_{m}\left( t\right) ...\right) \right) ^{2}  \label{mnZ}
\end{equation}%
However, some minimization functions may also include an additional term,
without any time derivative (see for example equation (\ref{dnp})), so that
the most general minimization function writes, for some function $g$:%
\begin{equation}
\sum_{i}\left( \frac{d\mathbf{A}_{i}\left( t\right) }{dt}-\sum_{j,k,l...}f%
\left( \mathbf{A}_{i}\left( t\right) ,\mathbf{A}_{j}\left( t\right) ,\mathbf{%
A}_{k}\left( t\right) ,\mathbf{\hat{A}}_{l}\left( t\right) ,\mathbf{\hat{A}}%
_{m}\left( t\right) ...\right) \right) ^{2}+\sum_{i}\sum_{j,k,l...}g\left( 
\mathbf{A}_{i}\left( t\right) ,\mathbf{A}_{j}\left( t\right) ,\mathbf{A}%
_{k}\left( t\right) ,\mathbf{\hat{A}}_{l}\left( t\right) ,\mathbf{\hat{A}}%
_{m}\left( t\right) ...\right)  \label{mNZ}
\end{equation}%
Ultimately, the sum $s$ of all the minimization functions is the sum of all
agents' squared deviation from the average dynamics within the system%
\footnote{%
The function $s$\ is related to the probabilistic approach, which associates
the probability $\exp \left( -s\right) $\ to the state of the system defined
by the $A_{i}\left( t\right) ,A_{j}\left( t\right) ,..$\ (see appendix 1)}.\
A generalisation of equation (\ref{mNZ}), in which agents interact at
different times, and its translation in term of field is presented in
appendix 1.

\subsubsection{Translation in terms of fields}

The translation itself can be divided into two relatively simple processes,
but varies slightly depending on the type of terms that appear in the
various minimization functions.

\paragraph{Indexed variables, without temporal derivative}

The terms in (\ref{mNZ}) that include indexed variables but no temporal
derivative terms are the easiest to translate.\ They are of the form:%
\begin{equation*}
\sum_{i}\sum_{j,k,l,m...}g\left( \mathbf{A}_{i}\left( t\right) ,\mathbf{A}%
_{j}\left( t\right) ,\mathbf{A}_{k}\left( t\right) ,\mathbf{\hat{A}}%
_{l}\left( t\right) ,\mathbf{\hat{A}}_{m}\left( t\right) ...\right)
\end{equation*}%
These terms describe the whole set of interactions both among and between
two groups of agents. Here, agents are characterized by their variables $%
\mathbf{A}_{i}\left( t\right) ,\mathbf{A}_{j}\left( t\right) ,\mathbf{A}%
_{k}\left( t\right) $... and $\mathbf{\hat{A}}_{l}\left( t\right) ,\mathbf{%
\hat{A}}_{m}\left( t\right) $... respectively, for instance in our model
firms and investors.

In the field translation, agents of type $\mathbf{A}_{i}\left( t\right) $
and $\mathbf{\hat{A}}_{l}\left( t\right) $ are described by a field $\Psi
\left( \mathbf{A}\right) $ and $\hat{\Psi}\left( \mathbf{\hat{A}}\right) $,
respectively.

In a first step, the variables indexed $i$ such as $\mathbf{A}_{i}\left(
t\right) $ are replaced by variables $\mathbf{A}$ in the expression of $g$.
The variables indexed $j$,$k$,$l$,$m...$, such as $\mathbf{A}_{j}\left(
t\right) $, $\mathbf{A}_{k}\left( t\right) $, $\mathbf{\hat{A}}_{l}\left(
t\right) ,\mathbf{\hat{A}}_{m}\left( t\right) $... are replaced by $\mathbf{A%
}^{\prime },\mathbf{A}^{\prime \prime }$, $\mathbf{\hat{A}}$, $\mathbf{\hat{A%
}}^{\prime }$ , and so on for all the indices in the function. This yields
the expression:

\begin{equation*}
\sum_{i}\sum_{j,k,l,m...}g\left( \mathbf{A},\mathbf{A}^{\prime },\mathbf{A}%
^{\prime \prime },\mathbf{\hat{A},\hat{A}}^{\prime }...\right)
\end{equation*}%
In a second step, each sum is replaced by a weighted integration symbol: 
\begin{eqnarray*}
\sum_{i} &\rightarrow &\int \left\vert \Psi \left( \mathbf{A}\right)
\right\vert ^{2}d\mathbf{A}\text{, }\sum_{j}\rightarrow \int \left\vert \Psi
\left( \mathbf{A}^{\prime }\right) \right\vert ^{2}d\mathbf{A}^{\prime }%
\text{, }\sum_{k}\rightarrow \int \left\vert \Psi \left( \mathbf{A}^{\prime
\prime }\right) \right\vert ^{2}d\mathbf{A}^{\prime \prime } \\
\sum_{l} &\rightarrow &\int \left\vert \hat{\Psi}\left( \mathbf{\hat{A}}%
\right) \right\vert ^{2}d\mathbf{\hat{A}}\text{, }\sum_{m}\rightarrow \int
\left\vert \hat{\Psi}\left( \mathbf{\hat{A}}^{\prime }\right) \right\vert
^{2}d\mathbf{\hat{A}}^{\prime }
\end{eqnarray*}%
which leads to the translation:%
\begin{eqnarray}
&&\sum_{i}\sum_{j}\sum_{j,k...}g\left( \mathbf{A}_{i}\left( t\right) ,%
\mathbf{A}_{j}\left( t\right) ,\mathbf{A}_{k}\left( t\right) ,\mathbf{\hat{A}%
}_{l}\left( t\right) ,\mathbf{\hat{A}}_{m}\left( t\right) ...\right)  \notag
\\
&\rightarrow &\int g\left( \mathbf{A},\mathbf{A}^{\prime },\mathbf{A}%
^{\prime \prime },\mathbf{\hat{A},\hat{A}}^{\prime }...\right) \left\vert
\Psi \left( \mathbf{A}\right) \right\vert ^{2}\left\vert \Psi \left( \mathbf{%
A}^{\prime }\right) \right\vert ^{2}\left\vert \Psi \left( \mathbf{A}%
^{\prime \prime }\right) \right\vert ^{2}d\mathbf{A}d\mathbf{A}^{\prime }d%
\mathbf{A}^{\prime \prime }\left\vert \hat{\Psi}\left( \mathbf{\hat{A}}%
\right) \right\vert ^{2}\left\vert \hat{\Psi}\left( \mathbf{\hat{A}}^{\prime
}\right) \right\vert ^{2}d\mathbf{\hat{A}}d\mathbf{\hat{A}}^{\prime }
\label{tln}
\end{eqnarray}

\paragraph{Variable with temporal derivative}

The terms in (\ref{mNZ}) that imply a variable temporal derivative are of
the form:%
\begin{equation}
\sum_{i}\left( \frac{d\mathbf{A}_{i}^{\left( \alpha \right) }\left( t\right) 
}{dt}-\sum_{j,k,l,m...}f^{\left( \alpha \right) }\left( \mathbf{A}_{i}\left(
t\right) ,\mathbf{A}_{j}\left( t\right) ,\mathbf{A}_{k}\left( t\right) ,%
\mathbf{\hat{A}}_{l}\left( t\right) ,\mathbf{\hat{A}}_{m}\left( t\right)
...\right) \right) ^{2}  \label{edr}
\end{equation}%
This particular form represents the dynamics of the $\alpha $-th coordinate
of a variable $\mathbf{A}_{i}\left( t\right) $ as a function of the other
agents.

The method of translation is similar to the above, but the time derivative
adds an additional operation.

In a first step, we translate the terms without derivative inside the
parenthesis:%
\begin{equation}
\sum_{j,k,l,m...}f^{\left( \alpha \right) }\left( \mathbf{A}_{i}\left(
t\right) ,\mathbf{A}_{j}\left( t\right) ,\mathbf{A}_{k}\left( t\right) ,%
\mathbf{\hat{A}}_{l}\left( t\right) ,\mathbf{\hat{A}}_{m}\left( t\right)
...\right)  \label{ntr}
\end{equation}%
This type of term has already been translated in the previous paragraph, but
since there is no sum over $i$ in (\ref{ntr}), there should be no integral
over $\mathbf{A}$\textbf{,} nor factor $\left\vert \Psi \left( \mathbf{A}%
\right) \right\vert ^{2}$.

The translation of (\ref{ntr}) is therefore, as before:%
\begin{equation}
\int f^{\left( \alpha \right) }\left( \mathbf{A},\mathbf{A}^{\prime },%
\mathbf{A}^{\prime \prime },\mathbf{\hat{A},\hat{A}}^{\prime }...\right)
\left\vert \Psi \left( \mathbf{A}^{\prime }\right) \right\vert
^{2}\left\vert \Psi \left( \mathbf{A}^{\prime \prime }\right) \right\vert
^{2}d\mathbf{A}^{\prime }d\mathbf{A}^{\prime \prime }\left\vert \hat{\Psi}%
\left( \mathbf{\hat{A}}\right) \right\vert ^{2}\left\vert \hat{\Psi}\left( 
\mathbf{\hat{A}}^{\prime }\right) \right\vert ^{2}d\mathbf{\hat{A}}d\mathbf{%
\hat{A}}^{\prime }  \label{trn}
\end{equation}%
A free variable $\mathbf{A}$ remains, which will be integrated later, when
we account for the external sum $\sum_{i}$. We will call $\Lambda (\mathbf{A}%
)$ the expression obtained:%
\begin{equation}
\Lambda (\mathbf{A})=\int f^{\left( \alpha \right) }\left( \mathbf{A},%
\mathbf{A}^{\prime },\mathbf{A}^{\prime \prime },\mathbf{\hat{A},\hat{A}}%
^{\prime }...\right) \left\vert \Psi \left( \mathbf{A}^{\prime }\right)
\right\vert ^{2}\left\vert \Psi \left( \mathbf{A}^{\prime \prime }\right)
\right\vert ^{2}d\mathbf{A}^{\prime }d\mathbf{A}^{\prime \prime }\left\vert 
\hat{\Psi}\left( \mathbf{\hat{A}}\right) \right\vert ^{2}\left\vert \hat{\Psi%
}\left( \mathbf{\hat{A}}^{\prime }\right) \right\vert ^{2}d\mathbf{\hat{A}}d%
\mathbf{\hat{A}}^{\prime }  \label{bdt}
\end{equation}%
In a second step, we account for the derivative in time by using field
gradients. To do so, and as a rule, we replace :%
\begin{equation}
\sum_{i}\left( \frac{d\mathbf{A}_{i}^{\left( \alpha \right) }\left( t\right) 
}{dt}-\sum_{j}\sum_{j,k...}f^{\left( \alpha \right) }\left( \mathbf{A}%
_{i}\left( t\right) ,\mathbf{A}_{j}\left( t\right) ,\mathbf{A}_{k}\left(
t\right) ,\mathbf{\hat{A}}_{l}\left( t\right) ,\mathbf{\hat{A}}_{m}\left(
t\right) ...\right) \right) ^{2}  \label{inco}
\end{equation}%
by:%
\begin{equation}
\int \Psi ^{\dag }\left( \mathbf{A}\right) \left( -\nabla _{\mathbf{A}%
^{\left( \alpha \right) }}\left( \frac{\sigma _{\mathbf{A}^{\left( \alpha
\right) }}^{2}}{2}\nabla _{\mathbf{A}^{\left( \alpha \right) }}+\Lambda (%
\mathbf{A})\right) \right) \Psi \left( \mathbf{A}\right) d\mathbf{A}
\label{Trl}
\end{equation}%
The variance $\sigma _{\mathbf{A}^{\left( \alpha \right) }}^{2}$ reflects
the probabilistic nature of the model which is hidden behind the field
formalism. This variance represents the characteristic level of uncertainty
of the system's dynamics. It is a parameter of the model. Note also that in (%
\ref{Trl}), the integral over $\mathbf{A}$ reappears at the end, along with
the square of the field $\left\vert \Psi \left( \mathbf{A}\right)
\right\vert ^{2}$.\ This square is split into two terms, $\Psi ^{\dag
}\left( \mathbf{A}\right) $ and $\Psi \left( \mathbf{A}\right) $, with a
gradient operator inserted in between.

\subsubsection{Gathering terms}

The field description is ultimately obtained by summing all the terms
translated above and introducing a time dependency. This sum is called the
action functional. It is the sum of terms of the form (\ref{tln}) and (\ref%
{Trl}), and is denoted $S\left( \Psi ,\Psi ^{\dag }\right) $.

\paragraph{Introducing time in the model}

Until now, no time variable was included in this model. We now introduce
one, written $\theta $ to distinguish it from the classical model
variables.\ We replace:%
\begin{eqnarray*}
\Psi \left( \mathbf{A}\right) &\rightarrow &\Psi \left( \mathbf{A},\theta
\right) \\
\hat{\Psi}\left( \mathbf{\hat{A}}\right) &\rightarrow &\hat{\Psi}\left( 
\mathbf{\hat{A}},\theta \right)
\end{eqnarray*}%
and introduce an additional contribution to $S\left( \Psi ,\Psi ^{\dag
}\right) $ (see GLW). The full action functional then becomes:%
\begin{eqnarray}
&&S\left( \Psi ,\Psi ^{\dag }\right) +\Psi ^{\dag }\left( \mathbf{A},\theta
\right) \left( -\nabla _{\theta }\left( \frac{\sigma _{\theta }^{2}}{2}%
\nabla _{\theta }-1\right) \right) \Psi \left( \mathbf{A},\theta \right)
\label{fct} \\
&&+\hat{\Psi}^{\dag }\left( \mathbf{\hat{A}},\theta \right) \left( -\nabla
_{\theta }\left( \frac{\sigma _{\theta }^{2}}{2}\nabla _{\theta }-1\right)
\right) \hat{\Psi}\left( \mathbf{\hat{A}},\theta \right) +\alpha \left\vert
\Psi \left( \mathbf{A}\right) \right\vert ^{2}+\alpha \left\vert \hat{\Psi}%
\left( \mathbf{\hat{A}}\right) \right\vert ^{2}  \notag
\end{eqnarray}%
where $\sigma _{\theta }^{2}$ is a variance term accounting for delays in
interactions, and $\frac{1}{\alpha }$ is a time scale describing the average
time span of interactions between agents. In practice $\sigma _{\theta
}^{2}<<1$ and $\alpha <<1$.

\paragraph{Remark about the introduction of the time variable}

As mentioned above, equation (\ref{fct}) is necessary to describe some
time-dependent processes.

However, including this time-variable in the model is not always necessary.
In this paper for instance, we are solely interested in the background
fields of the system. The background fields refer to a long-run,
stationary-type of equilibrium.\ As such, we can avoid introducing a time
variable in the model, and simply consider static fields, $\Psi \left( 
\mathbf{A}\right) $\ and $\hat{\Psi}\left( \mathbf{\hat{A}}\right) $.

We will later on include some time-dependent modifications in our background
fields. However, these modifications will reflect modifications in the
parameters, that must not be confused with the short-term $\theta $\
dependency.

\section{Application to the framework}

\subsection{Minimization functions}

In our model, the dynamics of the variable $X_{i}$ comes from the
minimization of the function:%
\begin{equation*}
\left( \frac{dX_{i}}{dt}-\nabla _{X}R\left( K_{i},X_{i}\right) H\left(
K_{i}\right) \right) ^{2}+\tau \sum_{j}\delta \left( X_{i}-X_{j}\right)
\end{equation*}%
We simply re-use this function. Since we are interested in the whole system,
we will sum over the whole set of agents, which yields the minimization
function for the capital allocation dynamics:

\begin{equation}
\sum_{i}\left( \frac{dX_{i}}{dt}-\nabla _{X}R\left( K_{i},X_{i}\right)
H\left( K_{i}\right) \right) ^{2}+\sum_{i}\tau \sum_{j}\delta \left(
X_{i}-X_{j}\right)  \label{minX}
\end{equation}%
On the contrary, the dynamics of the variables $K_{i},$ $\hat{K}_{i}$ et $%
\hat{X}_{i}$ are not the result of a minimization, but their associated
quadratic functions (\ref{mnZ}) can easily be found. These functions are
therefore:

for $K_{i}$, the minimization function for physical capital dynamics:

\begin{equation}
\sum_{i}\left( \frac{d}{dt}K_{i}+\frac{1}{\varepsilon }\left( K_{i}-\sum_{j}%
\frac{F_{2}\left( R\left( K_{i}\left( t\right) ,X_{i}\left( t\right) \right)
\right) G\left( X_{i}\left( t\right) -\hat{X}_{j}\right) }{%
\sum_{l}F_{2}\left( R\left( K_{l}\left( t\right) ,X_{l}\left( t\right)
\right) \right) G\left( X_{l}\left( t\right) -\hat{X}_{j}\right) }\hat{K}%
_{j}\left( t\right) \right) \right) ^{2}  \label{minK}
\end{equation}%
for $\hat{K}_{i}$, the minimization function for the financial capital
dynamics:%
\begin{equation}
\sum_{j}\left( \frac{d}{dt}\hat{K}_{j}-\frac{1}{\varepsilon }\left(
\sum_{i}\left( r_{i}+F_{1}\left( \frac{R\left( K_{i},X_{i}\right) }{%
\sum_{l}\delta \left( X_{l}-X_{i}\right) R\left( K_{l},X_{l}\right) },\frac{%
\dot{K}_{i}\left( t\right) }{K_{i}\left( t\right) }\right) \right) \frac{%
F_{2}\left( R\left( K_{i},X_{i}\right) \right) G\left( X_{i}-\hat{X}%
_{j}\right) }{\sum_{l}F_{2}\left( R\left( K_{l},X_{l}\right) \right) G\left(
X_{l}-\hat{X}_{j}\right) }\hat{K}_{j}\right) \right) ^{2}  \label{minKchap}
\end{equation}%
and for $\hat{X}_{i}$, the minimization function for financial capital
allocation: 
\begin{equation}
\sum_{i}\left( \frac{d}{dt}\hat{X}_{j}-\frac{1}{\sum_{i}\delta \left( X_{i}-%
\hat{X}_{j}\right) }\sum_{i}\left( \nabla _{\hat{X}}F_{0}\left( R\left(
K_{i},\hat{X}_{j}\right) \right) +\nu \nabla _{\hat{X}}F_{1}\left( \frac{%
R\left( K_{i},\hat{X}_{j}\right) }{\sum_{l}R\left( K_{l},X_{l}\right) }%
\right) \right) \right) ^{2}  \label{minXchap}
\end{equation}

\subsection{Translation in terms of fields}

We apply the general method developed above and translate the minimization
functions (\ref{minX}), (\ref{minK}), (\ref{minKchap}) and (\ref{minXchap})
in terms of fields. We start with producers, and translate first (\ref{minX}%
) and (\ref{minK}).

\subsubsection{The Real Economy}

In both\textbf{\ }capital allocation dynamics (\ref{minX}) and capital
accumulation dynamics (\ref{minK}), time derivatives appear.\ However, one
of them, equation (\ref{minX}), includes time-independent terms and is thus
of the form (\ref{mNZ}), the other, equation (\ref{minX}) is of the type (%
\ref{mnZ}).

\paragraph{Translation of the minimization function: Physical capital
allocation}

Let us start by translating in terms of fields the expression (\ref{minX}):

\begin{equation}
\sum_{i}\left( \left( \frac{dX_{i}}{dt}-\nabla _{X}R\left(
K_{i},X_{i}\right) H\left( K_{i}\right) \right) ^{2}+\tau \sum_{j}\delta
\left( X_{i}-X_{j}\right) \right)  \label{mnd}
\end{equation}%
To do so, we first consider the last term $\tau \sum_{i}\sum_{j}\delta
\left( X_{i}-X_{j}\right) $. This term contains no derivative. The form of
the translation is given by formula (\ref{tln}). Since the expression
contains two indices, both of them are summed.\ 

The first step of the translation is to replace $X_{i}$ and $X_{j}$ by two
variables $X$ et $X^{\prime }$, and substitute:

\begin{equation*}
\tau \delta \left( X_{i}-X_{j}\right) \rightarrow \tau \delta \left(
X-X^{\prime }\right)
\end{equation*}%
The sum over $i$ and the sum over $j$ are then replaced directly by the
integrals $\int \left\vert \Psi \left( K,X\right) \right\vert ^{2}d\left(
K,X\right) $, $\int \left\vert \Psi \left( K^{\prime },X^{\prime }\right)
\right\vert ^{2}d\left( K^{\prime },X^{\prime }\right) $, which leads to the
following translation:

\begin{eqnarray}
\tau \sum_{i}\sum_{j}\delta \left( X_{i}-X_{j}\right) &\rightarrow &\int
\left\vert \Psi \left( K,X\right) \right\vert ^{2}d\left( K,X\right) \int
\left\vert \Psi \left( K^{\prime },X^{\prime }\right) \right\vert
^{2}d\left( K^{\prime },X^{\prime }\right) \tau \delta \left( X-X^{\prime
}\right)  \notag \\
&=&\int \tau \left\vert \Psi \left( K,X\right) \right\vert ^{2}\left\vert
\Psi \left( K^{\prime },X\right) \right\vert ^{2}d\left( K,X\right)
dK^{\prime }  \label{tnl}
\end{eqnarray}%
To translate the first term in formula (\ref{mnd}): 
\begin{equation}
\sum_{i}\left( \frac{dX_{i}}{dt}-\nabla _{X}R\left( K_{i},X_{i}\right)
H\left( K_{i}\right) \right) ^{2}  \label{dnm}
\end{equation}%
We use the translation (\ref{Trl}) of a type-(\ref{inco}) expression. The
gradient term appearing in equation (\ref{Trl}) is $\nabla _{X}$. We thus
obtain the translation: 
\begin{eqnarray}
&&\sum_{i}\left( \frac{dX_{i}}{dt}-\nabla _{X}R\left( K_{i},X_{i}\right)
H\left( K_{i}\right) \right) ^{2}  \label{grl} \\
&\rightarrow &\int \Psi ^{\dag }\left( K,X\right) \left( -\nabla _{X}\left( 
\frac{\sigma _{X}^{2}}{2}\nabla _{X}+\Lambda (X,K)\right) \right) \Psi
\left( K,X\right) dKdX  \notag
\end{eqnarray}%
Note that the variance $\sigma _{X}^{2}$ reflects the probabilistic nature
of the model hidden behind the field formalism. This $\sigma _{X}^{2}$
represents the characteristic level of uncertainty of the sectors space
dynamics. It is a parameter of the model. The term $\Lambda (X,K)$ is the
translation of the term $-\nabla _{X}R\left( K_{i},X_{i}\right) H\left(
K_{i}\right) $ in the parenthesis of (\ref{dnm}). This term is a function of
one sole index "$i$". In that case, the term $\Lambda $ is simply obtained
by replacing $\left( K_{i},X_{i}\right) $ by $\left( K,X\right) $.\ We use
the translation (\ref{bdt}) of (\ref{ntr})-type term, so that $\Lambda $
writes:%
\begin{equation*}
\Lambda (X,K)=-\nabla _{X}R\left( K,X\right) H\left( K\right)
\end{equation*}%
and the translation of expression (\ref{dnm}) is: 
\begin{eqnarray}
&&\sum_{i}\left( \frac{dX_{i}}{dt}-\nabla _{X}R\left( K_{i},X_{i}\right)
H\left( K_{i}\right) \right) ^{2}  \label{ttn} \\
&\rightarrow &\int \Psi ^{\dag }\left( K,X\right) \left( -\nabla _{X}\left( 
\frac{\sigma _{X}^{2}}{2}\nabla _{X}-\nabla _{X}R\left( K,X\right) H\left(
K\right) \right) \right) \Psi \left( K,X\right) dKdX  \notag
\end{eqnarray}%
Using equations (\ref{tnl}) and (\ref{ttn}), the translation of (\ref{mnd})
is thus:%
\begin{eqnarray}
S_{1} &=&-\int \Psi ^{\dag }\left( K,X\right) \nabla _{X}\left( \frac{\sigma
_{X}^{2}}{2}\nabla _{X}-\nabla _{X}R\left( K,X\right) H\left( K\right)
\right) \Psi \left( K,X\right) dKdX  \label{sn} \\
&&+\tau \int \left\vert \Psi \left( K^{\prime },X\right) \right\vert
^{2}\left\vert \Psi \left( K,X\right) \right\vert ^{2}dK^{\prime }dKdX 
\notag
\end{eqnarray}

\paragraph{Translation of the minimization function: Physical capital}

We can now turn to the translation of the second equation (\ref{minK}),
which we rewrite for the sake of clarity as:

\begin{equation}
\sum_{i}\left( \frac{d}{dt}K_{i}+\frac{1}{\varepsilon }\left( K_{i}\left(
t\right) -\sum_{j}\frac{F_{2}\left( R\left( K_{i}\left( t\right)
,X_{i}\left( t\right) \right) \right) G\left( X_{i}\left( t\right) -\hat{X}%
_{j}\right) }{\sum_{l}F_{2}\left( R\left( K_{l}\left( t\right) ,X_{l}\left(
t\right) \right) \right) G\left( X_{l}\left( t\right) -\hat{X}_{j}\right) }%
\hat{K}_{j}\left( t\right) \right) \right) ^{2}  \label{mnn}
\end{equation}%
Once again, we use the translation (\ref{bdt}) of (\ref{ntr})-type term, and
start by building the field functional associated to the term inside the
square:%
\begin{equation*}
K_{i}\left( t\right) -\sum_{j}\frac{F_{2}\left( R\left( K_{i}\left( t\right)
,X_{i}\left( t\right) \right) \right) G\left( X_{i}\left( t\right) -\hat{X}%
_{j}\left( t\right) \right) }{\sum_{l}F_{2}\left( R\left( K_{l}\left(
t\right) ,X_{l}\left( t\right) \right) \right) G\left( X_{l}\left( t\right) -%
\hat{X}_{j}\left( t\right) \right) }\hat{K}_{j}\left( t\right)
\end{equation*}%
We replace:%
\begin{eqnarray*}
\left( K_{i}\left( t\right) ,X_{i}\left( t\right) \right) &\rightarrow
&\left( K,X\right) \\
\left( K_{l}\left( t\right) ,X_{l}\left( t\right) \right) &\rightarrow
&\left( K^{\prime },X^{\prime }\right) \\
\left( \hat{K}_{j}\left( t\right) ,\hat{X}_{j}\left( t\right) \right)
&\rightarrow &\left( \hat{K},\hat{X}\right)
\end{eqnarray*}%
and:%
\begin{equation}
K_{i}\left( t\right) -\sum_{j}\frac{F_{2}\left( R\left( K_{i}\left( t\right)
,X_{i}\left( t\right) \right) \right) G\left( X_{i}\left( t\right) -\hat{X}%
_{j}\right) }{\sum_{l}F_{2}\left( R\left( K_{l}\left( t\right) ,X_{l}\left(
t\right) \right) \right) G\left( X_{l}\left( t\right) -\hat{X}_{j}\right) }%
\hat{K}_{j}\left( t\right) \rightarrow K-\sum_{j}\frac{F_{2}\left( R\left(
K,X\right) \right) G\left( X-\hat{X}\right) }{\sum_{l}F_{2}\left( R\left(
K^{\prime },X^{\prime }\right) \right) G\left( X^{\prime }-\hat{X}\right) }%
\hat{K}  \label{tr}
\end{equation}%
The sum over $l$ is then replaced by an integral $\int \left\vert \Psi
\left( K^{\prime },X^{\prime }\right) \right\vert ^{2}d\left( K^{\prime
},X^{\prime }\right) $:%
\begin{eqnarray}
&&K_{i}\left( t\right) -\sum_{j}\frac{F_{2}\left( R\left( K_{i}\left(
t\right) ,X_{i}\left( t\right) \right) \right) G\left( X_{i}\left( t\right) -%
\hat{X}_{j}\right) }{\sum_{l}F_{2}\left( R\left( K_{l}\left( t\right)
,X_{l}\left( t\right) \right) \right) G\left( X_{l}\left( t\right) -\hat{X}%
_{j}\right) }\hat{K}_{j}\left( t\right)  \label{lrt} \\
&\rightarrow &K-\sum_{j}\frac{F_{2}\left( R\left( K,X\right) \right) G\left(
X-\hat{X}\right) }{\int \left\vert \Psi \left( K^{\prime },X^{\prime
}\right) \right\vert ^{2}d\left( K^{\prime },X^{\prime }\right) F_{2}\left(
R\left( K^{\prime },X^{\prime }\right) \right) G\left( X^{\prime }-\hat{X}%
_{j}\right) }\hat{K}  \notag
\end{eqnarray}%
Recall that investors' variables are denoted with an upper script $\symbol{94%
}$.

Finally, the sum over $j$ and the second field are replaced by $\int
\left\vert \hat{\Psi}\left( \hat{K},\hat{X}\right) \right\vert ^{2}d\left( 
\hat{K},\hat{X}\right) $. After introducing the characteristic factor $\frac{%
1}{\varepsilon }$ of the capital accumulation time scale (see (\ref{dnK})),
the translation becomes: 
\begin{eqnarray}
&&\frac{1}{\varepsilon }\left( K_{i}\left( t\right) -\sum_{j}\frac{%
F_{2}\left( R\left( K_{i}\left( t\right) ,X_{i}\left( t\right) \right)
\right) G\left( X_{i}\left( t\right) -\hat{X}_{j}\right) }{%
\sum_{l}F_{2}\left( R\left( K_{l}\left( t\right) ,X_{l}\left( t\right)
\right) \right) G\left( X_{l}\left( t\right) -\hat{X}_{j}\right) }\hat{K}%
_{j}\left( t\right) \right)  \notag \\
&\rightarrow &\frac{1}{\varepsilon }\left( K-\int \left\vert \hat{\Psi}%
\left( \hat{K},\hat{X}\right) \right\vert ^{2}d\left( \hat{K},\hat{X}\right) 
\frac{F_{2}\left( R\left( K,X\right) \right) G\left( X-\hat{X}\right) \hat{K}%
}{\int \left\vert \Psi \left( K^{\prime },X^{\prime }\right) \right\vert
^{2}d\left( K^{\prime },X^{\prime }\right) F_{2}\left( R\left( K^{\prime
},X^{\prime }\right) \right) G\left( X^{\prime }-\hat{X}\right) }\right) 
\notag \\
&\equiv &\Lambda \left( K,X\right)  \label{dml}
\end{eqnarray}%
Using the translation (\ref{Trl}) of (\ref{inco})-type term, we are led to
the translation of (\ref{mnn}). Since the square (\ref{mnn}) includes a
derivative $\frac{d}{dt}K_{i}$, the expression starts with a gradient with
respect to $K$, and we have:%
\begin{eqnarray}
&&\sum_{i}\left( \frac{d}{dt}K_{i}+\frac{1}{\varepsilon }\left(
K_{i}-\sum_{j}\frac{F_{2}\left( R\left( K_{i}\left( t\right) ,X_{i}\left(
t\right) \right) \right) G\left( X_{i}\left( t\right) -\hat{X}_{j}\right) }{%
\sum_{l}F_{2}\left( R\left( K_{l}\left( t\right) ,X_{l}\left( t\right)
\right) \right) G\left( X_{l}\left( t\right) -\hat{X}_{j}\right) }\hat{K}%
_{j}\left( t\right) \right) \right) ^{2}  \label{sq} \\
&\rightarrow &\int \Psi ^{\dag }\left( K,X\right) \left( -\nabla _{K}\left( 
\frac{\sigma _{K}^{2}}{2}\nabla _{K}+\Lambda \left( K,X\right) \right)
\right) \Psi \left( K,X\right) dKdX  \notag
\end{eqnarray}%
where, here again, the variance $\sigma _{K}^{2}$ reflects the probabilistic
nature of the model that is hidden behind the field formalism. Recall that
it represents the characteristic level of uncertainty in the dynamics of
capital.

Inserting result (\ref{dml}) in equation (\ref{sq}), the translation of (\ref%
{mnn)}) becomes:

\begin{equation}
S_{2}=-\int \Psi ^{\dag }\left( K,X\right) \nabla _{K}\left( \frac{\sigma
_{K}^{2}}{2}\nabla _{K}+\frac{1}{\varepsilon }\left( K-\int \frac{%
F_{2}\left( R\left( K,X\right) \right) G\left( X-\hat{X}\right) \hat{K}%
\left\Vert \hat{\Psi}\left( \hat{K},\hat{X}\right) \right\Vert ^{2}d\hat{K}d%
\hat{X}}{\int F_{2}\left( R\left( K,X\right) \right) G\left( X-\hat{X}%
\right) \left\Vert \Psi \left( K,X\right) \right\Vert ^{2}}\right) \right)
\Psi \left( K,X\right)  \label{sd}
\end{equation}

\subsubsection{Financial markets}

The functions to be translated are those of the financial capital dynamics (%
\ref{minKchap}) and of the financial capital allocation (\ref{minXchap}).\
Both expressions include a time derivative and are thus of type (\ref{edr}).
As for the real economy, the application of the translation rules is
straightforward.

\paragraph{Translation of the minimization function: Financial capital
dynamics}

We consider the function (\ref{minKchap}): 
\begin{equation}
\sum_{j}\left( \frac{d}{dt}\hat{K}_{j}-\frac{1}{\varepsilon }\left(
\sum_{i}\left( r_{i}+F_{1}\left( \frac{R\left( K_{i},X_{i}\right) }{%
\sum_{l}\delta \left( X_{l}-X_{i}\right) R\left( K_{l},X_{l}\right) },\frac{%
\dot{K}_{i}\left( t\right) }{K_{i}\left( t\right) }\right) \right) \frac{%
F_{2}\left( R\left( K_{i},X_{i}\right) \right) G\left( X_{i}-\hat{X}%
_{j}\right) }{\sum_{l}F_{2}\left( R\left( K_{l},X_{l}\right) \right) G\left(
X_{l}-\hat{X}_{j}\right) }\hat{K}_{j}\right) \right) ^{2}  \label{minKd}
\end{equation}%
which translates, using the general translation formula of expression (\ref%
{inco}) in (\ref{Trl}), into:%
\begin{equation*}
\int \hat{\Psi}^{\dag }\left( \hat{K},\hat{X}\right) \left( -\nabla _{\hat{K}%
}\left( \frac{\sigma _{\hat{K}}^{2}}{2}\nabla _{\hat{K}}+\Lambda \left( \hat{%
K},\hat{X}\right) \right) \right) \hat{\Psi}\left( \hat{K},\hat{X}\right) d%
\hat{K}d\hat{X}
\end{equation*}%
The function $\Lambda \left( \hat{K},\hat{X}\right) $ is obtained, as
before, by translating the term following the derivative in the function (%
\ref{minKd}):%
\begin{equation}
\frac{1}{\varepsilon }\sum_{i}\left( r_{i}+F_{1}\left( \frac{R\left(
K_{i},X_{i}\right) }{\sum_{l}\delta \left( X_{l}-X_{i}\right) R\left(
K_{l},X_{l}\right) },\frac{\dot{K}_{i}\left( t\right) }{K_{i}\left( t\right) 
}\right) \right) \frac{F_{2}\left( R\left( K_{i},X_{i}\right) \right)
G\left( X_{i}-\hat{X}_{j}\right) }{\sum_{l}F_{2}\left( R\left(
K_{l},X_{l}\right) \right) G\left( X_{l}-\hat{X}_{j}\right) }\hat{K}%
_{j}\rightarrow \Lambda \left( \hat{K},\hat{X}\right)  \label{rkt}
\end{equation}%
First, we use the price dynamics equation (\ref{pr}) at the zero-th order in
fluctuations to translate the capital dynamics $\frac{\dot{K}_{i}\left(
t\right) }{K_{i}\left( t\right) }$:%
\begin{eqnarray*}
\frac{\dot{K}_{i}\left( t\right) }{K_{i}\left( t\right) } &=&\sum_{j}\frac{%
F_{2}\left( R\left( K_{i}\left( t\right) ,X_{i}\left( t\right) \right)
\right) G\left( X_{i}\left( t\right) -\hat{X}_{j}\right) }{%
K_{i}\sum_{l}F_{2}\left( R\left( K_{l}\left( t\right) ,X_{l}\left( t\right)
\right) \right) G\left( X_{l}\left( t\right) -\hat{X}_{j}\right) }\hat{K}%
_{j}\left( t\right) -K_{i}\left( t\right) \\
&\rightarrow &\Gamma \left( K,X\right)
\end{eqnarray*}%
where:%
\begin{eqnarray}
\Gamma \left( K,X\right) &=&\frac{\int \frac{F_{2}\left( R\left( K,X\right)
\right) G\left( X-\hat{X}\right) }{\int F_{2}\left( R\left( K,X\right)
\right) G\left( X-\hat{X}\right) \left\Vert \Psi \left( K,X\right)
\right\Vert ^{2}}\hat{K}\left\Vert \hat{\Psi}\left( \hat{K},\hat{X}\right)
\right\Vert ^{2}d\left( \hat{K},\hat{X}\right) -K}{K}  \label{mg} \\
&=&\int \frac{F_{2}\left( R\left( K,X\right) \right) G\left( X-\hat{X}%
\right) }{K\int F_{2}\left( R\left( K,X\right) \right) G\left( X-\hat{X}%
\right) \left\Vert \Psi \left( K,X\right) \right\Vert ^{2}}\hat{K}\left\Vert 
\hat{\Psi}\left( \hat{K},\hat{X}\right) \right\Vert ^{2}d\left( \hat{K},\hat{%
X}\right) -1  \notag
\end{eqnarray}%
Then, using the translation (\ref{bdt}) of (\ref{ntr}), we translate
expression (\ref{rkt}) by replacing: 
\begin{eqnarray*}
\left( K_{i},X_{i}\right) &\rightarrow &\left( K,X\right) \\
\left( K_{l},X_{l}\right) &\rightarrow &\left( K^{\prime },X^{\prime }\right)
\\
\left( \hat{K}_{j},\hat{X}_{j}\right) &\rightarrow &\left( \hat{K},\hat{X}%
\right)
\end{eqnarray*}%
We also replace the sums by integrals times the appropriate square of field,
which yields:%
\begin{eqnarray*}
\Lambda \left( \hat{K},\hat{X}\right) &=&-\frac{\hat{K}}{\varepsilon }\int
\left( r\left( K,X\right) -\gamma \frac{\int K^{\prime }\left\Vert \Psi
\left( K^{\prime },X\right) \right\Vert ^{2}}{K}+F_{1}\left( \frac{R\left(
K,X\right) }{\int R\left( K^{\prime },X^{\prime }\right) \left\Vert \Psi
\left( K^{\prime },X^{\prime }\right) \right\Vert ^{2}d\left( K^{\prime
},X^{\prime }\right) },\Gamma \left( K,X\right) \right) \right) \\
&&\times \frac{F_{2}\left( R\left( K,X\right) \right) G\left( X-\hat{X}%
\right) }{\int F_{2}\left( R\left( K^{\prime },X^{\prime }\right) \right)
G\left( X^{\prime }-\hat{X}\right) \left\Vert \Psi \left( K^{\prime
},X^{\prime }\right) \right\Vert ^{2}d\left( K^{\prime },X^{\prime }\right) }%
\left\Vert \Psi \left( K,X\right) \right\Vert ^{2}d\left( K,X\right)
\end{eqnarray*}%
Ultimately, the translation of (\ref{minKchap}) is: 
\begin{eqnarray*}
S_{3} &=&-\int \hat{\Psi}^{\dag }\left( \hat{K},\hat{X}\right) \nabla _{\hat{%
K}}\left( \frac{\sigma _{\hat{K}}^{2}}{2}\nabla _{\hat{K}}-\frac{\hat{K}}{%
\varepsilon }\int \left( r\left( K,X\right) -\gamma \frac{\int K^{\prime
}\left\Vert \Psi \left( K^{\prime },X\right) \right\Vert ^{2}}{K}\right.
\right. \\
&&\left. +F_{1}\left( \frac{R\left( K,X\right) }{\int R\left( K^{\prime
},X^{\prime }\right) \left\Vert \Psi \left( K^{\prime },X^{\prime }\right)
\right\Vert ^{2}d\left( K^{\prime },X^{\prime }\right) },\Gamma \left(
K,X\right) \right) \right) \\
&&\times \left. \frac{F_{2}\left( R\left( K,X\right) \right) G\left( X-\hat{X%
}\right) }{\int F_{2}\left( R\left( K^{\prime },X^{\prime }\right) \right)
G\left( X^{\prime }-\hat{X}\right) \left\Vert \Psi \left( K^{\prime
},X^{\prime }\right) \right\Vert ^{2}d\left( K^{\prime },X^{\prime }\right) }%
\left\Vert \Psi \left( K,X\right) \right\Vert ^{2}d\left( K,X\right) \right) 
\hat{\Psi}\left( \hat{K},\hat{X}\right)
\end{eqnarray*}

\paragraph{Translation of the minimization function: Financial capital
allocation}

The translation of the function for financial capital allocation (\ref%
{minXchap}) follows the previous pattern. We obtain:%
\begin{eqnarray*}
S_{4} &=&-\int \hat{\Psi}^{\dag }\left( \hat{K},\hat{X}\right) \nabla _{\hat{%
X}}\left( \sigma _{\hat{X}}^{2}\nabla _{\hat{X}}\right. -\int \left( \nabla
_{\hat{X}}F_{0}\left( R\left( K,\hat{X}\right) \right) +\nu \nabla _{\hat{X}%
}F_{1}\left( \frac{R\left( K,\hat{X}\right) }{\int R\left( K^{\prime
},X^{\prime }\right) \left\Vert \Psi \left( K^{\prime },X^{\prime }\right)
\right\Vert ^{2}d\left( K^{\prime },X^{\prime }\right) }\right) \right) \\
&&\times \left. \frac{\left\Vert \Psi \left( K,\hat{X}\right) \right\Vert
^{2}dK}{\int \left\Vert \Psi \left( K^{\prime },\hat{X}\right) \right\Vert
^{2}dK^{\prime }}\right) \hat{\Psi}\left( \hat{K},\hat{X}\right)
\end{eqnarray*}

\subsection{Gathering contributions: the action functional}

Once these translations are performed, the action functional of the system
is described by the sum of all contributions:%
\begin{equation*}
S=S_{1}+S_{2}+S_{3}+S_{4}
\end{equation*}%
We write a compact form for the action functional $S$: 
\begin{eqnarray}
S &=&-\int \Psi ^{\dag }\left( K,X\right) \left( \nabla _{X}\left( \frac{%
\sigma _{X}^{2}}{2}\nabla _{X}-\nabla _{X}R\left( K,X\right) H\left(
K\right) \right) -\tau \left( \int \left\vert \Psi \left( K^{\prime
},X\right) \right\vert ^{2}dK^{\prime }\right) \right.  \label{fcn} \\
&&+\left. \nabla _{K}\left( \frac{\sigma _{K}^{2}}{2}\nabla _{K}+u\left(
K,X,\Psi ,\hat{\Psi}\right) \right) \right) \Psi \left( K,X\right) dKdX 
\notag \\
&&-\int \hat{\Psi}^{\dag }\left( \hat{K},\hat{X}\right) \left( \nabla _{\hat{%
K}}\left( \frac{\sigma _{\hat{K}}^{2}}{2}\nabla _{\hat{K}}-\hat{K}f\left( 
\hat{X},\Psi ,\hat{\Psi}\right) \right) +\nabla _{\hat{X}}\left( \frac{%
\sigma _{\hat{X}}^{2}}{2}\nabla _{\hat{X}}-g\left( K,X,\Psi ,\hat{\Psi}%
\right) \right) \right) \hat{\Psi}\left( \hat{K},\hat{X}\right)  \notag
\end{eqnarray}%
where each line corresponds to one $S_{i}$ and where, to simplify, we have
defined:%
\begin{eqnarray}
u\left( K,X,\Psi ,\hat{\Psi}\right) &=&\frac{1}{\varepsilon }\left( K-\int 
\frac{F_{2}\left( R\left( K,X\right) \right) G\left( X-\hat{X}\right) }{\int
F_{2}\left( R\left( K,X\right) \right) G\left( X-\hat{X}\right) \left\Vert
\Psi \left( K,X\right) \right\Vert ^{2}}\hat{K}\left\Vert \hat{\Psi}\left( 
\hat{K},\hat{X}\right) \right\Vert ^{2}d\hat{K}d\hat{X}\right)  \label{fcs}
\\
f\left( \hat{X},\Psi ,\hat{\Psi}\right) &=&\frac{1}{\varepsilon }\int \left(
r\left( K,X\right) -\frac{\gamma \int K^{\prime }\left\Vert \Psi \left(
K,X\right) \right\Vert ^{2}}{K}+F_{1}\left( \frac{R\left( K,X\right) }{\int
R\left( K^{\prime },X^{\prime }\right) \left\vert \Psi \left( K^{\prime
},X^{\prime }\right) \right\vert ^{2}dK^{\prime }dX^{\prime }},\Gamma \left(
K,X\right) \right) \right)  \notag \\
&&\times \frac{F_{2}\left( R\left( K,X\right) \right) G\left( X-\hat{X}%
\right) }{\int F_{2}\left( R\left( K^{\prime },X^{\prime }\right) \right)
G\left( X^{\prime }-\hat{X}\right) \left\Vert \Psi \left( K^{\prime
},X^{\prime }\right) \right\Vert ^{2}d\left( K^{\prime },X^{\prime }\right) }%
\left\vert \Psi \left( K,X\right) \right\vert ^{2}d\left( K,X\right)
\label{fcS} \\
g\left( K,\hat{X},\Psi ,\hat{\Psi}\right) &=&\int \left( \nabla _{\hat{X}%
}F_{0}\left( R\left( K,\hat{X}\right) \right) \right.  \label{fCS} \\
&&\left. +\nu \nabla _{\hat{X}}F_{1}\left( \frac{R\left( K,\hat{X}\right) }{%
\int R\left( K^{\prime },X^{\prime }\right) \left\Vert \Psi \left( K^{\prime
},X^{\prime }\right) \right\Vert ^{2}d\left( K^{\prime },X^{\prime }\right) }%
,\Gamma \left( K,X\right) \right) \right) \frac{\left\Vert \Psi \left( K,%
\hat{X}\right) \right\Vert ^{2}dK}{\int \left\Vert \Psi \left( K^{\prime },%
\hat{X}\right) \right\Vert ^{2}dK^{\prime }}  \notag
\end{eqnarray}%
The expression of \textbf{\ }$\Gamma \left( K,X\right) $ has been given in (%
\ref{mg}).

Recall that function $H\left( K_{X}\right) $\ encompasses the determinants
of the firms' mobility across the sector space. We will specify this
function below as a function of expected long term-returns and capital.

Function $u$\ describes the evolution of capital of a firm, located at $X$.
This dynamics depends on the relative value of a function $F_{2}$ that is
itself a function of the firms' expected returns $R\left( K,X\right) $.
Investors allocate their capital based on their expectations of the firms'
long-term returns.

Function $f$\ describes the returns of investors located at $\hat{X}$,\ and
investing in sector $X$\ a capital $K$. These returns depend on short-term
dividends $r\left( K,X\right) $, the field-equivalent cost of capital\textbf{%
\ }$\frac{\gamma \int K^{\prime }\left\Vert \Psi \left( K,X\right)
\right\Vert ^{2}}{K}$\textbf{,\ }and a function $F_{1}$ that depends on
firms' expected long-term stock valuations.\ These valuations themselves
depend on the relative attractivity of a firm expected long-term returns
vis-a-vis its competitors.

Function $g$\ describes investors' shifts across the sectors' space. They
are driven by the gradient of expected long-term returns and stocks
valuations, who themselves depend on the firms' relative expected long-term
returns.

Recall that we depart here from the general formalism: we do not introduce a
time variable in the present model. Indeed, as mentioned earlier, our
purpose is to find collective, or characteristic, configurations of the
system that, as such, can be considered static. It is only when we will
derive these configurations that a macro time scale will be introduced to
study how the evolution of the background states through time.

\section{Use of the field model}

Now that we have found the field action functional $S$, we can use field
theory to study the system of agents.\ This can be done at two levels.

\paragraph{At the collective level}

At the collective level, the background fields of the system, i.e. the
particular functions $\Psi \left( K,X\right) $\ and $\hat{\Psi}\left( \hat{K}%
,\hat{X}\right) $\ that minimize the functional $S$, can be computed. The
functions, squared, $\left\vert \Psi \left( K,X\right) \right\vert ^{2}$\
and $\left\vert \hat{\Psi}\left( \hat{K},\hat{X}\right) \right\vert ^{2}$,
represent the density of agents, both per sector and for a given capital $K$%
, in the collective state defined by $\Psi \left( K,X\right) $\ and $\hat{%
\Psi}\left( \hat{K},\hat{X}\right) $. Thus, the collective state determines,
for each sector and for a given capital $K$, the density of firms and the
density of investors.

Moreover, these two squared functions allow to compute various global
quantities of the system\ in the collective state $\Psi \left( K,X\right) $\
and $\hat{\Psi}\left( \hat{K},\hat{X}\right) $.

The sectors' number of producers $N\left( X\right) $\ and investors $\hat{N}%
\left( \hat{X}\right) $ are computed using the formula:%
\begin{eqnarray}
N\left( X\right) &=&\int \left\vert \Psi \left( K,X\right) \right\vert ^{2}dK
\label{Nx} \\
\hat{N}\left( \hat{X}\right) &=&\int \left\vert \hat{\Psi}\left( \hat{K},%
\hat{X}\right) \right\vert ^{2}d\hat{K}  \label{Nxh}
\end{eqnarray}%
The average values of total invested capital $\hat{K}_{X}$\ for each sector $%
X$ is:%
\begin{equation*}
\hat{K}_{\hat{X}}=\int \hat{K}\left\vert \hat{\Psi}\left( \hat{K},X\right)
\right\vert ^{2}d\hat{K}
\end{equation*}%
and the average invested capital per firm for sector $X$ is:\textbf{\ }%
\begin{equation}
K_{X}=\frac{\int \hat{K}\left\vert \hat{\Psi}\left( \hat{K},X\right)
\right\vert ^{2}d\hat{K}}{N\left( X\right) }  \label{kx}
\end{equation}%
Note that this $K_{X}$ is also equal to the average physical capital per
firm for sector $X$, i.e. :%
\begin{equation}
K_{X}=\frac{\int K\left\vert \Psi \left( K,X\right) \right\vert ^{2}dK}{%
N\left( X\right) }  \label{KX}
\end{equation}%
\textbf{\ }Indeed, given our assumptions, the total physical capital is
equal to the total capital invested:%
\begin{equation*}
\int K\left\vert \Psi \left( K,X\right) \right\vert ^{2}dK=\int \hat{K}%
\left\vert \hat{\Psi}\left( \hat{K},\hat{X}\right) \right\vert ^{2}d\hat{K}
\end{equation*}%
In the following, we will use alternately both expressions (\ref{kx}) or (%
\ref{KX}).

Ultimately, given a collective state, the distribution per sector of both
invested capital and capital per firm are given by $\frac{\left\vert \hat{%
\Psi}\left( \hat{K},X\right) \right\vert ^{2}}{\hat{N}\left( X\right) }$\
and $\frac{\left\vert \Psi \left( K,X\right) \right\vert ^{2}}{N\left(
X\right) }$\ respectively.

Gathering equations (\ref{Nx}), (\ref{Nxh}) and (\ref{kx}), we thus conclude
that each collective state is singularly determined by the collection of
data that characterizes each sector: the number of firms for each sector,
the number of investors for each sector, the average capital for each sector
and the density of distribution of capital in each sector.

All the above quantities allow to study the capital allocation among sectors
as well as its dependency in system parameters such as expected long-term
return, short-term return, or any parameters involved in the model.

\paragraph{At the individual level}

At the individual level, the field formalism allows to compute agents'
individual dynamics in the state defined by the background fields, through
the transition functions of the system. These transition functions are
themselves derived using the Green functions' formalism (see GLW). This
study is left for a subsequent work. In the following, we solve the system
for the background fields and compute the average associated quantities.
This "static" point of view, will be extended by introducing some
fluctuations in the expectations, leading to a dynamic of the average
capital at the macro-level.

\section{Resolution of the Model}

Now that our initial framework has been turned into a proper field
formalism, we can turn to the resolution of the model.\ More specifically,
we want to find the background fields of the system. We start with some
preliminary simplifications.

\subsection{Preliminaries}

To do so, we must find the configurations $\Psi \left( K,X\right) $ and $%
\hat{\Psi}\left( \hat{K},\hat{X}\right) $\ of the fields that minimize $S$.
To study the impact of financial on the real sector, we first minimize $%
S_{1}+S_{2}$, then solve for the field $\Psi \left( K,X\right) $ as a
function of the investors' variables. We then minimize $S_{3}+S_{4}$, and
find the minimal configuration of the investors' field $\hat{\Psi}\left( 
\hat{K},\hat{X}\right) $.

At this point, we can introduce a simplification and assume that investors
invest in only one sector. This translates into the following condition:%
\begin{equation}
G\left( X-\hat{X}\right) =\delta \left( X-\hat{X}\right)  \label{smp}
\end{equation}%
This simplification does not reduce the generality of our model: actually,
an investor acting in several sectors could be modelled as an aggregation of
several investors.\ Nor does it mean that investors should be static, since
they can still move from one sector to another.

The intermediate functions (\ref{fcs}), (\ref{fcS}) and (\ref{fCS}) involved
in the definition of the action functional (\ref{fcn}) thus become:%
\begin{eqnarray}
u\left( K,X,\Psi ,\hat{\Psi}\right) &=&\frac{1}{\varepsilon }\left( K-\int 
\frac{F_{2}\left( R\left( K,X\right) \right) }{\int F_{2}\left( R\left(
K^{\prime },X\right) \right) \left\Vert \Psi \left( K^{\prime },X\right)
\right\Vert ^{2}dK^{\prime }}\hat{K}\left\Vert \hat{\Psi}\left( \hat{K}%
,X\right) \right\Vert ^{2}d\hat{K}\right)  \label{fctt} \\
f\left( \hat{X},\Psi ,\hat{\Psi}\right) &=&\frac{1}{\varepsilon }\int \left(
r\left( K,X\right) -\gamma \frac{\int K^{\prime }\left\Vert \Psi \left(
K^{\prime },X\right) \right\Vert ^{2}}{K}+F_{1}\left( \frac{R\left(
K,X\right) }{\int R\left( K^{\prime },X^{\prime }\right) \left\Vert \Psi
\left( K^{\prime },X^{\prime }\right) \right\Vert ^{2}d\left( K^{\prime
},X^{\prime }\right) },\Gamma \left( K,X\right) \right) \right)  \notag \\
&&\times \frac{F_{2}\left( R\left( K,\hat{X}\right) \right) }{\int
F_{2}\left( R\left( K^{\prime },\hat{X}\right) \right) \left\Vert \Psi
\left( K^{\prime },\hat{X}\right) \right\Vert ^{2}dK^{\prime }}\left\Vert
\Psi \left( K,\hat{X}\right) \right\Vert ^{2}dK  \notag \\
g\left( K,\hat{X},\Psi ,\hat{\Psi}\right) &=&\int \left( \nabla _{\hat{X}%
}F_{0}\left( R\left( K,\hat{X}\right) \right) +\nu \nabla _{\hat{X}%
}F_{1}\left( \frac{R\left( K,\hat{X}\right) }{\int R\left( K^{\prime
},X^{\prime }\right) \left\Vert \Psi \left( K^{\prime },X^{\prime }\right)
\right\Vert ^{2}d\left( K^{\prime },X^{\prime }\right) }\right) \right) 
\frac{\left\Vert \Psi \left( K,\hat{X}\right) \right\Vert ^{2}dK}{\int
\left\Vert \Psi \left( K^{\prime },\hat{X}\right) \right\Vert ^{2}dK^{\prime
}}  \notag
\end{eqnarray}%
where:%
\begin{equation*}
\int \frac{F_{2}\left( R\left( K,X\right) \right) G\left( X-\hat{X}\right) }{%
K\int F_{2}\left( R\left( K,X\right) \right) G\left( X-\hat{X}\right)
\left\Vert \Psi \left( K,X\right) \right\Vert ^{2}}\hat{K}\left\Vert \hat{%
\Psi}\left( \hat{K},\hat{X}\right) \right\Vert ^{2}d\left( \hat{K},\hat{X}%
\right) -1
\end{equation*}

\subsection{\textbf{Background field for the real economy}}

\subsubsection{General formula}

We first compute\footnote{%
For detailed computations of this subsection, see appendix 2.} the field of
the real economy $\Psi \left( K,X\right) $ as a function of the field of the
financial sector $\hat{\Psi}\left( \hat{K},\hat{X}\right) $ by minimizing
the $\left( K,X\right) $ part of equation (\ref{fcn}): 
\begin{eqnarray}
S_{1}+S_{2} &=&-\int \Psi ^{\dag }\left( K,X\right) \left( \nabla _{X}\left( 
\frac{\sigma _{X}^{2}}{2}\nabla _{X}-\nabla _{X}R\left( K,X\right) H\left(
K\right) \right) -\tau \left( \int \left\vert \Psi \left( K^{\prime
},X\right) \right\vert ^{2}dK^{\prime }\right) \right. \\
&&+\left. \nabla _{K}\left( \frac{\sigma _{K}^{2}}{2}\nabla _{K}+u\left(
K,X,\Psi ,\hat{\Psi}\right) \right) \right) \Psi \left( K,X\right) dKdX 
\notag
\end{eqnarray}%
For relatively slow fluctuations in $X$, the background fields $\Psi \left(
K,X\right) $ and $\Psi ^{\dag }\left( K,X\right) $\ decompose as a product:%
\begin{eqnarray}
\Psi \left( K,X\right) &=&\exp \left( \int^{X}\frac{\nabla _{X}R\left(
X\right) }{\sigma _{X}^{2}}H\left( \frac{\int \hat{K}\left\Vert \hat{\Psi}%
\left( \hat{K},X\right) \right\Vert ^{2}d\hat{K}}{\left\Vert \Psi \left(
X\right) \right\Vert ^{2}}\right) \right)  \label{psf} \\
&&\times \exp \left( \int \left( K-\frac{F_{2}\left( R\left( K,X\right)
\right) K_{X}}{F_{2}\left( R\left( K_{X},X\right) \right) }\right) dK\right)
\Psi \left( X\right) \Psi _{1}\left( K-K_{X}\right)  \notag \\
\Psi ^{\dag }\left( K,X\right) &=&\exp \left( -\int^{X}\frac{\nabla
_{X}R\left( X\right) }{\sigma _{X}^{2}}H\left( \frac{\int \hat{K}\left\Vert 
\hat{\Psi}\left( \hat{K},X\right) \right\Vert ^{2}d\hat{K}}{\left\Vert \Psi
\left( X\right) \right\Vert ^{2}}\right) \right)  \label{fsp} \\
&&\times \exp \left( -\int \left( K-\frac{F_{2}\left( R\left( K,X\right)
\right) K_{X}}{F_{2}\left( R\left( K_{X},X\right) \right) }\right) dK\right)
\Psi ^{\dag }\left( X\right) \Psi _{1}^{\dag }\left( K-K_{X}\right)  \notag
\end{eqnarray}%
where $K_{X}$, the average invested capital per firm in sector $X$, is given
by:%
\begin{equation}
K_{X}=\frac{\int \hat{K}\left\Vert \hat{\Psi}\left( \hat{K},X\right)
\right\Vert ^{2}d\hat{K}}{\left\Vert \Psi \left( X\right) \right\Vert ^{2}}
\label{csc}
\end{equation}

\subsubsection{Determination of $\Psi \left( X\right) $ and $\left\Vert \Psi
\left( X\right) \right\Vert ^{2}$}

The function $\Psi \left( X\right) $, that arises in the definitions (\ref%
{psf}) and (\ref{fsp}) of the background fields $\Psi \left( K,X\right) $
and $\Psi ^{\dag }\left( K,X\right) $, minimizes:

\begin{eqnarray}
&&\int \Psi ^{\dag }\left( X\right) \left( -\frac{\sigma _{X}^{2}}{2}\nabla
_{X}^{2}+\frac{\left( \nabla _{X}R\left( X\right) H\left( K_{X}\right)
\right) ^{2}}{2\sigma _{X}^{2}}+\frac{\nabla _{X}^{2}R\left( K,X\right) }{2}%
H\left( K\right) +2\tau \left\vert \Psi \left( X\right) \right\vert
^{2}\right) \Psi \left( X\right)  \label{mnt} \\
&&+D\left( \left\Vert \Psi \right\Vert ^{2}\right) \left( \int \left\Vert
\Psi \left( X\right) \right\Vert ^{2}-N\right) +\int \mu \left( X\right)
\left\Vert \Psi \left( X\right) \right\Vert ^{2}  \notag
\end{eqnarray}%
where the constants $D\left( \left\Vert \Psi \right\Vert ^{2}\right) $ and $%
\mu \left( X\right) $ are Lagrange multipliers, and implement the
constraints: 
\begin{equation*}
\int \left\Vert \Psi \left( X\right) \right\Vert ^{2}=N
\end{equation*}%
where $N$ is the number of firms, and:%
\begin{equation*}
\left\Vert \Psi \left( X\right) \right\Vert ^{2}\geqslant 0
\end{equation*}%
Incidentally, note that, to keep track of the dependency of the Lagrange
multiplier in $\left\Vert \Psi \right\Vert ^{2}$ in the above, we have
chosen the notation $D\left( \left\Vert \Psi \right\Vert ^{2}\right) $.

The solution of the minimization problem for equation (\ref{mnt}) is a
function: 
\begin{equation*}
\Psi \left( X,\left( \nabla _{X}R\left( X\right) \right) ^{2},\frac{\hat{K}%
_{X}}{\hat{K}_{X,0}}\right)
\end{equation*}%
It depends on the parameters of the system: the average invested capital,
and the value of the expected return and its derivatives.\ The parameter $%
\hat{K}_{X,0}$ is a normalization factor. Solving equation (\ref{mnt}) in
the limit of small fluctuations $\sigma _{X}^{2}$ in $X$ , the minimization
of equation (\ref{mnt}) becomes: 
\begin{subequations}
\begin{equation}
D\left( \left\Vert \Psi \right\Vert ^{2}\right) =2\tau \left\Vert \Psi
\left( X\right) \right\Vert ^{2}+\frac{1}{2}\left( \left( \nabla _{X}R\left(
X\right) \right) ^{2}+\frac{\sigma _{X}^{2}\nabla _{X}^{2}R\left(
K_{X},X\right) }{H\left( K_{X}\right) }\right) H^{2}\left( \frac{\hat{K}_{X}%
}{\left\Vert \Psi \left( X\right) \right\Vert ^{2}}\right) \left( 1-\frac{%
H^{\prime }\left( \hat{K}_{X}\right) \hat{K}_{X}}{H\left( \hat{K}_{X}\right)
\left\Vert \Psi \left( X\right) \right\Vert ^{2}}\right)  \label{qnp}
\end{equation}%
where the Lagrange multiplier $D\left( \left\Vert \Psi \right\Vert
^{2}\right) $ satisfies: 
\end{subequations}
\begin{equation}
ND\left( \left\Vert \Psi \right\Vert ^{2}\right) =2\tau \int \left\vert \Psi
\left( X\right) \right\vert ^{4}+\frac{1}{2}\int \left( \nabla _{X}R\left(
X\right) H\left( \frac{\int \hat{K}\left\Vert \hat{\Psi}\left( \hat{K}%
,X\right) \right\Vert ^{2}d\hat{K}}{\left\Vert \Psi \left( X\right)
\right\Vert ^{2}}\right) \right) ^{2}\left\Vert \Psi \left( X\right)
\right\Vert ^{2}  \label{pnq}
\end{equation}%
The two equations above, (\ref{qnp}) and (\ref{pnq}),\ can be solved as a
function of the total capital invested in sector $X$, $\hat{K}_{X}$ for the
particular form of the function $H$. Two examples of such resolution are
given in appendix 2.2.

However a precise solution of (\ref{pnq}) is useless in the following, since
we will rather need the first equation, (\ref{qnp}), to be written as:%
\begin{equation}
\left\Vert \Psi \left( X\right) \right\Vert ^{2}=\left( 2\tau \right)
^{-1}\left( D\left( \left\Vert \Psi \right\Vert ^{2}\right) -\frac{1}{2}%
\left( \left( \nabla _{X}R\left( X\right) \right) ^{2}+\frac{\sigma
_{X}^{2}\nabla _{X}^{2}R\left( K_{X},X\right) }{H\left( K_{X}\right) }%
\right) H^{2}\left( K_{X}\right) \left( 1-\frac{H^{\prime }\left( \hat{K}%
_{X}\right) K_{X}}{H\left( \hat{K}_{X}\right) }\right) \right)  \label{psl}
\end{equation}%
This last formula will be used extensively in the sequel to compute $K_{X}$
the average physical capital per firm in sector $X$.

\subsubsection{Determination of $\Psi _{1}\left( K-K_{X}\right) $}

The function $\Psi _{1}\left( K-K_{X}\right) $, involved in the definitions (%
\ref{psf}) and (\ref{fsp}) of the background fields $\Psi \left( K,X\right) $
and $\Psi ^{\dag }\left( K,X\right) $, describes the fluctuations of capital
around an average value $K_{X}$. It satisfies the equation:%
\begin{equation}
-\nabla _{K}^{2}\Psi _{1}\left( K-K_{X}\right) +\left( K-\frac{F_{2}\left(
R\left( K,X\right) \right) K_{X}}{F_{2}\left( R\left( K_{X},X\right) \right) 
}\right) ^{2}\Psi _{1}\left( K-K_{X}\right) =0  \label{cpf}
\end{equation}%
with solution:%
\begin{equation}
\Psi _{1}\left( K-K_{X}\right) =\mathcal{N}\exp \left( -\left( K-\frac{%
F_{2}\left( R\left( K,X\right) \right) K_{X}}{F_{2}\left( R\left(
K_{X},X\right) \right) }\right) ^{2}\right)  \label{fpc}
\end{equation}%
where $\mathcal{N}$ is a normalization factor. The same formula holds for $%
\Psi _{1}^{\dag }\left( K-K_{X}\right) $.

\subsubsection{Interpretation of the formula and density of agents}

Formulas (\ref{psf}) and (\ref{fsp}) decompose in four terms:

The two exponential factors encapsulate the agents' average move towards
sector $X$ and capital accumulation at this point, respectively.

The two factors $\Psi \left( X\right) \Psi _{1}\left( K-K_{X}\right) $ and $%
\Psi ^{\dag }\left( X\right) \Psi _{1}^{\dag }\left( K-K_{X}\right) $
translate the fact that in the sector's space, motion is slower than capital
accumulation: capital accumulates as a function of the position $X$ of the
sector, through the capital allocated in this sector, $K_{X}$.\ The
repartition of agents in space is described by the squared field $\left\vert
\Psi \left( X\right) \right\vert ^{2}$ (see (\ref{Nx})).

Formula (\ref{fpc}) shows that the capital accumulated by a firm in a sector 
$X$ is centered around the average capital $K_{X}$ in this sector, weighted
by a factor $\frac{F_{2}\left( R\left( K,X\right) \right) }{F_{2}\left(
R\left( K_{X},X\right) \right) }$ that depends on the firm's expected
long-term return, and is relative to the average expected long-term return
of the whole sector $X$.\ The latter is described by the function $%
F_{2}\left( R\left( K_{X},X\right) \right) $\footnote{%
See discussion below equation (\ref{grandf2}).}.

Once the solution of (\ref{mnt}) found, it can be used alongside equation (%
\ref{fpc}) to compute $\left\Vert \Psi \left( K,X\right) \right\Vert ^{2}$
in the limit of relatively small fluctuations $\sigma _{X}^{2}$ in $X$. We
obtain:%
\begin{equation}
\left\Vert \Psi \left( K,X\right) \right\Vert ^{2}=\mathcal{N}\left\Vert
\Psi \left( X,\left( \nabla _{X}R\left( X\right) \right) ^{2},\frac{\hat{K}%
_{X}}{\hat{K}_{X,0}}\right) \right\Vert ^{2}\times \exp \left( -\left(
K-K_{X}\right) ^{2}\right)  \label{Psc}
\end{equation}%
Note that the form of the exponential in (\ref{Psc}) implies that the
average physical capital is given by $K_{X}$, i.e. the capital invested in
sector $X$,\ divided by the number of firms in this sector, which computes
the density of agents defined by a given position $X$\ and capital $K$.
Mathematically:%
\begin{equation*}
\int K\left\Vert \Psi \left( K,X\right) \right\Vert ^{2}d\hat{K}=\int \hat{K}%
\left\Vert \hat{\Psi}\left( \hat{K},X\right) \right\Vert ^{2}d\hat{K}
\end{equation*}

\subsection{B\textbf{ackground field\ }for the financial sector}

In this paragraph we compute the background field $\hat{\Psi}\left( \hat{K},%
\hat{X}\right) $\ representing the financial markets.

\subsubsection{Minimization of $S_{3}+S_{4}$}

Once the background fields for the real economy $\Psi \left( K,X\right) $
and $\Psi ^{\dag }\left( X,K\right) $ found, the minimization problem for $%
S_{3}+S_{4}$ can be considered. The computations are presented in appendix 3.

Given the form of $\Psi \left( X,K\right) $ and $\Psi ^{\dag }\left(
X,K\right) $ (\ref{psf}) and (\ref{fsp}), the field action for the
background field of the financial markets $\hat{\Psi}\left( \hat{K},\hat{X}%
\right) $ can be reduced to the expression (see appendix 3.1.1):%
\begin{equation}
S_{3}+S_{4}=-\int \hat{\Psi}^{\dag }\left( \hat{K},\hat{X}\right) \left(
\nabla _{\hat{K}}\left( \frac{\sigma _{\hat{K}}^{2}}{2}\nabla _{\hat{K}}-%
\hat{K}f\left( \hat{X},K_{\hat{X}}\right) \right) +\nabla _{\hat{X}}\left( 
\frac{\sigma _{\hat{X}}^{2}}{2}\nabla _{\hat{X}}-g\left( \hat{X},K_{\hat{X}%
}\right) \right) \right) \hat{\Psi}\left( \hat{K},\hat{X}\right)  \label{tsm}
\end{equation}%
with:

\begin{eqnarray}
f\left( \hat{X},K_{\hat{X}}\right) &=&\frac{1}{\varepsilon }\left( r\left(
K_{\hat{X}},\hat{X}\right) -\gamma \left\Vert \Psi \left( \hat{X}\right)
\right\Vert ^{2}+F_{1}\left( \frac{R\left( K_{\hat{X}},\hat{X}\right) }{\int
R\left( K_{X^{\prime }}^{\prime },X^{\prime }\right) \left\Vert \Psi \left(
X^{\prime }\right) \right\Vert ^{2}dX^{\prime }}\right) \right)  \label{fcf}
\\
g\left( \hat{X},K_{\hat{X}}\right) &=&\left( \frac{\nabla _{\hat{X}%
}F_{0}\left( R\left( K_{\hat{X}},\hat{X}\right) \right) }{\left\Vert \nabla
_{\hat{X}}R\left( K_{\hat{X}},\hat{X}\right) \right\Vert }+\nu \nabla _{\hat{%
X}}F_{1}\left( \frac{R\left( K_{\hat{X}},\hat{X}\right) }{\int R\left(
K_{X^{\prime }}^{\prime },X^{\prime }\right) \left\Vert \Psi \left(
X^{\prime }\right) \right\Vert ^{2}dX^{\prime }}\right) \right)  \label{fcg}
\end{eqnarray}%
The first order condition for (\ref{tsm}) is computed using a change of
variable (see appendix 3.1.2):%
\begin{equation*}
\hat{\Psi}\rightarrow \exp \left( \frac{1}{\sigma _{\hat{X}}^{2}}\int
g\left( \hat{X}\right) d\hat{X}+\frac{\hat{K}^{2}}{\sigma _{\hat{K}}^{2}}%
f\left( \hat{X}\right) \right) \hat{\Psi}
\end{equation*}%
which yields the following equations for $\hat{\Psi}$ and $\hat{\Psi}^{\dag
} $:%
\begin{eqnarray}
0 &=&\left( \frac{\sigma _{\hat{X}}^{2}}{2}\nabla _{\hat{X}}^{2}-\frac{1}{%
2\sigma _{\hat{X}}^{2}}\left( g\left( \hat{X},K_{\hat{X}}\right) \right)
^{2}-\frac{1}{2}\nabla _{\hat{X}}g\left( \hat{X},K_{\hat{X}}\right) \right) 
\hat{\Psi}  \label{hqn} \\
&&+\left( \nabla _{\hat{K}}\left( \frac{\sigma _{\hat{K}}^{2}}{2}\nabla _{%
\hat{K}}-\hat{K}f\left( \hat{X},K_{\hat{X}}\right) \right) -\hat{\lambda}%
\right) \hat{\Psi}-F\left( \hat{X},K_{\hat{X}}\right) \hat{K}\hat{\Psi} 
\notag
\end{eqnarray}%
\begin{eqnarray}
0 &=&\left( \frac{\sigma _{\hat{X}}^{2}}{2}\nabla _{\hat{X}}^{2}-\frac{1}{%
2\sigma _{\hat{X}}^{2}}\left( g\left( \hat{X},K_{\hat{X}}\right) \right)
^{2}-\frac{1}{2}\nabla _{\hat{X}}g\left( \hat{X},K_{\hat{X}}\right) \right) 
\hat{\Psi}^{\dag }  \label{nqh} \\
&&+\left( \left( \frac{\sigma _{\hat{K}}^{2}}{2}\nabla _{\hat{K}}+\hat{K}%
f\left( \hat{X},K_{\hat{X}}\right) \right) \nabla _{\hat{K}}-\hat{\lambda}%
\right) \hat{\Psi}-F\left( \hat{X},K_{\hat{X}}\right) \hat{K}\hat{\Psi}%
^{\dag }  \notag
\end{eqnarray}%
with:%
\begin{eqnarray}
F\left( \hat{X},K_{\hat{X}}\right) &=&\nabla _{K_{\hat{X}}}\left( \frac{%
\left( g\left( \hat{X},K_{\hat{X}}\right) \right) ^{2}}{2\sigma _{\hat{X}%
}^{2}}+\frac{1}{2}\nabla _{\hat{X}}g\left( \hat{X},K_{\hat{X}}\right)
+f\left( \hat{X},K_{\hat{X}}\right) \right) \frac{\left\Vert \hat{\Psi}%
\left( \hat{X}\right) \right\Vert ^{2}}{\left\Vert \Psi \left( \hat{X}%
\right) \right\Vert ^{2}}  \label{Fct} \\
&&+\frac{\nabla _{K_{\hat{X}}}f^{2}\left( \hat{X},K_{\hat{X}}\right) }{%
\sigma _{\hat{K}}^{2}\left\Vert \Psi \left( \hat{X}\right) \right\Vert ^{2}}%
\left\langle \hat{K}^{2}\right\rangle _{\hat{X}}  \notag
\end{eqnarray}%
where $\left\langle \hat{K}^{2}\right\rangle _{\hat{X}}$\ denotes the
average of $\hat{K}^{2}$\ in sector $\hat{X}$\ (see appendix 3.1.2) and\ $%
\left\Vert \hat{\Psi}\left( \hat{X}\right) \right\Vert ^{2}=\int \left\Vert 
\hat{\Psi}\left( \hat{X},\hat{K}\right) \right\Vert ^{2}d\hat{K}$.\ To close
the system of equations (\ref{hqn}) and (\ref{qnh}) for $\hat{\Psi}$\ and $%
\hat{\Psi}^{\dag }$,\ we add the constraint implemented by the Lagrange
multiplier $\hat{\lambda}$ :%
\begin{equation}
\int \left\Vert \hat{\Psi}\left( \hat{X},\hat{K}\right) \right\Vert ^{2}d%
\hat{X}d\hat{K}=\hat{N}  \label{nbg}
\end{equation}%
Note that the function $F\left( \hat{X},K_{\hat{X}}\right) $\ in (\ref{Fct}%
), which arises in the minimization equations (\ref{hqn}) and (\ref{qnh}),
describes the impact of individual variations on the collective state or
field $\hat{\Psi}$, and can be neglected in first approximation.

\subsubsection{Formula for the background fields}

The solutions of equations (\ref{hqn}) and (\ref{qnh})\ are computed in the
limit of small fluctuations in $\hat{X}$ (see appendix 3.1.3). We find an
infinite number of solutions parametrized by $\hat{\lambda}$:%
\begin{eqnarray}
\hat{\Psi}_{\hat{\lambda}}\left( \hat{X},\hat{K}\right) &=&\hat{\Psi}_{\hat{%
\lambda}}^{\left( 1\right) }\left( \hat{X},\hat{K}\right) \exp \left( \frac{1%
}{\sigma _{\hat{X}}^{2}}\int g\left( \hat{X}\right) d\hat{X}+\frac{\hat{K}%
^{2}}{\sigma _{\hat{K}}^{2}}f\left( \hat{X}\right) \right)  \label{psh} \\
&&\times D_{p\left( \hat{X},\hat{\lambda}\right) }\left( \frac{\left(
\left\vert f\left( \hat{X}\right) \right\vert \right) ^{\frac{1}{2}}}{\sigma
_{\hat{K}}}\left( \hat{K}+\frac{\sigma _{\hat{K}}^{2}F\left( \hat{X},K_{\hat{%
X}}\right) }{f^{2}\left( \hat{X}\right) }\right) \right)  \notag
\end{eqnarray}%
and:%
\begin{equation}
\hat{\Psi}_{\hat{\lambda}}^{\dag }\left( \hat{X},\hat{K}\right) =\hat{\Psi}_{%
\hat{\lambda}}^{\left( 1\right) \dag }\left( \hat{X},\hat{K}\right) \exp
\left( -\left( \frac{1}{\sigma _{\hat{X}}^{2}}\int g\left( \hat{X}\right) d%
\hat{X}+\frac{\hat{K}^{2}}{\sigma _{\hat{K}}^{2}}f\left( \hat{X}\right)
\right) \right) D_{p\left( \hat{X},\hat{\lambda}\right) }\left( \frac{\hat{K}%
}{\sigma _{\hat{K}}}\left( \left\vert f\left( \hat{X}\right) \right\vert
\right) ^{\frac{1}{2}}\right)  \label{hps}
\end{equation}%
with:%
\begin{equation}
p\left( \hat{X},\hat{\lambda}\right) =-\frac{\left( g\left( \hat{X}\right)
\right) ^{2}+\sigma _{\hat{X}}^{2}\left( f\left( \hat{X}\right) +\nabla _{%
\hat{X}}g\left( \hat{X},K_{\hat{X}}\right) -\frac{\sigma _{\hat{K}%
}^{2}F^{2}\left( \hat{X},K_{\hat{X}}\right) }{2f^{2}\left( \hat{X}\right) }+%
\hat{\lambda}\right) }{\sigma _{\hat{X}}^{2}\sqrt{f^{2}\left( \hat{X}\right) 
}}-\frac{1}{2}  \label{plb}
\end{equation}%
and where $D_{p}$ is the parabolic cylinder function with parameter $p$. The
fields $\hat{\Psi}_{\hat{\lambda}}^{\left( 1\right) }\left( \hat{X},\hat{K}%
\right) $ and $\hat{\Psi}_{\hat{\lambda}}^{\left( 1\right) \dag }\left( \hat{%
X},\hat{K}\right) $ are corrections in $\sigma _{X}^{2}$:%
\begin{equation}
\hat{\Psi}_{\hat{\lambda}}^{\left( 1\right) }\left( \hat{X}\right) =\sqrt{%
C\left( \hat{\lambda}\right) }\exp \left( -\int \frac{\left( \left( \hat{K}%
^{2}\mp \frac{1}{4}\left( \hat{K}+\frac{\sigma _{\hat{K}}^{2}F\left( \hat{X}%
,K_{\hat{X}}\right) }{f^{2}\left( \hat{X}\right) }\right) ^{2}\right)
f^{\prime }\left( \hat{X}\right) \right) ^{2}}{\sigma _{\hat{K}}^{2}\left( 2%
\hat{K}f\left( \hat{X}\right) -\left( \hat{K}+\frac{\sigma _{\hat{K}%
}^{2}F\left( \hat{X},K_{\hat{X}}\right) }{f^{2}\left( \hat{X}\right) }%
\right) \left\vert f\left( \hat{X}\right) \right\vert \right) }d\hat{K}%
\right)  \label{Psh}
\end{equation}%
and:

\begin{equation}
\hat{\Psi}^{\left( 1\right) \dag }\left( \hat{X}\right) =\sqrt{C\left( \hat{%
\lambda}\right) }\exp \left( \int \frac{\left( \left( \hat{K}^{2}\pm \frac{1%
}{4}\left( \hat{K}+\frac{\sigma _{\hat{K}}^{2}F\left( \hat{X},K_{\hat{X}%
}\right) }{f^{2}\left( \hat{X}\right) }\right) ^{2}\right) f^{\prime }\left( 
\hat{X}\right) \right) ^{2}}{\sigma _{\hat{K}}^{2}\left( 2\hat{K}f\left( 
\hat{X}\right) +\left( \hat{K}+\frac{\sigma _{\hat{K}}^{2}F\left( \hat{X},K_{%
\hat{X}}\right) }{f^{2}\left( \hat{X}\right) }\right) \left\vert f\left( 
\hat{X}\right) \right\vert \right) }d\hat{K}\right)  \label{Hps}
\end{equation}%
where the symbol $\pm $ stands for $sign\left( f\left( \hat{X}\right)
\right) $, $\mp $ for $-sign\left( f\left( \hat{X}\right) \right) $, and
where $C\left( \hat{\lambda}\right) $ is a function of $\hat{\lambda}$
obtained by imposing the normalization condition (\ref{nbg}) for $\hat{\Psi}%
_{\hat{\lambda}}\left( \hat{K},\hat{X}\right) $, i.e.:%
\begin{equation*}
\int \left\Vert \hat{\Psi}_{\hat{\lambda}}\left( \hat{K},\hat{X}\right)
\right\Vert ^{2}d\left( \hat{K},\hat{X}\right) =\hat{N}
\end{equation*}%
Appendix 3.1.4.2 shows that:%
\begin{equation}
C\left( \hat{\lambda}\right) \equiv C\left( \bar{p}\left( \hat{\lambda}%
\right) \right) \simeq \frac{\exp \left( -\frac{\sigma _{X}^{2}\sigma _{\hat{%
K}}^{2}\left( \frac{\left( \bar{p}\left( \hat{\lambda}\right) +\frac{1}{2}%
\right) f^{\prime }\left( X_{0}\right) }{f\left( \hat{X}_{0}\right) }\right)
^{2}}{96\left\vert f\left( \hat{X}_{0}\right) \right\vert }\right) \hat{N}%
\Gamma \left( -\bar{p}\left( \hat{\lambda}\right) \right) }{\left( \frac{%
\left\langle \left\vert f\left( \hat{X}\right) \right\vert \right\rangle }{%
\sigma _{\hat{K}}^{2}}\right) ^{-\frac{1}{2}}V_{r}\left( \func{Psi}\left( -%
\frac{\bar{p}\left( \hat{\lambda}\right) -1}{2}\right) -\func{Psi}\left( -%
\frac{\bar{p}\left( \hat{\lambda}\right) }{2}\right) \right) }
\end{equation}%
where:%
\begin{equation}
\bar{p}\left( \hat{\lambda}\right) =\left( -\frac{\frac{\left( g\left( \hat{X%
}_{0}\right) \right) ^{2}}{\sigma _{\hat{X}}^{2}}+\left( f\left( \hat{X}%
_{0}\right) +\frac{1}{2}\left\vert f\left( \hat{X}_{0}\right) \right\vert
+\nabla _{\hat{X}}g\left( \hat{X}_{0},K_{\hat{X}_{0}}\right) -\frac{\sigma _{%
\hat{K}}^{2}F^{2}\left( \hat{X}_{0},K_{\hat{X}_{0}}\right) }{2f^{2}\left( 
\hat{X}_{0}\right) }+\hat{\lambda}\right) }{\left\vert f\left( \hat{X}%
_{0}\right) \right\vert }\right)  \label{pBL}
\end{equation}%
and:%
\begin{equation}
\hat{X}_{0}=\arg \min_{\hat{X}}\left( \frac{\sigma _{X}^{2}\sigma _{\hat{K}%
}^{2}\left( \frac{\left( p\left( \hat{\lambda}\right) +\frac{1}{2}\right)
f^{\prime }\left( X\right) }{f\left( \hat{X}\right) }\right) ^{2}}{%
96\left\vert f\left( \hat{X}\right) \right\vert }\right)
\end{equation}%
The solutions (\ref{psh}) and (\ref{hps}) are thus a one-parameter $\hat{%
\lambda}$-family of solutions to (\ref{hqn}). In other words, there is an
infinite number of background fields, but in the following, only one will
eventually be relevant.

\subsubsection{Density of investors}

Finally, we find the density of investors at a given position $\hat{X}$\ and
capital $\hat{K}$\ in the background field or collective state $\hat{\Psi}_{%
\hat{\lambda}}\left( \hat{K},\hat{X}\right) $:%
\begin{eqnarray}
\left\Vert \hat{\Psi}_{\hat{\lambda}}\left( \hat{K},\hat{X}\right)
\right\Vert ^{2} &\simeq &C\left( \bar{p}\left( \hat{\lambda}\right) \right)
\exp \left( -\frac{\sigma _{X}^{2}\hat{K}^{4}\left( f^{\prime }\left(
X\right) \right) ^{2}}{96\sigma _{\hat{K}}^{2}\left\vert f\left( \hat{X}%
\right) \right\vert }\right)  \label{dns} \\
&&\times D_{p\left( \hat{X},\hat{\lambda}\right) }^{2}\left( \left( \frac{%
\left\vert f\left( \hat{X}\right) \right\vert }{\sigma _{\hat{K}}^{2}}%
\right) ^{\frac{1}{2}}\left( \hat{K}+\frac{\sigma _{\hat{K}}^{2}F\left( \hat{%
X},K_{\hat{X}}\right) }{f^{2}\left( \hat{X}\right) }\right) \right)  \notag
\end{eqnarray}%
Depending on the value of $\hat{\lambda}$, there is a multiplicity of
background fields, each contributing with its own weight $\left\Vert \hat{%
\Psi}_{\hat{\lambda}}\left( \hat{K},\hat{X}\right) \right\Vert ^{2}$,\
defined in (\ref{dns}), to the computation of the average capital invested.

\subsubsection{Estimation of $S_{3}+S_{4}$ for the background field}

We conclude this paragraph by computing an estimation of $S_{3}\left( \hat{%
\Psi}_{\hat{\lambda}}\left( \hat{K},\hat{X}\right) \right) +S_{4}\left( \hat{%
\Psi}_{\hat{\lambda}}\left( \hat{K},\hat{X}\right) \right) $\ for any
background field $\hat{\Psi}_{\hat{\lambda}}\left( \hat{K},\hat{X}\right) $
defined in (\ref{psh}).\ This estimation will be later used to discriminate
between the potential configurations $\hat{\Psi}_{\hat{\lambda}}\left( \hat{K%
},\hat{X}\right) $\ for the background field.

In order to do so, we multiply (\ref{hqn}) by $\hat{\Psi}_{\hat{\lambda}%
}^{\dagger }\left( \hat{K},\hat{X}\right) $\ on the left and integrate the
equation over $\hat{K}$\ and $\hat{X}$. It yields:%
\begin{equation*}
0=S_{3}\left( \hat{\Psi}_{\hat{\lambda}}\left( \hat{K},\hat{X}\right)
\right) +S_{4}\left( \hat{\Psi}_{\hat{\lambda}}\left( \hat{K},\hat{X}\right)
\right) -\hat{\lambda}\int \left\Vert \hat{\Psi}_{\hat{\lambda}}\left( \hat{K%
},\hat{X}\right) \right\Vert ^{2}d\hat{K}d\hat{X}-\int F\left( \hat{X},K_{%
\hat{X}}\right) \hat{K}\left\Vert \hat{\Psi}_{\hat{\lambda}}\left( \hat{K},%
\hat{X}\right) \right\Vert ^{2}d\hat{K}d\hat{X}
\end{equation*}

Using the constraint imposed on the number of investors:%
\begin{equation*}
\int \left\Vert \hat{\Psi}_{\hat{\lambda}}\left( \hat{K},\hat{X}\right)
\right\Vert ^{2}d\hat{K}=\hat{N}
\end{equation*}%
we find:%
\begin{equation}
S_{3}\left( \hat{\Psi}_{\hat{\lambda}}\left( \hat{K},\hat{X}\right) \right)
+S_{4}\left( \hat{\Psi}_{\hat{\lambda}}\left( \hat{K},\hat{X}\right) \right)
=\hat{\lambda}\hat{N}+\int F\left( \hat{X},K_{\hat{X}}\right) \hat{K}%
\left\Vert \hat{\Psi}_{\hat{\lambda}}\left( \hat{K},\hat{X}\right)
\right\Vert ^{2}d\hat{K}d\hat{X}  \label{smPR}
\end{equation}%
Then, estimating the second term of the rhs of (\ref{smPR}), we show in
appendix 3.1.4 that ultimately:%
\begin{equation}
S_{3}\left( \hat{\Psi}_{\hat{\lambda}}\left( \hat{K},\hat{X}\right) \right)
+S_{4}\left( \hat{\Psi}_{\hat{\lambda}}\left( \hat{K},\hat{X}\right) \right)
\simeq -\left( \left\vert \hat{\lambda}\right\vert -M\right) \hat{N}
\label{stM}
\end{equation}%
where $M$,\ the lowest bound for the eigenvalues $\left\vert \hat{\lambda}%
\right\vert $ will be defined in the next paragraph.

\subsection{Average capital invested}

\subsubsection{General equation}

Now that we have computed both the background fields for the real and the
financial sector of the economy, we can determine $K_{\hat{X}}$, the average
capital per firm in sector $\hat{X}$, in this environment.

To do so, we first rewrite the defining equation of $K_{\hat{X}}$, given in (%
\ref{csc}), as the following identity:%
\begin{equation}
K_{\hat{X}}\left\Vert \Psi \left( \hat{X}\right) \right\Vert ^{2}=\int \hat{K%
}\left\Vert \hat{\Psi}\left( \hat{K},\hat{X}\right) \right\Vert ^{2}d\hat{K}
\label{ctc}
\end{equation}%
This equation is actually an equation for $K_{\hat{X}}$. Actually, we have
expressed in equation (\ref{psl}) the squared background field $\left\Vert
\Psi \left( \hat{X}\right) \right\Vert ^{2}$ as a function of $K_{\hat{X}}$,
and the field $\hat{\Psi}\left( \hat{K},\hat{X}\right) $, and $\hat{\Psi}%
\left( \hat{K},\hat{X}\right) $ is itself a function of $K_{\hat{X}}$
through equation (\ref{dns}).

\subsubsection{Averaging over background fields}

Note that the equation (\ref{ctc}) above is defined for a given background
field $\hat{\Psi}\left( \hat{K},\hat{X}\right) $, whereas equation (\ref{dns}%
) in the previous paragraph yields an infinity of background fields $\hat{%
\Psi}_{\hat{\lambda}}\left( \hat{K},\hat{X}\right) $ indexed by the variable 
$\hat{\lambda}$. To account for this infinite number of solutions, we must
average the rhs of (\ref{ctc}) over $\hat{\lambda}$. The weight associated
to a particular value of $\hat{\lambda}$, which computes the probability of
the configuration $\hat{\Psi}_{\hat{\lambda}}$,is given by:%
\begin{equation}
\exp \left( -\left( S_{3}\left( \hat{\Psi}_{\hat{\lambda}}\right)
+S_{4}\left( \hat{\Psi}_{\hat{\lambda}}\right) \right) \right)
\label{weight}
\end{equation}%
Given the minimization equation (\ref{hqn}) that defines $\hat{\Psi}_{\hat{%
\lambda}}$, we have seen that (see (\ref{stM})):%
\begin{equation*}
S_{3}\left( \hat{\Psi}_{\hat{\lambda}}\right) +S_{4}\left( \hat{\Psi}_{\hat{%
\lambda}}\right) \simeq \left( \left\vert \hat{\lambda}\right\vert -M\right) 
\hat{N}
\end{equation*}%
where the last equality implements the constraint that the number of
investors should be $\hat{N}$. As a consequence, the weight (\ref{weight})
reduces to:%
\begin{equation*}
\exp \left( \hat{\lambda}\int \left\Vert \hat{\Psi}_{\hat{\lambda}}\left( 
\hat{K},\hat{X}\right) \right\Vert ^{2}\right) =\exp \left( -\left(
\left\vert \hat{\lambda}\right\vert -M\right) \hat{N}\right)
\end{equation*}%
and equation (\ref{ctc}) is replaced, up to a normalization factor, by:%
\begin{equation}
K_{X}\left\Vert \Psi \left( X\right) \right\Vert ^{2}=\int \hat{K}\int \exp
\left( -\left( \left\vert \hat{\lambda}\right\vert -M\right) \hat{N}\right)
\left\Vert \hat{\Psi}_{\hat{\lambda}}\left( \hat{K},X\right) \right\Vert
^{2}d\left\vert \hat{\lambda}\right\vert d\hat{K}  \label{ntg}
\end{equation}

\subsubsection{Computing the integral over $\left\vert \hat{\protect\lambda}%
\right\vert $}

Appendix 3.1.4.2 shows that the integral over $\left\vert \hat{\lambda}%
\right\vert $ has a lower bound $M$. This lower bound\footnote{%
This lower bound is reminiscent of the fact that the Lagrange multiplier $%
\lambda $ is the eigenvalue of the second order operator arising in equation
(\ref{nqh}), and that this operator is bounded from below.} is defined by:%
\begin{equation}
M=\max_{\hat{X}}A\left( \hat{X}\right)  \label{mqn}
\end{equation}%
with\footnote{%
In this formula, the term $\frac{\sigma _{\hat{K}}^{2}F^{2}\left( \hat{X},K_{%
\hat{X}}\right) }{2f^{2}\left( \hat{X}\right) }$ can be neglected in general
(see appendix 2). In the sequel, this term will often be omitted.
\par
{}}:%
\begin{equation}
A\left( \hat{X}\right) =\frac{\left( g\left( \hat{X}\right) \right) ^{2}}{%
\sigma _{\hat{X}}^{2}}+f\left( \hat{X}\right) +\frac{1}{2}\sqrt{f^{2}\left( 
\hat{X}\right) }+\nabla _{\hat{X}}g\left( \hat{X},K_{\hat{X}}\right) -\frac{%
\sigma _{\hat{K}}^{2}F^{2}\left( \hat{X},K_{\hat{X}}\right) }{2f^{2}\left( 
\hat{X}\right) }  \label{dft}
\end{equation}%
Incidentally, an interpretation of $A\left( \hat{X}\right) $ and its bound $%
M $\ will be given in section 8.2.1.

Once the parameter $M$\ found, the integral in (\ref{ntg}) becomes:%
\begin{equation}
K_{X}\left\Vert \Psi \left( X\right) \right\Vert ^{2}=\int \hat{K}%
\int_{\left\vert \hat{\lambda}\right\vert >M}\exp \left( -\left( \left\vert 
\hat{\lambda}\right\vert -M\right) \hat{N}\right) \left\Vert \hat{\Psi}_{%
\hat{\lambda}}\left( \hat{K},X\right) \right\Vert ^{2}d\left\vert \hat{%
\lambda}\right\vert d\hat{K}  \label{nTG}
\end{equation}%
\textbf{\ }However, since the number of investors $\hat{N}$\ is very large, $%
\hat{N}>>1$, the integral in (\ref{nTG}) is peaked at the minimum value for $%
\left\vert \hat{\lambda}\right\vert $, that is $M$, and $K_{X}$ satisfies:%
\begin{equation}
K_{X}\left\Vert \Psi \left( X\right) \right\Vert ^{2}=\int \hat{K}\left\Vert 
\hat{\Psi}_{p\left( -M\right) }\left( \hat{K},\hat{X}\right) \right\Vert
^{2}d\hat{K}  \label{qtl}
\end{equation}%
If we define $\bar{p}=\bar{p}\left( -M\right) $, $p=p\left( -M\right) $ and $%
\hat{\Psi}\left( \hat{K},\hat{X}\right) =\hat{\Psi}_{p\left( -M\right)
}\left( \hat{K},\hat{X}\right) $, we can rewrite equation (\ref{qtl}) as:%
\begin{equation}
K_{X}\left\Vert \Psi \left( X\right) \right\Vert ^{2}=\int \hat{K}\left\Vert 
\hat{\Psi}_{p}\left( \hat{K},\hat{X}\right) \right\Vert ^{2}d\hat{K}
\label{prK}
\end{equation}

\subsubsection{Final form of the capital equation}

Using the form of the financial background field (\ref{dns}) to compute the
integral in (\ref{prK}), the equation defining average capital at point $%
\hat{X}$ (see appendix 3.1.4.2) ultimately becomes:%
\begin{equation}
K_{\hat{X}}\left\Vert \Psi \left( \hat{X}\right) \right\Vert ^{2}\left\vert
f\left( \hat{X}\right) \right\vert =C\left( \bar{p}\right) \sigma _{\hat{K}%
}^{2}\hat{\Gamma}\left( p+\frac{1}{2}\right)  \label{qtk}
\end{equation}%
with:%
\begin{eqnarray}
\hat{\Gamma}\left( p+\frac{1}{2}\right) &=&\exp \left( -\frac{\sigma
_{X}^{2}\sigma _{\hat{K}}^{2}\left( p+\frac{1}{2}\right) ^{2}\left(
f^{\prime }\left( X\right) \right) ^{2}}{96\left\vert f\left( \hat{X}\right)
\right\vert ^{3}}\right)  \label{Gmh} \\
&&\times \left( \frac{\Gamma \left( -\frac{p+1}{2}\right) \Gamma \left( 
\frac{1-p}{2}\right) -\Gamma \left( -\frac{p}{2}\right) \Gamma \left( \frac{%
-p}{2}\right) }{2^{p+2}\Gamma \left( -p-1\right) \Gamma \left( -p\right) }+p%
\frac{\Gamma \left( -\frac{p}{2}\right) \Gamma \left( \frac{2-p}{2}\right)
-\Gamma \left( -\frac{p-1}{2}\right) \Gamma \left( -\frac{p-1}{2}\right) }{%
2^{p+1}\Gamma \left( -p\right) \Gamma \left( -p+1\right) }\right)  \notag
\end{eqnarray}%
and:%
\begin{equation}
p=\frac{M-\left( \left( g\left( \hat{X}\right) \right) ^{2}+\sigma _{\hat{X}%
}^{2}\left( f\left( \hat{X}\right) +\frac{1}{2}\sqrt{f^{2}\left( \hat{X}%
\right) }+\nabla _{\hat{X}}g\left( \hat{X},K_{\hat{X}}\right) -\frac{\sigma
_{\hat{K}}^{2}F^{2}\left( \hat{X},K_{\hat{X}}\right) }{2f^{2}\left( \hat{X}%
\right) }\right) \right) }{\sigma _{\hat{X}}^{2}\sqrt{f^{2}\left( \hat{X}%
\right) }}  \label{fRM}
\end{equation}%
It is this equation (\ref{qtk})\ that will be central to our following
computations. We will give in section 8.2.1 an interpretation of $p$\ in
terms of relative attractivity of a firm vis a vis its neighbours.

To conclude, note that for later purposes that the function $\hat{\Gamma}%
\left( p+\frac{1}{2}\right) $ is asymptotically given by:%
\begin{equation}
\hat{\Gamma}\left( p+\frac{1}{2}\right) \sim _{\infty }\alpha \exp \left( -%
\frac{\sigma _{X}^{2}\sigma _{\hat{K}}^{2}\left( p+\frac{1}{2}\right)
^{2}\left( f^{\prime }\left( X\right) \right) ^{2}}{96\left\vert f\left( 
\hat{X}\right) \right\vert ^{3}}\right) \Gamma \left( p+\frac{3}{2}\right)
\end{equation}%
with:%
\begin{equation*}
\alpha \simeq \frac{3}{2}
\end{equation*}%
Equation (\ref{qtk}) involves functions that have a general form, such as $f$%
, and complicated functions of the unknown variable $K_{\hat{X}}$\ (see
equation (\ref{plb})). As such, it cannot usually be solved analytically. We
will now provide several method to approach the solutions (\ref{qtk}).

\section{Finding the average capital in a given environment}

Except for particular cases\footnote{%
These particular cases will be studied in the following sections.}, equation
(\ref{qtk}) cannot be solved analytically. However, several complementary
approaches can be used to shed light on the behaviour of its solutions. The
first approach is the most general: we study the differential form of (\ref%
{qtk}) out of any assumptions on the parameter-functions. This allows to
compute, for each sector $\hat{X}$,\ the derivative of the average capital
per firm $K_{\hat{X}}$\ with respect to any parameter.\ In particular, the
derivative of the sector expected returns relative to its neighbours can be
computed, which shows the influence of the local environment on a sector.
However, this first approach does not yield the precise level of capital for
each sector. A second approach considers the expansion of (\ref{qtk}) around
particular values of capital, and a third approximates the resolution of
equation (\ref{qtk}) for standard forms of the parameter-functions.\
Combined, they confirm and precise our initial results.

We will check in the two last approaches that the equation (\ref{qtk}) for
average capital has several solutions, for a given set of parameters
functions, and that as such, background fields $\hat{\Psi}_{p}\left( \hat{K},%
\hat{X}\right) $\ are not unique. An infinite number of collective state may
arise, depending on some initial configuration. We will discuss this point
in section 10.

\subsection{First approach: differential form}

One way to better understand equation (\ref{qtk}) is to study its
differential form.

Assume at point $\hat{X}$ of the system, a variation $\delta Y\left( \hat{X}%
\right) $ for any parameter, in which the parameter $Y\left( \hat{X}\right) $
can be either $R\left( X\right) ,$ its gradient, or any parameter arising in
the definition of $f$ and $g$. This variation $\delta Y\left( \hat{X}\right) 
$ itself induces a variation $\delta K_{\hat{X}}$ in the average capital,
variation whose expression is obtained by differentiating equation (\ref{qtk}%
):

\begin{eqnarray}
\delta K_{\hat{X}} &=&\left( -\left( \frac{\frac{\partial f\left( \hat{X},K_{%
\hat{X}}\right) }{\partial K_{\hat{X}}}}{f\left( \hat{X},K_{\hat{X}}\right) }%
+\frac{\frac{\partial \left\Vert \Psi \left( \hat{X},K_{\hat{X}}\right)
\right\Vert ^{2}}{\partial K_{\hat{X}}}}{\left\Vert \Psi \left( \hat{X},K_{%
\hat{X}}\right) \right\Vert ^{2}}+l\left( \hat{X},K_{\hat{X}}\right) \right)
+k\left( p\right) \frac{\partial p}{\partial K_{\hat{X}}}\right) K_{\hat{X}%
}\delta K_{\hat{X}}  \label{rvd} \\
&&+\frac{\partial }{\partial Y\left( \hat{X}\right) }\left( \frac{\sigma _{%
\hat{K}}^{2}C\left( \bar{p}\right) 2\hat{\Gamma}\left( p+\frac{1}{2}\right) 
}{\left\vert f\left( \hat{X},K_{\hat{X}}\right) \right\vert \left\Vert \Psi
\left( \hat{X},K_{\hat{X}}\right) \right\Vert ^{2}}\right) \delta Y\left( 
\hat{X}\right)  \notag
\end{eqnarray}

\subsubsection{Developed form of the differential form (\protect\ref{rvd})}

For the later use of equation (\ref{rvd}) we give here its developed form
(see appendix 3.2.1):%
\begin{eqnarray}
\frac{\delta K_{\hat{X}}}{K_{\hat{X}}} &=&\left( k\left( p\right) \frac{%
\partial _{Y}\left( M-\left( \frac{\left( g\left( \hat{X},K_{\hat{X}}\right)
\right) ^{2}}{\sigma _{\hat{X}}^{2}}+\nabla _{\hat{X}}g\left( \hat{X},K_{%
\hat{X}}\right) -\frac{\sigma _{\hat{K}}^{2}F^{2}\left( \hat{X},K_{\hat{X}%
}\right) }{2f^{2}\left( \hat{X}\right) }\right) \right) }{f\left( \hat{X},K_{%
\hat{X}}\right) }\right.  \label{dtt} \\
&&\left. -\left( \frac{\frac{\partial f\left( \hat{X},K_{\hat{X}}\right) }{%
\partial Y}\left( 1+\left( p+\mathcal{H}\left( f\left( \hat{X},K_{\hat{X}%
}\right) \right) +\frac{1}{2}\right) k\left( p\right) \right) }{f\left( \hat{%
X},K_{\hat{X}}\right) }+\frac{\frac{\partial \left\Vert \Psi \left( \hat{X}%
,K_{\hat{X}}\right) \right\Vert ^{2}}{\partial Y}}{\left\Vert \Psi \left( 
\hat{X},K_{\hat{X}}\right) \right\Vert ^{2}}+m_{Y}\left( \hat{X},K_{\hat{X}%
}\right) \right) \right) \frac{\delta Y}{D}  \notag
\end{eqnarray}%
with $\mathcal{H}$ the Heaviside function, and where we have defined the
intermediate functions:%
\begin{eqnarray}
D &=&1+\left( \left( \frac{\frac{\partial f\left( \hat{X},K_{\hat{X}}\right) 
}{\partial K_{\hat{X}}}\left( 1+\left( p+\mathcal{H}\left( f\left( \hat{X}%
,K_{\hat{X}}\right) \right) +\frac{1}{2}\right) k\left( p\right) \right) }{%
f\left( \hat{X},K_{\hat{X}}\right) }+\frac{\frac{\partial \left\Vert \Psi
\left( \hat{X},K_{\hat{X}}\right) \right\Vert ^{2}}{\partial K_{\hat{X}}}}{%
\left\Vert \Psi \left( \hat{X},K_{\hat{X}}\right) \right\Vert ^{2}}+l\left( 
\hat{X},K_{\hat{X}}\right) \right) \right.  \label{D} \\
&&\left. -k\left( p\right) \frac{\partial _{K_{\hat{X}}}\left( M-\left( 
\frac{\left( g\left( \hat{X},K_{\hat{X}}\right) \right) ^{2}}{\sigma _{\hat{X%
}}^{2}}+\nabla _{\hat{X}}g\left( \hat{X},K_{\hat{X}}\right) -\frac{\sigma _{%
\hat{K}}^{2}F^{2}\left( \hat{X},K_{\hat{X}}\right) }{2f^{2}\left( \hat{X}%
\right) }\right) \right) }{f\left( \hat{X},K_{\hat{X}}\right) }\right) K_{%
\hat{X}}  \notag
\end{eqnarray}%
\begin{equation}
k\left( p\right) =\frac{\frac{d}{dp}\hat{\Gamma}\left( p+\frac{1}{2}\right) 
}{\hat{\Gamma}\left( p+\frac{1}{2}\right) }\sim _{\infty }\sqrt{\frac{p-%
\frac{1}{2}}{2}}-\frac{\sigma _{X}^{2}\sigma _{\hat{K}}^{2}\left( p+\frac{1}{%
2}\right) \left( f^{\prime }\left( X\right) \right) ^{2}}{48\left\vert
f\left( \hat{X}\right) \right\vert ^{3}}  \label{kdp}
\end{equation}%
\begin{eqnarray*}
l\left( \hat{X},K_{\hat{X}}\right) &=&\frac{\sigma _{X}^{2}\sigma _{\hat{K}%
}^{2}\left( \nabla _{K_{\hat{X}}}\left( f^{\prime }\left( \hat{X}\right)
\right) ^{2}\left\vert f\left( \hat{X}\right) \right\vert -3\left( \nabla
_{K_{\hat{X}}}\left\vert f\left( \hat{X}\right) \right\vert \right) \left(
f^{\prime }\left( \hat{X}\right) \right) ^{2}\right) \left( p+\frac{1}{2}%
\right) ^{2}}{120\left\vert f\left( \hat{X}\right) \right\vert ^{4}} \\
&&+\frac{\partial p}{\partial K_{\hat{X}}}\frac{\sigma _{X}^{2}\sigma _{\hat{%
K}}^{2}\left( p+\frac{1}{2}\right) \left( f^{\prime }\left( X\right) \right)
^{2}}{48\left\vert f\left( \hat{X}\right) \right\vert ^{3}}
\end{eqnarray*}%
and:

\begin{eqnarray*}
m_{Y}\left( \hat{X},K_{\hat{X}}\right) &=&\frac{\sigma _{X}^{2}\sigma _{\hat{%
K}}^{2}\left( \nabla _{Y}\left( f^{\prime }\left( \hat{X}\right) \right)
^{2}\left\vert f\left( \hat{X}\right) \right\vert -3\left( \nabla
_{Y}\left\vert f\left( \hat{X}\right) \right\vert \right) \left( f^{\prime
}\left( \hat{X}\right) \right) ^{2}\right) \left( p+\frac{1}{2}\right) ^{2}}{%
120\left\vert f\left( \hat{X}\right) \right\vert ^{4}} \\
&&+\nabla _{Y}p\frac{\sigma _{X}^{2}\sigma _{\hat{K}}^{2}\left( p+\frac{1}{2}%
\right) \left( f^{\prime }\left( X\right) \right) ^{2}}{48\left\vert f\left( 
\hat{X}\right) \right\vert ^{3}}
\end{eqnarray*}%
\textbf{\ }The interpretation of these intermediate functions will be given
later in section 10.2.1.\ 

Recall that equation (\ref{dtt}) is the differentiated version of the
equation defining the average capital $K_{\hat{X}}$, equation (\ref{qtk}).\
It will be used to compute the dependency of the average per firm in sector $%
\hat{X}$, $K_{\hat{X}}$, as a function of any parameter $Y(\hat{X})$. But
more fundamentally, it allows to understand the solutions of equation (\ref%
{qtk}) and their stability with respect to the parameters' variations.

\subsubsection{Local stability}

The differential form given by equation (\ref{rvd}), computes the effect of
a variation $\delta Y\left( \hat{X}\right) $ in the parameters on the
average capital $K_{\hat{X}}$. Actually, equation (\ref{rvd}) can be
understood as\ the fixed-point equation of a dynamical system through the
following mechanism\textbf{. }

Each variation $\delta Y\left( \hat{X}\right) $\ in the parameters impacts
the average capital, which must then be computed with the new parameters.
The first change induced is written $\delta K_{\hat{X}}^{\left( 1\right) }$:%
\begin{equation}
\delta K_{\hat{X}}^{\left( 1\right) }=\frac{\partial }{\partial Y\left( \hat{%
X}\right) }\left( \frac{\sigma _{\hat{K}}^{2}C\left( \bar{p}\right) 2\hat{%
\Gamma}\left( p+\frac{1}{2}\right) }{\left\vert f\left( \hat{X},K_{\hat{X}%
}\right) \right\vert \left\Vert \Psi \left( \hat{X},K_{\hat{X}}\right)
\right\Vert ^{2}}\right) \delta Y\left( \hat{X}\right)  \label{dk1}
\end{equation}
In a second step, the variation $\delta K_{\hat{X}}$ impacts the various
functions implied in (\ref{qtk}), and indirectly modifies $K_{\hat{X}}$
through the first term in the rhs of (\ref{rvd}):%
\begin{equation}
\left( -\left( \frac{\frac{\partial f\left( \hat{X},K_{\hat{X}}\right) }{%
\partial K_{\hat{X}}}}{f\left( \hat{X},K_{\hat{X}}\right) }+\frac{\frac{%
\partial \left\Vert \Psi \left( \hat{X},K_{\hat{X}}\right) \right\Vert ^{2}}{%
\partial K_{\hat{X}}}}{\left\Vert \Psi \left( \hat{X},K_{\hat{X}}\right)
\right\Vert ^{2}}\right) +k\left( p\right) \frac{\partial p}{\partial K_{%
\hat{X}}}\right) K_{\hat{X}}\delta K_{\hat{X}}^{\left( 1\right) }
\label{dk2}
\end{equation}%
These two effects combined, (\ref{dk1}) and (\ref{dk2}), yield the total
variation $\delta K_{\hat{X}}$.

Importantly, note that if we can interpret $\delta K_{\hat{X}}^{\left(
1\right) }$\ as a variation at time $t$, we can also infer from the indirect
effect (\ref{dk2}) that $\delta K_{\hat{X}}$\ is itself a variation at time $%
t+1$. Equation (\ref{rvd}) can thus be seen as the fixed point equation of a
dynamical system written:%
\begin{eqnarray}
\delta K_{\hat{X}}\left( t+1\right) &=&\left( -\left( \frac{\frac{\partial
f\left( \hat{X},K_{\hat{X}}\right) }{\partial K_{\hat{X}}}}{f\left( \hat{X}%
,K_{\hat{X}}\right) }+\frac{\frac{\partial \left\Vert \Psi \left( \hat{X},K_{%
\hat{X}}\right) \right\Vert ^{2}}{\partial K_{\hat{X}}}}{\left\Vert \Psi
\left( \hat{X},K_{\hat{X}}\right) \right\Vert ^{2}}+l\left( \hat{X},K_{\hat{X%
}}\right) \right) +k\left( p\right) \frac{\partial p}{\partial K_{\hat{X}}}%
\right) K_{\hat{X}}\delta K_{\hat{X}}\left( t\right)  \label{dts} \\
&&+\frac{\partial }{\partial Y\left( \hat{X},t\right) }\left( \frac{\sigma _{%
\hat{K}}^{2}C\left( \bar{p}\right) 2\hat{\Gamma}\left( p+\frac{1}{2}\right) 
}{\left\vert f\left( \hat{X},K_{\hat{X}}\right) \right\vert \left\Vert \Psi
\left( \hat{X},K_{\hat{X}}\right) \right\Vert ^{2}}\right) \delta Y\left( 
\hat{X},t\right)  \notag
\end{eqnarray}%
whose fixed point is the solution of (\ref{rvd}):%
\begin{equation}
\delta K_{\hat{X}}=\frac{\frac{\partial }{\partial Y\left( \hat{X}\right) }%
\left( \frac{\sigma _{\hat{K}}^{2}C\left( \bar{p}\right) 2\hat{\Gamma}\left(
p+\frac{1}{2}\right) }{\left\vert f\left( \hat{X},K_{\hat{X}}\right)
\right\vert \left\Vert \Psi \left( \hat{X},K_{\hat{X}}\right) \right\Vert
^{2}}\right) }{1+\left( \left( \frac{\frac{\partial f\left( \hat{X},K_{\hat{X%
}}\right) }{\partial K_{\hat{X}}}}{f\left( \hat{X},K_{\hat{X}}\right) }+%
\frac{\frac{\partial \left\Vert \Psi \left( \hat{X},K_{\hat{X}}\right)
\right\Vert ^{2}}{\partial K_{\hat{X}}}}{\left\Vert \Psi \left( \hat{X},K_{%
\hat{X}}\right) \right\Vert ^{2}}+l\left( \hat{X},K_{\hat{X}}\right) \right)
-k\left( p\right) \frac{\partial p}{\partial K_{\hat{X}}}\right) K_{\hat{X}}}%
\delta Y\left( \hat{X}\right)  \label{sol}
\end{equation}%
This solution (\ref{sol}) is stable when: 
\begin{subequations}
\begin{equation}
\left\vert k\left( p\right) \frac{\partial p}{\partial K_{\hat{X}}}-\left( 
\frac{\frac{\partial f\left( \hat{X},K_{\hat{X}}\right) }{\partial K_{\hat{X}%
}}}{f\left( \hat{X},K_{\hat{X}}\right) }+\frac{\frac{\partial \left\Vert
\Psi \left( \hat{X},K_{\hat{X}}\right) \right\Vert ^{2}}{\partial K_{\hat{X}}%
}}{\left\Vert \Psi \left( \hat{X},K_{\hat{X}}\right) \right\Vert ^{2}}%
+l\left( \hat{X},K_{\hat{X}}\right) \right) \right\vert <1  \label{sTB}
\end{equation}%
i.e. when $D$, defined in (\ref{D}), is positive, and unstable otherwise. So
that the stability of this average capital depends, in last analysis, on the
sign of $D$.

The notion of stability reveals the two types of solution in the model, the
solutions for the average capital per firm $K_{\hat{X}}$\ solution of (\ref%
{qtk}): the stable solutions $K_{\hat{X}}$\ can be considered as sector $%
\hat{X}$'s equilibrium averages.\ However unstable solutions must rather be
considered as thresholds\textbf{:} when $K_{\hat{X}}$\ is driven away from
this threshold, it may either converge toward a stable solution of (\ref{qtk}%
), or diverge towards $0$\ or infinity.

Introducing the dynamical system (\ref{dts}) may at first sight seem
artificial, but it is nonetheless coherent within the context of our field
model. Actually, this arbitrary variation $\delta K_{\hat{X}}^{\left(
1\right) }$\ induced by a change in parameter actually reveals a shift $%
\delta \hat{\Psi}\left( \hat{K},\hat{X}\right) $\ in the background state $%
\hat{\Psi}\left( \hat{K},\hat{X}\right) $. Since there is no reason for the
new configuration $\hat{\Psi}\left( \hat{K},\hat{X}\right) +\delta \hat{\Psi}%
\left( \hat{K},\hat{X}\right) $ to be a minimum of the action functional, we
must determine whether the system will settle on a slightly modified
background state with a different $K_{\hat{X}}$, or be driven towards a
different equilibrium. To do so, we must study the dynamics\ equation for $%
K_{\hat{X}}$\ (\ref{dts}).\ 

\subsubsection{Applications of the differential form}

Once the notion of stability understood, we can use equation (\ref{dtt}) to
compute the impact of the variation of any parameter $Y(\hat{X})$ on $\delta
K_{\hat{X}}$. Two applications are of particular interest to us.

\paragraph{Main application}

The main application of equation (\ref{dtt}) is to consider a parameter
denoted $Y(\hat{X})$, that encompasses the relative expected returns of
sector $X$ vis-\`{a}-vis its neighbouring sectors, and defined as: 
\end{subequations}
\begin{equation}
Y(\hat{X})=M-\left( \frac{\left( g\left( \hat{X},K_{\hat{X}}\right) \right)
^{2}}{\sigma _{\hat{X}}^{2}}+\nabla _{\hat{X}}g\left( \hat{X},K_{\hat{X}%
}\right) -\frac{\sigma _{\hat{K}}^{2}F^{2}\left( \hat{X},K_{\hat{X}}\right) 
}{2f^{2}\left( \hat{X}\right) }\right)  \label{fpR}
\end{equation}%
This parameter enters directly in (\ref{dtt}), the differentiated equation
for $K_{\hat{X}}$, through parameter (\ref{plb}). It is composed of three
terms.

The first term, $\frac{\left( g\left( \hat{X},K_{\hat{X}}\right) \right) ^{2}%
}{\sigma _{\hat{X}}^{2}}$,\ is directly proportional to the gradient of
expected long-term returns $\nabla R\left( K_{\hat{X}},\hat{X}\right) $%
\footnote{%
See the definition of the parameter function $g$, equation (\ref{fcg}).}.\
It is minimal for an extremum of the expected return $R\left( K_{\hat{X}},%
\hat{X}\right) $.

The second term, $\nabla _{\hat{X}}g\left( \hat{X},K_{\hat{X}}\right) $, is
proportional to the second derivative $\nabla ^{2}R\left( K_{\hat{X}},\hat{X}%
\right) $\ of $R\left( K_{\hat{X}},\hat{X}\right) ,$ and is minimal when
expected returns are maximum.

The third term, $-\frac{\sigma _{\hat{K}}^{2}F^{2}\left( \hat{X},K_{\hat{X}%
}\right) }{2f^{2}\left( \hat{X}\right) }$,\ is a corrective term. It can be
neglected in first approximation\footnote{%
See the discussion following equation (\ref{Fct}).}, but will be interpreted
in section.\textbf{\ }

As stated previously, the parameter $Y(\hat{X})$\ is thus a measure of the
expected long-term return of a sector relative to its neighbours: it is a
local maximum when $R\left( K_{\hat{X}},\hat{X}\right) $\ is itself a local
maximum.

Using (\ref{dtt}), we have:\textbf{\ }%
\begin{equation}
\frac{\delta K_{\hat{X}}}{K_{\hat{X}}}=-\frac{\frac{k\left( p\right) }{%
f\left( \hat{X},K_{\hat{X}}\right) }K_{\hat{X}}}{D}\delta Y\left( \hat{X}%
\right)  \label{drt}
\end{equation}%
Given equation (\ref{kdp}), $k\left( p\right) $ is positive at the first
order in $\sigma _{X}^{2}$. \ More precisely, using equation (\ref{kdp}):%
\begin{equation*}
k\left( p\right) \sim _{\infty }\sqrt{\frac{p-\frac{1}{2}}{2}}-\frac{\sigma
_{X}^{2}\sigma _{\hat{K}}^{2}\left( p+\frac{1}{2}\right) \left( f^{\prime
}\left( X\right) \right) ^{2}}{48\left\vert f\left( \hat{X}\right)
\right\vert ^{3}}
\end{equation*}%
along with equation (\ref{plb}), we can infer that $\sqrt{\frac{p-\frac{1}{2}%
}{2}}$\ is of order $\frac{1}{\sigma _{X}}$\ and $\frac{\sigma
_{X}^{2}\sigma _{\hat{K}}^{2}\left( p+\frac{1}{2}\right) \left( f^{\prime
}\left( X\right) \right) ^{2}}{48\left\vert f\left( \hat{X}\right)
\right\vert ^{3}}\sim 1$.

So that in a stable equilibrium, i.e. for $D>0$, equation (\ref{drt})
implies that the dependency of $K_{\hat{X}}$ in the parameter $A\left( \hat{X%
}\right) $ is negative:%
\begin{equation*}
\frac{\delta K_{\hat{X}}}{\delta A\left( \hat{X}\right) }<0
\end{equation*}%
We have seen above that $Y\left( \hat{X}\right) $\ is minimal for a maximum
expected long-term return $R\left( \hat{X},K_{\hat{X}}\right) $: when the
equilibrium is stable, capital accumulation is maximal for sectors that are
themselves a local maximum for $R\left( \hat{X},K_{\hat{X}}\right) $.

On the other hand, when the equilibrium is unstable, i.e. for $D<0$, the
capital $K_{\hat{X}}$\ is minimal for $R\left( \hat{X},K_{\hat{X}}\right) $\
maximal.

Actually, as seen above, in the instability range $D<0$\ ,the average
capital $K_{\hat{X}}$\ acts as a threshold. When, due to variations in the
system's parameters, the average capital per firm is shifted above the
threshold $K_{\hat{X}}$, \ capital will either move to the next stable
equilibrium, possibly zero, or tend to infinity. Our results show that when
the expected long-term return of a sector increases, the threshold $K_{\hat{X%
}}$ decreases, which favours capital accumulation.

\paragraph{Additional application}

A second use of equation (\ref{dtt}) is to consider $Y(\hat{X})$ as any
parameter-function involved in the definition of $f\left( \hat{X},K_{\hat{X}%
}\right) $ that may condition either real short-term returns or the
price-dividend ratio.

We can see that in this case, $Y(\hat{X})$ only impacts $f\left( \hat{X},K_{%
\hat{X}}\right) $, so that equation (\ref{dtt}) simplifies and yields: 
\begin{eqnarray}
\frac{\delta K_{\hat{X}}}{K_{\hat{X}}} &=&-\frac{m_{Y}\left( \hat{X},K_{\hat{%
X}}\right) }{D}\delta Y  \label{dkx} \\
&&-\frac{1}{D}\left( \frac{\frac{\partial f\left( \hat{X},K_{\hat{X}}\right) 
}{\partial Y}\left( 1+\left( p+H\left( f\left( \hat{X},K_{\hat{X}}\right)
\right) +\frac{1}{2}\right) k\left( p\right) \right) }{f\left( \hat{X},K_{%
\hat{X}}\right) }+\frac{\frac{\partial \left\Vert \Psi \left( \hat{X},K_{%
\hat{X}}\right) \right\Vert ^{2}}{\partial Y}}{\left\Vert \Psi \left( \hat{X}%
,K_{\hat{X}}\right) \right\Vert ^{2}}\right) \delta Y  \notag
\end{eqnarray}%
Incidentally, note that $p$ being proportional to $f^{-1}\left( \hat{X}%
\right) $, $m_{Y}\left( \hat{X},K_{\hat{X}}\right) $ rewrites: 
\begin{eqnarray}
-m_{Y}\left( \hat{X},K_{\hat{X}}\right) &=&\frac{\sigma _{X}^{2}\sigma _{%
\hat{K}}^{2}\left( 3\left( \nabla _{Y}\left\vert f\left( \hat{X}\right)
\right\vert \right) \left( f^{\prime }\left( \hat{X}\right) \right)
^{2}-\nabla _{Y}\left( f^{\prime }\left( \hat{X}\right) \right)
^{2}\left\vert f\left( \hat{X}\right) \right\vert \right) \left( p+\frac{1}{2%
}\right) ^{2}}{120\left\vert f\left( \hat{X}\right) \right\vert ^{4}}
\label{mx} \\
&&+\nabla _{Y}\left\vert f\left( \hat{X}\right) \right\vert \frac{\sigma
_{X}^{2}\sigma _{\hat{K}}^{2}p\left( p+\frac{1}{2}\right) \left( f^{\prime
}\left( X\right) \right) ^{2}}{48\left\vert f\left( \hat{X}\right)
\right\vert ^{4}}  \notag
\end{eqnarray}%
The first term in the rhs of (\ref{dkx}) is the impact of an increase in
investors' short-term returns.\ The second is the variation in capital
needed to maintain investors' overall returns.

The sign of $\frac{\delta K_{\hat{X}}}{K_{\hat{X}}}$ given by equation (\ref%
{dkx}) can be studied under two cases: the stable and the unstable
equilibrium.

Let us first consider the case of a stable equilibrium, i.e. $D>0$.

The first term in the rhs of (\ref{dkx}), the variation induced by an
increase in short-term returns, is in general positive for $f^{\prime
}\left( \hat{X}\right) $ proportional to $f\left( \hat{X}\right) $, that is
for instance when the function $f\left( \hat{X}\right) $, that describes
short-term returns and prices, depends on the variable $K_{\hat{X}}$ raised
to some arbitrary power.

Indeed in that case:%
\begin{equation*}
3\left( \nabla _{Y}\left\vert f\left( \hat{X}\right) \right\vert \right)
\left( f^{\prime }\left( \hat{X}\right) \right) ^{2}-\nabla _{Y}\left(
f^{\prime }\left( \hat{X}\right) \right) ^{2}\left\vert f\left( \hat{X}%
\right) \right\vert =\left( \nabla _{Y}\left\vert f\left( \hat{X}\right)
\right\vert \right) \left( f^{\prime }\left( \hat{X}\right) \right) ^{2}
\end{equation*}%
The second term in the rhs of (\ref{dkx}) is in general negative. When $%
\frac{\partial f\left( \hat{X},K_{\hat{X}}\right) }{\partial Y}>0$, i.e.
when returns are increasing in $Y$, a rise in $Y$\ increases returns and
decreases the capital needed to maintain these returns.\textbf{\ }Similarly,
when $\frac{\partial \left\Vert \Psi \left( \hat{X},K_{\hat{X}}\right)
\right\Vert ^{2}}{\partial Y}>0$, i.e. when the number of agents in sector $%
\hat{X}\ $is increasing in $Y$, a rise in $Y$\ increases the number of
agents that move towards point $\hat{X}$, and the average capital per firm
diminishes.

The net variation (\ref{dkx}) of $K_{\hat{X}}$ is the sum of these two
contributions. Considering an expansion of (\ref{dkx}) in powers of $\sigma
_{X}^{2}$, the first contribution $-m_{Y}\left( \hat{X},K_{\hat{X}}\right) $
is of magnitude $\left( \sigma _{X}^{2}\right) ^{-1}$, whereas the second is
proportional to $k\left( p\right) \sim $ $\left( \sigma _{X}\right) ^{-1}$.
The variation $\frac{\delta K_{\hat{X}}}{K_{\hat{X}}}$ is thus positive: $%
\frac{\delta K_{\hat{X}}}{K_{\hat{X}}}>0$. In most cases, a higher
short-term return, decomposed as a sum of dividend and price variation,
induces a higher average capital. This effect is magnified for larger levels
of capital: the third approach will confirm that, in most cases, the return $%
f\left( \hat{X}\right) $ is asymptotically a constant $c<<1$ when capital is
high: $K_{\hat{X}}>>1$.

Turning now to the case of an unstable equilibrium, i.e. $D<0$, the
variation $\frac{\delta K_{\hat{X}}}{K_{\hat{X}}}$ is negative: $\frac{%
\delta K_{\hat{X}}}{K_{\hat{X}}}<0$. In the instability range, and due to
this very instability, an increase in returns $f\left( \hat{X}\right) $
reduces the threshold of capital accumulation for low levels of capital.
When short-term returns $f\left( \hat{X}\right) $\ increase, a lower average
capital will trigger capital accumulation towards an equilibrium. Otherwise,
when average capital $K_{\hat{X}}$ is below this threshold, it will converge
toward $0$.

\subsection{Second approach: expansion around particular solutions}

The second approach to equation (\ref{qtk}) is to find the average capital
at some particular points, and then by first order expansion, the solutions
in the neighbourhood of these particular points. We choose as particular
points the values $\hat{X}$\ and $K_{\hat{X}}$\ that maximize $A\left( \hat{X%
}\right) $, with:\ 
\begin{equation}
A\left( \hat{X}\right) =\frac{\left( g\left( \hat{X},K_{\hat{X}_{M}}\right)
\right) ^{2}}{\sigma _{\hat{X}}^{2}}+f\left( \hat{X},K_{\hat{X}_{M}}\right) +%
\frac{1}{2}\sqrt{f^{2}\left( \hat{X},K_{\hat{X}_{M}}\right) }+\nabla _{\hat{X%
}}g\left( \hat{X},K_{\hat{X}_{M}}\right) -\frac{\sigma _{\hat{K}%
}^{2}F^{2}\left( \hat{X},K_{\hat{X}_{M}}\right) }{2f^{2}\left( \hat{X},K_{%
\hat{X}_{M}}\right) }  \label{dsc}
\end{equation}%
These points $\left( \hat{X}_{M},K_{\hat{X}_{M}}\right) $\ are such that%
\footnote{%
See equation (\ref{mqn}).}:%
\begin{equation}
A\left( \hat{X}_{M}\right) =M=\max_{\hat{X}}A\left( \hat{X}\right)
\label{mfl}
\end{equation}

\subsubsection{Interpretation of the particular points}

We can interpret these particular points $\hat{X}_{M}$\ by considering the
function $p$. We have already defined $p$ in (\ref{fRP}), as:%
\begin{equation*}
p=\frac{M-A\left( \hat{X}\right) }{f\left( \hat{X}\right) }\mathbf{\ }
\end{equation*}
Note that $p$\ is minimal and equal to $0$\ at points $\hat{X}_{M}$. The
function $p$\ measures sector $X$\ long-term attractivity relative to its
neighbouring sectors,\ normalized by its short run returns.

This can be showed by discarding the corrective term $\frac{\sigma _{\hat{K}%
}^{2}F^{2}\left( \hat{X},K_{\hat{X}_{M}}\right) }{2f^{2}\left( \hat{X},K_{%
\hat{X}_{M}}\right) }<<1$\ in (\ref{dsc}),\ and we have in first
approximation\footnote{%
We give an interpretation of $\frac{\sigma _{\hat{K}}^{2}F^{2}\left( \hat{X}%
,K_{\hat{X}}\right) }{2\sigma _{\hat{X}}^{2}\left( \sqrt{f^{2}\left( \hat{X}%
\right) }\right) ^{3}}$ below.}, up to a constant:%
\begin{equation}
p=\frac{M-A\left( \hat{X}\right) }{f\left( \hat{X}\right) }\simeq \frac{%
M-\left( \frac{\left( g\left( \hat{X},K_{\hat{X}_{M}}\right) \right) ^{2}}{%
\sigma _{\hat{X}}^{2}}+\nabla _{\hat{X}}g\left( \hat{X},K_{\hat{X}%
_{M}}\right) \right) }{f\left( \hat{X}\right) }+\frac{3}{2}  \label{rlt}
\end{equation}%
Recall that $g\left( \hat{X}\right) $\ is the investors' capital mobility at
sector $\hat{X}$.\ When capital mobility is high, i.e. $p$\ is low, sector $%
\hat{X}$\ is shunned by investors, who favour more attractive sectors. On
the contrary, when capital mobility is low, i.e. $p$\ is high, capital
accumulates at sector $\hat{X}$.

As a consequence, the particular points $\hat{X}_{M}$\ such that $p=0$\ are
relatively capital-deterrent sectors.

\subsubsection{General remarks}

Note that the parameter $p=\frac{M-A\left( \hat{X}\right) }{f\left( \hat{X}%
\right) }$\ is a relative version of parameter\ defined in (\ref{fpR}), $%
f\left( \hat{X}\right) $\ excluded: 
\begin{equation*}
Y(\hat{X})=M-\left( \frac{\left( g\left( \hat{X},K_{\hat{X}}\right) \right)
^{2}}{\sigma _{\hat{X}}^{2}}+\nabla _{\hat{X}}g\left( \hat{X},K_{\hat{X}%
}\right) \right)
\end{equation*}%
However both parameters have the same interpretation in terms of expected
long-term returns: both parameters are local maxima when long-term returns $%
R\left( K_{\hat{X}},\hat{X}\right) $\ are themselves local maxima\footnote{%
See discussion following equation (\ref{fpR}).}.

Note also that we can interpret the correction $\frac{\sigma _{\hat{K}%
}^{2}F^{2}\left( \hat{X},K_{\hat{X}}\right) }{2\sigma _{\hat{X}}^{2}\left( 
\sqrt{f^{2}\left( \hat{X}\right) }\right) ^{3}}$\ in $\frac{M-A\left( \hat{X}%
\right) }{f\left( \hat{X}\right) }$.

A close inspection of equation (\ref{Fct}) shows that this term contains
-squared- contributions of short-term returns, $f\left( \hat{X},K_{\hat{X}%
}\right) $,\ and the sector's relative attractivity, $\hat{X}$:\ $\frac{%
\left( g\left( \hat{X},K_{\hat{X}}\right) \right) ^{2}}{2\sigma _{\hat{X}%
}^{2}}+\frac{1}{2}\nabla _{\hat{X}}g\left( \hat{X},K_{\hat{X}}\right) $.
These contributions are both proportional to the gradient with respect to $%
K_{\hat{X}}$\ .

When this gradient is different from zero, i.e. when an increase in capital
may improve either the sector's relative attractivity or short-term returns,
the correction $\frac{\sigma _{\hat{K}}^{2}F^{2}\left( \hat{X},K_{\hat{X}%
}\right) }{2\sigma _{\hat{X}}^{2}\left( \sqrt{f^{2}\left( \hat{X}\right) }%
\right) ^{3}}$\ increases $\frac{A\left( \hat{X}\right) }{f\left( \hat{X}%
\right) },$\ and in turn $K_{\hat{X}}$,\ in most cases. This reflects the
tendency of the whole system to reach possibly stable configurations and
thus reduce capital discrepancies between close neighbours. Actually, the
derivation of the minimization equation in appendix 3.1.2 shows that the
term $F\left( \hat{X},K_{\hat{X}}\right) $\ arises as a backreaction of the
whole system with respect to modifications at one point of the thread.

\subsubsection{Simplification of equation (\protect\ref{qtk}) at particular
points}

Mathematically, the maximization (\ref{mqn}) is an equation on $\hat{X}$
with a set of solutions $\hat{X}_{M}$ and an associated value of $K_{\hat{X}%
_{M}}$. The maximum of $A\left( \hat{X}_{M}\right) $ is $M$. For such
points, the average capital equation (\ref{qtk}) simplifies. Actually, at
these points, the parameter $p$ is null: $p=0$.\ Consequently, given that: 
\begin{equation}
\hat{\Gamma}\left( \frac{1}{2}\right) =\exp \left( -\frac{\sigma
_{X}^{2}\sigma _{\hat{K}}^{2}\left( f^{\prime }\left( X,K_{\hat{X},M}\right)
\right) ^{2}}{384\left\vert f\left( \hat{X},K_{\hat{X},M}\right) \right\vert
^{3}}\right)  \label{Ghf}
\end{equation}%
equation (\ref{qtk}) at points $\left( \hat{X}_{M},K_{\hat{X}_{M}}\right) $
writes:%
\begin{equation}
K_{\hat{X},M}\left\vert f\left( \hat{X},K_{\hat{X},M}\right) \right\vert
\left\Vert \Psi \left( \hat{X},K_{\hat{X},M}\right) \right\Vert ^{2}\simeq
\sigma _{\hat{K}}^{2}C\left( \bar{p}\right) \exp \left( -\frac{\sigma
_{X}^{2}\sigma _{\hat{K}}^{2}\left( f^{\prime }\left( X,K_{\hat{X},M}\right)
\right) ^{2}}{384\left\vert f\left( \hat{X},K_{\hat{X},M}\right) \right\vert
^{3}}\right)  \label{kpt}
\end{equation}%
This equation shows that $K_{\hat{X},M}$ depends, through $\left\Vert \Psi
\left( \hat{X},K_{\hat{X},M}\right) \right\Vert ^{2}$\footnote{%
See equation (\ref{psl}).}, on the form of $R\left( \hat{X}\right) $ and the
return $f\left( \hat{X},K_{\hat{X},M}\right) $. Given the assumptions on $%
f\left( \hat{X},K_{\hat{X}}\right) $, equation (\ref{kpt}) has several
solutions. Actually, $\left\Vert \Psi \left( \hat{X},K_{\hat{X},M}\right)
\right\Vert ^{2}$ is decreasing in $K_{\hat{X}}$ (see (\ref{psrd})). If we
assume this is also true for $f\left( \hat{X},K_{\hat{X}}\right) $, equation
(\ref{qtk}) has two solutions.\ We give their forms and perform their
expansion below.

\subsubsection{Particular solutions for capital}

We have seen that equation (\ref{qtk}) for $p=0$, that defines average
capital per firm per sector, has in general two solutions. We can find an
approximate form for these solutions for some particular values of the
parameters. By now we will merely consider a power law for $f\left( \hat{X}%
\right) $:%
\begin{equation}
f\left( \hat{X}\right) \simeq B\left( X\right) K_{\hat{X}}^{\alpha }
\label{fbR}
\end{equation}%
The parameter $B\left( X\right) $ is the productivity in sector $X$, and
equation (\ref{fbR}) shows that the return $f\left( \hat{X}\right) $ is
increasing in $B\left( X\right) $.

The stable case corresponds to setting $\left\Vert \Psi \left( \hat{X}%
\right) \right\Vert ^{2}\simeq D$, i.e. given (\ref{psl}), $K_{\hat{X}%
}^{\alpha }<<D$. In this case, equation (\ref{qtk}) rewrites:

\begin{equation*}
DB\left( X\right) K_{\hat{X}}^{\alpha +1}=C\left( \bar{p}\right) \sigma _{%
\hat{K}}^{2}\exp \left( -\frac{\sigma _{X}^{2}\sigma _{\hat{K}}^{2}\left(
B^{\prime }\left( X\right) \right) ^{2}}{384\left( B\left( X\right) \right)
^{3}}K_{\hat{X}}^{-\alpha }\right)
\end{equation*}%
which has for solution:%
\begin{equation}
K_{\hat{X}}^{\alpha }=\left( \frac{DB\left( X\right) }{C\left( \bar{p}%
\right) \sigma _{\hat{K}}^{2}}\right) ^{-\frac{\alpha }{\alpha +1}}\exp
\left( W_{0}\left( -\frac{\sigma _{X}^{2}\sigma _{\hat{K}}^{2}\left(
B^{\prime }\left( X\right) \right) ^{2}\alpha }{384\left( B\left( X\right)
\right) ^{3}\left( \alpha +1\right) }\left( \frac{DB\left( X\right) }{%
C\left( \bar{p}\right) \sigma _{\hat{K}}^{2}}\right) ^{\frac{\alpha }{\alpha
+1}}\right) \right)  \label{Kms}
\end{equation}%
where $W_{0}$ is the Lambert $W$ function.

For $B\left( X\right) <<1$, we can check that $K_{\hat{X}}^{\alpha }$ is
increasing with $B\left( X\right) $, i.e. with short-term returns $f\left( 
\hat{X}\right) $\footnote{%
See equation (\ref{fbR}).}, which confirms the results found in approach
one: in the stable case, capital equilibrium increases with short-term
returns $f\left( \hat{X}\right) $.

The unstable case corresponds to a higher level of capital. Using (\ref{psl}%
), it amounts to considering in first approximation that:%
\begin{equation*}
D-\frac{1}{2}\left( \left( \nabla _{X}R\left( X\right) \right) ^{2}+\sigma
_{X}^{2}\frac{\nabla _{X}^{2}R\left( K_{X},X\right) }{H\left( K_{X}\right) }%
\right) H^{2}\left( K_{X}\right) <<1
\end{equation*}%
that is, $K_{X}$\ reaches a value such that:%
\begin{equation*}
\left\Vert \Psi \left( \hat{X},K_{\hat{X}}\right) \right\Vert ^{2}<<1
\end{equation*}%
and the capital is concentrated among a small group of agents. Using a power
law for $H^{2}\left( K_{X}\right) $:%
\begin{equation*}
H^{2}\left( K_{X}\right) =K_{X}^{\alpha }
\end{equation*}%
leads to write the solution (\ref{kpt}) at the first order in $D$:%
\begin{eqnarray}
K_{X}^{\alpha } &\simeq &\frac{2D}{\left( \nabla _{X}R\left( X\right)
\right) ^{2}+\sigma _{X}^{2}\frac{\nabla _{X}^{2}R\left( K_{X},X\right) }{%
H\left( K_{X}\right) }}  \label{sLT} \\
f\left( \hat{X}\right) &\simeq &B\left( X\right) K_{\hat{X}}^{\alpha } 
\notag
\end{eqnarray}%
Including corrections of order $\frac{1}{D}$\ to formula (\ref{sLT}) yields
the approximate solution to (\ref{kpt}):%
\begin{eqnarray}
K_{\hat{X}}^{\alpha } &\simeq &\frac{2D}{\left( \nabla _{X}R\left( X\right)
\right) ^{2}+\sigma _{X}^{2}\frac{\nabla _{X}^{2}R\left( K_{X},X\right) }{%
H\left( K_{X}\right) }}  \label{Kmn} \\
&&-\left( \frac{\left( \nabla _{X}R\left( X\right) \right) ^{2}+\sigma
_{X}^{2}\frac{\nabla _{X}^{2}R\left( K_{X},X\right) }{H\left( K_{X}\right) }%
}{2D}\right) ^{\frac{1}{\alpha }}\frac{\sigma _{\hat{K}}^{2}C\left( \bar{p}%
\right) }{DB\left( X\right) }  \notag \\
&&\times \exp \left( -\frac{\sigma _{X}^{2}\sigma _{\hat{K}}^{2}\left(
B^{\prime }\left( X\right) \right) ^{2}}{768D\left( B\left( X\right) \right)
^{3}}\left( \left( \nabla _{X}R\left( X\right) \right) ^{2}+\sigma _{X}^{2}%
\frac{\nabla _{X}^{2}R\left( K_{X},X\right) }{H\left( K_{X}\right) }\right)
\right)  \notag
\end{eqnarray}%
The previous (in)stability analysis applies. In the range $B\left( X\right)
<<1$, when $f\left( \hat{X}\right) $\ increases, or which is equivalent, $%
B\left( X\right) $ increases, average capital must reduce to preserve the
possibility of unstable equilibria. Likewise, equilibrium capital is higher
when expected returns $R\left( X\right) $ are minimal. When expected returns
increase, the threshold defined by the unstable equilibrium decreases.

\subsubsection{Expansion around particular solutions for $p=0$}

To better understand the behaviour of the solutions of equation (\ref{qtk}),
we expand this equation around the points $\left( \hat{X},K_{\hat{X}%
,M}\right) $ that solve equation (\ref{qtk}). Appendix 3.2.2.2 computes this
expansion at the second order around $\hat{X}_{M}$ and $K_{\hat{X},M}$. We
find:

\begin{eqnarray}
\left( K_{\hat{X}}-K_{\hat{X},M}\right) &=&\frac{1}{D}\left( \sigma
_{X}^{2}\sigma _{\hat{K}}^{2}\frac{3\left( f^{\prime }\left( \hat{X}\right)
\right) ^{3}-2f^{\prime }\left( X\right) f^{\prime \prime }\left( \hat{X}%
\right) \left\vert f\left( \hat{X}\right) \right\vert }{120\left\vert
f\left( \hat{X}\right) \right\vert ^{4}}\right.  \label{pnm} \\
&&\left. -\frac{\frac{\partial f\left( \hat{X},K_{\hat{X}}\right) }{\partial 
\hat{X}}}{f\left( \hat{X},K_{\hat{X}}\right) }-\frac{\frac{\partial
\left\Vert \Psi \left( \hat{X},K_{\hat{X}}\right) \right\Vert ^{2}}{\partial 
\hat{X}}}{\left\Vert \Psi \left( \hat{X},K_{\hat{X}}\right) \right\Vert ^{2}}%
\right) _{K_{\hat{X},M}}\left( \hat{X}-\hat{X}_{M}\right)  \notag \\
&&+\frac{1}{D}\frac{b}{2}\left( \hat{X}-\hat{X}_{M}\right) \nabla _{\hat{X}%
}^{2}\left( \frac{M-A\left( \hat{X}\right) }{f\left( \hat{X}\right) }\right)
_{K_{\hat{X},M}}\left( \hat{X}-\hat{X}_{M}\right)  \notag
\end{eqnarray}%
with $\frac{A\left( \hat{X}\right) }{f\left( \hat{X}\right) }$\ given in
formula (\ref{rlt}):%
\begin{equation*}
D=\left( 1+\frac{\frac{\partial f\left( \hat{X},K_{\hat{X}}\right) }{%
\partial K_{\hat{X}}}}{f\left( \hat{X},K_{\hat{X}}\right) }+\frac{\frac{%
\partial \left\Vert \Psi \left( \hat{X},K_{\hat{X}}\right) \right\Vert ^{2}}{%
\partial K_{\hat{X}}}}{\left\Vert \Psi \left( \hat{X},K_{\hat{X}}\right)
\right\Vert ^{2}}+\frac{\sigma _{X}^{2}\sigma _{\hat{K}}^{2}\left( \nabla
_{K_{\hat{X}}}\left( f^{\prime }\left( \hat{X}\right) \right) ^{2}\left\vert
f\left( \hat{X}\right) \right\vert -3\left( \nabla _{K_{\hat{X}}}\left\vert
f\left( \hat{X}\right) \right\vert \right) \left( f^{\prime }\left( \hat{X}%
\right) \right) ^{2}\right) }{120\left\vert f\left( \hat{X}\right)
\right\vert ^{4}}\right) _{K_{\hat{X}_{M}}}
\end{equation*}%
and $b=\left( 2-\ln 2-\gamma _{0}\right) $, with $\gamma _{0}$ the
Euler-Mascheroni constant.

As in the first approach, $D>0$ corresponds to a stable equilibrium, and $%
D<0 $ to an unstable one. The expansion (\ref{pnm}) describes the local
variations of $K_{\hat{X}}$ in the neighbourhood of the points $K_{\hat{X}%
,M} $. This approximation (\ref{pnm}) suffices to understand the role of the
parameters of the system.

We consider the case of stable equilibria, i.e. $D>0$. Note that under
unstable equilibria, $D<0$, the interpretations are inverted, since $K_{\hat{%
X}}$ is interpreted as a threshold\footnote{%
See the first approach.}.

The equation (\ref{pnm}), that expands average capital at sector $\hat{X}%
_{M} $,\ is composed of a first order and a second order contributions.

The\textbf{\ }first order part in the expansion (\ref{pnm}) writes: 
\begin{equation}
\frac{1}{D}\left( \sigma _{X}^{2}\sigma _{\hat{K}}^{2}\frac{3\left(
f^{\prime }\left( \hat{X}\right) \right) ^{3}-2f^{\prime }\left( X\right)
f^{\prime \prime }\left( \hat{X}\right) \left\vert f\left( \hat{X}\right)
\right\vert }{120\left\vert f\left( \hat{X}\right) \right\vert ^{4}}-\frac{%
\frac{\partial f\left( \hat{X},K_{\hat{X}}\right) }{\partial \hat{X}}}{%
f\left( \hat{X},K_{\hat{X}}\right) }-\frac{\frac{\partial \left\Vert \Psi
\left( \hat{X},K_{\hat{X}}\right) \right\Vert ^{2}}{\partial \hat{X}}}{%
\left\Vert \Psi \left( \hat{X},K_{\hat{X}}\right) \right\Vert ^{2}}\right)
_{K_{\hat{X},M}}\left( \hat{X}-\hat{X}_{M}\right)  \label{pnM}
\end{equation}%
It represents the variation of equilibrium capital as a function of its
position. It is decomposed in three contributions:

For $f^{\prime }\left( \hat{X}\right) >0$, the second contribution in (\ref%
{pnM}):

\begin{equation*}
-\frac{\frac{\partial f\left( \hat{X},K_{\hat{X}}\right) }{\partial \hat{X}}%
\left( \hat{X}-\hat{X}_{M}\right) }{f\left( \hat{X},K_{\hat{X}}\right) }
\end{equation*}%
is positive. It represents the decrease in capital needed to reach
equilibrium. Actually, the return is higher at point $\hat{X}$ than at $\hat{%
X}_{M}$: a lower capital will yield the same overall return at point $\hat{X}
$. On the contrary, the first contribution in (\ref{pnM}): 
\begin{equation*}
\frac{\sigma _{X}^{2}\sigma _{\hat{K}}^{2}\frac{3\left( f^{\prime }\left( 
\hat{X}\right) \right) ^{3}-2f^{\prime }\left( X\right) f^{\prime \prime
}\left( \hat{X}\right) \left\vert f\left( \hat{X}\right) \right\vert }{%
120\left\vert f\left( \hat{X}\right) \right\vert ^{4}}\left( \hat{X}-\hat{X}%
_{M}\right) }{D}
\end{equation*}%
describes the "net" variation of capital due to a variation in $f\left(
X\right) $. When returns are decreasing, i.e. when $f^{\prime }\left( \hat{X}%
\right) >0$ and $f^{\prime \prime }\left( \hat{X}\right) <0$, this first
contribution has the sign of $f^{\prime }\left( \hat{X}\right) $. An
increase in returns attracts capital.

The third term in (\ref{pnM}):%
\begin{equation*}
-\frac{1}{D}\left( \frac{\frac{\partial \left\Vert \Psi \left( \hat{X},K_{%
\hat{X}}\right) \right\Vert ^{2}}{\partial \hat{X}}}{\left\Vert \Psi \left( 
\hat{X},K_{\hat{X}}\right) \right\Vert ^{2}}\right) _{K_{\hat{X},M}}\left( 
\hat{X}-\hat{X}_{M}\right)
\end{equation*}%
represents the number effect.\ Actually, when: 
\begin{equation*}
\frac{\frac{\partial \left\Vert \Psi \left( \hat{X},K_{\hat{X}}\right)
\right\Vert ^{2}}{\partial \hat{X}}}{\left\Vert \Psi \left( \hat{X},K_{\hat{X%
}}\right) \right\Vert ^{2}}>0
\end{equation*}%
the number of agents is higher at $\hat{X}$ than at $\hat{X}_{M}$: the
average capital per agent is reduced.

The second order contribution in (\ref{pnm}) represents the effect of the
neighbouring sector space on each sector. Given the first order condition (%
\ref{mfl}): 
\begin{equation*}
\nabla _{\hat{X}}^{2}\left( \frac{M-A\left( \hat{X}\right) }{f\left( \hat{X}%
\right) }\right) _{K_{\hat{X},M}}=\left( \frac{\nabla _{\hat{X}}^{2}\left(
M-A\left( \hat{X}\right) \right) }{f\left( \hat{X}\right) }\right) _{K_{\hat{%
X},M}}
\end{equation*}%
and since $A\left( \hat{X}_{M}\right) $ is a maximum, we have:%
\begin{equation*}
\left( \hat{X}-\hat{X}_{M}\right) \nabla _{\hat{X}}^{2}\left( \frac{%
M-A\left( \hat{X}\right) }{f\left( \hat{X}\right) }\right) _{K_{\hat{X}%
,M}}\left( \hat{X}-\hat{X}_{M}\right) >0
\end{equation*}

When $f\left( \hat{X}\right) $ is constant, $A\left( \hat{X}_{M}\right) $\
is a local maximum, and $K_{\hat{X}_{M}}$\ is a minimum. To put it
differently, $K_{\hat{X}}$ is a decreasing function of $A\left( \hat{X}%
\right) $.\ This is in line with the definition of $A\left( \hat{X}\right) $%
\footnote{%
See discussions after equations (\ref{mqn}) and (\ref{dsc}).}, which
measures the relative attractiveness of sector $\hat{X}$'s\ neighbours: the
higher $A\left( \hat{X}\right) $, the lower the incentive for capital to
stay in sector $\hat{X}$.

\subsection{Third approach:\ solving for standard parameter functions}

A third approach computes the approximate solutions for the average capital
per firm per sector $X$, which is given by equation (\ref{qtk}). To do so,\
we choose some general forms for the three parameters functions arising in
the definition of the action functional, $f$,\ $g$ and$\ H\left(
K_{X}\right) $, given by equations (\ref{fcf}), (\ref{fcg}) and (\ref{psl})
respectively. Recall that $f$\ defines short-term returns, including
dividend and expected long-term price variations. The function $g$\
describes investors' mobility in the sector space.\ The function $H\left(
K_{X}\right) $, which is involved in the background field for firms,
describes firms' moves\textbf{\ }in the sectors space.

Once these parameter functions chosen, the approximate solutions of equation
(\ref{qtk}) for average capital per firm per sector can be found. The second
approach has already shown that this equation has in general several
solutions. To find these solutions, we thus examine relevant ranges for $%
K_{X}$, namely $K_{X}>>1$, $K_{X}<<1$\ and the intermediate range $\infty
>K_{X}>1$, and find the solutions for each $K_{X}$ within these various
ranges.\textbf{\ }

\subsubsection{Choice of parameter functions}

Our choices for the parameter functions $f$,\ $g$ and$\ H^{2}\left(
K_{X}\right) $ are the following.

\paragraph{Function $H^{2}\left( K_{X}\right) $}

We can choose for $H^{2}\left( K_{X}\right) $ a power function of $K_{X}$,
so that equation (\ref{psl}) rewrites:%
\begin{equation}
\left\Vert \Psi \left( X\right) \right\Vert ^{2}\simeq \frac{D\left(
\left\Vert \Psi \right\Vert ^{2}\right) -\frac{F}{2\sigma _{X}^{2}}\left(
\left( \nabla _{X}R\left( X\right) \right) ^{2}+\frac{2\sigma _{X}^{2}\nabla
_{X}^{2}R\left( K_{X},X\right) }{H\left( K_{X}\right) }\right) K_{X}^{\eta }%
}{2\tau }\equiv D-L\left( X\right) \left( \nabla _{X}R\left( X\right)
\right) ^{2}K_{X}^{\eta }  \label{rsp}
\end{equation}

\paragraph{Function $f$}

To determine the function $f$, we must first assume a form for $r\left(
K,X\right) $, the physical capital marginal returns, and for $F_{1}$, the
function that measures the impact of expected long-term return on investment
choices.

Assuming the production functions are of Cobb-Douglas type, i.e. $B\left(
X\right) K^{\alpha }$ with $B\left( X\right) $ a productivity factor, we
have for $r\left( K,X\right) $:%
\begin{equation}
r\left( K,X\right) =\frac{\partial r\left( K,X\right) }{\partial K}=\alpha
B\left( X\right) K^{\alpha -1}  \label{rkX}
\end{equation}

For function $F_{1}$, the simplest choice would be a linear form:%
\begin{equation*}
F_{1}\left( \frac{R\left( K_{\hat{X}},\hat{X}\right) }{\int R\left(
K_{X^{\prime }}^{\prime },X^{\prime }\right) \left\Vert \Psi \left(
X^{\prime }\right) \right\Vert ^{2}dX^{\prime }}\right) \simeq F_{1}\left( 
\frac{R\left( K_{\hat{X}},\hat{X}\right) }{\left\langle K_{\hat{X}}^{\alpha
}\right\rangle \left\langle R\left( \hat{X}\right) \right\rangle }\right)
=b\left( \frac{K_{\hat{X}}^{\alpha }R\left( \hat{X}\right) }{\left\langle
K_{X}^{\alpha }\right\rangle \left\langle R\left( X\right) \right\rangle }%
-1\right)
\end{equation*}%
where, for any function $u\left( \hat{X}\right) $, $\left\langle u\left( 
\hat{X}\right) \right\rangle $\ denotes its average over the sector space,\
and $b$\ an arbitrary parameter.

However, when capital$\ K_{\hat{X}}^{\alpha }\rightarrow \infty $ and is
concentrated at $\hat{X}$, we have $\left\langle K_{X}^{\alpha
}\right\rangle \simeq \frac{K_{\hat{X}}^{\alpha }}{N^{\alpha }\left(
X\right) }$, so that $\frac{K_{\hat{X}}^{\alpha }R\left( \hat{X}\right) }{%
\left\langle K_{X}^{\alpha }\right\rangle \left\langle R\left( X\right)
\right\rangle }\rightarrow \frac{N^{\alpha }\left( X\right) R\left( \hat{X}%
\right) }{\left\langle R\left( X\right) \right\rangle }>>1$. To impose some
bound on moves in the sector space we rather choose:%
\begin{equation}
F_{1}\left( \frac{R\left( K_{\hat{X}},\hat{X}\right) }{\left\langle K_{\hat{X%
}}^{\alpha }\right\rangle \left\langle R\left( \hat{X}\right) \right\rangle }%
\right) \simeq b\arctan \left( \frac{K_{\hat{X}}^{\alpha }R\left( \hat{X}%
\right) }{\left\langle K_{X}^{\alpha }\right\rangle \left\langle R\left(
X\right) \right\rangle }-1\right)  \label{prf}
\end{equation}%
so that $F_{1}\left( \frac{R\left( K_{\hat{X}},\hat{X}\right) }{\left\langle
K_{\hat{X}}^{\alpha }\right\rangle \left\langle R\left( \hat{X}\right)
\right\rangle }\right) >0$ when$\ \frac{K_{\hat{X}}^{\alpha }R\left( \hat{X}%
\right) }{\left\langle K_{X}^{\alpha }\right\rangle \left\langle R\left(
X\right) \right\rangle }>1$.

Given the above assumptions, the general formula for $f$\ given in equation (%
\ref{fcf}) rewrites:%
\begin{equation}
f\left( \hat{X},\Psi ,\hat{\Psi}\right) =\frac{1}{\varepsilon }\left(
r\left( \hat{X}\right) K_{\hat{X}}^{\alpha -1}-\gamma \left\Vert \Psi \left( 
\hat{X}\right) \right\Vert ^{2}+b\arctan \left( \frac{K_{\hat{X}}^{\alpha
}R\left( \hat{X}\right) }{\left\langle K_{X}^{\alpha }\right\rangle
\left\langle R\left( X\right) \right\rangle }-1\right) \right)  \label{Spt}
\end{equation}%
This general formula can be approximated for $\frac{K_{\hat{X}}^{\alpha
}R\left( \hat{X}\right) }{\left\langle K_{X}^{\alpha }\right\rangle
\left\langle R\left( X\right) \right\rangle }\simeq 1$, when average capital
in sector $\hat{X}$ is close to the average capital of the whole space,
which is usually the case.

Using our choices (\ref{rsp}), (\ref{rkX}) and (\ref{prf}) for $\left\Vert
\Psi \left( X\right) \right\Vert ^{2}$\ $r\left( \hat{X}\right) $\ and $%
F_{1} $\ respectively, the equation (\ref{fcf}) for $f\left( \hat{X},\Psi ,%
\hat{\Psi}\right) $ becomes:%
\begin{equation*}
f\left( \hat{X},\Psi ,\hat{\Psi}\right) =\frac{1}{\varepsilon }\left( \left(
r\left( \hat{X}\right) +\frac{bR\left( \hat{X}\right) K_{\hat{X}}^{\alpha }}{%
\left\langle K_{\hat{X}}^{\alpha }\right\rangle \left\langle R\left( \hat{X}%
\right) \right\rangle }\right) +\gamma L\left( \hat{X}\right) K_{X}^{\eta
}-\gamma D-b\right)
\end{equation*}%
We may assume without impairing the results that $\eta =\alpha $. We thus
have:%
\begin{eqnarray}
f\left( \hat{X},\Psi ,\hat{\Psi}\right) &=&\frac{1}{\varepsilon }\left(
\left( \frac{r\left( \hat{X}\right) }{K_{\hat{X}}^{\alpha }}+\frac{bR\left( 
\hat{X}\right) }{\left\langle K_{\hat{X}}^{\alpha }\right\rangle
\left\langle R\left( \hat{X}\right) \right\rangle }+\gamma L\left( \hat{X}%
\right) \right) K_{\hat{X}}^{\alpha }-\gamma D-b\right)  \label{stp} \\
&\equiv &B_{1}\left( \hat{X}\right) K_{\hat{X}}^{\alpha -1}+B_{2}\left( \hat{%
X}\right) K_{\hat{X}}^{\alpha }-C\left( \hat{X}\right)  \notag
\end{eqnarray}

where:%
\begin{eqnarray*}
B_{1}\left( \hat{X}\right) &=&\frac{\alpha B\left( \hat{X}\right) }{%
\varepsilon } \\
B_{2}\left( \hat{X}\right) &=&\frac{bR\left( \hat{X}\right) }{\varepsilon
\left\langle K_{\hat{X}}^{\alpha }\right\rangle \left\langle R\left( \hat{X}%
\right) \right\rangle }+\frac{\gamma }{\varepsilon } \\
C\left( \hat{X}\right) &=&\gamma D+b
\end{eqnarray*}

\paragraph{Function $g$}

To determine the form of function $g$, equation (\ref{fcg}), we must first
choose a form for the function $F_{0}$.\ 

We assume that: 
\begin{equation}
F_{0}\left( R\left( \hat{X},K_{\hat{X}}\right) \right) =a\arctan \left( K_{%
\hat{X}}^{\alpha }R\left( \hat{X}\right) \right)  \label{zrg}
\end{equation}%
where is $a$\ an arbitrary constant.

Combined to our assumption for $F_{1}$, (\ref{prf}), the formula (\ref{fcg})
for $g$ can be written:%
\begin{equation}
g\left( \hat{X},\Psi ,\hat{\Psi}\right) =a\nabla _{\hat{X}}\arctan \left( K_{%
\hat{X}}^{\alpha }R\left( \hat{X}\right) \right) +b\nabla _{\hat{X}}\arctan
\left( \frac{K_{\hat{X}}^{\alpha }R\left( \hat{X}\right) }{\left\langle
K_{X}^{\alpha }\right\rangle \left\langle R\left( X\right) \right\rangle }%
-1\right)  \label{tsp}
\end{equation}%
where the $\arctan $ function ensures that the velocity in the sector space $%
g$ increases with capital and is maximal when average capital per firm in
sector $\hat{X}$ tends to infinity, i.e. $K_{\hat{X}}^{\alpha }\rightarrow
\infty $.

This general formula, equation (\ref{tsp}), can be approximated for $\frac{%
K_{\hat{X}}^{\alpha }R\left( \hat{X}\right) }{\left\langle K_{X}^{\alpha
}\right\rangle \left\langle R\left( X\right) \right\rangle }\simeq 1$, when
average capital in sector $\hat{X}$ is close to the average capital of the
whole space. It then reduces to:%
\begin{equation}
g\left( \hat{X},\Psi ,\hat{\Psi}\right) \simeq \frac{K_{\hat{X}}^{\alpha }}{%
\left\langle K_{\hat{X}}^{\alpha }\right\rangle }\nabla _{\hat{X}}R\left( 
\hat{X}\right) \left( 1+\frac{b}{\left\langle R\left( \hat{X}\right)
\right\rangle }\right) \equiv \nabla _{\hat{X}}R\left( \hat{X}\right)
A\left( \hat{X}\right) K_{\hat{X}}^{\alpha }  \label{gtp}
\end{equation}%
which in turn allows to approximate the gradient of $g$, $\nabla _{\hat{X}%
}g\left( \hat{X},\Psi ,\hat{\Psi}\right) $, by: 
\begin{equation}
\nabla _{\hat{X}}g\left( \hat{X},\Psi ,\hat{\Psi}\right) \simeq \frac{\nabla
_{\hat{X}}^{2}R\left( \hat{X}\right) }{\left\langle K_{\hat{X}}^{\alpha
}\right\rangle }\left( 1+\frac{b}{\left\langle K_{\hat{X}}^{\alpha
}\right\rangle \left\langle R\left( \hat{X}\right) \right\rangle }\right) K_{%
\hat{X}}^{\alpha }\equiv \nabla _{\hat{X}}^{2}R\left( \hat{X}\right) A\left( 
\hat{X}\right) K_{\hat{X}}^{\alpha }  \label{ptg}
\end{equation}

\subsubsection{Solutions for the average capital}

Now that the forms of the particular functions have been chosen, we can find
the approximate solutions to (\ref{qtk}) in the ranges of average capital : $%
K_{X}>>1$, $K_{X}>>>1$, $K_{X}<1$\ and the intermediate case $\infty
>K_{X}>1 $.

Besides, we only consider the case in which returns are positive\footnote{%
Solutions for negative returns, $f<0$, are discussed below.}, $f>0$.\ 

The asymptotic form of equation (\ref{qtk}) which determines average capital
per firm per sector, writes: 
\begin{equation}
K_{\hat{X}}\left\Vert \Psi \left( \hat{X}\right) \right\Vert ^{2}\left\vert
f\left( \hat{X}\right) \right\vert =C\left( \bar{p}\right) \sigma _{\hat{K}%
}^{2}\exp \left( -\frac{\sigma _{X}^{2}\sigma _{\hat{K}}^{2}\left( p+\frac{1%
}{2}\right) ^{2}\left( f^{\prime }\left( X\right) \right) ^{2}}{96\left\vert
f\left( \hat{X}\right) \right\vert ^{3}}\right) \Gamma \left( p+\frac{3}{2}%
\right)  \label{qtc}
\end{equation}%
where:%
\begin{equation}
p\simeq \frac{M-\left( \left( g\left( \hat{X}\right) \right) ^{2}+\sigma _{%
\hat{X}}^{2}\left( f\left( \hat{X}\right) +\nabla _{\hat{X}}g\left( \hat{X}%
,K_{\hat{X}}\right) \right) \right) }{\sigma _{\hat{X}}^{2}\sqrt{f^{2}\left( 
\hat{X}\right) }}  \label{xpr}
\end{equation}

This equation (\ref{qtc}) has several solutions depending on the range of
average capital considered.

\paragraph{Case 1: $K_{\hat{X}}>>1$}

Under the assumption average capital $K_{\hat{X}}$ is large, but not
excessively so, we can assume in first approximation that:%
\begin{equation}
\left\Vert \Psi \left( \hat{X}\right) \right\Vert ^{2}\simeq D  \label{Bts}
\end{equation}%
From the expressions (\ref{stp}) and (\ref{tsp}), for $f\left( \hat{X},\Psi ,%
\hat{\Psi}\right) $ and $g\left( \hat{X},\Psi ,\hat{\Psi}\right) $
respectively,\ we know that $f\left( \hat{X}\right) $ is independent of $K_{%
\hat{X}}$, and that $g\left( \hat{X}\right) $ is proportional to $\nabla _{%
\hat{X}}R\left( \hat{X}\right) $. These functions rewrite:%
\begin{eqnarray}
f\left( \hat{X}\right) &\equiv &c-\frac{d}{K_{\hat{X}}^{\alpha }R\left( \hat{%
X}\right) }-\gamma D  \label{ffpr} \\
f^{\prime }\left( \hat{X}\right) &\simeq &\frac{d\nabla _{\hat{X}}R\left( 
\hat{X}\right) }{K_{\hat{X}}^{\alpha }R^{2}\left( \hat{X}\right) }  \notag
\end{eqnarray}%
and:%
\begin{eqnarray}
g\left( \hat{X}\right) &\simeq &-\frac{\nabla _{\hat{X}}R\left( \hat{X}%
\right) f}{K_{\hat{X}}^{\alpha }R\left( \hat{X}\right) }  \label{prgg} \\
\nabla _{\hat{X}}g\left( \hat{X}\right) &\simeq &-\frac{\nabla _{\hat{X}%
}^{2}R\left( \hat{X}\right) f}{K_{\hat{X}}^{\alpha }R\left( \hat{X}\right) }
\notag
\end{eqnarray}

We can moreover include the constant $\alpha $\ in the definition of $%
C\left( \bar{p}\right) $, so that equation (\ref{qtc}) becomes:%
\begin{equation}
K_{\hat{X}}D\left\vert f\left( \hat{X}\right) \right\vert =C\left( \bar{p}%
\right) \sigma _{\hat{K}}^{2}\exp \left( -\frac{\sigma _{X}^{2}\sigma _{\hat{%
K}}^{2}\left( p+\frac{1}{2}\right) ^{2}\left( f^{\prime }\left( X\right)
\right) ^{2}}{96\left\vert f\left( \hat{X}\right) \right\vert ^{3}}\right)
\Gamma \left( p+\frac{3}{2}\right)  \label{llqg}
\end{equation}%
Appendix 3.2.3.2 shows that the solution of (\ref{llqj}) is:%
\begin{equation}
K_{\hat{X}}^{\alpha }=\frac{C\left( \bar{p}\right) \sigma _{\hat{K}%
}^{2}\Gamma \left( \frac{M}{c}\right) }{D\left( c-\gamma D\right) }+\frac{d}{%
\left( c-\gamma D\right) R\left( \hat{X}\right) }\left( 1+M\func{Psi}\left( 
\frac{M}{c}\right) \left( 1+\frac{\nabla _{\hat{X}}^{2}R\left( \hat{X}%
\right) f}{M\left( c-\gamma D\right) }\right) \right)  \label{sshk}
\end{equation}%
where the function $\func{Psi}$\ is defined as $\func{Psi}\left( x\right) =%
\frac{\Gamma ^{\prime }\left( x\right) }{\Gamma \left( x\right) }$. This
solution holds for $K_{\hat{X}}>>1$, but only when $\left( c-\gamma D\right)
>0$ and $\frac{C\left( \bar{p}\right) \sigma _{\hat{K}}^{2}\Gamma \left( 
\frac{M}{c}\right) }{D\left( c-\gamma D\right) }>>1$.\ Formula (\ref{sshk})
shows that the dependency of $K_{\hat{X}}^{\alpha }$ in $R\left( \hat{X}%
\right) $ in turns\ depends on the sign of the last term $1+M\func{Psi}%
\left( \frac{M}{c}\right) \left( 1+\frac{\nabla _{\hat{X}}^{2}R\left( \hat{X}%
\right) f}{M\left( c-\gamma D\right) }\right) $.

When:%
\begin{equation*}
1+M\func{Psi}\left( \frac{M}{c}\right) \left( 1+\frac{\nabla _{\hat{X}%
}^{2}R\left( \hat{X}\right) f}{M\left( c-\gamma D\right) }\right) >0
\end{equation*}%
average capital in sector $\hat{X}$, $K_{\hat{X}}^{\alpha }$, is decreasing
in $R\left( \hat{X}\right) $ and $f\left( \hat{X}\right) \simeq c-\gamma D$.
Given the results of section 8.1.3 about stability, this implies that the
equilibrium is unstable.

When:%
\begin{equation*}
1+M\func{Psi}\left( \frac{M}{c}\right) \left( 1+\frac{\nabla _{\hat{X}%
}^{2}R\left( \hat{X}\right) f}{M\left( c-\gamma D\right) }\right) <0
\end{equation*}%
a stable equilibrium, i.e. $K_{\hat{X}}^{\alpha }$ increasing with $R\left( 
\hat{X}\right) $ and $f\left( \hat{X}\right) $\ is possible. This arises for 
$\nabla _{\hat{X}}^{2}R\left( \hat{X}\right) <<0$, for instance when $%
R\left( \hat{X}\right) $ is maximum. In such a case, an increase in $R\left( 
\hat{X}\right) $ induces a higher number $\left\Vert \Psi \left( \hat{X}%
\right) \right\Vert ^{2}$ of firms, without impairing average capital per
firm.

\paragraph{Case 2: $K_{\hat{X}}>>>1$}

When average capital per firm in sector $\hat{X}$ is very high, and higher
than case 1, $K_{\hat{X}}>>>1$, factor $L\left( \hat{X}\right) $ can be
discarded and we can assume that, in first approximation:%
\begin{equation}
\left\Vert \Psi \left( \hat{X}\right) \right\Vert ^{2}\simeq D-\left( \left(
\nabla _{X}R\left( X\right) \right) ^{2}+\frac{2\sigma _{X}^{2}\nabla
_{X}^{2}R\left( K_{X},X\right) }{H\left( K_{X}\right) }\right) K_{\hat{X}%
}^{\alpha }<<1  \label{nlt}
\end{equation}%
The function $f\left( \hat{X},\Psi ,\hat{\Psi}\right) $ given in expression (%
\ref{stp}) can then be rewritten:%
\begin{equation*}
f\left( \hat{X}\right) \simeq b\left( \frac{\pi }{2}-\frac{\left\langle
K_{X}^{\alpha }\right\rangle \left\langle R\left( X\right) \right\rangle }{%
K_{\hat{X}}^{\alpha }R\left( \hat{X}\right) }\right) \equiv c-\frac{d}{K_{%
\hat{X}}^{\alpha }R\left( \hat{X}\right) }\simeq c
\end{equation*}%
The expressions found in the first case, $K_{\hat{X}}>>1$, for $f^{\prime
}\left( \hat{X}\right) $, $g\left( \hat{X}\right) $ and $\nabla _{\hat{X}%
}g\left( \hat{X}\right) $, equations (\ref{ffpr}) and (\ref{prgg})
respectively, are still valid.

Appendix 3.2.3.2 solves equation (\ref{qtc}) given these assumptions.\ Two
cases arise:

For $\left( \nabla _{\hat{X}}R\left( \hat{X}\right) \right) ^{2}\neq 0$, we
find:%
\begin{eqnarray}
K_{\hat{X}}^{\alpha } &\simeq &\frac{D}{\left( \nabla _{\hat{X}}R\left( \hat{%
X}\right) \right) ^{2}}-\frac{C\left( \bar{p}\right) \sigma _{\hat{K}}^{2}%
\sqrt{\frac{M-c}{c}}}{\left( \nabla _{\hat{X}}R\left( \hat{X}\right) \right)
^{2\left( 1-\frac{1}{\alpha }\right) }D^{\frac{1}{\alpha }}c}  \label{mgk} \\
&&-\frac{d}{R\left( \hat{X}\right) }\frac{\left( \nabla _{\hat{X}}R\left( 
\hat{X}\right) \right) ^{\frac{2}{\alpha }}C\left( \bar{p}\right) \sigma _{%
\hat{K}}^{2}\left( \sqrt{\frac{M-c}{c}}+\frac{\frac{M}{c}+\nabla _{\hat{X}%
}^{2}R\left( \hat{X}\right) \frac{f}{d}}{2\sqrt{\frac{M-c}{c}}}\right) }{%
c^{2}D^{1+\frac{1}{\alpha }}\left( 1-\frac{\left( \nabla _{\hat{X}}R\left( 
\hat{X}\right) \right) ^{\frac{2}{\alpha }}C\left( \bar{p}\right) \sigma _{%
\hat{K}}^{2}}{cD^{1+\frac{1}{\alpha }}}\sqrt{\frac{M-c}{c}}\right) }  \notag
\end{eqnarray}%
which shows that $K_{\hat{X}}^{\alpha }$ is increasing in $f\left( \hat{X}%
\right) $ and $R\left( \hat{X}\right) $ for $K_{\hat{X}}^{\alpha }$ large $%
\sqrt{f^{2}\left( \hat{X}\right) }\simeq c<<1$ and $D>>1$.\ As explained in
section 8.1.3, this corresponds to a stable local equilibrium.

For $\left( \nabla _{\hat{X}}R\left( \hat{X}\right) \right) ^{2}\rightarrow
0 $, we must come back to (\ref{rsp}), and replace:%
\begin{equation*}
\left( \nabla _{X}R\left( X\right) \right) ^{2}\rightarrow \left( \nabla
_{X}R\left( X\right) \right) ^{2}+\frac{2\sigma _{X}^{2}\nabla
_{X}^{2}R\left( K_{X},X\right) }{H\left( K_{X}\right) }
\end{equation*}%
There are two possibilities:

When $\nabla _{X}^{2}R\left( K_{X},X\right) <0$, average capital is given by:%
\begin{equation*}
K_{\hat{X}}=\left( \frac{C\left( \bar{p}\right) \sigma _{\hat{K}}^{2}}{%
\left\vert \nabla _{X}^{2}R\left( K_{X},X\right) \right\vert c}\Gamma \left( 
\frac{M-\nabla _{\hat{X}}g\left( \hat{X},K_{\hat{X}}\right) }{c}\right)
\right) ^{\frac{2}{3\alpha }}
\end{equation*}%
This case is unstable. Actually, $K_{X}$ is decreasing in $c$, i.e. in $%
f\left( \hat{X}\right) $. When returns increase, an equilibrium arises only
for a relatively low average capital. Otherwise, capital tends to accumulate
infinitely. When the sector's expected returns are at a local maximum, the
pattern of accumulation becomes unstable. Note that an equilibrium with $K_{%
\hat{X}}>>>1$ is merely possible for $c<<1$.\ Otherwise, there is no
equilibrium for $R\left( K_{X},X\right) $ maximum.

When $\nabla _{X}^{2}R\left( K_{X},X\right) >0$, average capital is given by:%
\begin{equation*}
K_{\hat{X}}^{\alpha }\simeq \left( \frac{D}{\sigma _{X}^{2}\nabla
_{X}^{2}R\left( K_{X},X\right) }\right) ^{2}
\end{equation*}%
and points such that $\nabla _{X}^{2}R\left( K_{X},X\right) >0$ and $\nabla
_{X}R\left( X\right) =0$ are minima of $R\left( X\right) $. This equilibrium
may exist only when capital (\ref{drsp}) is sufficiently high to compensate
for the low expected returns, i.e. to match the condition:%
\begin{equation*}
\frac{K_{\hat{X}}^{\alpha }R\left( \hat{X}\right) }{\left\langle
K_{X}^{\alpha }\right\rangle \left\langle R\left( X\right) \right\rangle }%
-1>0
\end{equation*}%
This equilibrium is thus unlikely and may be discarded in general.

\paragraph{Case 3: $K_{\hat{X}}<<1$}

When average physical capital per firm in sector $\hat{X}$ is very low, we
can use our assumptions about $g\left( \hat{X},\Psi ,\hat{\Psi}\right) $ and 
$\nabla _{\hat{X}}g\left( \hat{X},\Psi ,\hat{\Psi}\right) $, equations (\ref%
{gtp}) and (\ref{ptg}), and assume that:%
\begin{equation}
f\left( \hat{X}\right) \simeq B_{1}\left( \hat{X}\right) K_{\hat{X}}^{\alpha
-1}>>1  \label{srt}
\end{equation}%
and: 
\begin{equation*}
g\left( \hat{X}\right) \simeq 0
\end{equation*}%
and moreover that:%
\begin{equation*}
\left\Vert \Psi \left( \hat{X}\right) \right\Vert ^{2}=D-L\left( X\right)
\left( \nabla _{X}R\left( X\right) \right) ^{2}K_{\hat{X}}^{\alpha }\simeq D
\end{equation*}%
For these conditions, the solution of (\ref{qtc}) is locally stable.

Moreover, the conditions $K_{\hat{X}}<<1$ and the defining equation (\ref%
{stp}) for $f$ imply that $f>0$, and that for $\alpha <1$:%
\begin{equation*}
\frac{\sigma _{X}^{2}\sigma _{\hat{K}}^{2}\left( p+\frac{1}{2}\right)
^{2}\left( f^{\prime }\left( X\right) \right) ^{2}}{96\left\vert f\left( 
\hat{X}\right) \right\vert ^{3}}<<1
\end{equation*}%
Under these assumptions, equation (\ref{qtc}) reduces to:%
\begin{equation}
K_{\hat{X}}D\left\vert f\left( \hat{X}\right) \right\vert \simeq C\left( 
\bar{p}\right) \sigma _{\hat{K}}^{2}\hat{\Gamma}\left( p+\frac{1}{2}\right)
\label{qtC}
\end{equation}%
This equation (\ref{qtC}) can be approximated.\ Actually, using formula (\ref%
{xpr}) for $p$\ yields:%
\begin{equation*}
p+\frac{1}{2}=\frac{M-\left( \frac{\left( g\left( \hat{X}\right) \right) ^{2}%
}{\sigma _{\hat{X}}^{2}}+\nabla _{\hat{X}}g\left( \hat{X},K_{\hat{X}}\right)
\right) }{\sqrt{f^{2}\left( \hat{X}\right) }}-1\simeq -1
\end{equation*}%
and an expansion of $\hat{\Gamma}\left( p+\frac{1}{2}\right) $\ around the
value $p+\frac{1}{2}=-1$\ writes:%
\begin{equation*}
\hat{\Gamma}\left( p+\frac{1}{2}\right) \simeq \hat{\Gamma}\left( -1\right) +%
\hat{\Gamma}^{\prime }\left( -1\right) \frac{M-\left( \frac{\left( g\left( 
\hat{X}\right) \right) ^{2}}{\sigma _{\hat{X}}^{2}}+\nabla _{\hat{X}}g\left( 
\hat{X},K_{\hat{X}}\right) \right) }{\sqrt{f^{2}\left( \hat{X}\right) }}
\end{equation*}%
As a consequence, when returns are large, i.e. $f\left( \hat{X}\right) >>1$,
equation (\ref{qtc}) writes:%
\begin{equation*}
K_{\hat{X}}\left( B_{1}\left( \hat{X}\right) K_{\hat{X}}^{\alpha -1}\right)
\simeq \frac{C\left( \bar{p}\right) \sigma _{\hat{K}}^{2}}{D}\left( \hat{%
\Gamma}\left( -1\right) +\hat{\Gamma}^{\prime }\left( -1\right) \frac{%
M-\left( \frac{\left( g\left( \hat{X}\right) \right) ^{2}}{\sigma _{\hat{X}%
}^{2}}+\nabla _{\hat{X}}g\left( \hat{X},K_{\hat{X}}\right) \right) }{%
B_{1}\left( \hat{X}\right) K_{\hat{X}}^{\alpha -1}}\right)
\end{equation*}%
with first order solution\footnote{%
Given our hypotheses, $D>>1$\ , which implies that $K_{\hat{X}}<<1$,\ as
needed.}:%
\begin{equation}
K_{\hat{X}}=\left( \frac{C\left( \bar{p}\right) \sigma _{\hat{K}}^{2}\hat{%
\Gamma}\left( -1\right) }{DB_{1}\left( \hat{X}\right) }\right) ^{\frac{1}{%
\alpha }}+\frac{\frac{C\left( \bar{p}\right) \sigma _{\hat{K}}^{2}}{D}\hat{%
\Gamma}^{\prime }\left( -1\right) \left( M-\left( \frac{\left( g\left( \hat{X%
}\right) \right) ^{2}}{\sigma _{\hat{X}}^{2}}+\nabla _{\hat{X}}g\left( \hat{X%
},K_{\hat{X}}\right) \right) \right) }{B_{1}^{\frac{1}{\alpha }}\left( \hat{X%
}\right) \left( \frac{C\left( \bar{p}\right) \sigma _{\hat{K}}^{2}\hat{\Gamma%
}\left( -1\right) }{D}\right) ^{1-\frac{1}{\alpha }}}  \label{rst}
\end{equation}

Equation (\ref{rst}) shows that average capital $K_{\hat{X}}$\ increases
with $M-\left( \frac{\left( g\left( \hat{X}\right) \right) ^{2}}{\sigma _{%
\hat{X}}^{2}}+\nabla _{\hat{X}}g\left( \hat{X},K_{\hat{X}}\right) \right) $:
when expected long-term returns increase, more capital is allocated to the
sector. Equation (\ref{srt}) also shows that average capital $K_{\hat{X}}$\
is maximal when returns $R\left( \hat{X}\right) $\ are at a local maximum,
i.e. when $\frac{\left( g\left( \hat{X}\right) \right) ^{2}}{\sigma _{\hat{X}%
}^{2}}=0$\ and $\nabla _{\hat{X}}g\left( \hat{X},K_{\hat{X}}\right) <0$.

Inversely, the same equations (\ref{rst}) and (\ref{srt}) show that average
capital $K_{\hat{X}}$ is decreasing in $f\left( \hat{X}\right) $. The
equilibrium is unstable. When average capital is very low, i.e. $K_{\hat{X}%
}<<1$, which is the case studied here, marginal returns are high.\ Any
increase in capital above the threshold widely increases returns, which
drives capital towards the next stable equilibrium, with higher $K_{\hat{X}}$%
. Recall that in this unstable equilibrium, $K_{\hat{X}}$ must be seen as a
threshold.\ The rise in $f\left( \hat{X}\right) $ reduces the threshold $K_{%
\hat{X}}$, which favours\ capital accumulation and increases the average
capital $K_{\hat{X}}$.

This case is thus an exception: the dependency of $K_{\hat{X}}$\ in $R\left( 
\hat{X}\right) $\ is stable, but the dependency in $f\left( \hat{X}\right) $%
\ is unstable. This saddle path type of instability may lead the sector,
either towards a higher level of capital (case 4 below) or towards $0$.
where the sector disappears.

\paragraph{Case 4: $\infty >>K_{\hat{X}}>1$ intermediate case}

To solve equation (\ref{qtc}) in this general case, we define an
intermediate variable $W$ given by:%
\begin{equation}
W=\sqrt{\frac{\sigma _{X}^{2}\sigma _{\hat{K}}^{2}\left( f^{\prime }\left(
X\right) \right) ^{2}}{96\left\vert f\left( \hat{X}\right) \right\vert ^{3}}}%
\left( p+\frac{1}{2}-\frac{48\left\vert f\left( \hat{X}\right) \right\vert
^{3}}{\sigma _{X}^{2}\sigma _{\hat{K}}^{2}\left( f^{\prime }\left( X\right)
\right) ^{2}}\left( \ln \left( \bar{p}+\frac{1}{2}\right) -1\right) \right)
\label{kdw}
\end{equation}%
so that equation (\ref{qtc}) rewrites: 
\begin{eqnarray}
&&K_{\hat{X}}\left\Vert \Psi \left( \hat{X}\right) \right\Vert ^{2}\left( 
\frac{\sigma _{X}^{2}\left( f^{\prime }\left( X\right) \right)
^{2}\left\vert f\left( \hat{X}\right) \right\vert \exp \left( -\frac{%
96\left\vert f\left( \hat{X}\right) \right\vert ^{3}}{\sigma _{X}^{2}\sigma
_{\hat{K}}^{2}\left( f^{\prime }\left( X\right) \right) ^{2}}\left( \ln
\left( \bar{p}+\frac{1}{2}\right) -1\right) ^{2}\right) }{96\left( \sigma _{%
\hat{K}}^{2}\right) ^{3}}\right) ^{\frac{1}{4}}  \label{qtcw} \\
&=&C\left( \bar{p}\right) \exp \left( -W^{2}\right) \sqrt{W+2\sqrt{\frac{%
96\left\vert f\left( \hat{X}\right) \right\vert ^{3}}{\sigma _{X}^{2}\sigma
_{\hat{K}}^{2}\left( f^{\prime }\left( X\right) \right) ^{2}}}\left( \ln
\left( \bar{p}+\frac{1}{2}\right) -1\right) }  \notag
\end{eqnarray}%
where $\bar{p}$, the average value of parameter $p$, is defined by $\bar{p}=%
\bar{p}\left( -M\right) $, and is given by equation (\ref{pBL}) when $%
\lambda =-M$.

Equation (\ref{qtcw}) can be solved for $K_{\hat{X}}$\ under the following
simplifying assumptions: 
\begin{equation}
f\left( \hat{X}\right) \simeq B_{2}\left( X\right) K_{\hat{X}}^{\alpha }
\label{fRP}
\end{equation}%
and:%
\begin{equation*}
\left\Vert \Psi \left( \hat{X}\right) \right\Vert ^{2}\simeq D
\end{equation*}%
Eventually, appendix 3.2.3.2 shows that for $\sigma _{X}^{2}<<1$:

\begin{eqnarray}
K_{\hat{X}}^{\alpha } &=&\left( \frac{8C\left( \bar{p}\right) }{D}\sqrt{%
\frac{3\sigma _{\hat{K}}^{2}\left\vert B_{2}\left( X\right) \right\vert }{%
\sigma _{X}^{2}\left( B_{2}^{\prime }\left( X\right) \right) ^{2}}\left( \ln
\left( \bar{p}+\frac{1}{2}\right) -1\right) }\right) ^{\frac{2\alpha }{%
1+\alpha }}  \label{kfcw} \\
&&\times \exp \left( -W_{0}\left( -\frac{48\alpha }{1+\alpha }\left( \sqrt{%
\frac{3\sigma _{\hat{K}}^{2}}{\sigma _{X}^{2}}}\frac{8C\left( \bar{p}\right) 
}{D}\right) ^{\frac{2\alpha }{1+\alpha }}\frac{\left\vert B_{2}\left(
X\right) \right\vert ^{3+\frac{\alpha }{1+\alpha }}}{\sigma _{X}^{2}\sigma _{%
\hat{K}}^{2}\left( B_{2}^{\prime }\left( X\right) \right) ^{2+\frac{2\alpha 
}{1+\alpha }}}\left( \ln \left( \bar{p}+\frac{1}{2}\right) -1\right) ^{2+%
\frac{\alpha }{1+\alpha }}\right) \right)  \notag
\end{eqnarray}%
where $W_{0}$ is the Lambert $W$ function.

In first approximation, equation (\ref{kfcw}) implies that $K_{\hat{X}%
}^{\alpha }$\ is an increasing function of $B_{2}\left( X\right) $. Given
our simplifying assumption (\ref{fRP}), average capital is higher in high
short-term returns sectors.

Moreover, $K_{\hat{X}}^{\alpha }$\ is a decreasing function of $\left(
\nabla _{\hat{X}}R\left( \hat{X}\right) \right) ^{2}$\ and $\nabla _{\hat{X}%
}^{2}R\left( \hat{X}\right) $: capital accumulation is locally maximal when
expected returns $R\left( \hat{X}\right) $ of sector $\hat{X}$\ are at a
local maxima, i.e. $\left( \nabla _{\hat{X}}R\left( \hat{X}\right) \right)
^{2}=0$\ and $\nabla _{\hat{X}}^{2}R\left( \hat{X}\right) <0$.

Thus, in the intermediate case, the average values $K_{\hat{X}}$ are stable.
In addition, both short-term and long term returns matter in the
intermediate range.

\subsubsection{Case $f<0$}

In the four cases above, we have only considered the case where a sector $%
\hat{X}$\ short-term returns are positive $f\left( \hat{X}\right) >0$. We
can nonetheless extend our analysis to the case $f\left( \hat{X}\right) <0$.

In such a case, the equilibria, whether stable or unstable, defined in cases
1, 2 with $K_{\hat{X}}>>1$, and 4 with $K_{\hat{X}}>1$, are still valid, and
capital allocation relies on expectations of high long-term returns. If we
consider that $f\left( \hat{X}\right) <0$\ is an extreme case, where
expectations of large future profits must offset short-term losses. However,
such equilibria become unsustainable when $R\left( \hat{X}\right) $\
decreases to such an extent that it does not compensate for the loss $%
f\left( \hat{X}\right) $\textbf{. }Case 3, $K_{\hat{X}}<1$\ is the only case
that is no longer possible when $f\left( \hat{X}\right) <0$, since the
returns that matter in this case are dividends. If they turn negative, the
equilibrium is no longer sustainable.

\section{\textbf{Finding average capital in a dynamic environment}}

So far, we have determined and studied the dependency in parameters of
average capital per firm and per sector. However parameters may vary over
time, and so should average capital values. We thus introduce a macro time
scale and design a dynamic model that involves average capital and time
varying expectations in long-term returns.

\subsection{A\textbf{verage capital} and long-run expected returns}

We consider the dynamics for $K_{\hat{X}}$ generated by modifications in
parameters. Assuming that some time-dependent parameters modify expected
long-term returns $R\left( X\right) $, average capital $K_{\hat{X}}$ becomes
a function of the time variable $\theta $. To find the evolution over time
of the average physical capital per firm in sector $\hat{X}$, $K_{\hat{X}}$,
can be found by defining the equation for $K_{\hat{X}}$, (\ref{qtk}), and
compute its variation with respect to $\theta $, using the fact that the
functions $\left\Vert \Psi \left( \hat{X}\right) \right\Vert ^{2}$\ and $%
\hat{\Gamma}\left( p+\frac{1}{2}\right) $\ both depend on time $\theta $\
through $K_{\hat{X}}$\ and $R\left( X\right) $. The variations of these two
functions with respect to the two dynamical variables $K_{\hat{X}}$ and $%
R\left( X\right) $ are computed in appendix 4.1.\ We show that, when $%
C\left( \bar{p}\right) $ constant, the variation of (\ref{qtk}) writes:%
\begin{equation}
k\frac{\nabla _{\theta }K_{\hat{X}}}{K_{\hat{X}}}+l\frac{\nabla _{\theta
}R\left( \hat{X}\right) }{R\left( \hat{X}\right) }-2m\frac{\nabla _{\hat{X}%
}\nabla _{\theta }R\left( \hat{X}\right) }{\nabla _{\hat{X}}R\left( \hat{X}%
\right) }+n\frac{\nabla _{\hat{X}}^{2}\nabla _{\theta }R\left( \hat{X}%
\right) }{\nabla _{\hat{X}}^{2}R\left( \hat{X}\right) }=-C_{3}\left( p,\hat{X%
}\right) \frac{\nabla _{\theta }r\left( \hat{X}\right) }{f\left( \hat{X}%
\right) }  \label{krv}
\end{equation}%
with:%
\begin{eqnarray}
k &=&1-\eta \left( 1-\frac{\gamma C_{3}\left( p,\hat{X}\right) }{\left\vert
f\left( \hat{X}\right) \right\vert }\right) \frac{D-\left\Vert \Psi \left( 
\hat{X}\right) \right\Vert ^{2}}{\left\Vert \Psi \left( \hat{X}\right)
\right\Vert ^{2}}\frac{D^{\frac{1}{\alpha }}c}{\left( \nabla _{\hat{X}%
}R\left( \hat{X}\right) \right) ^{\frac{2}{\alpha }}C\left( \bar{p}\right)
\sigma _{\hat{K}}^{2}\sqrt{\frac{M-c}{c}}}  \label{cFT} \\
&&+\frac{\alpha \left( 2\frac{g^{2}\left( \hat{X}\right) }{\sigma _{\hat{X}%
}^{2}}+\nabla _{\hat{X}}g\left( \hat{X}\right) \right) C_{2}\left( p,\hat{X}%
\right) -\left( 1-\alpha \right) C_{3}\left( p,\hat{X}\right) }{\left\vert
f\left( \hat{X}\right) \right\vert }  \notag \\
l &=&\frac{\varsigma F_{1}\left( R\left( K_{\hat{X}},\hat{X}\right) \right)
C_{3}\left( p,\hat{X}\right) }{f\left( \hat{X}\right) }  \notag \\
m &=&\left( 1-\frac{\gamma C_{3}\left( p,\hat{X}\right) }{f\left( \hat{X}%
\right) }\right) \frac{D-\left\Vert \Psi \left( \hat{X}\right) \right\Vert
^{2}}{\left\Vert \Psi \left( \hat{X}\right) \right\Vert ^{2}}-\frac{%
g^{2}\left( \hat{X}\right) C_{2}\left( p,\hat{X}\right) }{\sigma _{\hat{X}%
}^{2}}  \notag \\
n &=&\frac{\nabla _{\hat{X}}g\left( \hat{X}\right) C_{2}\left( p,\hat{X}%
\right) }{\left\vert f\left( \hat{X}\right) \right\vert }  \notag
\end{eqnarray}%
and:%
\begin{eqnarray*}
C_{1}\left( p,\hat{X}\right) &=&\frac{\sigma _{X}^{2}\sigma _{\hat{K}%
}^{2}\left( p+\frac{1}{2}\right) ^{2}\left( f^{\prime }\left( X\right)
\right) ^{2}}{96\left\vert f\left( \hat{X}\right) \right\vert ^{3}} \\
C_{2}\left( p,\hat{X}\right) &=&\ln \left( p+\frac{1}{2}\right) -\frac{%
2C_{1}\left( p,\hat{X}\right) }{p+\frac{1}{2}} \\
C_{3}\left( p,\hat{X}\right) &=&1-C_{1}\left( p,\hat{X}\right) +\left( p+%
\frac{3}{2}\right) C_{2}\left( p,\hat{X}\right)
\end{eqnarray*}%
To make the system self-consistent, and since $K_{\hat{X}}$ already depends
on $R$, we merely need to introduce an endogenous dynamics for $R$.

To do so, we assume that $R$ depends on $K_{\hat{X}},\hat{X}$ and $\nabla
_{\theta }K_{\hat{X}}$, and that this dependency has the form of a diffusion
process (see appendix 4.2). This leads to write $R$ as a function $R\left(
K_{\hat{X}},\hat{X},\nabla _{\theta }K_{\hat{X}}\right) $. The variation of $%
R$ is of the form:%
\begin{eqnarray}
\nabla _{\theta }R\left( \theta ,\hat{X}\right) &=&a_{0}\left( \hat{X}%
\right) \nabla _{\theta }K_{\hat{X}}+b\left( \hat{X}\right) \nabla _{\hat{X}%
}^{2}\nabla _{\theta }K_{\hat{X}}+c\left( \hat{X}\right) \nabla _{\theta
}\left( \nabla _{\theta }K_{\hat{X}}\right) +d\left( \hat{X}\right) \nabla
_{\theta }^{2}\left( \nabla _{\theta }K_{\hat{X}}\right)  \label{rvn} \\
&&+f\left( \hat{X}\right) \nabla _{\hat{X}}^{2}\left( \nabla _{\theta
}R\left( \theta ,\hat{X}\right) \right) +h\left( \hat{X}\right) \nabla
_{\theta }^{2}\left( \nabla _{\theta }R\left( \theta ,\hat{X}\right) \right)
\notag \\
&&+u\left( \hat{X}\right) \nabla _{\hat{X}}\nabla _{\theta }\left( \nabla
_{\theta }K_{\hat{X}}\right) +v\left( \hat{X}\right) \nabla _{\hat{X}}\nabla
_{\theta }\left( \nabla _{\theta }R\left( \theta ,\hat{X}\right) \right) 
\notag
\end{eqnarray}%
We can also assume that the coefficients in the expansion are slowly
varying, since they are obtained by computing averages.

This dynamics corresponds to a diffusion process: expected returns in one
sector depend on the variations of capital and returns in neighbouring
sectors.

To find the intrinsic dynamics for $K_{\hat{X}}$, we assume that the
exogenous variation $\frac{\nabla _{\theta }r\left( \hat{X}\right) }{r\left(
K_{\hat{X}},\hat{X}\right) }$ is null, and that the system of equations (\ref%
{krv}) and (\ref{rvn}) yields the dynamics for $\nabla _{\theta }K_{\hat{X}}$
and $\nabla _{\theta }R\left( \theta ,\hat{X}\right) $. Approximating these
dynamics to the first order in derivatives, we find:%
\begin{eqnarray}
&&0=\left( 
\begin{array}{cc}
\frac{k}{K_{\hat{X}}} & \frac{l}{R\left( \hat{X}\right) } \\ 
-a_{0}\left( \hat{X}\right) & 1%
\end{array}%
\right) \left( 
\begin{array}{c}
\nabla _{\theta }K_{\hat{X}} \\ 
\nabla _{\theta }R%
\end{array}%
\right)  \label{dqk} \\
&&-\left( 
\begin{array}{cc}
0 & \frac{2m}{\nabla _{\hat{X}}R\left( \hat{X}\right) }\nabla _{\hat{X}} \\ 
a\left( \hat{X}\right) \nabla _{\hat{X}}+c\left( \hat{X}\right) \nabla
_{\theta } & e\left( \hat{X}\right) \nabla _{\hat{X}}+g\left( \hat{X}\right)
\nabla _{\theta }%
\end{array}%
\right) \left( 
\begin{array}{c}
\nabla _{\theta }K_{\hat{X}} \\ 
\nabla _{\theta }R%
\end{array}%
\right)  \notag \\
&&-\left( 
\begin{array}{cc}
0 & -\frac{n}{\nabla _{\hat{X}}^{2}R\left( \hat{X}\right) }\nabla _{\hat{X}%
}^{2} \\ 
d\left( \hat{X}\right) \nabla _{\theta }^{2}+b\left( \hat{X}\right) \nabla _{%
\hat{X}}^{2}+u\left( \hat{X}\right) \nabla _{\hat{X}}\nabla _{\theta } & 
e\left( \hat{X}\right) \nabla _{\theta }^{2}+f\left( \hat{X}\right) \nabla _{%
\hat{X}}^{2}+v\nabla _{\hat{X}}\nabla _{\theta }%
\end{array}%
\right) \left( 
\begin{array}{c}
\nabla _{\theta }K_{\hat{X}} \\ 
\nabla _{\theta }R%
\end{array}%
\right)  \notag
\end{eqnarray}

\subsection{Oscillatory solutions}

We look for solutions of (\ref{dqk}) of the type:%
\begin{equation}
\left( 
\begin{array}{c}
\nabla _{\theta }K_{\hat{X}} \\ 
\nabla _{\theta }R\left( \hat{X}\right)%
\end{array}%
\right) =\exp \left( i\Omega \left( \hat{X}\right) \theta +iG\left( \hat{X}%
\right) \hat{X}\right) \left( 
\begin{array}{c}
\nabla _{\theta }K_{0} \\ 
\nabla _{\theta }R_{0}%
\end{array}%
\right)  \label{dqK}
\end{equation}%
with slowly varying $G\left( \hat{X}\right) $ and $\Omega \left( \hat{X}%
\right) $.\ We are then led to the relation between $\Omega \left( \hat{X}%
\right) $ and $G\left( \hat{X}\right) $: 
\begin{eqnarray*}
0 &=&\frac{k}{K_{\hat{X}}}\left( 1-ieG-ig\Omega \right) +\left( \frac{l}{%
R\left( \hat{X}\right) }-i\frac{2m}{\nabla _{\hat{X}}R\left( \hat{X}\right) }%
G\right) \left( a_{0}+iaG+ic\Omega \right) \\
&&-\frac{l}{R\left( \hat{X}\right) }\left( d\Omega ^{2}+bG^{2}+u\Omega
G\right) +\frac{k}{K_{\hat{X}}}\left( e\Omega ^{2}+fG^{2}+v\Omega G\right)
\end{eqnarray*}%
In the sequel, we will limit ourselves to the first order terms which yields
the expression for $\Omega $ (see appendix 4.3):%
\begin{eqnarray}
&&\Omega =\frac{\frac{lc}{R\left( \hat{X}\right) }\left( \frac{2ma_{0}}{%
\nabla _{\hat{X}}R\left( \hat{X}\right) }\right) G-\frac{2mc}{\nabla _{\hat{X%
}}R\left( \hat{X}\right) }G\left( \frac{k}{K_{\hat{X}}}+\frac{a_{0}l}{%
R\left( \hat{X}\right) }\right) }{\left( \frac{lc}{R\left( \hat{X}\right) }%
\right) ^{2}+\left( \frac{2mc}{\nabla _{\hat{X}}R\left( \hat{X}\right) }%
G\right) ^{2}}  \label{slg} \\
&&+i\frac{\frac{lc}{R\left( \hat{X}\right) }\left( \frac{k}{K_{\hat{X}}}+%
\frac{a_{0}l}{R\left( \hat{X}\right) }\right) +\frac{2mc}{\nabla _{\hat{X}%
}R\left( \hat{X}\right) }\left( \frac{2ma_{0}}{\nabla _{\hat{X}}R\left( \hat{%
X}\right) }\right) G^{2}}{\left( \frac{lc}{R\left( \hat{X}\right) }\right)
^{2}+\left( \frac{2mc}{\nabla _{\hat{X}}R\left( \hat{X}\right) }G\right) ^{2}%
}  \notag
\end{eqnarray}%
Once the frequencies (\ref{slg}) of oscillations found, we can determine
their condition of stability. When: 
\begin{equation}
\frac{lc}{R\left( \hat{X}\right) }\left( \frac{k}{K_{\hat{X}}}+\frac{a_{0}l}{%
R\left( \hat{X}\right) }\right) +\frac{4m^{2}ca_{0}}{\left( \nabla _{\hat{X}%
}R\left( \hat{X}\right) \right) ^{2}}G^{2}>0  \label{sln}
\end{equation}%
oscillations are dampened and return to the steady state. Otherwise,
oscillations are diverging: the system settles on another steady state, i.e.
another background state. Appendix 4.4 studies the condition (\ref{sln}) as
a function of the parameter functions $f\left( \hat{X}\right) $\ and $%
R\left( \hat{X}\right) $, the level of average capital $K_{\hat{X}}$, and
the coefficients arising in the expectations formations. The results are
presented in the next section.

\section{Interpretations of the results}

Let us now gather and interpret our results. We will discuss the
determinants of capital accumulation, its dependency in the parameters, the
several patterns of accumulation and the stability of the system, and detail
the dynamic system including endogenized expectations. We present a
synthesis of the main results obtained at the end of the section.

\subsection{Average capital in a given environment}

Capital accumulation and the stability of a configuration both depend on
several parameters. We will describe the determinants of capital
accumulation, its patterns, its dependency in parameters, before studying
the density of firms and investors per sector.

\subsubsection{Determinants of Capital accumulation}

Average capital in sector $\hat{X}$ is determined short-term returns, $%
f\left( \hat{X}\right) $, which decompose into dividends and price
fluctuations. Average capital is also determined by expected long-term
returns, $R\left( \hat{X}\right) $,$\ $which encompass the gross prospects
of the firm. Short term returns and long-term returns are not fully
independent: within short-term returns, price fluctuations are driven by
expected long-term returns.\textbf{\ }

A third determinant of average capital is the environment of sector $\hat{X}$%
, i.e. on the expected long-term return of neighbouring sectors.
Mathematically, this means that the derivatives of the expected long-term
return $R\left( \hat{X}\right) $ of a sector,\ along the sectors space do
matter for capital accumulation: average capital is determined by $\nabla _{%
\hat{X}}R\left( \hat{X}\right) $, $\nabla _{\hat{X}}^{2}R\left( \hat{X}%
\right) $\ and $\nabla _{\hat{X}}f\left( \hat{X}\right) $. Throughout the
derivation of the model, we have seen that the dependency of sector $\hat{X}$%
's capital accumulation\ in neighbouring sectors can be measured by the
parameter defined in (\ref{fpR}), $Y(\hat{X})$:%
\begin{equation}
Y(\hat{X})=M-\left( \frac{\left( g\left( \hat{X}\right) \right) ^{2}}{\sigma
_{\hat{X}}^{2}}+\nabla _{\hat{X}}g\left( \hat{X},K_{\hat{X}}\right) -\frac{%
\sigma _{\hat{K}}^{2}F^{2}\left( \hat{X},K_{\hat{X}}\right) }{2f^{2}\left( 
\hat{X}\right) }\right)  \label{nPT}
\end{equation}%
or, alternately, its value normalized by short-term returns, defined in (\ref%
{fRM}), up to a constant:%
\begin{equation}
p=\frac{M-\left( \frac{\left( g\left( \hat{X},K_{\hat{X}_{M}}\right) \right)
^{2}}{\sigma _{\hat{X}}^{2}}+\nabla _{\hat{X}}g\left( \hat{X},K_{\hat{X}%
_{M}}\right) -\frac{\sigma _{\hat{K}}^{2}F^{2}\left( \hat{X},K_{\hat{X}%
}\right) }{2f^{2}\left( \hat{X}\right) }\right) }{f\left( \hat{X}\right) }+%
\frac{3}{2}  \label{nPR}
\end{equation}

From these equations, we can see that both $Y(\hat{X})$ and $p$ depend on
the gradients of long-term returns $R\left( \hat{X}\right) $ through the
function $g\left( \hat{X}\right) $, capital mobility at sector $\hat{X}$.\ 

This function $g\left( \hat{X}\right) $, which depicts investors' propensity
to seek higher returns across sectors, is indeed proportional to $\nabla _{%
\hat{X}}R\left( \hat{X}\right) $. The gradient of $g$, $\nabla _{\hat{X}}g$,
is proportional to $\nabla _{\hat{X}}^{2}R\left( \hat{X}\right) $: it
measures the position of the sector relative to its neighbours.\ At a
maximum, $\nabla _{\hat{X}}^{2}R\left( \hat{X}\right) <0$, at a minimum, $%
\nabla _{\hat{X}}^{2}R\left( \hat{X}\right) >0$.

The last term, $\frac{\sigma _{\hat{K}}^{2}F^{2}\left( \hat{X},K_{\hat{X}%
}\right) }{2f^{2}\left( \hat{X}\right) }$, involved in the definition of $Y(%
\hat{X})$ and $p$ is a smoothing factor between neighbours' sectors. It will
be discussed below.

The parameters $Y(\hat{X})$ and $p$ defined in (\ref{nPT}) and (\ref{nPR})
represent the relative attractivity of a sector vis-a-vis its neighbours.
The parameter $p$ is normalized by short-term returns. It computes the ratio
of relative attractivity to short-term returns and allows to consider these
two variables separately.

\ Both parameters, $Y(\hat{X})$\ and $p$,\ are maximal when $R\left( \hat{X}%
\right) $\ is maximum (see section 8.2.1 and 8.2.2). More generally, the
higher $Y(\hat{X})$\ and $p$, the more attractive is sector $\hat{X}$\
relative to its neighbours.

\subsubsection{Patterns of capital accumulation}

Section 8.2 has showed that, for each sector $\hat{X}$, the equation (\ref%
{qtk}) for average capital per firm per sector has, in general, several
solutions: these potential equilibria depend on the parameter-functions \ $%
f\left( \hat{X}\right) $, $R\left( \hat{X}\right) $ and $Y\left( \hat{X}%
\right) $, and on firms' densities $\left\Vert \Psi \left( X,K_{X}\right)
\right\Vert ^{2}$. Each of these parameters influence eachothers. Three
patterns of capital accumulation arise.\ Some will be deemed stable, others
unstable with respect to some variations in parameters, a notion that will
be detailed later on.

\paragraph{First pattern: low capital, high short-term returns driven by
dividends only.}

In this pattern, the low level of capital implies that firms' returns rely
on short term returns $f\left( \hat{X}\right) $\ through dividends rather
than on expected long-term returns. Here, for low capital, dividends are
driven by a high marginal productivity. Capital accumulation depends mainly
on short-term returns $f\left( \hat{X}\right) $. Such sectors are in general
stable. Agents are scarce: there is a niche effect. Changes in parameters
impact $f\left( \hat{X}\right) $ through a change in marginal returns are
adjusted for by a change in the number of firms.

\paragraph{Second pattern: intermediate to high level of capital, short-term
returns, long-term expectations.}

Capital accumulation increases with any parameter that increases short-term
returns - dividends and stock prices - or long-term returns through relative
attractivity $Y\left( \hat{X}\right) $, as detailed above: locally, the
higher the sector relative attractivity, the higher capital accumulation.
This is the most standard pattern of capital allocation.

\paragraph{Third pattern: high capital, long-term returns and relative
attractivity\ }

In these sectors, capital accumulation depends on high expected long-term
returns, themselves sustained by high levels of capital. Sectors with
maximal expected returns, i.e. maximal attractivity, dominate their
neighbours and may accumulate extremely high levels of capital. However,
such equilibria are unstable (see section 8.1): in a dynamic perspective,
all else equal, capital could grow indefinitely.

Combined, the two last patterns show that, in sectors where expected
long-term returns are maximal, two outcomes may arise: a stable pattern of
high capital, and an unstable pattern.

A last particular case arise in the third pattern.\ It is a limit case of
our model, where low expected returns do not deter extremely high capital.
However this case is unsustainable.

\subsubsection{Dependency in parameters}

The stability of the configuration influences the dependency in the
parameters of the system.

Recall that in the stable case (patterns 1,2 and partly 3), average values
can be understood as equilibrium values (see section 8.1), but that unstable
equilibria rather describe potential thresholds, not actual ones.

In\ locally stable configurations, average capital is increasing in
short-term returns, expected long-term returns, and relative attractivity of
the sector, $f\left( \hat{X}\right) $, $R\left( \hat{X}\right) $ and $%
Y\left( \hat{X}\right) $ respectively. The higher the returns, the higher
the capital accumulation.

For unstable equilibria, on the contrary, average capital is decreasing in
these variables: an increase in short-term returns or expected long-term
returns facilitates capital accumulation and reduces the threshold. When
capital moves below these thresholds, the sector average capital move toward
the next stable equilibrium, whereas when capital exceeds the threshold, the
sector's average capital per firm grows indefinitely.

This form of instability will be discussed below, and will later lead to
consider the dynamics aspect of the average values of capital per sector at
macro-time scale.

Lastly, we can compare the relative effect of parameters' variation on
neighbouring sectors. An equal increase of long-term returns per unit of
capital, $R\left( \hat{X}\right) $,\ in two close sectors favours the best
capitalised sector, since its total returns' expectations are higher.
Capital flows thus increase inequality between neighbouring sectors. On the
other hand, an increase in\ global productivity impacts short-turn returns, $%
f\left( \hat{X}\right) $, of least capitalized sectors and lures in
investors.

\subsubsection{Density of firms per sector}

Formula (\ref{psl}) defines the density of firms per sector.\ Its central
parameter depends on expected long-term returns. For any given level of
capital, the number of firms in sector $X$ is given by (\ref{psl}):%
\begin{equation}
\left\Vert \Psi \left( X\right) \right\Vert ^{2}=\left( 2\tau \right)
^{-1}\left( D\left( \left\Vert \Psi \right\Vert ^{2}\right) -\frac{1}{2}%
\left( \left( \nabla _{X}R\left( X\right) \right) ^{2}+\frac{\sigma
_{X}^{2}\nabla _{X}^{2}R\left( K_{X},X\right) }{H\left( K_{X}\right) }%
\right) H^{2}\left( K_{X}\right) \left( 1-\frac{H^{\prime }\left( \hat{K}%
_{X}\right) K_{X}}{H\left( \hat{K}_{X}\right) }\right) \right)  \label{pSL}
\end{equation}%
and this function is decreasing in: 
\begin{equation*}
\left( \nabla _{X}R\left( X\right) \right) ^{2}+\frac{\sigma _{X}^{2}\nabla
_{X}^{2}R\left( K_{X},X\right) }{H\left( K_{X}\right) }
\end{equation*}%
Recall that $\nabla _{X}R\left( X\right) $ is the gradient of expected
long-term returns relative to the sectors space, and that $\nabla
_{X}^{2}R\left( K_{X},X\right) $ is the Laplacian, i.e. the generalisation
of the second derivative of $R\left( K_{X},X\right) $ with respect to the
sectors' space.

When $\nabla _{X}R\left( X\right) \neq 0$, the sector is only 'transitory".\
The sector $X$\ is on a slope: neighbouring sectors have lower expected
returns, others have higher ones. Firms head towards the sectors with higher
returns. The higher the slope $\nabla _{X}R\left( X\right) $, the faster
firms leave the sector.

For $\nabla _{X}R\left( X\right) =0$ and $\nabla _{X}^{2}R\left(
K_{X},X\right) <0$, sector $X$'s returns $R\left( X\right) $ are at a local
maximum.\ Formula (\ref{pSL}) shows that the number of firms in the sector
is maximal, since firms tend to accumulate capital in most profitable
sectors. Thus, at these points, there are both a large number of firms and
high level of capital per firm. Yet competition ensures that some firms do
remain in sectors with low, or even minimal expected returns. We will see
that these equilibria are unstable.

For $\nabla _{X}R\left( X\right) =0$ and $\nabla _{X}^{2}R\left(
K_{X},X\right) >0$, the density of firms is much lower than in the case where%
\textbf{\ }$\nabla _{X}^{2}R\left( K_{X},X\right) <0$. Potential incomers
are crowded out by competitors with higher capital. Competition also implies
that, for given returns $R\left( X\right) $, the number of firms in a sector 
$X$\ decreases with the average level of capital $K_{X}$. Equilibria in
sector $X$\ with high level of capital per firm $K_{X}$\ have a relatively
low number of firms: firms with high capital deter incomers.

\subsubsection{Density of investors per sector}

Formula (\ref{dns}) shows that the average number of investors at sector $%
\hat{X}$ is an increasing function of short-term returns,\ $f\left( \hat{X}%
\right) $, and the sector $\hat{X}$'s relative long-term attractivity, $p$,
defined in equation (\ref{fRM}). All else equal, an increase in short-term
returns or an improvement of the sector's relative long-term attractivity
increases the number of investors and, through them, the disposable capital
for firms.

The density of investors in sector $\hat{X}$ increases with its relative
attractivity $p$. \ which can be written as:%
\begin{equation*}
p=\frac{M-\left( \left( g\left( \hat{X}\right) \right) ^{2}+\sigma _{\hat{X}%
}^{2}\left( f\left( \hat{X}\right) +\nabla _{\hat{X}}g\left( \hat{X},K_{\hat{%
X}}\right) \right) \right) }{\sigma _{\hat{X}}^{2}\sqrt{f^{2}\left( \hat{X}%
\right) }}+\frac{\sigma _{\hat{K}}^{2}F^{2}\left( \hat{X},K_{\hat{X}}\right) 
}{2\sigma _{\hat{X}}^{2}\left( \sqrt{f^{2}\left( \hat{X}\right) }\right) ^{3}%
}
\end{equation*}%
where the first term is the attractivity of sector's $\hat{X}$ relative to
its neighbours, normalized by short-term returns $f\left( \hat{X}\right) $,
and the second is a smoothing term that reduces differences between sectors:
it increases when the relative attractivity with respect to $K_{\hat{X}}$
decreases. The number of investors and capital will increase in sectors\
that are in the neighbourhood of significantly more attractive sectors, i.e.
with higher average capital and number of investors.

\subsection{(In)Stability}

Two sources of instability may arise in the model. One is local, and stems
from solutions for average capital per firm per sector, equation (\ref{qtk}%
).\ Another is global, and stems from the constraint imposed, in the model,
on the total number of investors.

\subsubsection{Local stability of capital accumulation patterns}

\paragraph{Mechanisms of local instability}

The equation (\ref{qtk}) defining average capital per firm per sector can
itself be seen as the stable point of a dynamical equation with varying
parameters\footnote{%
The definitions of the parameters have been given after equation (\ref{dtt}).%
} (see (\ref{dts})):%
\begin{eqnarray}
\delta K_{\hat{X}}\left( t+1\right) &=&\left( -\left( \frac{\frac{\partial
f\left( \hat{X},K_{\hat{X}}\right) }{\partial K_{\hat{X}}}}{f\left( \hat{X}%
,K_{\hat{X}}\right) }+\frac{\frac{\partial \left\Vert \Psi \left( \hat{X},K_{%
\hat{X}}\right) \right\Vert ^{2}}{\partial K_{\hat{X}}}}{\left\Vert \Psi
\left( \hat{X},K_{\hat{X}}\right) \right\Vert ^{2}}+l\left( \hat{X},K_{\hat{X%
}}\right) \right) +k\left( p\right) \frac{\partial p}{\partial K_{\hat{X}}}%
\right) K_{\hat{X}}\delta K_{\hat{X}}\left( t\right)  \notag \\
&&+\frac{\partial }{\partial Y\left( \hat{X},t\right) }\left( \frac{\sigma _{%
\hat{K}}^{2}C\left( \bar{p}\right) 2\hat{\Gamma}\left( p+\frac{1}{2}\right) 
}{\left\vert f\left( \hat{X},K_{\hat{X}}\right) \right\vert \left\Vert \Psi
\left( \hat{X},K_{\hat{X}}\right) \right\Vert ^{2}}\right) \delta Y\left( 
\hat{X},t\right)
\end{eqnarray}%
Capital accumulation is potentially instable for some sectors (see (\ref{sTB}%
)), when:%
\begin{equation}
\left\vert B\left( \hat{X}\right) \right\vert \equiv \left\vert k\left(
p\right) \frac{\partial p}{\partial K_{\hat{X}}}-\left( \frac{\frac{\partial
f\left( \hat{X},K_{\hat{X}}\right) }{\partial K_{\hat{X}}}}{f\left( \hat{X}%
,K_{\hat{X}}\right) }+\frac{\frac{\partial \left\Vert \Psi \left( \hat{X},K_{%
\hat{X}}\right) \right\Vert ^{2}}{\partial K_{\hat{X}}}}{\left\Vert \Psi
\left( \hat{X},K_{\hat{X}}\right) \right\Vert ^{2}}+l\left( \hat{X},K_{\hat{X%
}}\right) \right) \right\vert >1  \label{nNT}
\end{equation}%
When this inequality holds, the equilibrium is not a steady state: a
variation $\delta Y\left( \hat{X},t\right) $ drives the system away from
potential equilibria. This variation shifts the average capital per firm $K_{%
\hat{X}}$ via the four terms in $\left\vert B\left( \hat{X}\right)
\right\vert $:

The first term $\frac{\partial f\left( \hat{X},K_{\hat{X}}\right) }{\partial
K_{\hat{X}}}$ captures the relative variation of short-term returns,
dividends and price fluctuations.

The second term is the variation in the number of firms moving in, or out
of, sector $K_{\hat{X}}$.

The two last term $l\left( \hat{X},K_{\hat{X}}\right) $ and $k\left(
p\right) \frac{\partial p}{\partial K_{\hat{X}}}$ compute the modification
in the background field induced by the indirect change $\delta Y\left( \hat{X%
},t\right) $.

The term $l\left( \hat{X},K_{\hat{X}}\right) $ is the direct variation
induced by a modification of $f\left( \hat{X}\right) $.

The term $k\left( p\right) \frac{\partial p}{\partial K_{\hat{X}}}$ is the
variation induced by a modification of $p\left( \hat{X}\right) $, the
relative position of sector $\hat{X}$ in the space of return, which depends
on the shape of the returns around $\hat{X}$.

Thus, the modification of one parameter affects the system as a whole, and
reshapes the collective state through modifications of the background field.

The sum of these four modifications magnifies or dampens any initial
modification and, depending on the amplitude of $\left\vert B\left( \hat{X}%
\right) \right\vert $, determines the stability of the system.

\paragraph{Consequences of local instability}

Instability can arise in two cases\footnote{%
We assume physical capital returns are Cobb-Douglas.}:

When\textbf{\ }$B\left( \hat{X}\right) <-1$\textbf{, }$K_{\hat{X}%
}\rightarrow 0$ and the short-term return becomes infinite\textbf{\ }$%
f\left( \hat{X}\right) \rightarrow \infty $\textbf{\ . }In such a case, the
background field becomes null at sector $\hat{X}$\textbf{, }$\left\Vert \hat{%
\Psi}\left( \hat{X},\hat{K}\right) \right\Vert ^{2}\rightarrow 0$, i.e.
firms desert the sector\textbf{.} This corresponds to the extreme case of a
niche effect. Investors and firms are scarce and returns are high, since,
for a very low capital marginal productivity is mathematically high. However
the total capital involved in this case is negligible, and does not impact
the system globally.

When $B\left( \hat{X}\right) >1$, $K_{\hat{X}}\rightarrow \infty $ and $%
f\left( \hat{X}\right) \rightarrow c$ for some constant $c<<1$.

Given the definition of the parameters, we can assume that for $K_{\hat{X}%
}\rightarrow \infty $, $\frac{\partial f\left( \hat{X},K_{\hat{X}}\right) }{%
\partial K_{\hat{X}}}\rightarrow 0$ so that $\frac{\partial p}{\partial K_{%
\hat{X}}}\rightarrow 0$. Moreover, $l\left( \hat{X},K_{\hat{X}}\right)
\rightarrow 0$. As a consequence, (\ref{nNT}) becomes:%
\begin{equation*}
\frac{\frac{\partial \left\Vert \Psi \left( \hat{X},K_{\hat{X}}\right)
\right\Vert ^{2}}{\partial K_{\hat{X}}}}{\left\Vert \Psi \left( \hat{X},K_{%
\hat{X}}\right) \right\Vert ^{2}}>1
\end{equation*}%
which implies that $\left\Vert \Psi \left( \hat{X},K_{\hat{X}}\right)
\right\Vert ^{2}$ behaves approximatively as: 
\begin{equation*}
\left\vert \hat{\Psi}\left( \hat{X},\hat{K}\right) \right\vert ^{2}\simeq
\exp \left( rK_{\hat{X}}\right)
\end{equation*}%
with $r>1$.

Since $K_{\hat{X}}\rightarrow \infty $, the background field $\left\vert 
\hat{\Psi}\left( \hat{X},\hat{K}\right) \right\vert ^{2}\rightarrow \infty $
and the number of firms grows indefinitely. The number of agents is fixed:
they move towards sector $\hat{X}$, until some maximum number of agents is
reached, and: 
\begin{equation*}
\left\vert \hat{\Psi}\left( \hat{X},\hat{K}\right) \right\vert
^{2}=\left\vert \hat{\Psi}\left( \hat{X},\hat{K}\right) \right\vert _{\max
}^{2}>>1
\end{equation*}%
and the capital at sector $\hat{X}$ reaches:%
\begin{equation*}
K_{\hat{X}}\simeq K_{\max }=\frac{\ln \left( \left\vert \hat{\Psi}\left( 
\hat{X},\hat{K}\right) \right\vert _{\max }^{2}\right) }{r}
\end{equation*}%
So that when a point\ exists where $B\left( \hat{X}\right) >1$, the total
remaining capital available $V\left\langle K\right\rangle $ is reduced to $%
V\left\langle K\right\rangle -K_{\max }$, with $V$ the volume of the sector
space and $\left\langle K\right\rangle $ the average physical capital in the
space. Average capital available reduces to $\left\langle K\right\rangle -%
\frac{K_{\max }}{V}$, which directly impacts other points in the sectors'
space, since capital accumulation in one point depends explicitly on the
function:%
\begin{equation*}
F_{1}\left( \frac{R\left( K_{\hat{X}},\hat{X}\right) }{\int R\left(
K_{X^{\prime }}^{\prime },X^{\prime }\right) \left\Vert \Psi \left(
X^{\prime }\right) \right\Vert ^{2}dX^{\prime }}\right)
\end{equation*}%
The function $F_{1}$ depends on $\left\langle K\right\rangle -\frac{K_{\max }%
}{V}$, which in turn modifies the function $f\left( \hat{X}\right) $ and $%
\left\vert B\left( \hat{X}\right) \right\vert $ over the whole space: some
points move over the threshold $B\left( \hat{X}\right) >1$, others below $%
B\left( \hat{X}\right) <-1$. Some sectors experience a capital increase,
others disappear. As a consequence, if a stable situation finally emerges,
the sector space may be considered as a reduced one: some sectors disappear,
and only sectors with positive capital remain. Ultimately, the state defined
by $\hat{\Psi}\left( \hat{X},\hat{K}\right) $ should be transformed into a
field $\hat{\Psi}_{\text{red}}\left( \hat{X},\hat{K}\right) $ of stable
equilibria on this reduced space: the model should be transformed ultimately
in another one.

\subsubsection{Global stability}

A second source of instability of the system arises outside of the equations
for average capital per firm per sector, (\ref{qtk}), and its differential
version, (\ref{rvd}).\ It stems from the sectors' space expected long-term
returns. It is induced by the minimization equations (\ref{hqn}) and (\ref%
{nqh}), and is a source of global instability for the background field.

\paragraph{Description of global instability}

In these equations, the Lagrange multiplier $\hat{\lambda}$ is the
eigenvalue of a second-order differential equation. Because there exist an
infinite number of eigenvalues $\hat{\lambda}$, there are an infinite number
of local minimum background fields $\Psi \left( \hat{X},K_{\hat{X}}\right) $%
.\ But the most likely minimum, given in (\ref{mqn}), is obtained for $\hat{%
\lambda}=M$ (see appendix 2).

Yet\textbf{\ }$\hat{\lambda}$ is also the Lagrange multiplier that
implements the constraint of a fixed number $N$\ of agents.\ 

Since the number of investors is computed by:%
\begin{equation*}
\int \left\Vert \Psi \left( \hat{X},K_{\hat{X}}\right) \right\Vert
^{2}d\left( \hat{X},K_{\hat{X}}\right)
\end{equation*}%
the constraint implemented by $\hat{\lambda}$ is: 
\begin{equation}
\hat{N}=\int \left\Vert \Psi \left( \hat{X},K_{\hat{X}}\right) \right\Vert
^{2}d\left( \hat{X},K_{\hat{X}}\right)  \label{grandn}
\end{equation}%
since this constraint runs over the whole space, it is a global property of
the system.

Yet equations (\ref{hqn}) and (\ref{nqh}), the minimization equations
defining $\Psi \left( \hat{X},K_{\hat{X}}\right) $, may also be viewed as a
set of local minimization equations at each point $\hat{X}$ of the sector
space. Considered individually, each provide a lower minimum that could be
reached separately for each $\hat{X}$. In other words, provided each
sector's number of agents is fixed independently from the rest of the
system, a stable background field could be reached at every point.

However, our global constraint rules out this set of local minimizations.\
The solutions of (\ref{hqn}) and (\ref{nqh}) are thus a local minimum for
the sole points $\hat{X}$\ such that $M$: the lowest value of $\hat{\lambda}$
is reached at $\hat{X}$, such that:%
\begin{equation}
Y\left( \hat{X}\right) =\frac{\left( g\left( \hat{X}\right) \right) ^{2}}{%
\sigma _{\hat{X}}^{2}}+f\left( \hat{X}\right) +\frac{1}{2}\sqrt{f^{2}\left( 
\hat{X}\right) }+\nabla _{\hat{X}}g\left( \hat{X},K_{\hat{X}}\right) -\frac{%
\sigma _{\hat{K}}^{2}F^{2}\left( \hat{X},K_{\hat{X}}\right) }{2f^{2}\left( 
\hat{X}\right) }=M  \label{stl}
\end{equation}%
For points $\hat{X}$ that do not satisfy (\ref{stl}), the solution $\Psi
\left( \hat{X},K_{\hat{X}}\right) $ and $\Psi ^{\dag }\left( \hat{X},K_{\hat{%
X}}\right) $\ of (\ref{hqn}) and (\ref{nqh}), with $\hat{\lambda}=M$ are not
a global minimum, but merely a local one. Any perturbation $\delta \Psi
\left( \hat{X},K_{\hat{X}}\right) $ in the parameters destabilize the
system: the equilibrium is unstable.

The stability of both the background field and the potential equilibria are
thus determined by $Y\left( \hat{X}\right) $, the sector space's overall
shape of returns and expectations. An homogeneous shape, a space such that $%
Y\left( \hat{X}\right) $, presents small deviations around $M$ and is more
background-stable than an heterogeneous space.

More importantly, the background fields and associated average capital must
be understood as potential, not actual long-run equilibria: the whole system
is better described as a dynamical system, which will be defined in section
9, between potential backgrounds where time enters as a macro-variable. We
consider the results of the background field's dynamical behaviour in
section 10.3.

\paragraph{Reducing agents mobility and removing instability}

As mentioned above, an homogeneous shape is a space such that $Y\left( \hat{X%
}\right) $ presents small deviations around $M$. In an heterogeneous shape,
the space presents large differences in $Y\left( \hat{X}\right) $. We find
that homogeneous shapes are more background-stable than heterogeneous ones.
This partly results from the global constraint (\ref{grandn}) imposed on the
number of agents in the model, which ensures that the number of financial
agents in the system is fixed over the whole sector space.

Relaxing this constraint fully would render the number of agents in sectors
independent.\ The associated background field of each sector could, at each
point, adjust to be minimum and stabilize the system.

To do so, we replace equation (\ref{hqn}), the minimization equation, by a
set of independent equations with independent Lagrange multipliers $\hat{%
\lambda}_{\hat{X}}$ for each sector $\hat{X}$: 
\begin{eqnarray}
0 &=&\left( \frac{\sigma _{\hat{X}}^{2}}{2}\nabla _{\hat{X}}^{2}-\frac{1}{%
2\sigma _{\hat{X}}^{2}}\left( g\left( \hat{X},K_{\hat{X}}\right) \right)
^{2}-\frac{1}{2}\nabla _{\hat{X}}g\left( \hat{X},K_{\hat{X}}\right) \right) 
\hat{\Psi}  \label{mN} \\
&&+\left( \nabla _{\hat{K}}\left( \frac{\sigma _{\hat{K}}^{2}}{2}\nabla _{%
\hat{K}}-\hat{K}f\left( \hat{X},K_{\hat{X}}\right) \right) -\hat{\lambda}_{%
\hat{X}}\right) \hat{\Psi}-F\left( \hat{X},K_{\hat{X}}\right) \hat{K}\hat{%
\Psi}  \notag
\end{eqnarray}%
For each $\hat{X}$, the minimum configuration is reached by setting: 
\begin{equation*}
\hat{\lambda}_{\hat{X}}=\frac{\left( g\left( \hat{X}\right) \right) ^{2}}{%
\sigma _{\hat{X}}^{2}}+f\left( \hat{X}\right) +\frac{1}{2}\sqrt{f^{2}\left( 
\hat{X}\right) }+\nabla _{\hat{X}}g\left( \hat{X},K_{\hat{X}}\right) -\frac{%
\sigma _{\hat{K}}^{2}F^{2}\left( \hat{X},K_{\hat{X}}\right) }{2f^{2}\left( 
\hat{X}\right) }
\end{equation*}%
which is similar to the Lagrange multiplier (\ref{mqn}) of the minimization
equation for the background field, stripped of the maximum condition, and
where the average capital equation (\ref{qtk}) is replaced by\footnote{%
Expression (\ref{Ghf}) is used to compute $\hat{\Gamma}\left( \frac{1}{2}%
\right) $.}:%
\begin{equation}
K_{\hat{X}}\left\Vert \Psi \left( \hat{X}\right) \right\Vert ^{2}\left\vert
f\left( \hat{X}\right) \right\vert =C\left( \bar{p}\right) \sigma _{\hat{K}%
}^{2}\hat{\Gamma}\left( \frac{1}{2}\right) =C\left( \bar{p}\right) \sigma _{%
\hat{K}}^{2}\exp \left( -\frac{\sigma _{X}^{2}\sigma _{\hat{K}}^{2}\left(
f^{\prime }\left( X,K_{\hat{X}}\right) \right) ^{2}}{384\left\vert f\left( 
\hat{X},K_{\hat{X}}\right) \right\vert ^{3}}\right)  \label{gbd}
\end{equation}%
This equation is identical to (\ref{kpt}) and has thus at least one locally
stable solution. The solutions are computed in (\ref{Kms}) and (\ref{Kmn}).

Solutions to (\ref{gbd}) do no longer directly depend on the relative
characteristics of a particular sector, but rather on the returns at point $%
f\left( \hat{X}\right) $ and on the number of firms in the sector, $%
\left\Vert \Psi \left( \hat{X}\right) \right\Vert ^{2}$. Yet this dependency
is only indirect, through the firms' density at sector $\hat{X}$, $%
\left\Vert \Psi \left( \hat{X}\right) \right\Vert ^{2}$, and this quantity
does not vary much in the sector space.

An intermediate situation between (\ref{qtk}) and (\ref{gbd}) could also be
considered: it would be to assume a constant number of agents in some
regions of the sector space.

Alternatively, limiting the number of investors per sector can be achieved
through some public regulation to maintain a constant flow of investment in
the sector.

\subsection{Average capital in a dynamic environment}

Recall that we have extended our model in section 9 by endogenizing the
expected long-term returns $R\left( \hat{X}\right) $. We have found a
dynamic equation (\ref{dqk}) for the two variables $\left( K_{\hat{X}%
},R\left( \hat{X}\right) \right) $ system, and shown that this system has
oscillatory solutions (\ref{dqK}):

\begin{equation}
\left( 
\begin{array}{c}
\nabla _{\theta }K_{\hat{X}} \\ 
\nabla _{\theta }R\left( \hat{X}\right)%
\end{array}%
\right) =\exp \left( i\Omega \left( \hat{X}\right) \theta +iG\left( \hat{X}%
\right) \hat{X}\right) \left( 
\begin{array}{c}
\nabla _{\theta }K_{0} \\ 
\nabla _{\theta }R_{0}%
\end{array}%
\right)
\end{equation}%
where the frequencies $\Omega \left( \hat{X}\right) $\ satisfy (\ref{slg}):%
\begin{eqnarray}
&&\Omega =\frac{\frac{lc}{R\left( \hat{X}\right) }\left( \frac{2ma_{0}}{%
\nabla _{\hat{X}}R\left( \hat{X}\right) }\right) G-\frac{2mc}{\nabla _{\hat{X%
}}R\left( \hat{X}\right) }G\left( \frac{k}{K_{\hat{X}}}+\frac{a_{0}l}{%
R\left( \hat{X}\right) }\right) }{\left( \frac{lc}{R\left( \hat{X}\right) }%
\right) ^{2}+\left( \frac{2mc}{\nabla _{\hat{X}}R\left( \hat{X}\right) }%
G\right) ^{2}}  \label{slG} \\
&&+i\frac{\frac{lc}{R\left( \hat{X}\right) }\left( \frac{k}{K_{\hat{X}}}+%
\frac{a_{0}l}{R\left( \hat{X}\right) }\right) +\frac{2mc}{\nabla _{\hat{X}%
}R\left( \hat{X}\right) }\left( \frac{2ma_{0}}{\nabla _{\hat{X}}R\left( \hat{%
X}\right) }\right) G^{2}}{\left( \frac{lc}{R\left( \hat{X}\right) }\right)
^{2}+\left( \frac{2mc}{\nabla _{\hat{X}}R\left( \hat{X}\right) }G\right) ^{2}%
}  \notag
\end{eqnarray}%
The coefficients arising in the frequency equation, given in (\ref{cFT}) and
(\ref{rvn}), depend on $K_{\hat{X}}$\ and on the form of the expectations'
formation (\ref{rvn}).

We will now study the frequency $\Omega $\ in equation (\ref{slG}) as a
function of both average capital $K_{\hat{X}}$\ and the form of expectations%
\footnote{%
See appendix 3.}. We are mainly interested in the imaginary part of (\ref%
{slg}), whose sign determines whether the dynamics is stable.

We have seen in (\ref{sln}) that the condition for stable oscillations is:%
\begin{equation}
\frac{lc}{R\left( \hat{X}\right) }\left( \frac{k}{K_{\hat{X}}}+\frac{a_{0}l}{%
R\left( \hat{X}\right) }\right) +\frac{4m^{2}ca_{0}}{\left( \nabla _{\hat{X}%
}R\left( \hat{X}\right) \right) ^{2}}G^{2}>0  \label{sLN}
\end{equation}%
When this condition is satisfied, oscillations at point $\hat{X}$\ dampen
and the system returns to its equilibrium, i.e. the background state.
Otherwise, diverging oscillations lead the system to the next background
state.

To investigate these two possibilities, the most important parameter is the
coefficient $c$\ in equation (\ref{sLN}) that has been defined in equation (%
\ref{rvn}).\ Its values determine two relevant forms of expectations.
Equation (\ref{rvn}) shows that, when $c>0$, expectations are highly
reactive to variations in capital, i.e. expected long-term returns depend
positively on the variations of average capital $K_{\hat{X}}$. When $c<0$,
expectations are moderately reactive to variations in capital, i.e. expected
long-term returns depend negatively on the variations of average capital $K_{%
\hat{X}}$.

\subsubsection{Case 1: $K_{\hat{X}}<<1$}

When average capital is very low, for $K_{\hat{X}}<<1$, the dominant
coefficient in the equation defining the frequencies of oscillations (\ref%
{slg}) is $\left\vert \frac{k}{K_{\hat{X}}}\right\vert >>l>>1$, so that the
solution of (\ref{slg}) is stable if:%
\begin{equation}
\frac{cl}{R\left( \hat{X}\right) }\frac{k}{K_{\hat{X}}}>0  \label{Sgg}
\end{equation}%
Appendix 4.4 shows that for $K_{\hat{X}}<<1$ we have $k<0$. As a
consequence, equation (\ref{Sgg}) implies that, for $K_{\hat{X}}\leqslant 1$
and $c>0$, i.e. for highly reactive expectations, oscillations are unstable,
whereas for $c<0$, i.e. moderately reactive expectations, they are stable.

This is in line with the notion of stable average capital defined above.
Actually, for such stable values, there is a positive relation between
variations of $R\left( \hat{X}\right) $ and average capital.

In addition, for $c>0$, expected long-term returns depend positively on the
variations of average capital $K_{\hat{X}}$. This creates an amplification
in the dynamics of these two variables.

On the contrary, for $K_{\hat{X}}\leqslant 1$ and $c<0$, the expected
long-term returns depend negatively on the variations of average capital $K_{%
\hat{X}}$. Stabilization occurs with dampening oscillations. This shows that
for expectations moderately reactive to variations of capital, some
equilibria with relative low capital are possible and resilient to
oscillations in expectations, a niche effect may exist for some sectors.

\subsubsection{Case 2: $K_{\hat{X}}>>1$}

For $K_{\hat{X}}>>1$, on the contrary, both in stable and unstable
equilibrium with high average capital, oscillations are dampening for $c>0$
and explosive for $c<0$.

When $c>0$, expectations are highly reactive and this leads to an
amplification between variations of capital and return expectations.

In the stable case, high average capital depends on this amplification by
expectations: fluctuations that would otherwise be destabilizing for low
capital sectors may stabilize or maintain stable sectors with high level of
capital. This does not mean that these sectors become attraction points,%
\textbf{\ }but rather that a large reactivity between expectations and
capital will allow for their intrinsic high level of capital to consolidate.

In the unstable case, due to the specificity of the equilibrium for $K_{\hat{%
X}}>>1$, the mechanism of stabilization is as follows: an initial increase
in $K_{\hat{X}}$\ implies an increase in the expected long-term return. But
in turn, the negative relation between $K_{\hat{X}}$\ and the variations of $%
R\left( \hat{X}\right) $\ lowers the average value of the average capital.
Thus, an increase in equilibrium capital $K_{\hat{X}}$\ improves sector $%
\hat{X}$'s profitability, which in turn lowers the entry's threshold in this
sector and ultimately reduces the potential equilibrium level of capital.
Put differently, an initial rise in the sector's average capital impacts
with amplification the sector's expected return, which reduces the potential
average equilibrium by an amount that offsets the initial rise in capital.

For $c<0$, moderately reactive expectations impair the dampening
oscillations mechanism that arise for $c>0$. For instance in the unstable
case, an initial rise in the threshold $K_{\hat{X}}$, impacts moderately
sectors' expected returns, without offsetting the initial rise in capital.

\subsubsection{Case 3: Intermediate values $\infty >K_{\hat{X}}>1$}

For intermediate values of capital, several possibilities arise.

When $c<0$, the oscillations are stable if: 
\begin{equation}
\frac{a_{0}}{R\left( \hat{X}\right) }-\frac{1-\alpha }{\varsigma K_{\hat{X}%
}F_{1}\left( R\left( K_{\hat{X}},\hat{X}\right) \right) }<0  \label{stc}
\end{equation}%
and:%
\begin{equation}
G^{2}<\frac{l^{2}\left( \nabla _{\hat{X}}R\left( \hat{X}\right) \right) ^{2}%
}{4a_{0}R\left( \hat{X}\right) }\left( \frac{\varsigma \left( \nabla _{\hat{X%
}}R\left( \hat{X}\right) F_{1}\left( R\left( K_{\hat{X}},\hat{X}\right)
\right) \left\Vert \Psi \left( \hat{X}\right) \right\Vert ^{2}\right) }{%
\gamma \left( D-\left\Vert \Psi \left( \hat{X}\right) \right\Vert
^{2}\right) }\right) ^{2}\left\vert \frac{a_{0}}{R\left( \hat{X}\right) }-%
\frac{1-\alpha }{\varsigma F_{1}\left( R\left( K_{\hat{X}},\hat{X}\right)
\right) }\right\vert  \label{sTC}
\end{equation}

The constant $\varsigma $\ is irrelevant here, although it arises in
appendix 3 to estimate short-term returns, and the function $F_{1}$,\
defined in (\ref{pr}), determines the stock's prices evolution. The
coefficient $\alpha $ is the Cobb-Douglas power arising in the dividend part
of short-term returns. The constant $D$, defined in (\ref{psl}), determines
the relation between number of firms and average capital at sector $\hat{X}$.

Conditions (\ref{stc}) and (\ref{sTC}) correspond to the case of relatively
low capital for which a stability in the oscillations may be reached when
expectations are moderately reactive to variation in capital.

For $c>0$ the oscillations are stable if: 
\begin{equation*}
\frac{a_{0}}{R\left( \hat{X}\right) }-\frac{1-\alpha }{\varsigma K_{\hat{X}%
}F_{1}\left( R\left( K_{\hat{X}},\hat{X}\right) \right) }>0
\end{equation*}%
or if:%
\begin{equation*}
\frac{a_{0}}{R\left( \hat{X}\right) }-\frac{1-\alpha }{\varsigma K_{\hat{X}%
}F_{1}\left( R\left( K_{\hat{X}},\hat{X}\right) \right) }<0
\end{equation*}%
and:%
\begin{equation*}
G^{2}>\frac{l^{2}\left( \nabla _{\hat{X}}R\left( \hat{X}\right) \right) ^{2}%
}{4a_{0}R\left( \hat{X}\right) }\left( \frac{\varsigma \left( \nabla _{\hat{X%
}}R\left( \hat{X}\right) F_{1}\left( R\left( K_{\hat{X}},\hat{X}\right)
\right) \left\Vert \Psi \left( \hat{X}\right) \right\Vert ^{2}\right) }{%
\gamma \left( D-\left\Vert \Psi \left( \hat{X}\right) \right\Vert
^{2}\right) }\right) ^{2}\left\vert \frac{a_{0}}{R\left( \hat{X}\right) }-%
\frac{1-\alpha }{\varsigma F_{1}\left( R\left( K_{\hat{X}},\hat{X}\right)
\right) }\right\vert
\end{equation*}%
We recover the large average capital case. A relatively high reactivity of
expectations to fluctuations in capital allows to maintain the capital at
its equilibrium value.

\subsubsection{two types of oscillations}

These results show that the threshold's values between dampening and
explosive oscillations depend on the sectors and on the parameters of the
system. Interactions between moderately reactive expectations and capital
favour patterns 1 and 2, i.e low to high capital sectors, and to impair
pattern 3, i.e. very high capital sectors.

On the other hand, very reactive expectations favour pattern 3 and impair
patterns 1 and 2. \ Actually, in this case, oscillations are relatively
weaker for high capital sectors, which leads to a reallocation of capital.
If, for instance, expected long-term returns decrease in the neighbourhood
of a high capital sector, capital will be reallocated from neighbours to the
considered sector and will stabilize the high capital sector.

Recall moreover, that extreme cases of pattern 3, i.e. both maximal capital
and returns, represent moving thresholds that repulse low-capital firms and
allow already high-capital firms to grow indefinitely. These thresholds and
their oscillations should generate highly global instability: oscillations
constantly drive firms below the threshold out of the sector.

To conclude, let us stress that, as in the static case, the dynamics of
average capital and expected returns can itself be seen as a dynamics for
the system's background fields or collective states. As a consequence, the
background fields are themselves subject to fluctuations.\ Moreover, since
oscillations around a collective state may destabilize the patterns in some
sectors, fluctuations may ultimately switch the collective state and modify
the repartition of patterns across the sectors.

\subsection{Synthesis of the results}

Let us now synthetize our results.

1. In our model, when firms reallocate their capital, they tend to do so in
sectors with relatively higher long-term returns. The speed at which they
reallocate depends on their capital endowment, but can be crowded out by
their competitors. The higher the firm's capital, te greater the chance to
overcome competitors. Eventually, highest capital firms concentrate in
highest expected long-term return sectors, while the rest locate in
neighbours sectors, and possibly least expected return sectors.

2. Financial capital allocation depends on short-term returns, dividends and
price fluctuations, and expected long-term returns. However, since price
fluctuations are driven by expected long-term returns, short and long-term
returns are not independent. Financial capital allocation also depends on
the sector's relative attractivity, which measure the expected returns of a
sector relative to its neighbours. However financial capital is volatile.
High short-term returns are an incentive, but the relative attractivity of
sectors lures investors. Financial capital allocation thus depends on the
ratio of sectors' relative attractivity to short-term returns. Since this
ratio depends on expectations, it is subject to fluctuations, which in turn
impact the collective state.

3. Capital allocation of firms and investors differ and interact. They also
impact the form of the background field and the average values of capital
per sector. Average capital per firm per sector depends on short-term
returns, both dividends, that are driven by marginal productivity, and stock
prices fluctuations, and on expectations of long-term returns.

4. For each sector, three patterns of accumulation emerge. They depend on
three parameters of the model, short-term returns, expected long-term
returns and the sector's relative attractivity. In the first pattern, the
dividend component of short-term returns is determinant for sectors with
small number of firms and low capital. \ In the second pattern, both short
and long-term returns drive intermediate-to-high capital in the sector. In
the third pattern, higher expectations of long-term returns drive massive
inputs of capital in sectors. In this pattern, firms with maximal expected
return can theoretically accumulate without bound. Practically this
accumulation stops when there is no more capital available.

5. These values of average capital are stationary results: agents move and
accumulate but, in average, density of firms and average capital per sector
are constant.

6. In a dynamic perspective, the patterns of accumulation are in general
stable: small deviations from the average values induce the system to return
to its initial values. Only sectors in the third pattern with maximal
expected returns are unstable. The average value of capital per firm located
at these sectors is a threshold: firms with capital below this threshold
move toward the next stable equilibrium, firms with capital above accumulate
indefinitely.

7. The results above, for average values of capital, hold for a given
background field.\ However, for each sector, three possible patterns of
capital accumulation may exist. This combination of various accumulation
patterns in each sector yields an infinite number of possible collective
states.\ It does not follow that these various possibilities are free:
sector patterns depend on the relative attractivity of both the sector and
its neighbours'. There are also constraints: massive inflows of capital are
only driven by high expected long-term returns, while niche effects merely
occur for relatively high productivity firms. However, from relatively
homogeneous levels of capital to largely heterogeneous patterns of
accumulation between sectors, a potentially infinite range of collective
states may exist. So that when parameters vary, given collective states may
switch to another: a change in expectations or parameters may, for instance,
induce variations in average capital, and in turn induce changes in sectors'
patterns of capital accumulation. To study these possible switches, we
introduce a dynamic interaction between average capital and expected
long-term\ returns, now endogenized.

8. The main characteristic of this dynamic interaction depends both on the
patterns of accumulation and expectations formation. Two types of
expectations are relevant: expectations of long-term returns, that react
positively to any variation in capital, i. e. highly reactive expectations,
and expectations reacting negatively to this variation, i.e. moderately
reactive. In this dynamics, average capital and expectations present some
oscillatory patterns that may dampen equilibria or drive them towards other
equilibria. \ Expectations highly reactive to capital variations stabilize
high-capital configurations.\ They drive low-to-moderate capital sectors
towards zero or higher capital, depending on their initial conditions.
Inversely, expectations moderately reactive to capital variations stabilize
low-to moderate capital configurations, and drive high-capital sectors
towards lower capital equilibria.

\section{Discussion}

The use of statistical field theory has led us to describe a microeconomic
framework in terms of collective states. These collective states are
composed of sectors themselves made up of a large number of firms. Recall
that our notion of firm is versatile: a single company could be modelled as
a set of independent firms. Similarly, the notion of sector merely refers to
a group of entities with similar activities.

Each collective state is singularly determined by the collection of data
that characterizes each sector: number of firms, number of investors,
average capital and density of distribution of capital. Mathematically, the
collective state is described by what we have called a background field. Any
change in this collection of data would describe another collective state.\
However collective states do not change at the slightest variation of one of
these data: they deal with theoretical averages over long-term periods, not
instantaneous empirical averages. Nor are collective states arbitrary: they
directly result from agents' interactions and are the most probable stable
states of the system obtained by minimization conditions. Other states
exist, but they are unstable.

The collective states describe the possible background states of the economy
considered that eventually condition the agents' individual dynamics. They
depend on the parameters of the model, short-term and long-term returns,
relative attractiveness of the sector, and any parameter conditioning these
three quantities. Their multiplicity stems from the multiple possibility of
patterns in each sector. For instance, pattern-3, stable and unstable, are
more present in the US than in the UK stock markets, where pattern 1 and 2
dominate.

A particular collective state can be described by its distribution into
patterns of capital accumulation - type 1, 2 or 3 - across sectors. Each
sector has its own pattern of accumulation, and the distribution in patterns
is directly conditioned by the economic constraints imposed on the system.\
Type-3 patterns appear in sectors locally more attractive in the long-term.
It is this relative attractivity that determines the sector's capital.
Patterns 1 or 2, that are relatively less attractive sectors, lure in
capital with dividends and expected returns. Besides, sectors are connected
and benefit from the relative attractiveness of their neighbours: this
smoothing effect between sectors materialise in mergers and acquisitions.

The selection of a particular collective state and its sectoral patterns is
ultimately determined by exogenous conditions. Structural changes, such as
an extra-loose monetary policy or the choice of a pension system are
external conditions that modify collective states. \textbf{\ }

Collective states are not static. Their dynamics depend on short-term and
long-term return functions, that are exogenous, and more broadly on a whole
landscape of technological and economic conditions. But as a system, they
also present an internal dynamics. We have considered these two types of
variations in the paper.

First, exogeneous modifications in the parameters change collective states.
Any modification in expectations or, more generally, structural changes in
economic and/or monetary conditions, may change expected returns and in turn
the collective state. Unstable type-3 sectors are particularly sensitive to
these changes in long-term growth, inflation and interest rates. Higher
expectations in these sectors attract investment, which in turn increase
expectations. This seemingly endless expected growth spirals until outlook
flattens or deteriorates. A typical example would be the growth model of a
company such as Amazon, whose ever-broadening product ranges has fuelled
higher expected long-term returns and stock prices increase. Type-1 and -2
sectors attract capital through dividends and, although only partially and
for high capital type-2 sectors, expected returns. Under higher
expectations, these sectors are relatively less attractive than nearby
type-3 sector.\ They may nonetheless survive in the long-term provided their
short-term returns and dividends are high enough. This may be done by
cutting costs or investment, at the expense of future growth. Moreover,
advert signalling may emerge: an increase in dividends can be interpreted as
faltering growth prospects. Conversely, any increase in long-term
uncertainty impact expected returns and drive sector-3 capital towards
other\ patterns. External shocks, inflation and monetary policy impact
expectations, reduce long-term investment and either drive capital out of
sectors 3 to sector-1 or -2, or favour other pattern-3 sectors.

Second, a deviation of capital from its collective state equilibrium value
in one sector may itself initiate oscillations in the entire system.
Actually, a temporary deviation from the given collective state implies an
unstable redistribution of capital, expectations and returns. This generates
interactions between sectors, reallocation of capital and global
oscillations. These oscillations can dampen, or alternately drive the system
towards a new collective state. There are thus potential transitions between
collective states. This dynamics occurs at a slower, larger time-scale than
that of market fluctuations.\ In the long-run, when transitions occur, both
sectors' averages and patterns may have changed: patterns 2 may morph in,
say, pattern 3 stable or unstable, sectors may simply disappear. Concretely,
any significant modification in average capital in a sector could induce
oscillations and initiate a transition.

Moreover, once endogenous expectations are introduced, they react to
variations in capital, collective states of mixed 1-2-3-patterns are
difficult to maintain. Highly reactive expectations favour patterns 3:
expected returns magnify capital accumulation at the expense of other
patterns. Mildly reactive expectations favour patterns 1 and 2: their
oscillations, that are actually induced by uncertainties, dampen.\ Type-3
sectors on the contrary experience strong fluctuations in capital :
attracting capital is less effective with fading expectations. The threshold
in capital accumulation shifts upwards and least-profitable firms are ousted
out of the sector.\ The recent evolution in performances between value and
growth investment strategies exemplifies these shifts in investors'
sentiment between expected growth and real returns. In periods of
uncertainties, fluctuations affect capital accumulation in growth sectors,
today's tech companies, and strengthen more dividend-driven investments.
Note however that the most profitable and best capitalized firms, that
remain above the threshold, maintain relatively high levels of capital.\
Here our versatile notion of firms proves convenient: any firm that
accumulates enough capital to be able to buy back, in periods of volatility,
its own stocks is actually acting as an autonomous investor. When volatility
is high, the most likely investors for the best capitalized firms are, first
and foremost, the best capitalized companies themselves. They act so to
speak as pools of closely held investors.\ In other words, provided firms
have high enough capital, they can always cushion the impact of price
fluctuations and adverse shocks through buybacks. Similarly, they also could
choose to acquire companies in their sector or neighbouring sectors.

Fluctuations in financial expectations impose their pace to the real
economy.\ Actually, expected returns are both exogenous and endogenous.
Because they are exogenous, they may change quickly: expected returns, that
theoretically should reflect long-term perspectives, actually rely on
short-term sentiments: new information, changes in global economic outlook,
adverse shocks regularly happen and modifies the long-term expectations.
Capital moves from sectors to sectors quickly. But expected returns are also
endogenous.\ Being expectations, they react, either highly or mildly, to
changes within the system. \ When high levels of capital seek to maximize
returns, we can suspect that expectations will react strongly to capital
changes. The combination of expectations both highly sensitive to exogenous
conditions and highly reactive to variations in capital imply that large
fluctuations of capital in the system. Creating or inflating expectations
attracts capital, at times unduly.\ When this cannot be done, the sole
remaining tool to reduce capital outflows is dividend policy, which may be
done at the expense of labour force, capital expenditures and future growth.

\section{Conclusion}

We have studied the impact of financial capital on physical capital
allocation and shown that collective states distribute sectors into several
patterns of accumulation. All else equal, sectors with highest expected
returns and capital may, through expectations, indefinitely attract capital
at the expense of other sectors. This expansion is nevertheless unstable
since adverse changes in expectations drive capital away.

At a macro timescale, the system can be globally described by oscillations
between average capital and expected long-term returns, depending on the
sectors' patterns. These oscillations, that can be either dampening or
explosive, may change sectors patterns and explain switches from one
collective state to another. Markets, supposedly the most efficient
ressource allocation mechanism, add in a context of uncertainties impose
their fluctuations to those of the real economy. This should render the role
of Central Banks, or any kind of regulation, crucial to the good functioning
of these markets.\pagebreak

\section*{Appendix 1 From large number of agents to field formalism}

This appendix summarizes the most useful steps of the method developed in
Gosselin, Lotz and Wambst (2017, 2020, 2021),\ to switch from the
probabilistic description of the model to the field theoretic formalism and
summarizes the translation of a generalisation of (\ref{mNZ}) involving
different time variables. By convention and unless otherwise mentioned, the
symbol $\int $\ refers to all the variables involved.

\subsection*{A1.1 Probabilistic formalism}

The probabilistic formalism for a system with $N$ identical economic agents
in interaction is based on the minimization functions described in the text.
Classically, the dynamics derives through the optimization problem of these
functions. The probabilitic formalism relies on the contrary on the fact,
that, due to uncertainties, shocks... agents do not optimize fully these
functions. Moreover, given the large number of agents, there may be some
discrepency between agents minimization functions, and this fact may be
translated in an uncertainty of behaviour around some average minimization,
or objective funtion.

We thus assume that each agent chooses for his action a path randomly
distributed around the optimal path. The agent's behavior can be described
as a weight that is an exponential of the intertemporal utility, that
concentrates the probability around the optimal path. This feature models
some internal uncertainty as well as non-measurable shocks. Gathering all
agents, it yields a probabilistic description of the system in terms of a
probabilistic weight.

In general, this weight includes utility functions and internalizes
forward-looking behaviors, such as intertemporal budget constraints and
interactions among agents. These interactions may for instance arise through
constraints, since income flows depend on other agents demand. The
probabilistic description then allows to compute the transition functions of
the system, and in turn compute the probability for a system to evolve from
an initial state to a final state within a given time span. They have the
form of Euclidean path integrals.

In the context of the present paper, we have seen that the minimization
functions for the system considered in this work have the form:%
\begin{equation}
\int dt\left( \sum_{i}\left( \frac{d\mathbf{A}_{i}\left( t\right) }{dt}%
-\sum_{j,k,l...}f\left( \mathbf{A}_{i}\left( t\right) ,\mathbf{A}_{j}\left(
t\right) ,\mathbf{A}_{k}\left( t\right) ,\mathbf{A}_{l}\left( t\right)
...\right) \right) ^{2}+\sum_{i}\left( \sum_{j,k,l...}g\left( \mathbf{A}%
_{i}\left( t\right) ,\mathbf{A}_{j}\left( t\right) ,\mathbf{A}_{k}\left(
t\right) ,\mathbf{A}_{l}\left( t\right) ...\right) \right) \right)
\label{mnz}
\end{equation}%
This minimization of this function will yield a dynamic equation for $N$
agents in interaction described by a set of dynamic variables $\mathbf{A}%
_{i}\left( t\right) $ during a given timespan $T$.

The probabilistic description is straightforwardly obtained from (\ref{mnz}%
). The probability associated to a configuration $\left( \mathbf{A}%
_{i}\left( t\right) \right) _{\substack{ i=1,...,N  \\ 0\leqslant t\leqslant
T }}$ \ is directly given by:%
\begin{eqnarray}
&&\mathcal{N}\exp \left( -\frac{1}{\sigma ^{2}}\int dt\left( \sum_{i}\left( 
\frac{d\mathbf{A}_{i}\left( t\right) }{dt}-\sum_{j,k,l...}f\left( \mathbf{A}%
_{i}\left( t\right) ,\mathbf{A}_{j}\left( t\right) ,\mathbf{A}_{k}\left(
t\right) ,\mathbf{A}_{l}\left( t\right) ...\right) \right) ^{2}\right.
\right.  \label{prz} \\
&&\left. \left. +\sum_{i}\left( \sum_{j,k,l...}g\left( \mathbf{A}_{i}\left(
t\right) ,\mathbf{A}_{j}\left( t\right) ,\mathbf{A}_{k}\left( t\right) ,%
\mathbf{A}_{l}\left( t\right) ...\right) \right) \right) \right)  \notag
\end{eqnarray}%
where $\mathcal{N}$ is a normalization factor and $\sigma ^{2}$ is a
variance whose magnitude describes the amplitude of deviations around the
optimal path.

As in the paper, the system is in general modelled by several equations, and
thus, several minimization function. The overall system is thus described by
several functions, and the minimization function of the whole system is
described by the set of functions:%
\begin{equation}
\int dt\left( \sum_{i}\left( \frac{d\mathbf{A}_{i}\left( t\right) }{dt}%
-\sum_{j,k,l...}f^{\left( \alpha \right) }\left( \mathbf{A}_{i}\left(
t\right) ,\mathbf{A}_{j}\left( t\right) ,\mathbf{A}_{k}\left( t\right) ,%
\mathbf{A}_{l}\left( t\right) ...\right) \right) ^{2}+\sum_{i}\left(
\sum_{j,k,l...}g^{\left( \alpha \right) }\left( \mathbf{A}_{i}\left(
t\right) ,\mathbf{A}_{j}\left( t\right) ,\mathbf{A}_{k}\left( t\right) ,%
\mathbf{A}_{l}\left( t\right) ...\right) \right) \right)  \label{znm}
\end{equation}%
where $\alpha $ runs over the set equations describing the system's
dynamics. The associated weight is then:%
\begin{eqnarray}
&&\mathcal{N}\exp \left( -\left( \sum_{i,\alpha }\frac{1}{\sigma _{\alpha
}^{2}}\int dt\left( \frac{d\mathbf{A}_{i}\left( t\right) }{dt}%
-\sum_{j,k,l...}f^{\left( \alpha \right) }\left( \mathbf{A}_{i}\left(
t\right) ,\mathbf{A}_{j}\left( t\right) ,\mathbf{A}_{k}\left( t\right) ,%
\mathbf{A}_{l}\left( t\right) ...\right) \right) ^{2}\right. \right.
\label{pnz} \\
&&\left. \left. +\sum_{i,\alpha }\left( \sum_{j,k,l...}g^{\left( \alpha
\right) }\left( \mathbf{A}_{i}\left( t\right) ,\mathbf{A}_{j}\left( t\right)
,\mathbf{A}_{k}\left( t\right) ,\mathbf{A}_{l}\left( t\right) ...\right)
\right) \right) \right)  \notag
\end{eqnarray}

The appearance of the sum of minimization functions in the probabilitic
weight (\ref{pnz}) translates the hypothesis that the deviations with
respect to the optimization of the functions (\ref{znm}) are assumed to be
independent.

For a large number of agents, the system described by (\ref{pnz}) involves a
large number of variables $K_{i}\left( t\right) $, $P_{i}\left( t\right) $
and $X_{i}\left( t\right) $\ that are difficult to handle. To overcome this
difficulty, we consider the space $H$\ of complex functions defined on the
space of a single agent's actions. The space $H$ describes the collective
behaviour of the system. Each function $\Psi $ of $H$ encodes a particular
state of the system. We then associate to\ each function $\Psi $ of $H$ a
statistical weight, i.e. a probability describing the state encoded in $\Psi 
$. This probability is written $\exp \left( -S\left( \Psi \right) \right) $,
where $S\left( \Psi \right) $ is a functional, i.e. the function of the
function $\Psi $. The form of $S\left( \Psi \right) $ is derived directly
from the form of (\ref{pnz}) as detailed in the text. As seen from (\ref{pnz}%
), this translation can in fact be directly obtained from the sum of
"classical" minimization functions weighted by the factors $\frac{1}{\sigma
_{\alpha }^{2}}$:%
\begin{equation*}
\sum_{i,\alpha }\frac{1}{\sigma _{\alpha }^{2}}\int dt\left( \frac{d\mathbf{A%
}_{i}\left( t\right) }{dt}-\sum_{j,k,l...}f^{\left( \alpha \right) }\left( 
\mathbf{A}_{i}\left( t\right) ,\mathbf{A}_{j}\left( t\right) ,\mathbf{A}%
_{k}\left( t\right) ,\mathbf{A}_{l}\left( t\right) ...\right) \right)
^{2}+\sum_{i,\alpha }\left( \sum_{j,k,l...}g^{\left( \alpha \right) }\left( 
\mathbf{A}_{i}\left( t\right) ,\mathbf{A}_{j}\left( t\right) ,\mathbf{A}%
_{k}\left( t\right) ,\mathbf{A}_{l}\left( t\right) ...\right) \right)
\end{equation*}%
This is this shortcut we used in the text.

\subsection*{A1.2 Interactions between agents at different times}

A straightforward generalisation of (\ref{mNZ}) involve agents interactions
at different times. The terms considered have the form:%
\begin{eqnarray}
&&\sum_{i}\left( \frac{d\mathbf{A}_{i}\left( t\right) }{dt}%
-\sum_{j,k,l...}\int f\left( \mathbf{A}_{i}\left( t_{i}\right) ,\mathbf{A}%
_{j}\left( t_{j}\right) ,\mathbf{A}_{k}\left( t_{k}\right) ,\mathbf{A}%
_{l}\left( t_{l}\right) ...,\mathbf{t}\right) \mathbf{dt}\right) ^{2}
\label{gR} \\
&&+\sum_{i}\sum_{j,k,l...}\int g\left( \mathbf{A}_{i}\left( t_{i}\right) ,%
\mathbf{A}_{j}\left( t_{j}\right) ,\mathbf{A}_{k}\left( t_{k}\right) ,%
\mathbf{A}_{l}\left( t_{l}\right) ...,\mathbf{t}\right) \mathbf{dt}  \notag
\end{eqnarray}%
where $\mathbf{t}$\ stands for $\left( t_{i},t_{j},t_{k},t_{l}...\right) $
and $\mathbf{dt}$ stands for $dt_{i}dt_{j}dt_{k}dt_{l}...$

The translation is straightforward. We introduce a time variable $\theta $
on the field side and the fields write $\left\vert \Psi \left( \mathbf{A}%
,\theta \right) \right\vert ^{2}$ and $\left\vert \hat{\Psi}\left( \mathbf{%
\hat{A}},\hat{\theta}\right) \right\vert ^{2}$. The second term in (\ref{gR}%
) becomes: 
\begin{eqnarray}
&&\sum_{i}\sum_{j}\sum_{j,k...}\int g\left( \mathbf{A}_{i}\left(
t_{i}\right) ,\mathbf{A}_{j}\left( t_{j}\right) ,\mathbf{A}_{k}\left(
t_{k}\right) ,\mathbf{A}_{l}\left( t_{l}\right) ...,\mathbf{t}\right) 
\mathbf{dt}  \notag \\
&\rightarrow &\int g\left( \mathbf{A},\mathbf{A}^{\prime },\mathbf{A}%
^{\prime \prime },\mathbf{\hat{A},\hat{A}}^{\prime }...,\mathbf{\theta ,\hat{%
\theta}}\right) \left\vert \Psi \left( \mathbf{A},\theta \right) \right\vert
^{2}\left\vert \Psi \left( \mathbf{A}^{\prime },\theta ^{\prime }\right)
\right\vert ^{2}\left\vert \Psi \left( \mathbf{A}^{\prime \prime },\theta
^{\prime \prime }\right) \right\vert ^{2}d\mathbf{A}d\mathbf{A}^{\prime }d%
\mathbf{A}^{\prime \prime } \\
&&\times \left\vert \hat{\Psi}\left( \mathbf{\hat{A}},\hat{\theta}\right)
\right\vert ^{2}\left\vert \hat{\Psi}\left( \mathbf{\hat{A}}^{\prime },\hat{%
\theta}^{\prime }\right) \right\vert ^{2}d\mathbf{\hat{A}}d\mathbf{\hat{A}}%
^{\prime }\mathbf{d\theta d\hat{\theta}}  \notag
\end{eqnarray}%
where $\mathbf{\theta }$ and $\mathbf{\hat{\theta}}$ are the multivariables $%
\left( \theta ,\theta ^{\prime },\theta ^{\prime \prime }...\right) $ and $%
\left( \hat{\theta},\hat{\theta}^{\prime }...\right) $ respectively and $%
\mathbf{d\theta d\hat{\theta}}$ stands for $d\theta d\theta ^{\prime
}d\theta ^{\prime \prime }...$ and $d\hat{\theta}d\hat{\theta}^{\prime }...$

Similarly, the first term in (\ref{gR}) translates as:%
\begin{eqnarray}
&&\sum_{i}\left( \frac{d\mathbf{A}_{i}\left( t\right) }{dt}%
-\sum_{j,k,l...}\int f\left( \mathbf{A}_{i}\left( t_{i}\right) ,\mathbf{A}%
_{j}\left( t_{j}\right) ,\mathbf{A}_{k}\left( t_{k}\right) ,\mathbf{A}%
_{l}\left( t_{l}\right) ...,\mathbf{t}\right) \mathbf{dt}\right) ^{2} \\
&\rightarrow &\int \Psi ^{\dag }\left( \mathbf{A},\theta \right) \left(
-\nabla _{\mathbf{A}^{\left( \alpha \right) }}\left( \frac{\sigma _{\mathbf{A%
}^{\left( \alpha \right) }}^{2}}{2}\nabla _{\mathbf{A}^{\left( \alpha
\right) }}+\Lambda (\mathbf{A},\theta )\right) \right) \Psi \left( \mathbf{A}%
,\theta \right) d\mathbf{A}d\theta
\end{eqnarray}%
by:%
\begin{eqnarray}
\Lambda (\mathbf{A},\theta ) &=&\int f^{\left( \alpha \right) }\left( 
\mathbf{A},\mathbf{A}^{\prime },\mathbf{A}^{\prime \prime },\mathbf{\hat{A},%
\hat{A}}^{\prime }...,\mathbf{\theta ,\hat{\theta}}\right) \left\vert \Psi
\left( \mathbf{A}^{\prime },\theta ^{\prime }\right) \right\vert
^{2}\left\vert \Psi \left( \mathbf{A}^{\prime \prime },\theta ^{\prime
\prime }\right) \right\vert ^{2}d\mathbf{A}^{\prime }d\mathbf{A}^{\prime
\prime } \\
&&\times \left\vert \hat{\Psi}\left( \mathbf{\hat{A}},\theta \right)
\right\vert ^{2}\left\vert \hat{\Psi}\left( \mathbf{\hat{A}}^{\prime
},\theta ^{\prime \prime }\right) \right\vert ^{2}d\mathbf{\hat{A}}d\mathbf{%
\hat{A}}^{\prime }\mathbf{d\bar{\theta}d\hat{\theta}}  \notag
\end{eqnarray}%
with $\mathbf{d\bar{\theta}}=d\theta ^{\prime }d\theta ^{\prime \prime }$.

Ultimately, as in the text, additional terms (\ref{fct}):%
\begin{eqnarray}
&&\Psi ^{\dag }\left( \mathbf{A},\theta \right) \left( -\nabla _{\theta
}\left( \frac{\sigma _{\theta }^{2}}{2}\nabla _{\theta }-1\right) \right)
\Psi \left( \mathbf{A},\theta \right) \\
&&+\hat{\Psi}^{\dag }\left( \mathbf{\hat{A}},\theta \right) \left( -\nabla
_{\theta }\left( \frac{\sigma _{\theta }^{2}}{2}\nabla _{\theta }-1\right)
\right) \hat{\Psi}\left( \mathbf{\hat{A}},\theta \right) +\alpha \left\vert
\Psi \left( \mathbf{A}\right) \right\vert ^{2}+\alpha \left\vert \hat{\Psi}%
\left( \mathbf{\hat{A}}\right) \right\vert ^{2}  \notag
\end{eqnarray}%
are included to the action functional to take into account for the time
variable.

\section*{Appendix 2 expression of $\Psi \left( K,X\right) $ as function of
financial variables}

\subsection*{A2.1 \textbf{Finding }$\Psi \left( K,X\right) $: principle}

In this paragraph, we give the principle of resolution for $\Psi \left(
K,X\right) $ for an arbitrary function $H$. The full resolution for some
particular cases is given below. Given a particular state $\hat{\Psi}$, we
aim at minimizing the action functional $S_{1}+S_{2}+S_{3}+S_{4}$. However,
given our assumptions, the action functional $S_{3}+S_{4}$ depends on $\Psi
\left( K,X\right) $, through average quantities, and moreover, we have
assumed that physical capital dynamics depends on financial accumulation. As
a consequence, we can neglect, in first approximation, the impact of $\Psi
\left( K,X\right) $ on $S_{3}+S_{4}$ and consider rather the minimization of 
$S_{1}+S_{2}$ which is given by:

\begin{eqnarray}
S_{1}+S_{2} &=&-\int \Psi ^{\dag }\left( K,X\right) \left( \nabla _{X}\left( 
\frac{\sigma _{X}^{2}}{2}\nabla _{X}-\nabla _{X}R\left( K,X\right) H\left(
K\right) \right) -\tau \left( \int \left\vert \Psi \left( K^{\prime
},X\right) \right\vert ^{2}dK^{\prime }\right) \right.  \label{prt} \\
&&+\left. \nabla _{K}\left( \frac{\sigma _{K}^{2}}{2}\nabla _{K}+u\left(
K,X,\Psi ,\hat{\Psi}\right) \right) \right) \Psi \left( K,X\right) dKdX 
\notag
\end{eqnarray}%
with:%
\begin{equation}
u\left( K,X,\Psi ,\hat{\Psi}\right) =\frac{1}{\varepsilon }\left( K-\int 
\frac{F_{2}\left( R\left( K,X\right) \right) G\left( X-\hat{X}\right) }{\int
F_{2}\left( R\left( K,X\right) \right) G\left( X-\hat{X}\right) \left\Vert
\Psi \left( K,X\right) \right\Vert ^{2}}\hat{K}\left\Vert \hat{\Psi}\left( 
\hat{K},\hat{X}\right) \right\Vert ^{2}d\hat{K}d\hat{X}\right)  \label{lpr}
\end{equation}%
and:%
\begin{equation*}
\Gamma \left( K,X\right) =\int \frac{F_{2}\left( R\left( K,X\right) \right)
G\left( X-\hat{X}\right) }{K\int F_{2}\left( R\left( K,X\right) \right)
G\left( X-\hat{X}\right) \left\Vert \Psi \left( K,X\right) \right\Vert ^{2}}%
\hat{K}\left\Vert \hat{\Psi}\left( \hat{K},\hat{X}\right) \right\Vert
^{2}d\left( \hat{K},\hat{X}\right) -1
\end{equation*}%
This is done in two steps. First, we find $\Psi \left( X\right) $, the
background field for $X$ when $K$ determined by $X$. We then find the
corrections to the particular cases considered and compute $\Psi \left(
K,X\right) $.

\subsubsection*{A2.1.1 \textbf{Particular case}: $K$ determined by $X$}

A simplification arises, assuming $K$ adaptating to $X$. We assume that in
first approximation $K$ is a function of $X$, written $K_{X}$: 
\begin{equation}
K=K_{X}=\int \frac{F_{2}\left( R\left( K_{X},X\right) \right) G\left( X-\hat{%
X}\right) }{\int F_{2}\left( R\left( K_{X^{\prime }}^{\prime },X^{\prime
}\right) \right) G\left( X^{\prime }-\hat{X}\right) \left\Vert \Psi \left(
X^{\prime }\right) \right\Vert ^{2}dX^{\prime }}\hat{K}\left\Vert \hat{\Psi}%
\left( \hat{K},\hat{X}\right) \right\Vert ^{2}d\left( \hat{K},\hat{X}\right)
\label{kvr}
\end{equation}%
This means that for any sector $X$, the capital of all agents in this sector
are equal. At the individual level, this corresponds to set $\frac{d}{dt}%
K_{i}\left( t\right) =0$. The level of capital adapts faster than the motion
in sector space and reaches quickly its equilibrium value. Incindently, (\ref%
{kvr}) implies that $\Gamma \left( K,X\right) =0$. Actually, using (\ref{kvr}%
):%
\begin{eqnarray*}
\Gamma \left( K,X\right) &=&\int \frac{F_{2}\left( R\left( K,X\right)
\right) G\left( X-\hat{X}\right) }{K\int F_{2}\left( R\left( K,X\right)
\right) G\left( X-\hat{X}\right) \left\Vert \Psi \left( K,X\right)
\right\Vert ^{2}}\hat{K}\left\Vert \hat{\Psi}\left( \hat{K},\hat{X}\right)
\right\Vert ^{2}d\left( \hat{K},\hat{X}\right) -1 \\
&=&\int \frac{F_{2}\left( R\left( K,X\right) \right) G\left( X-\hat{X}%
\right) \hat{K}\left\Vert \hat{\Psi}\left( \hat{K},X\right) \right\Vert ^{2}d%
\hat{K}}{\int F_{2}\left( R\left( K_{X},X\right) \right) G\left( X-\hat{X}%
\right) \hat{K}\left\Vert \hat{\Psi}\left( \hat{K},\hat{X}\right)
\right\Vert ^{2}d\left( \hat{K},\hat{X}\right) }-1 \\
&=&0
\end{eqnarray*}

\paragraph*{A2.1.1.1 Justification of approximation (\protect\ref{kvr})}

Approximation (\ref{kvr}) justifies in the following way. When $F_{2}$ is
slowly varying with $K$, we perform the following change of variable in (\ref%
{prt}):%
\begin{eqnarray*}
\Psi &\rightarrow &\Psi \exp \left( -\frac{\int u\left( K,X,\Psi ,\hat{\Psi}%
\right) dK}{\sigma _{K}^{2}}\right) \simeq \Psi \exp \left( -\frac{1}{%
2\sigma _{K}^{2}}\varepsilon u^{2}\left( K,X,\Psi ,\hat{\Psi}\right) \right)
\\
\Psi ^{\dag } &\rightarrow &\Psi ^{\dag }\exp \left( \frac{1}{\sigma _{K}^{2}%
}\int u\left( K,X,\Psi ,\hat{\Psi}\right) dK\right) \simeq \Psi ^{\dag }\exp
\left( -\frac{1}{2\sigma _{K}^{2}}\varepsilon u^{2}\left( K,X,\Psi ,\hat{\Psi%
}\right) \right)
\end{eqnarray*}%
and this replaces $S_{2}$ in (\ref{prt}) by:%
\begin{equation}
-\int \Psi ^{\dag }\left( K,X\right) \left( \frac{\sigma _{K}^{2}}{2}\nabla
_{K}^{2}-\frac{u^{2}}{2\sigma _{K}^{2}}\left( K,X,\Psi ,\hat{\Psi}\right) +%
\frac{1}{2}\nabla _{K}u\left( K,X,\Psi ,\hat{\Psi}\right) \right) \Psi
\left( K,X\right) dKdX  \label{trd}
\end{equation}%
The change of variable modifies $S_{1}$ in (\ref{prt}). Actually, the
derivative $\nabla _{X}$ acts on $\exp \left( -\frac{1}{2\sigma _{K}^{2}}%
u^{2}\left( K,X,\Psi ,\hat{\Psi}\right) \right) $ and the term:%
\begin{equation*}
-\int \Psi ^{\dag }\left( K,X\right) \nabla _{X}\left( \frac{\sigma _{X}^{2}%
}{2}\nabla _{X}-\nabla _{X}R\left( K,X\right) H\left( K\right) \right) \Psi
\left( K,X\right) dKdX
\end{equation*}%
becomes:%
\begin{eqnarray}
&&-\int \Psi ^{\dag }\left( X\right) \nabla _{X}\left( \frac{\sigma _{X}^{2}%
}{2}\nabla _{X}-\nabla _{X}R\left( K,X\right) H\left( K\right) \right) \Psi
\left( X\right) dKdX  \label{trf} \\
&&+\varepsilon \int \Psi ^{\dag }\left( K,X\right) \left( \frac{\sigma
_{X}^{2}}{2\sigma _{K}^{2}}u\nabla _{X}u\right) \nabla _{X}\Psi \left(
K,X\right) dKdX+\varepsilon \int \Psi ^{\dag }\left( K,X\right) \left( \frac{%
\sigma _{X}^{2}}{2\sigma _{K}^{2}}\left( \left( \nabla _{X}u\right)
^{2}+u\nabla _{X}^{2}u\right) \right) \Psi \left( K,X\right) dKdX  \notag \\
&&-\int \Psi ^{\dag }\left( K,X\right) \left( \varepsilon \frac{u\nabla _{X}u%
}{\sigma _{K}^{2}}\nabla _{X}R\left( K,X\right) H\left( K\right)
+\varepsilon ^{2}\frac{\sigma _{X}^{2}}{2\sigma _{K}^{4}}\left( u\nabla
_{X}u\right) ^{2}\right) \Psi \left( K,X\right) dKdX  \notag
\end{eqnarray}%
Using that $u$ is of order $\frac{1}{\varepsilon }$ (see(\ref{lpr})), the
minimum of $S_{1}+S_{2}$ is obtained when the potential:%
\begin{eqnarray}
&&\int \Psi ^{\dag }\left( K,X\right) \left( \frac{u^{2}}{2\sigma _{K}^{2}}-%
\frac{1}{2}\nabla _{K}u\right) \Psi \left( K,X\right) dKdX  \label{ptl} \\
&&+\varepsilon \int \Psi ^{\dag }\left( K,X\right) \left( \frac{\sigma
_{X}^{2}}{2\sigma _{K}^{2}}\left( \left( \nabla _{X}u\right) ^{2}+u\nabla
_{X}^{2}u\right) \right) \Psi \left( K,X\right) dKdX  \notag \\
&&-\int \Psi ^{\dag }\left( K,X\right) \left( \varepsilon \frac{u\nabla _{X}u%
}{\sigma _{K}^{2}}\nabla _{X}R\left( K,X\right) H\left( K\right)
+\varepsilon ^{2}\frac{\sigma _{X}^{2}}{2\sigma _{K}^{4}}\left( u\nabla
_{X}u\right) ^{2}\right) \Psi \left( K,X\right) dKdX  \notag
\end{eqnarray}%
is nul. The dominant term in (\ref{ptl}) for $\varepsilon <<1$ is: 
\begin{equation}
\int \Psi ^{\dag }\left( K,X\right) \left( \frac{u^{2}}{2\sigma _{K}^{2}}%
-\varepsilon ^{2}\frac{\sigma _{X}^{2}}{2\sigma _{K}^{4}}\left( u\nabla
_{X}u\right) ^{2}\right) \Psi \left( K,X\right) dKdX  \label{trm}
\end{equation}%
For $\sigma _{X}^{2}<<\sigma _{K}^{2}$ it implies that the minimum for $%
S_{1}+S_{2}$ is obtained for: 
\begin{equation*}
u\left( K,X,\Psi ,\hat{\Psi}\right) \simeq 0
\end{equation*}%
with solution (\ref{kvr}).

\paragraph*{A2.1.1.2 Rewriting the action $S_{1}+S_{2}$}

With our choice $G\left( X-\hat{X}\right) =\delta \left( X-\hat{X}\right) $
we find:%
\begin{equation}
K_{X}=\frac{\int \hat{K}\left\Vert \hat{\Psi}\left( \hat{K},X\right)
\right\Vert ^{2}d\hat{K}}{\left\Vert \Psi \left( X\right) \right\Vert ^{2}}
\label{xK}
\end{equation}%
and $\Psi \left( K,X\right) $ becomes a function $\Psi \left( X\right) $:%
\begin{equation*}
\Psi \left( K,X\right) \rightarrow \Psi \left( X\right)
\end{equation*}%
To find the action for $\Psi \left( X\right) $ we evaluate (\ref{ptl}) using 
$u\left( K_{X},X,\Psi ,\hat{\Psi}\right) =0$, and compute the first term in (%
\ref{trm}) for $\Psi \left( X\right) =\Psi \left( K_{X},X\right) \delta
\left( u\right) $ by replacing:%
\begin{equation*}
\delta \left( u\right) \rightarrow \frac{\exp \left( -\varepsilon
u^{2}\right) }{\sqrt{2\pi }\varepsilon }
\end{equation*}%
We obtain: 
\begin{eqnarray*}
-\int \Psi ^{\dag }\left( K,X\right) \left( \frac{\sigma _{K}^{2}}{2}\nabla
_{K}^{2}\right) \Psi \left( K,X\right) dKdX &=&\frac{\sigma _{K}^{2}}{2}\int
\left\vert \Psi \left( X\right) \right\vert ^{2}dX\int \frac{\exp \left(
-\varepsilon u^{2}\right) }{\sqrt{2\pi \varepsilon }}\nabla _{K}^{2}\frac{%
\exp \left( -\varepsilon u^{2}\right) }{\sqrt{2\pi \varepsilon }}dK \\
&\simeq &\frac{\sigma _{K}^{2}}{2\varepsilon }\int \left\vert \Psi \left(
X\right) \right\vert ^{2}dX
\end{eqnarray*}%
and the action $S_{1}$ restricted to the variable $X$\ is given by:%
\begin{eqnarray*}
S_{1} &=&\int \Psi ^{\dag }\left( X\right) \left( -\nabla _{X}\left( \frac{%
\sigma _{X}^{2}}{2}\nabla _{X}-\left( \nabla _{X}R\left( X\right) H\left(
K_{X}\right) \right) \right) +\tau \left\vert \Psi \left( X\right)
\right\vert ^{2}\right) \Psi \left( X\right) \\
&&+\int \Psi ^{\dag }\left( K,X\right) \left( \frac{\sigma _{X}^{2}}{4\sigma
_{K}^{2}}\left( \nabla _{X}u\left( K_{X},X,\Psi ,\hat{\Psi}\right) \right)
^{2}\right) \Psi \left( K,X\right) dKdX \\
&&+\int \left( \frac{\sigma _{K}^{2}}{2\varepsilon }-\frac{1}{2}\nabla
_{K}u\left( K_{X},X,\Psi ,\hat{\Psi}\right) \right) \left\vert \Psi \left(
X\right) \right\vert ^{2}dX
\end{eqnarray*}%
In our order of appromation $\nabla _{K}u\left( K_{X},X,\Psi ,\hat{\Psi}%
\right) \simeq \varepsilon $. Ultimately, for $\sigma _{X}^{2}<<\sigma
_{K}^{2}$, action $S_{1}$ reduces to:

\begin{equation}
S_{1}=\int \Psi ^{\dag }\left( X\right) \left( -\nabla _{X}\left( \frac{%
\sigma _{X}^{2}}{2}\nabla _{X}-\left( \nabla _{X}R\left( X\right) H\left(
K_{X}\right) \right) \right) +\tau \left\vert \Psi \left( X\right)
\right\vert ^{2}+\frac{\sigma _{K}^{2}-1}{2\varepsilon }\right) \Psi \left(
X\right)  \label{cnt}
\end{equation}%
and we look for $\Psi \left( X\right) $ minimizing (\ref{cnt}).

\paragraph*{A2.1.1.3 Minimization of (\protect\ref{cnt})}

To minimize (\ref{cnt}), we assume for the sake of simplicity, that for $%
i\neq j$:%
\begin{equation*}
\left\vert \nabla _{X_{i}}\nabla _{X_{j}}R\left( X\right) \right\vert
<<\left\vert \nabla _{X_{i}}^{2}R\left( X\right) \right\vert
\end{equation*}%
which is the case for example if $R\left( X\right) $ is a function with
separated variables : $R\left( X\right) =\sum R_{i}\left( X_{i}\right) $.
This can be also realized if locally, one chooses the variables $X_{i}$ to
diagonalize $\nabla _{X_{i}}\nabla _{X_{j}}R\left( X\right) $ at some points
in the sector space.

We then perform the change of variables:%
\begin{equation*}
\exp \left( \int^{X}\frac{\nabla _{X}R\left( X\right) }{\sigma
_{X}^{2}\left\Vert \nabla _{X}R\left( X\right) \right\Vert }H\left(
K_{X}\right) \right) \Psi \left( X\right) \rightarrow \Psi \left( X\right)
\end{equation*}%
and:%
\begin{equation*}
\exp \left( -\int^{X}\nabla _{X}R\left( X\right) H\left( K_{X}\right)
\right) \Psi ^{\dag }\left( X\right) \rightarrow \Psi ^{\dag }\left( X\right)
\end{equation*}%
so that (\ref{cnt}) becomes:%
\begin{equation}
\int \Psi ^{\dag }\left( X\right) \left( -\frac{\sigma _{X}^{2}}{2}\nabla
_{X}^{2}+\frac{1}{2\sigma _{X}^{2}}\left( \nabla _{X}R\left( X\right)
H\left( K_{X}\right) \right) ^{2}+\frac{\nabla _{X}^{2}R\left(
K_{X},X\right) }{2}H\left( K_{X}\right) +\tau \left\vert \Psi \left(
X\right) \right\vert ^{2}+\frac{\sigma _{K}^{2}-1}{2\varepsilon }\right)
\Psi \left( X\right)  \label{ntc}
\end{equation}%
which is of second order in derivatives with a potential:%
\begin{equation*}
\tau \left\Vert \Psi \left( X\right) \right\Vert ^{4}+\frac{1}{2\sigma
_{X}^{2}}\int \left( \nabla _{X}R\left( X\right) H\left( K_{X}\right)
\right) ^{2}\left\Vert \Psi \left( X\right) \right\Vert ^{2}
\end{equation*}%
We assume the number of agents fixed equal to $N$. We have to minimize (\ref%
{ntc}) with the contraint $\left\Vert \Psi \left( X\right) \right\Vert
^{2}\geqslant 0$ and $\int \left\Vert \Psi \left( X\right) \right\Vert
^{2}=N $. We thus replace (\ref{ntc}) by:%
\begin{eqnarray}
&&\int \Psi ^{\dag }\left( X\right) \left( -\frac{\sigma _{X}^{2}\nabla
_{X}^{2}}{2}+\frac{\left( \nabla _{X}R\left( X\right) H\left( K_{X}\right)
\right) ^{2}}{2\sigma _{X}^{2}}+\frac{\nabla _{X}^{2}R\left( K_{X},X\right) 
}{2}H\left( K_{X}\right) +\tau \left\vert \Psi \left( X\right) \right\vert
^{2}+\frac{\sigma _{K}^{2}-1}{2\varepsilon }\right) \Psi \left( X\right) 
\notag \\
&&+D\left( \left\Vert \Psi \right\Vert ^{2}\right) \left( \int \left\Vert
\Psi \left( X\right) \right\Vert ^{2}-N\right) +\int \mu \left( X\right)
\left\Vert \Psi \left( X\right) \right\Vert ^{2}  \label{ntC}
\end{eqnarray}%
we have written $D\left( \left\Vert \Psi \right\Vert ^{2}\right) $ the
Lagrange multiplier for $\int \left\Vert \Psi \left( X\right) \right\Vert
^{2}$,$\ $to keep track of its dependency multiplier in $\left\Vert \Psi
\right\Vert ^{2}$. By a redefinition $D\left( \left\Vert \Psi \right\Vert
^{2}\right) -\frac{\sigma _{K}^{2}-1}{2\varepsilon }\rightarrow D\left(
\left\Vert \Psi \right\Vert ^{2}\right) $, $\frac{D\left( \left\Vert \Psi
\right\Vert ^{2}\right) }{D\left( \left\Vert \Psi \right\Vert ^{2}\right) -%
\frac{\sigma _{K}^{2}}{2\varepsilon }}N\rightarrow N$ we can write (\ref{ntC}%
) as: 
\begin{eqnarray}
&&\int \Psi ^{\dag }\left( X\right) \left( -\frac{\sigma _{X}^{2}}{2}\nabla
_{X}^{2}+\frac{1}{2\sigma _{X}^{2}}\left( \nabla _{X}R\left( X\right)
H\left( K_{X}\right) \right) ^{2}+\frac{H\left( K_{X}\right) \nabla
_{X}^{2}R\left( K_{X},X\right) }{2}+\tau \left\vert \Psi \left( X\right)
\right\vert ^{2}\right) \Psi \left( X\right)  \label{rdc} \\
&&+D\left( \left\Vert \Psi \right\Vert ^{2}\right) \left( \int \left\Vert
\Psi \left( X\right) \right\Vert ^{2}-N\right) +\int \mu \left( X\right)
\left\Vert \Psi \left( X\right) \right\Vert ^{2}  \notag
\end{eqnarray}%
Introducing the change of variable for $\nabla _{X}R\left( X\right) $ for
the sake of simplicity:%
\begin{equation}
\left( \nabla _{X}R\left( X\right) \right) ^{2}+\sigma _{X}^{2}\frac{\nabla
_{X}^{2}R\left( K_{X},X\right) }{H\left( K_{X}\right) }\rightarrow \left(
\nabla _{X}R\left( X\right) \right) ^{2}  \label{dfg}
\end{equation}%
the minimization of the potential yields, for $\sigma _{X}^{2}<<1$:%
\begin{eqnarray}
&&iD\left( \left\Vert \Psi \right\Vert ^{2}\right) +\mu \left( X\right)
\label{bnq} \\
&=&2\tau \left\Vert \Psi \left( X\right) \right\Vert ^{2}-\frac{H^{\prime
}\left( \frac{\int \hat{K}\left\Vert \hat{\Psi}\left( \hat{K},X\right)
\right\Vert ^{2}d\hat{K}}{\left\Vert \Psi \left( X\right) \right\Vert ^{2}}%
\right) }{2\sigma _{X}^{2}H\left( \frac{\int \hat{K}\left\Vert \hat{\Psi}%
\left( \hat{K},X\right) \right\Vert ^{2}d\hat{K}}{\left\Vert \Psi \left(
X\right) \right\Vert ^{2}}\right) }  \notag \\
&&\times \left( \nabla _{X}R\left( X\right) H\left( \frac{\int \hat{K}%
\left\Vert \hat{\Psi}\left( \hat{K},X\right) \right\Vert ^{2}d\hat{K}}{%
\left\Vert \Psi \left( X\right) \right\Vert ^{2}}\right) \right) ^{2}\frac{%
\int \hat{K}\left\Vert \hat{\Psi}\left( \hat{K},X\right) \right\Vert ^{2}d%
\hat{K}}{\left\Vert \Psi \left( X\right) \right\Vert ^{4}}\left\Vert \Psi
\left( X\right) \right\Vert ^{2}  \notag \\
&&+\frac{1}{2\sigma _{X}^{2}}\left( \nabla _{X}R\left( X\right) H\left( 
\frac{\int \hat{K}\left\Vert \hat{\Psi}\left( \hat{K},X\right) \right\Vert
^{2}d\hat{K}}{\left\Vert \Psi \left( X\right) \right\Vert ^{2}}\right)
\right) ^{2}  \notag
\end{eqnarray}%
Moreover, multiplying (\ref{bnq}) by $\left\Vert \Psi \left( X\right)
\right\Vert ^{2}$ and integrating yields:

\begin{eqnarray}
D\left( \left\Vert \Psi \right\Vert ^{2}\right) N &=&2\tau \int \left\vert
\Psi \left( X\right) \right\vert ^{4}  \label{tgn} \\
&&-\int \frac{H^{\prime }\left( \frac{\int \hat{K}\left\Vert \hat{\Psi}%
\left( \hat{K},X\right) \right\Vert ^{2}d\hat{K}}{\left\Vert \Psi \left(
X\right) \right\Vert ^{2}}\right) }{2\sigma _{X}^{2}H\left( \frac{\int \hat{K%
}\left\Vert \hat{\Psi}\left( \hat{K},X\right) \right\Vert ^{2}d\hat{K}}{%
\left\Vert \Psi \left( X\right) \right\Vert ^{2}}\right) }\left( \nabla
_{X}R\left( X\right) H\left( \frac{\int \hat{K}\left\Vert \hat{\Psi}\left( 
\hat{K},X\right) \right\Vert ^{2}d\hat{K}}{\left\Vert \Psi \left( X\right)
\right\Vert ^{2}}\right) \right) ^{2}\int \hat{K}\left\Vert \hat{\Psi}\left( 
\hat{K},X\right) \right\Vert ^{2}d\hat{K}  \notag \\
&&+\frac{1}{2\sigma _{X}^{2}}\int \left( \nabla _{X}R\left( X\right) H\left( 
\frac{\int \hat{K}\left\Vert \hat{\Psi}\left( \hat{K},X\right) \right\Vert
^{2}d\hat{K}}{\left\Vert \Psi \left( X\right) \right\Vert ^{2}}\right)
\right) ^{2}\left\Vert \Psi \left( X\right) \right\Vert ^{2}  \notag \\
&\simeq &2\tau \int \left\vert \Psi \left( X\right) \right\vert ^{4}+\frac{1%
}{2\sigma _{X}^{2}}\int \left( \nabla _{X}R\left( X\right) H\left( \frac{%
\int \hat{K}\left\Vert \hat{\Psi}\left( \hat{K},X\right) \right\Vert ^{2}d%
\hat{K}}{\left\Vert \Psi \left( X\right) \right\Vert ^{2}}\right) \right)
^{2}\left\Vert \Psi \left( X\right) \right\Vert ^{2}  \notag
\end{eqnarray}%
Note that in first approximation, for $H^{\prime }<<1$, (\ref{bnq}) and (\ref%
{tgn}) become:%
\begin{equation}
D\left( \left\Vert \Psi \right\Vert ^{2}\right) +\mu \left( X\right) =2\tau
\left\Vert \Psi \left( X\right) \right\Vert ^{2}+\frac{1}{2\sigma _{X}^{2}}%
\left( \nabla _{X}R\left( X\right) \right) ^{2}H^{2}\left( \frac{\int \hat{K}%
\left\Vert \hat{\Psi}\left( \hat{K},X\right) \right\Vert ^{2}d\hat{K}}{%
\left\Vert \Psi \left( X\right) \right\Vert ^{2}}\right)  \label{psn}
\end{equation}%
and:%
\begin{equation}
ND\left( \left\Vert \Psi \right\Vert ^{2}\right) =2\tau \int \left\vert \Psi
\left( X\right) \right\vert ^{4}+\frac{1}{2\sigma _{X}^{2}}\int \left(
\nabla _{X}R\left( X\right) H\left( \frac{\int \hat{K}\left\Vert \hat{\Psi}%
\left( \hat{K},X\right) \right\Vert ^{2}d\hat{K}}{\left\Vert \Psi \left(
X\right) \right\Vert ^{2}}\right) \right) ^{2}\left\Vert \Psi \left(
X\right) \right\Vert ^{2}  \label{snp}
\end{equation}

\paragraph*{A2.1.1.4 Resolution of (\protect\ref{psn}) and (\protect\ref{snp}%
)}

Two cases arise in the resolution:

\subparagraph{Case 1: $\left\Vert \Psi \left( X\right) \right\Vert ^{2}>0$}

For $\left\Vert \Psi \left( X\right) \right\Vert ^{2}>0$, (\ref{bnq})
writes: 
\begin{subequations}
\begin{equation}
D\left( \left\Vert \Psi \right\Vert ^{2}\right) =2\tau \left\Vert \Psi
\left( X\right) \right\Vert ^{2}+\frac{1}{2\sigma _{X}^{2}}\left( \nabla
_{X}R\left( X\right) \right) ^{2}H^{2}\left( \frac{\hat{K}_{X}}{\left\Vert
\Psi \left( X\right) \right\Vert ^{2}}\right) \left( 1-\frac{H^{\prime
}\left( \hat{K}_{X}\right) }{H\left( \hat{K}_{X}\right) }\frac{\hat{K}_{X}}{%
\left\Vert \Psi \left( X\right) \right\Vert ^{2}}\right)  \label{nbq}
\end{equation}%
with: 
\end{subequations}
\begin{equation}
\hat{K}_{X}=\int \hat{K}\left\Vert \hat{\Psi}\left( \hat{K},X\right)
\right\Vert ^{2}d\hat{K}=K_{X}\left\Vert \Psi \left( X\right) \right\Vert
^{2}  \label{Kx}
\end{equation}%
Note that restoring the initial variable:%
\begin{equation}
\left( \nabla _{X}R\left( X\right) \right) ^{2}\rightarrow \left( \nabla
_{X}R\left( X\right) \right) ^{2}+\sigma _{X}^{2}\frac{\nabla
_{X}^{2}R\left( K_{X},X\right) }{H\left( K_{X}\right) }  \label{stg}
\end{equation}%
yields (\ref{psl}) in the text.

Given the setup, we can assume that%
\begin{equation*}
H^{2}\left( \frac{\hat{K}_{X}}{\left\Vert \Psi \left( X\right) \right\Vert
^{2}}\right) \left( 1-\frac{H^{\prime }\left( \hat{K}_{X}\right) }{H\left( 
\hat{K}_{X}\right) }\frac{\hat{K}_{X}}{\left\Vert \Psi \left( X\right)
\right\Vert ^{2}}\right)
\end{equation*}%
is a decreasing function of $\left\Vert \Psi \left( X\right) \right\Vert
^{2} $. Assume a minimum $\Psi _{0}\left( X\right) $ for the right hand side
of (\ref{nbq}). It leads to a condition for $D\left( \left\Vert \Psi
\right\Vert ^{2}\right) $:%
\begin{equation}
D\left( \left\Vert \Psi \right\Vert ^{2}\right) >2\tau \left\Vert \Psi
_{0}\left( X\right) \right\Vert ^{2}+\frac{1}{2\sigma _{X}^{2}}\left( \nabla
_{X}R\left( X\right) \right) ^{2}H^{2}\left( \frac{\hat{K}_{X}}{\left\Vert
\Psi _{0}\left( X\right) \right\Vert ^{2}}\right) \left( 1-\frac{H^{\prime
}\left( \hat{K}_{X}\right) }{H\left( \hat{K}_{X}\right) }\frac{\hat{K}_{X}}{%
\left\Vert \Psi _{0}\left( X\right) \right\Vert ^{2}}\right)  \label{cd}
\end{equation}%
and the solution of (\ref{nbq}) writes:%
\begin{equation}
\left\Vert \Psi \left( X,\left( \nabla _{X}R\left( X\right) \right) ^{2},%
\frac{\hat{K}_{X}}{\hat{K}_{X,0}}\right) \right\Vert ^{2}  \label{sp}
\end{equation}%
where $\hat{K}_{X,0}$ is a constant representing some average to normalize $%
\frac{\hat{K}_{X}}{\hat{K}_{X,0}}$ as a dimensionless number.

\subparagraph{Case 2 $\left\Vert \Psi \left( X\right) \right\Vert ^{2}=0$}

On the other hand, if:%
\begin{equation}
D\left( \left\Vert \Psi \right\Vert ^{2}\right) <2\tau \left\Vert \Psi
_{0}\left( X\right) \right\Vert ^{2}+\frac{1}{2\sigma _{X}^{2}}\left( \nabla
_{X}R\left( X\right) \right) ^{2}H^{2}\left( \frac{\hat{K}_{X}}{\left\Vert
\Psi _{0}\left( X\right) \right\Vert ^{2}}\right) \left( 1-\frac{H^{\prime
}\left( \hat{K}_{X}\right) }{H\left( \hat{K}_{X}\right) }\frac{\hat{K}_{X}}{%
\left\Vert \Psi _{0}\left( X\right) \right\Vert ^{2}}\right)  \label{dc}
\end{equation}%
the solution of (\ref{nbq}) is $\left\Vert \Psi \left( X\right) \right\Vert
^{2}=0$

\subparagraph{Gathering both cases}

The value of $\left\Vert \Psi \right\Vert ^{2}$ thus depends on the
conditions (\ref{cd}) and (\ref{dc}). To compute the value of $D\left(
\left\Vert \Psi \right\Vert ^{2}\right) $ we integrate (\ref{nbq}) over $%
V/V_{0}$ with $V_{0}$ locus where $\left\Vert \Psi \left( X\right)
\right\Vert ^{2}=0$. $V_{0}$ will be then defined by (\ref{dc}) once $%
D\left( \left\Vert \Psi \right\Vert ^{2}\right) $ found. For $H$ slowly
varying, we can replace $\frac{\hat{K}_{X}}{\left\Vert \Psi \left( X\right)
\right\Vert ^{2}}$ by:%
\begin{equation*}
\frac{\int \hat{K}\left\Vert \hat{\Psi}\left( \hat{K},X\right) \right\Vert
^{2}d\hat{K}dX}{\int \left\Vert \Psi \left( X\right) \right\Vert ^{2}dX}=%
\frac{\int \hat{K}\left\Vert \hat{\Psi}\left( \hat{K},X\right) \right\Vert
^{2}d\hat{K}dX}{N}
\end{equation*}%
so that the integration of (\ref{dc}) over $X$ yields:

\begin{eqnarray*}
D\left( \left\Vert \Psi \right\Vert ^{2}\right) \left( V-V_{0}\right)
&\simeq &2\tau N+\frac{1}{2\sigma _{X}^{2}}\int \left( \nabla _{X}R\left(
X\right) \right) ^{2}H^{2}\left( \frac{\int \hat{K}\left\Vert \hat{\Psi}%
\left( \hat{K},X\right) \right\Vert ^{2}d\hat{K}dX}{N}\right) \\
&&\times \left( 1-\frac{H^{\prime }\left( \frac{\int \hat{K}\left\Vert \hat{%
\Psi}\left( \hat{K},X\right) \right\Vert ^{2}d\hat{K}dX}{N}\right) }{H\left( 
\frac{\int \hat{K}\left\Vert \hat{\Psi}\left( \hat{K},X\right) \right\Vert
^{2}d\hat{K}dX}{N}\right) }\frac{\int \hat{K}\left\Vert \hat{\Psi}\left( 
\hat{K},X\right) \right\Vert ^{2}d\hat{K}dX}{N}\right) \\
&=&2\tau N+\frac{1}{2\sigma _{X}^{2}}\left( \nabla _{X}R\left( X\right)
\right) ^{2}H^{2}\left( \frac{\left\langle \hat{K}\right\rangle }{N}\right)
\left( 1-\frac{H^{\prime }\left( \frac{\left\langle \hat{K}\right\rangle }{N}%
\right) }{H\left( \frac{\left\langle \hat{K}\right\rangle }{N}\right) }\frac{%
\left\langle \hat{K}\right\rangle }{N}\right)
\end{eqnarray*}%
As a consequence:%
\begin{equation*}
D\left( \left\Vert \Psi \right\Vert ^{2}\right) \simeq 2\tau \frac{N}{V-V_{0}%
}+\frac{1}{2\sigma _{X}^{2}}\left\langle \left( \nabla _{X}R\left( X\right)
\right) ^{2}\right\rangle _{V/V_{0}}H^{2}\left( \frac{\left\langle \hat{K}%
\right\rangle }{N}\right) \left( 1-\frac{H^{\prime }\left( \frac{%
\left\langle \hat{K}\right\rangle }{N}\right) }{H\left( \frac{\left\langle 
\hat{K}\right\rangle }{N}\right) }\frac{\left\langle \hat{K}\right\rangle }{N%
}\right)
\end{equation*}%
and $V_{0}$ is defined by (\ref{dc}):%
\begin{eqnarray}
&&2\tau \frac{N}{V-V_{0}}+\frac{1}{2\sigma _{X}^{2}}\left\langle \left(
\nabla _{X}R\left( X\right) \right) ^{2}\right\rangle _{V/V_{0}}H^{2}\left( 
\frac{\left\langle \hat{K}\right\rangle }{N}\right) \left( 1-\frac{H^{\prime
}\left( \frac{\left\langle \hat{K}\right\rangle }{N}\right) }{H\left( \frac{%
\left\langle \hat{K}\right\rangle }{N}\right) }\frac{\left\langle \hat{K}%
\right\rangle }{N}\right)  \label{qvn} \\
&<&2\tau \left\Vert \Psi _{0}\left( X\right) \right\Vert ^{2}+\frac{1}{%
2\sigma _{X}^{2}}\left( \nabla _{X}R\left( X\right) \right) ^{2}H^{2}\left( 
\frac{\hat{K}_{X}}{\left\Vert \Psi _{0}\left( X\right) \right\Vert ^{2}}%
\right) \left( 1-\frac{H^{\prime }\left( \hat{K}_{X}\right) }{H\left( \hat{K}%
_{X}\right) }\frac{\hat{K}_{X}}{\left\Vert \Psi _{0}\left( X\right)
\right\Vert ^{2}}\right)  \notag
\end{eqnarray}%
On $V/V_{0}$, $\left\Vert \Psi \right\Vert ^{2}$ is given by (\ref{sp}) and
on $V_{0}$, $\left\Vert \Psi \right\Vert ^{2}=0$.

Below, we give explicitely the form of $\Psi \left( X\right) $ form two
different form of the function $H$.

\subsubsection*{A2.1.2 \textbf{Introducing the }$K$ dependency}

\paragraph*{A2.1.2.1 First order condition}

To go beyond approximation (\ref{kvr}) and solve for the field $\Psi \left(
K,X\right) $ that minimizes (\ref{prt}), we come back to the full system for 
$K$ and $X$: 
\begin{eqnarray}
&&\int \Psi ^{\dag }\left( K,X\right) \left( \left( -\nabla _{X}\left( \frac{%
\sigma _{X}^{2}}{2}\nabla _{X}-\left( \frac{\nabla _{X}R\left( K,X\right) }{%
\left\Vert \nabla _{X}R\left( K,X\right) \right\Vert }\right) H\left(
K\right) +\tau \left\vert \Psi \left( K,X\right) \right\vert ^{2}\right)
\right) \right.  \label{cmk} \\
&&\left. -\nabla _{K}\left( \frac{\sigma _{K}^{2}}{2}\nabla _{K}+u\left(
K,X,\Psi ,\hat{\Psi}\right) \right) -\frac{1}{2}\nabla _{K}u\left( K,X,\Psi ,%
\hat{\Psi}\right) \right) \Psi \left( K,X\right)  \notag
\end{eqnarray}%
with $u\left( K,X,\Psi ,\hat{\Psi}\right) $ given by (\ref{lpr}). We then
look for a minimum of (\ref{cmk}) of the form:%
\begin{equation}
\Psi \left( K,X\right) =\Psi \left( X\right) \Psi _{1}\left( K-K_{X}\right)
\label{dcp}
\end{equation}%
with $K_{X}$ given in (\ref{xK}):%
\begin{equation}
K_{X}=\frac{\int \hat{K}\left\Vert \hat{\Psi}\left( \hat{K},X\right)
\right\Vert ^{2}d\hat{K}}{\left\Vert \Psi \left( X\right) \right\Vert ^{2}}
\label{xkdd}
\end{equation}%
and $\Psi _{1}$ peaked around $0$ and of norm $1$. When $H\left( K\right) $
is slowly varying around $K_{X}$, the minimization of (\ref{cmk}) for $\Psi
_{1}\left( K-K_{X}\right) $ writes:%
\begin{equation}
\nabla _{K}\left( \frac{\sigma _{K}^{2}}{2}\nabla _{K}+u\left( K,X,\Psi ,%
\hat{\Psi}\right) +\frac{1}{2}\nabla _{K}u\left( K,X,\Psi ,\hat{\Psi}\right)
\right) \Psi _{1}\left( K-K_{X}\right) =0  \label{qnx}
\end{equation}%
Then, using that, in first approximation:%
\begin{equation*}
\int F_{2}\left( R\left( K^{\prime },X\right) \right) \left\Vert \Psi \left(
K^{\prime },X\right) \right\Vert ^{2}dK^{\prime }\simeq F_{2}\left( R\left(
K_{X},X\right) \right) \left\Vert \Psi \left( X\right) \right\Vert ^{2}
\end{equation*}%
Equation (\ref{qnx}) becomes:%
\begin{equation}
\nabla _{K}\left( \frac{\sigma _{K}^{2}}{2}\nabla _{K}+K-\frac{F_{2}\left(
R\left( K,X\right) \right) K_{X}}{F_{2}\left( R\left( K_{X},X\right) \right) 
}\right) \Psi _{1}\left( K-K_{X}\right) =0  \label{tsl}
\end{equation}

\paragraph*{A2.1.2.2 Solving (\protect\ref{tsl})}

To solve the first order condition (\ref{tsl}) we perform the change of
variable:%
\begin{equation*}
\Psi _{1}\left( K-K_{X}\right) \rightarrow \exp \left( \frac{1}{\sigma
_{K}^{2}}\int \left[ K-\frac{F_{2}\left( R\left( K,X\right) \right) K_{X}}{%
F_{2}\left( R\left( K_{X},X\right) \right) }\right] dK\right) \Psi
_{1}\left( K-K_{X}\right)
\end{equation*}%
and (\ref{tsl}) is transformed into%
\begin{equation}
-\frac{\sigma _{K}^{2}}{2}\nabla _{K}^{2}\Psi _{1}\left( K-K_{X}\right) +%
\frac{1}{2\sigma _{K}^{2}}\left( K-\frac{F_{2}\left( R\left( K,X\right)
\right) K_{X}}{F_{2}\left( R\left( K_{X},X\right) \right) }\right) ^{2}\Psi
_{1}\left( K-K_{X}\right) =0  \label{scn}
\end{equation}%
This equation can be solved by implementing the constraint:%
\begin{equation*}
\int \left\Vert \Psi _{1}\left( K-K_{X}\right) \right\Vert ^{2}=1
\end{equation*}%
and we find:%
\begin{eqnarray*}
&&\Psi _{1}\left( K-K_{X}\right) \simeq \mathcal{N}\exp \left( -\frac{1}{%
\sigma _{K}^{2}}\left( K-\frac{F_{2}\left( R\left( K,X\right) \right) K_{X}}{%
F_{2}\left( R\left( K_{X},X\right) \right) }\right) ^{2}\right) \\
&\simeq &\mathcal{N}\exp \left( -\frac{1}{\sigma _{K}^{2}}\left(
K-K_{X}-\left( K-K_{X}\right) \frac{\partial _{K}R\left( K_{X},X\right)
F_{2}^{\prime }\left( R\left( K_{X},X\right) \right) }{F_{2}\left( R\left(
K_{X},X\right) \right) }K_{X}\right) ^{2}\right) \\
&=&\mathcal{N}\exp \left( -\frac{1}{\sigma _{K}^{2}}\left( 1-\frac{\partial
_{K}R\left( K_{X},X\right) F_{2}^{\prime }\left( R\left( K,X\right) \right) 
}{F_{2}\left( R\left( K_{X},X\right) \right) }K_{X}\right) ^{2}\left(
K-K_{X}\right) ^{2}\right)
\end{eqnarray*}%
with the normalization factor $\mathcal{N}$\ given by:%
\begin{equation*}
\mathcal{N}=\sqrt{\frac{c}{\sigma _{K}^{2}\left( 1-\frac{\partial
_{K}R\left( K_{X},X\right) F_{2}^{\prime }\left( R\left( K,X\right) \right) 
}{F_{2}\left( R\left( K_{X},X\right) \right) }K_{X}\right) ^{2}}}
\end{equation*}

\paragraph*{A2.1.2.3 Expression for the density of firms $\left\Vert \Psi
\left( K,X\right) \right\Vert ^{2}$}

Having found $\Psi _{1}$, and using (\ref{sp}) and (\ref{dcp}) we obtain the
expression for $\left\Vert \Psi \left( K,X\right) \right\Vert ^{2}$:%
\begin{eqnarray}
\left\Vert \Psi \left( K,X\right) \right\Vert ^{2} &=&\mathcal{N}\left\Vert
\Psi \right\Vert ^{2}\left( X,\left( \nabla _{X}R\left( X\right) \right)
^{2},\frac{\hat{K}_{X}}{\hat{K}_{X,0}}\right)  \label{PSc} \\
&&\times \exp \left( -\frac{1}{\sigma _{K}^{2}}\left( K-\frac{F_{2}\left(
R\left( K,X\right) \right) }{F_{2}\left( R\left( K_{X},X\right) \right)
\left\Vert \Psi \left( X\right) \right\Vert ^{2}}\int \hat{K}\left\Vert \hat{%
\Psi}\left( \hat{K},X\right) \right\Vert ^{2}d\hat{K}\right) ^{2}\right) 
\notag \\
&=&\left\Vert \Psi \right\Vert ^{2}\left( X,\left( \nabla _{X}R\left(
X\right) \right) ^{2},\frac{\hat{K}_{X}}{\hat{K}_{X,0}}\right) \frac{c\exp
\left( -\frac{1}{\sigma _{K}^{2}}\left( 1-\frac{\partial _{K}R\left(
K_{X},X\right) F_{2}^{\prime }\left( R\left( K,X\right) \right) }{%
F_{2}\left( R\left( K_{X},X\right) \right) }K_{X}\right) ^{2}\left(
K-K_{X}\right) ^{2}\right) }{\frac{1}{\sigma _{K}^{2}}\left( 1-\frac{%
\partial _{K}R\left( K_{X},X\right) F_{2}^{\prime }\left( R\left( K,X\right)
\right) }{F_{2}\left( R\left( K_{X},X\right) \right) }K_{X}\right) ^{2}} 
\notag
\end{eqnarray}%
for $X\in V/V_{0}$ and $\left\Vert \Psi \left( K,X\right) \right\Vert ^{2}=0$
otherwise.

As stated in the text, note that the form of the exponential in (\ref{PSc})
implies that:%
\begin{equation*}
\int K\left\Vert \Psi \left( K,X\right) \right\Vert ^{2}d\hat{K}=\int \hat{K}%
\left\Vert \hat{\Psi}\left( \hat{K},X\right) \right\Vert ^{2}d\hat{K}
\end{equation*}

\subsection*{A2.2 \textbf{Examples}}

\subsubsection*{A2.2.1 Example 1}

We compute $\Psi \left( K,X\right) $ for the specific function: 
\begin{equation*}
H\left( y\right) =\left( \frac{y}{1+y}\right) ^{\varsigma }\text{, }%
H^{\prime }\left( y\right) =\varsigma \frac{\left( \frac{y}{y+1}\right)
^{\varsigma }}{y\left( y+1\right) }
\end{equation*}%
We use the simplified equations (\ref{psn}) and (\ref{snp}) that yield:%
\begin{equation*}
D\left( \left\Vert \Psi \right\Vert ^{2}\right) +\mu \left( X\right) =\tau
\left\Vert \Psi \left( X\right) \right\Vert ^{2}+\frac{\frac{1}{\sigma
_{X}^{2}}\left( \nabla _{X}R\left( X\right) \right) ^{2}\left( \left( \frac{%
\int \hat{K}\left\Vert \hat{\Psi}\left( \hat{K},X\right) \right\Vert ^{2}d%
\hat{K}}{\left\Vert \Psi \left( X\right) \right\Vert ^{2}}\right)
^{\varsigma }\right) ^{2}\left( 1-\varsigma \frac{1}{\left( \frac{\int \hat{K%
}\left\Vert \hat{\Psi}\left( \hat{K},X\right) \right\Vert ^{2}d\hat{K}}{%
\left\Vert \Psi \left( X\right) \right\Vert ^{2}}+\left\langle \hat{K}%
\right\rangle \right) }\right) }{\left( \left\langle \hat{K}\right\rangle +%
\frac{\int \hat{K}\left\Vert \hat{\Psi}\left( \hat{K},X\right) \right\Vert
^{2}d\hat{K}}{\left\Vert \Psi \left( X\right) \right\Vert ^{2}}\right)
^{2\varsigma }}
\end{equation*}%
or equivalently:%
\begin{eqnarray*}
D\left( \left\Vert \Psi \right\Vert ^{2}\right) +\mu \left( X\right) &=&\tau
\left\Vert \Psi \left( X\right) \right\Vert ^{2} \\
&&+\frac{\frac{1}{\sigma _{X}^{2}}\left( \nabla _{X}R\left( X\right) \right)
^{2}\left( \int \hat{K}\left\Vert \hat{\Psi}\left( \hat{K},X\right)
\right\Vert ^{2}d\hat{K}\right) ^{2\varsigma }}{\left( \left\langle \hat{K}%
\right\rangle \left\Vert \Psi \left( X\right) \right\Vert ^{2}+\int \hat{K}%
\left\Vert \hat{\Psi}\left( \hat{K},X\right) \right\Vert ^{2}d\hat{K}\right)
^{2\varsigma +1}} \\
&&\times \left( \int \hat{K}\left\Vert \hat{\Psi}\left( \hat{K},X\right)
\right\Vert ^{2}d\hat{K}+\left( 1-\varsigma \right) \left\langle \hat{K}%
\right\rangle \left\Vert \Psi \left( X\right) \right\Vert ^{2}\right)
\end{eqnarray*}%
For $\varsigma \simeq \frac{1}{2}$, this reduces to:%
\begin{equation*}
D\left( \left\Vert \Psi \right\Vert ^{2}\right) +\mu \left( X\right) =\tau
\left\Vert \Psi \left( X\right) \right\Vert ^{2}+\frac{\frac{1}{\sigma
_{X}^{2}}\left( \nabla _{X}R\left( X\right) \right) ^{2}\hat{K}_{X}\left( 
\hat{K}_{X}+\frac{1}{2}\left\langle \hat{K}\right\rangle \left\Vert \Psi
\left( X\right) \right\Vert ^{2}\right) }{\left( \left\langle \hat{K}%
\right\rangle \left\Vert \Psi \left( X\right) \right\Vert ^{2}+\hat{K}%
_{X}\right) ^{2}}
\end{equation*}%
and for $\left\langle \hat{K}\right\rangle \left\Vert \Psi \left( X\right)
\right\Vert ^{2}<<\hat{K}_{X}$ this becomes:%
\begin{equation}
D\left( \left\Vert \Psi \right\Vert ^{2}\right) +\mu \left( X\right) \simeq
\tau \left\Vert \Psi \left( X\right) \right\Vert ^{2}+\frac{\frac{1}{\sigma
_{X}^{2}}\left( \nabla _{X}R\left( X\right) \right) ^{2}\hat{K}_{X}}{\left(
\left\langle \hat{K}\right\rangle \left\Vert \Psi \left( X\right)
\right\Vert ^{2}+\hat{K}_{X}\right) }  \label{prn}
\end{equation}%
Two cases arise.

When $\frac{1}{\sigma _{X}^{2}}\left( \nabla _{X}R\left( X\right) \right)
^{2}<<\tau $: 
\begin{eqnarray}
\left\Vert \Psi \left( X\right) \right\Vert ^{2} &=&\frac{\left( D\left(
\left\Vert \Psi \right\Vert ^{2}\right) -\tau \frac{\hat{K}_{X}}{%
\left\langle \hat{K}\right\rangle }\right) +\sqrt{\left( D\left( \left\Vert
\Psi \right\Vert ^{2}\right) -\tau \frac{\hat{K}_{X}}{\left\langle \hat{K}%
\right\rangle }\right) ^{2}-4\tau \frac{\hat{K}_{X}}{\left\langle \hat{K}%
\right\rangle }\left( \frac{\left( \nabla _{X}R\left( X\right) \right) ^{2}}{%
\sigma _{X}^{2}}-D\left( \left\Vert \Psi \right\Vert ^{2}\right) \right) }}{%
2\tau }  \label{psp} \\
&=&\frac{4\tau \frac{\hat{K}_{X}}{\left\langle \hat{K}\right\rangle }\left( 
\frac{1}{\sigma _{X}^{2}}\left( \nabla _{X}R\left( X\right) \right)
^{2}-D\left( \left\Vert \Psi \right\Vert ^{2}\right) \right) }{2\tau \left(
\left( D\left( \left\Vert \Psi \right\Vert ^{2}\right) -\tau \frac{\hat{K}%
_{X}}{\left\langle \hat{K}\right\rangle }\right) -\sqrt{\left( D\left(
\left\Vert \Psi \right\Vert ^{2}\right) -\tau \frac{\hat{K}_{X}}{%
\left\langle \hat{K}\right\rangle }\right) ^{2}-4\tau \frac{\hat{K}_{X}}{%
\left\langle \hat{K}\right\rangle }\left( \frac{\left( \nabla _{X}R\left(
X\right) \right) ^{2}}{\sigma _{X}^{2}}-D\left( \left\Vert \Psi \right\Vert
^{2}\right) \right) }\right) }  \notag
\end{eqnarray}%
This is positive on the set: 
\begin{equation}
\left\{ \left( D\left( \left\Vert \Psi \right\Vert ^{2}\right) -\tau \frac{%
\hat{K}_{X}}{\left\langle \hat{K}\right\rangle }\right) >0\right\} \cup
\left\{ \frac{1}{\sigma _{X}^{2}}\left( \nabla _{X}R\left( X\right) \right)
^{2}-D\left( \left\Vert \Psi \right\Vert ^{2}\right) <0\right\}  \label{dbc}
\end{equation}%
To detail these two conditions, we write (\ref{prn}) for $\left\Vert \Psi
\left( X\right) \right\Vert ^{2}>0$:%
\begin{equation*}
D\left( \left\Vert \Psi \right\Vert ^{2}\right) \simeq \tau \left\Vert \Psi
\left( X\right) \right\Vert ^{2}+\frac{\frac{1}{\sigma _{X}^{2}}\left(
\nabla _{X}R\left( X\right) \right) ^{2}\frac{\hat{K}_{X}}{\left\langle \hat{%
K}\right\rangle }}{\left( \left\Vert \Psi \left( X\right) \right\Vert ^{2}+%
\frac{\hat{K}_{X}}{\left\langle \hat{K}\right\rangle }\right) }
\end{equation*}%
which is equivalent to:%
\begin{equation*}
\frac{\frac{1}{\sigma _{X}^{2}}\left( \nabla _{X}R\left( X\right) \right)
^{2}-D\left( \left\Vert \Psi \right\Vert ^{2}\right) }{\left( \left\Vert
\Psi \left( X\right) \right\Vert ^{2}+\frac{\hat{K}_{X}}{\left\langle \hat{K}%
\right\rangle }\right) }\frac{\hat{K}_{X}}{\left\langle \hat{K}\right\rangle 
}=\frac{-\tau \left\Vert \Psi \left( X\right) \right\Vert ^{2}+D\left(
\left\Vert \Psi \right\Vert ^{2}\right) -\tau \frac{\hat{K}_{X}}{%
\left\langle \hat{K}\right\rangle }}{\left( \left\Vert \Psi \left( X\right)
\right\Vert ^{2}+\frac{\hat{K}_{X}}{\left\langle \hat{K}\right\rangle }%
\right) }\left\Vert \Psi \left( X\right) \right\Vert ^{2}
\end{equation*}%
Then, we have the implication:%
\begin{equation}
\frac{1}{\sigma _{X}^{2}}\left( \nabla _{X}R\left( X\right) \right)
^{2}-D\left( \left\Vert \Psi \right\Vert ^{2}\right) >0\Rightarrow D\left(
\left\Vert \Psi \right\Vert ^{2}\right) -\tau \frac{\hat{K}_{X}}{%
\left\langle \hat{K}\right\rangle }>0  \label{mpn}
\end{equation}%
This implies that (\ref{dbc}) is always satisfied, and formula (\ref{psp})
is valid for all $X$.

The second case arises when $\frac{1}{\sigma _{X}^{2}}\left( \nabla
_{X}R\left( X\right) \right) ^{2}<<\tau $. In this case, the solution is:%
\begin{equation*}
\left\Vert \Psi \left( X\right) \right\Vert ^{2}=\frac{\left( D\left(
\left\Vert \Psi \right\Vert ^{2}\right) -\tau \frac{\hat{K}_{X}}{%
\left\langle \hat{K}\right\rangle }\right) -\sqrt{\left( D\left( \left\Vert
\Psi \right\Vert ^{2}\right) -\tau \frac{\hat{K}_{X}}{\left\langle \hat{K}%
\right\rangle }\right) ^{2}-4\tau \frac{\hat{K}_{X}}{\left\langle \hat{K}%
\right\rangle }\left( \frac{\left( \nabla _{X}R\left( X\right) \right) ^{2}}{%
\sigma _{X}^{2}}-D\left( \left\Vert \Psi \right\Vert ^{2}\right) \right) }}{%
2\tau }
\end{equation*}%
This solution is valid, i.e. $\left\Vert \Psi \left( X\right) \right\Vert
^{2}>0$, under the conditions: 
\begin{equation}
\left\{ D\left( \left\Vert \Psi \right\Vert ^{2}\right) -\tau \frac{\hat{K}%
_{X}}{\left\langle \hat{K}\right\rangle }>0\right\} \cap \left\{ \frac{1}{%
\sigma _{X}^{2}}\left( \nabla _{X}R\left( X\right) \right) ^{2}-D\left(
\left\Vert \Psi \right\Vert ^{2}\right) >0\right\}  \label{cdt}
\end{equation}%
and $\left\Vert \Psi \right\Vert ^{2}=0$ for:%
\begin{equation*}
\left\{ D\left( \left\Vert \Psi \right\Vert ^{2}\right) -\tau \frac{\hat{K}%
_{X}}{\left\langle \hat{K}\right\rangle }<0\right\} \cup \left\{ \frac{1}{%
\sigma _{X}^{2}}\left( \nabla _{X}R\left( X\right) \right) ^{2}-D\left(
\left\Vert \Psi \right\Vert ^{2}\right) <0\right\}
\end{equation*}%
To detail these two conditions, we use the implication (\ref{mpn}) that is
equivalent to:%
\begin{equation*}
D\left( \left\Vert \Psi \right\Vert ^{2}\right) -\tau \frac{\hat{K}_{X}}{%
\left\langle \hat{K}\right\rangle }<0\Rightarrow \frac{1}{\sigma _{X}^{2}}%
\left( \nabla _{X}R\left( X\right) \right) ^{2}-D\left( \left\Vert \Psi
\right\Vert ^{2}\right) <0
\end{equation*}%
As a consequence, $\left\Vert \Psi \left( X\right) \right\Vert ^{2}=0$ only
if:%
\begin{equation}
\frac{1}{\sigma _{X}^{2}}\left( \nabla _{X}R\left( X\right) \right)
^{2}-D\left( \left\Vert \Psi \right\Vert ^{2}\right) <0  \label{ncl}
\end{equation}%
We find $D\left( \left\Vert \Psi \right\Vert ^{2}\right) $ by integration
of: 
\begin{equation}
D\left( \left\Vert \Psi \right\Vert ^{2}\right) +\mu \left( X\right) \simeq
\tau \left\Vert \Psi \left( X\right) \right\Vert ^{2}+\frac{\frac{1}{\sigma
_{X}^{2}}\left( \nabla _{X}R\left( X\right) \right) ^{2}\hat{K}_{X}}{\left(
\left\langle \hat{K}\right\rangle \left\Vert \Psi \left( X\right)
\right\Vert ^{2}+\hat{K}_{X}\right) }  \label{psd}
\end{equation}%
and this leads to:%
\begin{eqnarray*}
\int_{V/V_{0}}D\left( \left\Vert \Psi \right\Vert ^{2}\right) &\simeq &\tau
N+\int_{V/V_{0}}\frac{\frac{1}{\sigma _{X}^{2}}\left( \nabla _{X}R\left(
X\right) \right) ^{2}\frac{\hat{K}_{X}}{\left\langle \hat{K}\right\rangle }}{%
\left( \left\Vert \Psi \left( X\right) \right\Vert ^{2}+\frac{\hat{K}_{X}}{%
\left\langle \hat{K}\right\rangle }\right) } \\
&\simeq &\tau N+\frac{1}{2}\int_{V/V_{0}}\frac{1}{\sigma _{X}^{2}}\left(
\nabla _{X}R\left( X\right) \right) ^{2}=\tau N+\frac{1}{2}\left(
V-V_{0}\right) \left\langle \frac{1}{\sigma _{X}^{2}}\left( \nabla
_{X}R\left( X\right) \right) ^{2}\right\rangle _{V/V_{0}}
\end{eqnarray*}%
we thus have:%
\begin{equation}
D\left( \left\Vert \Psi \right\Vert ^{2}\right) \simeq \frac{\tau N}{V-V_{0}}%
+\frac{1}{2}\left\langle \frac{1}{\sigma _{X}^{2}}\left( \nabla _{X}R\left(
X\right) \right) ^{2}\right\rangle _{V/V_{0}}  \label{cdn}
\end{equation}%
and $V_{0}$ is defined using (\ref{ncl}). It is the set of points $X$\ such
that :

\begin{equation}
\frac{\tau N}{V-V_{0}}+\frac{1}{2}\left\langle \frac{1}{\sigma _{X}^{2}}%
\left( \nabla _{X}R\left( X\right) \right) ^{2}\right\rangle _{V/V_{0}}-%
\frac{1}{\sigma _{X}^{2}}\left( \nabla _{X}R\left( X\right) \right) ^{2}>0
\label{bnc}
\end{equation}%
Similarly, the set $V/V_{0}$ is defined by: 
\begin{equation}
\frac{\tau N}{V-V_{0}}+\frac{1}{2}\left\langle \frac{1}{\sigma _{X}^{2}}%
\left( \nabla _{X}R\left( X\right) \right) ^{2}\right\rangle _{V/V_{0}}-%
\frac{1}{\sigma _{X}^{2}}\left( \nabla _{X}R\left( X\right) \right) ^{2}<0
\label{cnb}
\end{equation}%
To each function $R\left( X\right) $ and any $d>0$, we associate two
functions that depend on the form of $\frac{1}{\sigma _{X}^{2}}\left( \nabla
_{X}R\left( X\right) \right) ^{2}$ over the whole space. First, $v\left(
V-V_{0}\right) $ is a decreasing function of $V-V_{0}$, defined by:%
\begin{equation}
V\left( \frac{1}{\sigma _{X}^{2}}\left( \nabla _{X}R\left( X\right) \right)
^{2}>v\left( V-V_{0}\right) \right) =V-V_{0}  \label{vft}
\end{equation}%
Second, for every $d\geqslant 0$, the function $h\left( d\right) $ is given
by: 
\begin{equation}
h\left( d\right) =\frac{1}{\int_{\nabla _{X}R\left( X\right) >d}dX}%
\int_{\nabla _{X}R\left( X\right) >d}\frac{1}{\sigma _{X}^{2}}\left( \nabla
_{X}R\left( X\right) \right) ^{2}dX  \label{hcf}
\end{equation}%
This is an increasing function of $d$.

Thus, we can rewrite (\ref{cnb}) as:%
\begin{equation}
\frac{\tau N}{V-V_{0}}+\frac{1}{2}\left\langle \frac{1}{\sigma _{X}^{2}}%
\left( \nabla _{X}R\left( X\right) \right) ^{2}\right\rangle
_{V/V_{0}}=v\left( V-V_{0}\right)  \label{rmf}
\end{equation}%
and moreover, by integration of (\ref{cnb}) over $V/V_{0}$:%
\begin{equation}
\left\langle \frac{1}{\sigma _{X}^{2}}\left( \nabla _{X}R\left( X\right)
\right) ^{2}\right\rangle _{V/V_{0}}=h\left( \frac{\tau N}{V-V_{0}}+\frac{1}{%
2}\left\langle \frac{1}{\sigma _{X}^{2}}\left( \nabla _{X}R\left( X\right)
\right) ^{2}\right\rangle _{V/V_{0}}\right)  \label{frm}
\end{equation}%
Equations (\ref{rmf}) and (\ref{frm}) combine as:%
\begin{equation}
2\left( v\left( V-V_{0}\right) -\frac{\tau N}{V-V_{0}}\right) =h\left(
v\left( V-V_{0}\right) \right)  \label{lvt}
\end{equation}%
which is an equation depending on the form of $R\left( X\right) $. If it has
a solution, the set on which $\left\Vert \Psi \left( X\right) \right\Vert
^{2}=0$ is defined by:%
\begin{equation*}
\frac{1}{\sigma _{X}^{2}}\left( \nabla _{X}R\left( X\right) \right)
^{2}<v\left( V-V_{0}\right)
\end{equation*}%
and $D\left( \left\Vert \Psi \right\Vert ^{2}\right) $ is given by 
\begin{equation*}
D\left( \left\Vert \Psi \right\Vert ^{2}\right) \simeq v\left( V-V_{0}\right)
\end{equation*}%
Once the solution of (\ref{lvt}) is known, the constant $D\left( \left\Vert
\Psi \right\Vert ^{2}\right) $ is given by (\ref{cdn}) and:

\begin{equation}
\left\Vert \Psi \left( X\right) \right\Vert ^{2}=\frac{2\frac{\hat{K}_{X}}{%
\left\langle \hat{K}\right\rangle }\left( \frac{1}{\sigma _{X}^{2}}\left(
\nabla _{X}R\left( X\right) \right) ^{2}-D\left( \left\Vert \Psi \right\Vert
^{2}\right) \right) }{D\left( \left\Vert \Psi \right\Vert ^{2}\right) -\tau 
\frac{\hat{K}_{X}}{\left\langle \hat{K}\right\rangle }+\sqrt{\left( D\left(
\left\Vert \Psi \right\Vert ^{2}\right) -\tau \frac{\hat{K}_{X}}{%
\left\langle \hat{K}\right\rangle }\right) ^{2}-4\tau \frac{\hat{K}_{X}}{%
\left\langle \hat{K}\right\rangle }\left( \frac{1}{\sigma _{X}^{2}}\left(
\nabla _{X}R\left( X\right) \right) ^{2}-D\left( \left\Vert \Psi \right\Vert
^{2}\right) \right) }}  \label{spp}
\end{equation}%
for $X\in V/V_{0}$.

\subsubsection*{A2.2.2 Example 2}

We choose $H\left( y\right) =y$ and equations (\ref{psn}) and (\ref{snp})
yield:%
\begin{equation*}
D\left( \left\Vert \Psi \right\Vert ^{2}\right) \simeq \tau \left\Vert \Psi
\left( X\right) \right\Vert ^{2}+\frac{1}{\sigma _{X}^{2}}\left( \nabla
_{X}R\left( X\right) \right) ^{2}\frac{\hat{K}_{X}}{\left\Vert \Psi \left(
X\right) \right\Vert ^{2}}
\end{equation*}%
If:%
\begin{equation}
D\left( \left\Vert \Psi \right\Vert ^{2}\right) >2\sqrt{\tau \frac{1}{\sigma
_{X}^{2}}\left( \nabla _{X}R\left( X\right) \right) ^{2}\hat{K}_{X}}
\label{dcn}
\end{equation}

then:%
\begin{equation*}
\left\Vert \Psi \left( X\right) \right\Vert ^{2}=\frac{1}{2\tau }\left(
D\left( \left\Vert \Psi \right\Vert ^{2}\right) -\sqrt{\left( D\left(
\left\Vert \Psi \right\Vert ^{2}\right) \right) ^{2}-4\hat{K}_{X}\frac{1}{%
\sigma _{X}^{2}}\left( \nabla _{X}R\left( X\right) \right) ^{2}\tau }\right)
>0
\end{equation*}%
To solve (\ref{cdn}) and to find $V_{0}$, we compute $D\left( \left\Vert
\Psi \right\Vert ^{2}\right) $ by integrating (\ref{psd}) and (\ref{cdn}) is
still valid:%
\begin{equation}
D\left( \left\Vert \Psi \right\Vert ^{2}\right) \simeq \frac{\tau N}{V-V_{0}}%
+\frac{1}{2}\left\langle \frac{1}{\sigma _{X}^{2}}\left( \nabla _{X}R\left(
X\right) \right) ^{2}\right\rangle _{V/V_{0}}  \label{dnc}
\end{equation}%
We proceed as in the previous paragraph to find $D\left( \left\Vert \Psi
\right\Vert ^{2}\right) $ and $V_{0}$. Using (\ref{dnc}), (\ref{dcn})
becomes:%
\begin{equation}
\frac{1}{4\tau \hat{K}_{X}}\left( \frac{\tau N}{V-V_{0}}+\frac{1}{2}%
\left\langle \frac{1}{\sigma _{X}^{2}}\left( \nabla _{X}R\left( X\right)
\right) ^{2}\right\rangle _{V/V_{0}}\right) ^{2}>\frac{1}{\sigma _{X}^{2}}%
\left( \nabla _{X}R\left( X\right) \right) ^{2}  \label{qln}
\end{equation}%
Definitions (\ref{vft}) and (\ref{hcf}) allow to rewrite (\ref{dnc}) and (%
\ref{qln}):%
\begin{equation*}
\frac{1}{4\tau \hat{K}_{X}}\left( \frac{\tau N}{V-V_{0}}+\frac{1}{2}%
\left\langle \frac{1}{\sigma _{X}^{2}}\left( \nabla _{X}R\left( X\right)
\right) ^{2}\right\rangle _{V/V_{0}}\right) ^{2}=v\left( V-V_{0}\right)
\end{equation*}%
\begin{equation*}
\left\langle \frac{1}{\sigma _{X}^{2}}\left( \nabla _{X}R\left( X\right)
\right) ^{2}\right\rangle _{V/V_{0}}=h\left( v\left( V-V_{0}\right) \right)
\end{equation*}%
that reduce to an equation for $V-V_{0}$:%
\begin{equation*}
2\left( 2\sqrt{\tau v\left( V-V_{0}\right) \hat{K}_{X}}-\frac{\tau N}{V-V_{0}%
}\right) =h\left( v\left( V-V_{0}\right) \right)
\end{equation*}%
If it has a solution, the set on which $\left\Vert \Psi \left( X\right)
\right\Vert ^{2}=0$ is defined by:%
\begin{equation*}
\frac{1}{\sigma _{X}^{2}}\left( \nabla _{X}R\left( X\right) \right)
^{2}<v\left( V-V_{0}\right)
\end{equation*}%
and $D\left( \left\Vert \Psi \right\Vert ^{2}\right) $ is given by 
\begin{equation*}
D\left( \left\Vert \Psi \right\Vert ^{2}\right) \simeq 2\sqrt{\tau v\left(
V-V_{0}\right) \hat{K}_{X}}
\end{equation*}

\section*{Appendix 3. Computation of the background field $\hat{\Psi}\left( 
\hat{K},\hat{X}\right) $ and average capital $\hat{K}_{X}$}

\subsection*{A3.1 System for $\hat{\Psi}\left( \hat{K},\hat{X}\right) $}

\subsubsection*{A3.1.1 \textbf{Replacing quantities depending on }$\left(
K,X\right) $}

\textbf{\ }Having found $\Psi \left( K,X\right) $, we can rewrite an action
functional for $\hat{\Psi}\left( \hat{K},\hat{X}\right) $. To do so, we
first replace the quantities depending on $\Psi \left( K,X\right) $ in the
action (\ref{fcn}). Given the form of this function we can use the
approximation $K\simeq K_{X}$: at the collective level, the relevant
quantity, from the point of view of investors are the sectors.

Using that:%
\begin{equation*}
\frac{R\left( K,X\right) }{\int R\left( K^{\prime },X^{\prime }\right)
\left\Vert \Psi \left( K^{\prime },X^{\prime }\right) \right\Vert
^{2}d\left( K^{\prime },X^{\prime }\right) }\simeq \frac{R\left( K,X\right) 
}{\int R\left( K_{X^{\prime }}^{\prime },X^{\prime }\right) \left\Vert \Psi
\left( X^{\prime }\right) \right\Vert ^{2}dX^{\prime }}
\end{equation*}%
we first start by rewriting $F_{1}$ and we have: \textbf{\ }%
\begin{equation*}
F_{1}\left( \frac{R\left( K,X\right) }{\int R\left( K^{\prime },X^{\prime
}\right) \left\Vert \Psi \left( K^{\prime },X^{\prime }\right) \right\Vert
^{2}d\left( K^{\prime },X^{\prime }\right) },\Gamma \left( K,X\right)
\right) \simeq F_{1}\left( \frac{R\left( K_{X},X\right) }{\int R\left(
K_{X^{\prime }}^{\prime },X^{\prime }\right) \left\Vert \Psi \left(
X^{\prime }\right) \right\Vert ^{2}dX^{\prime }},\Gamma \left( K,X\right)
\right)
\end{equation*}%
As explained in appendix 1, when $K\simeq K_{X}$, we also have:%
\begin{equation*}
\Gamma \left( K,X\right) =\int \frac{F_{2}\left( R\left( K,X\right) \right) 
}{K_{X}F_{2}\left( R\left( K_{X},X\right) \right) \left\Vert \Psi \left(
K_{X},X\right) \right\Vert }\hat{K}\left\Vert \hat{\Psi}\left( \hat{K}%
,X\right) \right\Vert ^{2}d\hat{K}-1=0
\end{equation*}%
Then, we rewrite the expression involving $F_{2}$ in (\ref{fcn}): 
\begin{eqnarray*}
\frac{F_{2}\left( R\left( K,\hat{X}\right) \right) }{\int F_{2}\left(
R\left( K^{\prime },\hat{X}\right) \right) \left\Vert \Psi \left( K^{\prime
},\hat{X}\right) \right\Vert ^{2}dK^{\prime }}\left\Vert \Psi \left( K,\hat{X%
}\right) \right\Vert ^{2} &\simeq &\frac{F_{2}\left( R\left( K,\hat{X}%
\right) \right) }{F_{2}\left( R\left( K_{\hat{X}},\hat{X}\right) \right)
\left\Vert \Psi \left( \hat{X}\right) \right\Vert ^{2}}\left\Vert \Psi
\left( K,\hat{X}\right) \right\Vert ^{2} \\
&=&\frac{F_{2}\left( R\left( K,\hat{X}\right) \right) \left\Vert \Psi
_{0}\left( K-K_{\hat{X}}\right) \right\Vert ^{2}}{F_{2}\left( R\left( K_{%
\hat{X}},\hat{X}\right) \right) }
\end{eqnarray*}%
and the $\hat{\Psi}\left( \hat{K},\hat{X}\right) $ part of the action
functional (\ref{fcn}) writes:%
\begin{eqnarray}
&&S_{3}+S_{4}=-\int \hat{\Psi}^{\dag }\left( \hat{K},\hat{X}\right) \left(
\nabla _{\hat{K}}\left( \frac{\sigma _{\hat{K}}^{2}}{2}\nabla _{\hat{K}}-%
\hat{K}f\left( K,X,\Psi ,\hat{\Psi}\right) \right) \right.  \label{mts} \\
&&\left. +\nabla _{\hat{X}}\left( \frac{\sigma _{\hat{X}}^{2}}{2}\nabla _{%
\hat{X}}-g\left( K,X,\Psi ,\hat{\Psi}\right) \right) \right) \hat{\Psi}%
\left( \hat{K},\hat{X}\right)  \notag
\end{eqnarray}%
where:%
\begin{eqnarray}
f\left( \hat{X},\Psi ,\hat{\Psi}\right) &=&\frac{1}{\varepsilon }\int \left(
\nabla _{K}R\left( K,X\right) -\gamma \frac{\int K^{\prime }\left\Vert \Psi
\left( K^{\prime },X\right) \right\Vert ^{2}}{K}+F_{1}\left( \frac{R\left(
K,X\right) }{\int R\left( K^{\prime },X^{\prime }\right) \left\Vert \Psi
\left( K^{\prime },X^{\prime }\right) \right\Vert ^{2}d\left( K^{\prime
},X^{\prime }\right) }\right) \right)  \notag \\
&&\times \frac{F_{2}\left( R\left( K,\hat{X}\right) \right) \left\Vert \Psi
_{0}\left( K-K_{\hat{X}}\right) \right\Vert ^{2}}{F_{2}\left( R\left( K_{%
\hat{X}},\hat{X}\right) \right) }dK  \label{hfr} \\
g\left( \hat{X},\Psi ,\hat{\Psi}\right) &=&\int \left( \frac{\nabla _{\hat{X}%
}F_{0}\left( R\left( K,\hat{X}\right) \right) }{\left\Vert \nabla _{\hat{X}%
}R\left( K,\hat{X}\right) \right\Vert }+\nu \nabla _{\hat{X}}F_{1}\left( 
\frac{R\left( K,\hat{X}\right) }{\int R\left( K^{\prime },X^{\prime }\right)
\left\Vert \Psi \left( K^{\prime },X^{\prime }\right) \right\Vert
^{2}d\left( K^{\prime },X^{\prime }\right) },\Gamma \left( K,X\right)
\right) \right)  \notag \\
&&\times \frac{\left\Vert \Psi \left( K,\hat{X}\right) \right\Vert ^{2}dK}{%
\int \left\Vert \Psi \left( K^{\prime },\hat{X}\right) \right\Vert
^{2}dK^{\prime }}  \label{hgr}
\end{eqnarray}%
An other simplification arises for the function $F_{2}\left( R\left( K,\hat{X%
}\right) \right) $. Actually:

\begin{eqnarray*}
&&\frac{F_{2}\left( R\left( K,\hat{X}\right) \right) }{\int F_{2}\left(
R\left( K^{\prime },\hat{X}\right) \right) \left\Vert \Psi \left( K^{\prime
},\hat{X}\right) \right\Vert ^{2}dK^{\prime }}\left\Vert \Psi \left( K,\hat{X%
}\right) \right\Vert ^{2} \\
&\simeq &\frac{F_{2}\left( R\left( K,\hat{X}\right) \right) }{\int
F_{2}\left( R\left( K_{\hat{X}},\hat{X}\right) \right) \left\Vert \Psi
\left( \hat{X}\right) \right\Vert ^{2}}\left\Vert \Psi \left( K,\hat{X}%
\right) \right\Vert ^{2} \\
&\simeq &\frac{F_{2}\left( R\left( K,\hat{X}\right) \right) }{F_{2}\left(
R\left( K_{\hat{X}},\hat{X}\right) \right) }\left\Vert \Psi \left( K-K_{\hat{%
X}}\right) \right\Vert ^{2}
\end{eqnarray*}%
and by integration in (\ref{hfr}) and (\ref{hgr}), we have:%
\begin{eqnarray}
f\left( \hat{X},\Psi ,\hat{\Psi}\right) &=&\frac{1}{\varepsilon }\left(
r\left( K_{\hat{X}},\hat{X}\right) -\gamma \left\Vert \Psi \left( \hat{X}%
\right) \right\Vert ^{2}+F_{1}\left( \frac{R\left( K_{\hat{X}},\hat{X}%
\right) }{\int R\left( K_{X^{\prime }}^{\prime },X^{\prime }\right)
\left\Vert \Psi \left( X^{\prime }\right) \right\Vert ^{2}dX^{\prime }}%
\right) \right)  \label{ncf} \\
g\left( \hat{X},\Psi ,\hat{\Psi}\right) &=&\frac{\nabla _{\hat{X}%
}F_{0}\left( R\left( K_{\hat{X}},\hat{X}\right) \right) }{\left\Vert \nabla
_{\hat{X}}R\left( K_{\hat{X}},\hat{X}\right) \right\Vert }+\nu \nabla _{\hat{%
X}}F_{1}\left( \frac{R\left( K_{\hat{X}},\hat{X}\right) }{\int R\left(
K_{X^{\prime }}^{\prime },X^{\prime }\right) \left\Vert \Psi \left(
X^{\prime }\right) \right\Vert ^{2}dX^{\prime }}\right)  \notag
\end{eqnarray}%
In the sequel, for the sake of simplicity, we will write $f\left( \hat{X}%
\right) $ and $g\left( \hat{X}\right) $ for $f\left( \hat{X},K_{\hat{X}%
}\right) $ and $g\left( \hat{X},K_{\hat{X}}\right) $ respectively. We then
perform the following change of variable in (\ref{mts}): 
\begin{eqnarray*}
\hat{\Psi} &\rightarrow &\exp \left( \frac{1}{\sigma _{\hat{X}}^{2}}\int
g\left( \hat{X}\right) d\hat{X}\right) \hat{\Psi} \\
\hat{\Psi}^{\dag } &\rightarrow &\exp \left( \frac{1}{\sigma _{\hat{X}}^{2}}%
\int g\left( \hat{X}\right) d\hat{X}\right) \hat{\Psi}^{\dag }
\end{eqnarray*}%
so that (\ref{mts}) becomes:%
\begin{eqnarray}
&&S_{3}+S_{4}=-\int \hat{\Psi}^{\dag }\left( \frac{\sigma _{\hat{X}}^{2}}{2}%
\nabla _{\hat{X}}^{2}-\frac{1}{2\sigma _{\hat{X}}^{2}}\left( g\left( \hat{X}%
,K_{\hat{X}}\right) \right) ^{2}-\frac{1}{2}\nabla _{\hat{X}}g\left( \hat{X}%
,K_{\hat{X}}\right) \right) \hat{\Psi}  \label{stm} \\
&&-\int \hat{\Psi}^{\dag }\left( \nabla _{\hat{K}}\left( \frac{\sigma _{\hat{%
K}}^{2}}{2}\nabla _{\hat{K}}-\hat{K}f\left( \hat{X},K_{\hat{X}}\right)
\right) \right) \hat{\Psi}  \notag
\end{eqnarray}%
This action functional for $\hat{\Psi}$ will be minimized in the next
paragraph. Note that we should also include to (\ref{stm}), the action
functional $S_{1}+S_{2}$ evaluated at the background field $\Psi $, since
this ones depends on $\hat{\Psi}$. However, we have seen that at the
background field $\Psi $, for $K\simeq K_{X}$, $u\left( K,X,\Psi ,\hat{\Psi}%
\right) \simeq 0$ and the action functional $S_{1}+S_{2}$ defined in (\ref%
{prt}) reduces to:

\begin{equation}
S_{1}+S_{2}\simeq \int \Psi ^{\dag }\left( X\right) \left( -\nabla
_{X}\left( \frac{\sigma _{X}^{2}}{2}\nabla _{X}-\left( \nabla _{X}R\left(
X\right) H\left( K_{X}\right) \right) \right) +\tau \left\vert \Psi \left(
X\right) \right\vert ^{2}+\frac{\sigma _{K}^{2}-1}{2\varepsilon }\right)
\Psi \left( X\right)
\end{equation}%
and this depends on through $K_{X}$. Then, due to the first order condition
for $\Psi \left( X\right) $, one has:%
\begin{equation*}
\frac{\delta }{\delta \hat{\Psi}}\left( S_{1}+S_{2}\right) =\frac{\delta
K_{X}}{\delta \hat{\Psi}}\frac{\partial }{\partial K_{X}}\left(
S_{1}+S_{2}\right)
\end{equation*}%
We have assumed previously that $H\left( K_{X}\right) $ is slowly varying.
Moreover, due to is definition:%
\begin{equation*}
\frac{\delta K_{X}}{\delta \hat{\Psi}\left( \hat{K},X\right) }=\frac{\hat{K}%
}{\left\Vert \Psi \left( X\right) \right\Vert ^{2}}
\end{equation*}%
In most of the cases, this reduces to:%
\begin{equation*}
\frac{\delta K_{X}}{\delta \hat{\Psi}\left( \hat{K},X\right) }\simeq \frac{%
\hat{K}}{D\left( \left\Vert \Psi \right\Vert ^{2}\right) }<<\hat{K}
\end{equation*}%
As a consequence, we can assume that $\frac{\delta }{\delta \hat{\Psi}}%
\left( S_{1}+S_{2}\right) $ will be negligible with respect to the other
quantities in the minimization with respect to $\hat{\Psi}\left( \hat{K}%
,X\right) $. The rationale for this approximation is the following. The
field action $S_{1}+S_{2}$ for $\Psi \left( X\right) $ depends on the global
quantity $\int \hat{K}\left\Vert \hat{\Psi}\left( \hat{K},X\right)
\right\Vert ^{2}d\hat{K}$ that represents the total investment in sector $X$%
. While minimizing the field action $S_{1}+S_{2}$ with respect to $\hat{\Psi}%
\left( \hat{K},X\right) $, we compute the change in this action with respect
to an individual variation $\hat{\Psi}\left( \hat{K},X\right) $, and the
impact of this variation is, as a consequence, negligible.

\subsubsection*{A3.1.2 \textbf{Minimization for} $\hat{\Psi}\left( \hat{K},%
\hat{X}\right) $}

Adding the Lagrange multiplier $\hat{\lambda}$ implementing the constraint $%
\int \left\Vert \hat{\Psi}\left( \hat{K},\hat{X}\right) \right\Vert ^{2}=%
\hat{N}$ , the minimization of (\ref{stm}) with the functions given by (\ref%
{ncf}) leads to the first order conditions:

\begin{eqnarray}
0 &=&\left( \frac{\sigma _{\hat{X}}^{2}\nabla _{\hat{X}}^{2}}{2}-\frac{%
\left( g\left( \hat{X},K_{\hat{X}}\right) \right) ^{2}}{2\sigma _{\hat{X}%
}^{2}}-\frac{\nabla _{\hat{X}}g\left( \hat{X},K_{\hat{X}}\right) }{2}\right) 
\hat{\Psi}+\nabla _{\hat{K}}\left( \frac{\sigma _{\hat{K}}^{2}\nabla _{\hat{K%
}}}{2}-\hat{K}f\left( \hat{X},K_{\hat{X}}\right) -\hat{\lambda}\right) \hat{%
\Psi}  \label{drf} \\
&&-\left( \int \hat{\Psi}^{\dag }\frac{\delta }{\delta \hat{\Psi}^{\dag }}%
\left( \frac{1}{2\sigma _{\hat{X}}^{2}}\left( g\left( \hat{X},K_{\hat{X}%
}\right) \right) ^{2}+\frac{1}{2}\nabla _{\hat{X}}g\left( \hat{X},K_{\hat{X}%
}\right) \right) \hat{\Psi}\right) -\left( \int \hat{\Psi}^{\dag }\nabla _{%
\hat{K}}\frac{\delta }{\delta \hat{\Psi}^{\dag }}\left( \hat{K}f\left( \hat{X%
},K_{\hat{X}}\right) \right) \hat{\Psi}\right)  \notag
\end{eqnarray}%
Using that:%
\begin{equation*}
\frac{\delta }{\delta \hat{\Psi}^{\dag }}K_{\hat{X}}=\frac{\hat{K}}{%
\left\Vert \Psi \left( \hat{X}\right) \right\Vert ^{2}}\hat{\Psi}
\end{equation*}%
equation (\ref{drf}) becomes: 
\begin{eqnarray}
0 &=&\left( \frac{\sigma _{\hat{X}}^{2}}{2}\nabla _{\hat{X}}^{2}-\frac{1}{%
2\sigma _{\hat{X}}^{2}}\left( g\left( \hat{X},K_{\hat{X}}\right) \right)
^{2}-\frac{1}{2}\nabla _{\hat{X}}g\left( \hat{X},K_{\hat{X}}\right) \right) 
\hat{\Psi}  \label{fdr} \\
&&+\left( \nabla _{\hat{K}}\left( \frac{\sigma _{\hat{K}}^{2}}{2}\nabla _{%
\hat{K}}-\hat{K}f\left( \hat{X},K_{\hat{X}}\right) \right) -\hat{\lambda}%
\right) \hat{\Psi}-F\left( \hat{X},K_{\hat{X}}\right) \hat{K}\hat{\Psi} 
\notag
\end{eqnarray}%
with:%
\begin{equation}
F\left( \hat{X},K_{\hat{X}}\right) =\frac{\left\langle \nabla _{K_{\hat{X}%
}}\left( \frac{\left( g\left( \hat{X},K_{\hat{X}}\right) \right) ^{2}}{%
2\sigma _{\hat{X}}^{2}}+\frac{1}{2}\nabla _{\hat{X}}g\left( \hat{X},K_{\hat{X%
}}\right) \right) \right\rangle }{\left\Vert \Psi \left( \hat{X}\right)
\right\Vert ^{2}}+\frac{\left\langle \nabla _{\hat{K}}\left( \hat{K}\nabla
_{K_{\hat{X}}}f\left( \hat{X},K_{\hat{X}}\right) \right) \right\rangle }{%
\left\Vert \Psi \left( \hat{X}\right) \right\Vert ^{2}}  \label{dnf}
\end{equation}%
The brackets in (\ref{dnf}) are given by:%
\begin{eqnarray}
&&\left\langle \nabla _{K_{\hat{X}}}\left( \frac{\left( g\left( \hat{X},K_{%
\hat{X}}\right) \right) ^{2}}{2\sigma _{\hat{X}}^{2}}+\frac{1}{2}\nabla _{%
\hat{X}}g\left( \hat{X},K_{\hat{X}}\right) \right) \right\rangle  \notag \\
&=&\int \hat{\Psi}^{\dag }\left( \hat{X},\hat{K}\right) \nabla _{K_{\hat{X}%
}}\left( \frac{\left( g\left( \hat{X},K_{\hat{X}}\right) \right) ^{2}}{%
2\sigma _{\hat{X}}^{2}}+\frac{1}{2}\nabla _{\hat{X}}g\left( \hat{X},K_{\hat{X%
}}\right) \right) \hat{\Psi}\left( \hat{X},\hat{K}\right) d\hat{K}  \notag \\
&\equiv &\nabla _{K_{\hat{X}}}\left( \frac{\left( g\left( \hat{X},K_{\hat{X}%
}\right) \right) ^{2}}{2\sigma _{\hat{X}}^{2}}+\frac{1}{2}\nabla _{\hat{X}%
}g\left( \hat{X},K_{\hat{X}}\right) \right) \left\Vert \hat{\Psi}\left( \hat{%
X}\right) \right\Vert ^{2}  \notag \\
&&\left\langle \nabla _{\hat{K}}\left( \hat{K}\nabla _{K_{\hat{X}}}f\left( 
\hat{X},K_{\hat{X}}\right) \right) \right\rangle  \notag \\
&=&\int \hat{\Psi}^{\dag }\left( \hat{X},K_{\hat{X}}\right) \nabla _{\hat{K}%
}\left( \hat{K}\nabla _{K_{\hat{X}}}f\left( \hat{X},K_{\hat{X}}\right)
\right) \hat{\Psi}\left( \hat{X},K_{\hat{X}}\right) d\hat{K}  \notag \\
&=&-\nabla _{K_{\hat{X}}}f\left( \hat{X},K_{\hat{X}}\right) \int \left( \hat{%
K}\nabla _{\hat{K}}\left\Vert \hat{\Psi}\left( \hat{X},K_{\hat{X}}\right)
\right\Vert ^{2}-\frac{2\hat{K}^{2}}{\sigma _{\hat{K}}^{2}}f\left( \hat{X}%
\right) \left\Vert \hat{\Psi}\left( \hat{X},K_{\hat{X}}\right) \right\Vert
^{2}\right) d\hat{K}  \notag \\
&=&\nabla _{K_{\hat{X}}}f\left( \hat{X},K_{\hat{X}}\right) \left\Vert \hat{%
\Psi}\left( \hat{X}\right) \right\Vert ^{2}+\frac{\nabla _{K_{\hat{X}%
}}f^{2}\left( \hat{X},K_{\hat{X}}\right) }{\sigma _{\hat{K}}^{2}}%
\left\langle \hat{K}^{2}\right\rangle _{\hat{X}}  \label{smt}
\end{eqnarray}%
Where the average $\left\langle \hat{K}^{2}\right\rangle _{\hat{X}}$ is
defined by:%
\begin{equation*}
\left\langle \hat{K}^{2}\right\rangle _{\hat{X}}=\int \left\Vert \hat{\Psi}%
\left( \hat{X},\hat{K}\right) \right\Vert ^{2}d\hat{K}
\end{equation*}%
The previous expression (\ref{smt}) for $F\left( \hat{X},K_{\hat{X}}\right) $
can also be rewritten as: 
\begin{eqnarray}
F\left( \hat{X},K_{\hat{X}}\right) &=&\frac{\left\langle \nabla _{K_{\hat{X}%
}}\left( \frac{\left( g\left( \hat{X},K_{\hat{X}}\right) \right) ^{2}}{%
2\sigma _{\hat{X}}^{2}}+\frac{1}{2}\nabla _{\hat{X}}g\left( \hat{X},K_{\hat{X%
}}\right) \right) \right\rangle }{\left\Vert \Psi \left( \hat{X}\right)
\right\Vert ^{2}}+\frac{\left\langle \nabla _{\hat{K}}\left( \hat{K}\nabla
_{K_{\hat{X}}}f\left( \hat{X},K_{\hat{X}}\right) \right) \right\rangle }{%
\left\Vert \Psi \left( \hat{X}\right) \right\Vert ^{2}}  \label{frl} \\
&=&\nabla _{K_{\hat{X}}}\left( \frac{\left( g\left( \hat{X},K_{\hat{X}%
}\right) \right) ^{2}}{2\sigma _{\hat{X}}^{2}}+\frac{1}{2}\nabla _{\hat{X}%
}g\left( \hat{X},K_{\hat{X}}\right) +f\left( \hat{X},K_{\hat{X}}\right)
\right) \frac{\left\Vert \hat{\Psi}\left( \hat{X}\right) \right\Vert ^{2}}{%
\left\Vert \Psi \left( \hat{X}\right) \right\Vert ^{2}}  \notag \\
&&+\frac{\nabla _{K_{\hat{X}}}f^{2}\left( \hat{X},K_{\hat{X}}\right) }{%
\sigma _{\hat{K}}^{2}\left\Vert \Psi \left( \hat{X}\right) \right\Vert ^{2}}%
\left\langle \hat{K}^{2}\right\rangle _{\hat{X}}  \notag
\end{eqnarray}%
It will be useful to rewrite the last term as:%
\begin{equation}
\frac{\nabla _{K_{\hat{X}}}f^{2}\left( \hat{X},K_{\hat{X}}\right) }{\sigma _{%
\hat{K}}^{2}\left\Vert \Psi \left( \hat{X}\right) \right\Vert ^{2}}%
\left\langle \hat{K}^{2}\right\rangle _{\hat{X}}\simeq \frac{\nabla _{K_{%
\hat{X}}}f^{2}\left( \hat{X},K_{\hat{X}}\right) }{\sigma _{\hat{K}}^{2}}%
\left\langle \hat{K}\right\rangle _{\hat{X}}^{2}=\frac{\nabla _{K_{\hat{X}%
}}f^{2}\left( \hat{X},K_{\hat{X}}\right) }{\sigma _{\hat{K}}^{2}}\frac{%
\left\Vert \Psi \left( \hat{X}\right) \right\Vert ^{2}}{\left\Vert \hat{\Psi}%
\left( \hat{X}\right) \right\Vert ^{2}}  \label{lfr}
\end{equation}%
As a consequence:%
\begin{eqnarray}
F\left( \hat{X},K_{\hat{X}}\right) &=&\nabla _{K_{\hat{X}}}\left( \frac{%
\left( g\left( \hat{X},K_{\hat{X}}\right) \right) ^{2}}{2\sigma _{\hat{X}%
}^{2}}+\frac{1}{2}\nabla _{\hat{X}}g\left( \hat{X},K_{\hat{X}}\right)
+f\left( \hat{X},K_{\hat{X}}\right) \right) \frac{\left\Vert \hat{\Psi}%
\left( \hat{X}\right) \right\Vert ^{2}}{\left\Vert \Psi \left( \hat{X}%
\right) \right\Vert ^{2}}  \label{fnd} \\
&&+\frac{\nabla _{K_{\hat{X}}}f^{2}\left( \hat{X},K_{\hat{X}}\right) }{%
\sigma _{\hat{K}}^{2}}\frac{\left\Vert \Psi \left( \hat{X}\right)
\right\Vert ^{2}}{\left\Vert \hat{\Psi}\left( \hat{X}\right) \right\Vert ^{2}%
}  \notag
\end{eqnarray}%
We also have an equation for $\hat{\Psi}^{\dag }$ similar to (\ref{fdr}):%
\begin{eqnarray}
0 &=&\left( \frac{\sigma _{\hat{X}}^{2}}{2}\nabla _{\hat{X}}^{2}-\frac{1}{%
2\sigma _{\hat{X}}^{2}}\left( g\left( \hat{X},K_{\hat{X}}\right) \right)
^{2}-\frac{1}{2}\nabla _{\hat{X}}g\left( \hat{X},K_{\hat{X}}\right) \right) 
\hat{\Psi}^{\dag }  \label{fdp} \\
&&+\left( \left( \frac{\sigma _{\hat{K}}^{2}}{2}\nabla _{\hat{K}}+\hat{K}%
f\left( \hat{X},K_{\hat{X}}\right) \right) \nabla _{\hat{K}}-\hat{\lambda}%
\right) \hat{\Psi}-F\left( \hat{X},K_{\hat{X}}\right) \hat{K}\hat{\Psi}%
^{\dag }  \notag
\end{eqnarray}

\subsubsection*{A3.1.3 Resolution of (\protect\ref{fdr})}

\paragraph{A3.1.3.1 \textbf{zeroth order in} $\protect\sigma _{X}^{2}$}

We consider $\sigma _{X}^{2}<<1$ (which means that fluctuation in $X<<$
fluctuation in $K$). Thus (\ref{fdr}) writes at the lowest order:

\begin{equation}
\left( \nabla _{\hat{K}}\left( \frac{\sigma _{\hat{K}}^{2}}{2}\nabla _{\hat{K%
}}-\hat{K}f\left( \hat{X},K_{\hat{X}}\right) \right) -\frac{\left( g\left( 
\hat{X}\right) \right) ^{2}}{2\sigma _{\hat{X}}^{2}}-\frac{\nabla _{\hat{X}%
}g\left( \hat{X},K_{\hat{X}}\right) }{2}-F\left( \hat{X},K_{\hat{X}}\right) 
\hat{K}-\hat{\lambda}\right) \hat{\Psi}=0  \label{knq}
\end{equation}%
Performing the change of variable:%
\begin{equation*}
\hat{\Psi}\rightarrow \exp \left( \frac{\hat{K}^{2}}{\sigma _{\hat{K}}^{2}}%
f\left( \hat{X}\right) \right) \hat{\Psi}
\end{equation*}%
leads to the equation for $\hat{K}$: 
\begin{equation}
\frac{\sigma _{\hat{K}}^{2}}{2}\nabla _{\hat{K}}^{2}\hat{\Psi}-\left( \frac{%
\hat{K}^{2}}{2\sigma _{\hat{K}}^{2}}f^{2}\left( \hat{X}\right) +F\left( \hat{%
X},K_{\hat{X}}\right) \hat{K}+\frac{1}{2}f\left( \hat{X},K_{\hat{X}}\right) +%
\frac{\left( g\left( \hat{X}\right) \right) ^{2}}{2\sigma _{\hat{X}}^{2}}+%
\frac{1}{2}\nabla _{\hat{X}}g\left( \hat{X},K_{\hat{X}}\right) +\hat{\lambda}%
\right) \hat{\Psi}\simeq 0  \label{nqk}
\end{equation}%
This equation can be normalized by dividing by $f^{2}\left( \hat{X}\right) $%
: 
\begin{equation*}
\frac{\sigma _{\hat{K}}^{2}\nabla _{\hat{K}}^{2}\hat{\Psi}}{2f^{2}\left( 
\hat{X}\right) }-\left( \frac{\hat{K}^{2}}{2\sigma _{\hat{K}}^{2}}+\frac{%
F\left( \hat{X},K_{\hat{X}}\right) \hat{K}}{f^{2}\left( \hat{X}\right) }+%
\frac{\frac{f\left( \hat{X},K_{\hat{X}}\right) }{2}+\frac{\left( g\left( 
\hat{X}\right) \right) ^{2}}{2\sigma _{\hat{X}}^{2}}+\frac{1}{2}\nabla _{%
\hat{X}}g\left( \hat{X},K_{\hat{X}}\right) +\hat{\lambda}}{f^{2}\left( \hat{X%
}\right) }\right) \hat{\Psi}\simeq 0
\end{equation*}%
We then define:%
\begin{equation*}
y=\frac{\hat{K}+\frac{\sigma _{\hat{K}}^{2}F\left( \hat{X},K_{\hat{X}%
}\right) }{f^{2}\left( \hat{X}\right) }}{\sqrt{\sigma _{\hat{K}}^{2}}}\left(
f^{2}\left( \hat{X}\right) \right) ^{\frac{1}{4}}
\end{equation*}%
and (\ref{fdr}) is transformed into:%
\begin{equation}
\nabla _{y}^{2}\hat{\Psi}-\left( \frac{y^{2}}{4}+\frac{\left( g\left( \hat{X}%
\right) \right) ^{2}+\sigma _{\hat{X}}^{2}\left( f\left( \hat{X}\right)
+\nabla _{\hat{X}}g\left( \hat{X},K_{\hat{X}}\right) -\frac{\sigma _{\hat{K}%
}^{2}F^{2}\left( \hat{X},K_{\hat{X}}\right) }{2f^{2}\left( \hat{X}\right) }+%
\hat{\lambda}\right) }{\sigma _{\hat{X}}^{2}\sqrt{f^{2}\left( \hat{X}\right) 
}}\right) \Psi \simeq 0  \label{lgm}
\end{equation}%
Solutions of (\ref{lgm}) are obtained by rewriting (\ref{lgm}):%
\begin{equation*}
\hat{\Psi}^{\prime \prime }+\left( p\left( \hat{X},\hat{\lambda}\right) +%
\frac{1}{2}-\frac{1}{4}y^{2}\right) \hat{\Psi}
\end{equation*}%
$\allowbreak \allowbreak $where:%
\begin{equation}
p\left( \hat{X},\hat{\lambda}\right) =-\frac{\left( g\left( \hat{X}\right)
\right) ^{2}+\sigma _{\hat{X}}^{2}\left( f\left( \hat{X}\right) +\nabla _{%
\hat{X}}g\left( \hat{X},K_{\hat{X}}\right) -\frac{\sigma _{\hat{K}%
}^{2}F^{2}\left( \hat{X},K_{\hat{X}}\right) }{2f^{2}\left( \hat{X}\right) }+%
\hat{\lambda}\right) }{\sigma _{\hat{X}}^{2}\sqrt{f^{2}\left( \hat{X}\right) 
}}-\frac{1}{2}  \label{pxd}
\end{equation}%
The solution of (\ref{lgm}) is thus:%
\begin{equation}
\hat{\Psi}_{\hat{\lambda},C}^{\left( 0\right) }\left( \hat{X},\hat{K}\right)
=\sqrt{C}D_{p\left( \hat{X},\hat{\lambda}\right) }\left( \left( \left\vert
f\left( \hat{X}\right) \right\vert \right) ^{\frac{1}{2}}\frac{\left( \hat{K}%
+\frac{\sigma _{\hat{K}}^{2}F\left( \hat{X},K_{\hat{X}}\right) }{f^{2}\left( 
\hat{X}\right) }\right) }{\sigma _{\hat{K}}}\right)  \label{slt}
\end{equation}%
where $D_{p}$ denotes the parabolic cylinder function with parameter $p$ and 
$C$ is a normalization constant that will be computed as a function of $%
\lambda $ using the constraint $\int \left\Vert \hat{\Psi}\left( \hat{K},%
\hat{X}\right) \right\Vert ^{2}=\hat{N}$.

A similar equation to (\ref{knq}) can be obtained for $\hat{\Psi}^{\dag }$.
The equivalent of (\ref{drf}) is (\ref{fdp}):%
\begin{eqnarray}
0 &=&\left( \frac{\sigma _{\hat{X}}^{2}}{2}\nabla _{\hat{X}}^{2}-\frac{1}{%
2\sigma _{\hat{X}}^{2}}\left( g\left( \hat{X},K_{\hat{X}}\right) \right)
^{2}-\frac{1}{2}\nabla _{\hat{X}}g\left( \hat{X},K_{\hat{X}}\right) \right) 
\hat{\Psi}  \label{cjt} \\
&&+\left( \left( \frac{\sigma _{\hat{K}}^{2}}{2}\nabla _{\hat{K}}+\hat{K}%
f\left( \hat{X},K_{\hat{X}}\right) \right) \nabla _{\hat{K}}-\hat{\lambda}%
\right) \hat{\Psi}-F\left( \hat{X},K_{\hat{X}}\right) \hat{K}\hat{\Psi} 
\notag
\end{eqnarray}%
The change of variable:%
\begin{equation*}
\hat{\Psi}^{\dag }\rightarrow \exp \left( -\frac{\hat{K}^{2}}{\sigma _{\hat{K%
}}^{2}}f\left( \hat{X}\right) \right) \hat{\Psi}^{\dag }
\end{equation*}%
and the approximation $\sigma _{\hat{X}}^{2}<<1$ lead ultimately to:%
\begin{equation}
\frac{\sigma _{\hat{K}}^{2}}{2}\nabla _{\hat{K}}^{2}\hat{\Psi}^{\dag
}-\left( \frac{\hat{K}^{2}}{2\sigma _{\hat{K}}^{2}}f^{2}\left( \hat{X}%
\right) +\frac{1}{2}\nabla _{\hat{X}}f\left( \hat{X},K_{\hat{X}}\right) +%
\frac{1}{2\sigma _{\hat{X}}^{2}}\left( g\left( \hat{X}\right) \right) ^{2}+%
\frac{1}{2}\nabla _{\hat{X}}g\left( \hat{X},K_{\hat{X}}\right) +F\left( \hat{%
X},K_{\hat{X}}\right) +\hat{\lambda}\right) \hat{\Psi}^{\dag }\simeq 0
\label{fbm}
\end{equation}%
which is the same equation as (\ref{nqk}). As a consequence, the solutions
of (\ref{fbm}) write:%
\begin{equation}
\hat{\Psi}_{\lambda ,C}^{\left( 0\right) \dag }\left( \hat{X},\hat{K}\right)
=\hat{\Psi}_{\lambda ,C}^{\left( 0\right) }\left( \hat{X},\hat{K}\right) =%
\sqrt{C}D_{p\left( \hat{X},\hat{\lambda}\right) }\left( \left( \left\vert
f\left( \hat{X}\right) \right\vert \right) ^{\frac{1}{2}}\frac{\left( \hat{K}%
+\frac{\sigma _{\hat{K}}^{2}F\left( \hat{X},K_{\hat{X}}\right) }{f^{2}\left( 
\hat{X}\right) }\right) }{\sigma _{\hat{K}}}\right)  \label{tls}
\end{equation}%
To conclude this section, we detail the expressions for $\frac{\sigma _{\hat{%
K}}^{2}F\left( \hat{X},K_{\hat{X}}\right) }{f^{2}\left( \hat{X}\right) }$
and $\frac{\sigma _{\hat{K}}^{2}F^{2}\left( \hat{X},K_{\hat{X}}\right) }{%
2f^{2}\left( \hat{X}\right) }$. \ Given the expression for $F\left( \hat{X}%
,K_{\hat{X}}\right) $ in (\ref{fnd}), the term $\frac{\sigma _{\hat{K}%
}^{2}F\left( \hat{X},K_{\hat{X}}\right) }{f^{2}\left( \hat{X}\right) }$
arising in (\ref{slt}) and (\ref{tls}) 
\begin{eqnarray}
\frac{\sigma _{\hat{K}}^{2}F\left( \hat{X},K_{\hat{X}}\right) }{f^{2}\left( 
\hat{X},K_{\hat{X}}\right) } &=&\frac{\sigma _{\hat{K}}^{2}}{f^{2}\left( 
\hat{X}\right) }\nabla _{K_{\hat{X}}}\left( \frac{\left( g\left( \hat{X},K_{%
\hat{X}}\right) \right) ^{2}}{2\sigma _{\hat{X}}^{2}}+\frac{1}{2}\nabla _{%
\hat{X}}g\left( \hat{X},K_{\hat{X}}\right) +f\left( \hat{X},K_{\hat{X}%
}\right) \right) \frac{\left\Vert \hat{\Psi}\left( \hat{X}\right)
\right\Vert ^{2}}{\left\Vert \Psi \left( \hat{X}\right) \right\Vert ^{2}} 
\notag \\
&&+\frac{\nabla _{K_{\hat{X}}}f^{2}\left( \hat{X},K_{\hat{X}}\right) }{%
f^{2}\left( \hat{X},K_{\hat{X}}\right) }\frac{\left\Vert \Psi \left( \hat{X}%
\right) \right\Vert ^{2}}{\left\Vert \hat{\Psi}\left( \hat{X}\right)
\right\Vert ^{2}}  \notag \\
&\simeq &\frac{\nabla _{K_{\hat{X}}}f\left( \hat{X},K_{\hat{X}}\right) }{%
f\left( \hat{X},K_{\hat{X}}\right) }\frac{\left\Vert \Psi \left( \hat{X}%
\right) \right\Vert ^{2}}{\left\Vert \hat{\Psi}\left( \hat{X}\right)
\right\Vert ^{2}}  \label{tmd}
\end{eqnarray}%
$\frac{\sigma _{\hat{K}}^{2}F^{2}\left( \hat{X},K_{\hat{X}}\right) }{%
2f^{2}\left( \hat{X}\right) }$ arising in the definition (\ref{pxd}) of $%
p\left( \hat{X},\hat{\lambda}\right) $ is equal to: 
\begin{eqnarray}
\frac{\sigma _{\hat{K}}^{2}F^{2}\left( \hat{X},K_{\hat{X}}\right) }{%
2f^{2}\left( \hat{X}\right) } &=&\frac{\sigma _{\hat{K}}^{2}}{2}\left(
\left( \frac{\nabla _{K_{\hat{X}}}\left( g\left( \hat{X},K_{\hat{X}}\right)
\right) ^{2}+\sigma _{\hat{X}}^{2}\left( \nabla _{\hat{X}}^{2}g\left( \hat{X}%
,K_{\hat{X}}\right) +\nabla _{K_{\hat{X}}}f\left( \hat{X},K_{\hat{X}}\right)
\right) }{2\sigma _{\hat{X}}^{2}f\left( \hat{X},K_{\hat{X}}\right) }\right) 
\frac{\left\Vert \hat{\Psi}\left( \hat{X}\right) \right\Vert ^{2}}{%
\left\Vert \Psi \left( \hat{X}\right) \right\Vert ^{2}}\right.  \notag \\
&&\left. +2\nabla _{K_{\hat{X}}}f\left( \hat{X},K_{\hat{X}}\right) \frac{%
\left\Vert \Psi \left( \hat{X}\right) \right\Vert ^{2}}{\left\Vert \hat{\Psi}%
\left( \hat{X}\right) \right\Vert ^{2}}\right) ^{2}  \label{kmd}
\end{eqnarray}%
and this simplifies as:%
\begin{equation}
\frac{\sigma _{\hat{K}}^{2}F^{2}\left( \hat{X},K_{\hat{X}}\right) }{%
2f^{2}\left( \hat{X}\right) }\simeq 2\sigma _{\hat{K}}^{2}\left( \nabla _{K_{%
\hat{X}}}f\left( \hat{X},K_{\hat{X}}\right) \frac{\left\Vert \Psi \left( 
\hat{X}\right) \right\Vert ^{2}}{\left\Vert \hat{\Psi}\left( \hat{X}\right)
\right\Vert ^{2}}\right) ^{2}  \label{kmdp}
\end{equation}%
since:%
\begin{eqnarray*}
&&\frac{\nabla _{K_{\hat{X}}}\left( g\left( \hat{X},K_{\hat{X}}\right)
\right) ^{2}+\sigma _{\hat{X}}^{2}\left( \nabla _{\hat{X}}^{2}g\left( \hat{X}%
,K_{\hat{X}}\right) +\nabla _{K_{\hat{X}}}f\left( \hat{X},K_{\hat{X}}\right)
\right) }{2\sigma _{\hat{X}}^{2}f\left( \hat{X},K_{\hat{X}}\right) }\frac{%
\left\Vert \hat{\Psi}\left( \hat{X}\right) \right\Vert ^{2}}{\left\Vert \Psi
\left( \hat{X}\right) \right\Vert ^{2}} \\
&\sim &\frac{\left( g\left( \hat{X},K_{\hat{X}}\right) \right) ^{2}+\sigma _{%
\hat{X}}^{2}\left( \nabla _{\hat{X}}^{2}g\left( \hat{X},K_{\hat{X}}\right)
+\nabla _{K_{\hat{X}}}f\left( \hat{X},K_{\hat{X}}\right) \right) }{2\sigma _{%
\hat{X}}^{2}f\left( \hat{X},K_{\hat{X}}\right) }\left( \frac{\left\Vert \hat{%
\Psi}\left( \hat{X}\right) \right\Vert ^{2}}{K_{\hat{X}}\left\Vert \Psi
\left( \hat{X}\right) \right\Vert ^{2}}\right) <<1
\end{eqnarray*}

\paragraph*{A3.1.3.2 \textbf{Corrections in} $\protect\sigma _{X}^{2}$:}

To introduce the corrections in $\sigma _{X}^{2}$ in (\ref{fdr}) we factor
the solution as: 
\begin{eqnarray*}
\hat{\Psi}_{\lambda ,C}\left( \hat{K},\hat{X}\right) &=&\sqrt{C}\exp \left( 
\frac{\hat{K}^{2}}{\sigma _{\hat{K}}^{2}}f\left( \hat{X}\right) \right)
D_{p\left( \hat{X},\hat{\lambda}\right) }\left( \left( \frac{\left\vert
f\left( \hat{X}\right) \right\vert }{\sigma _{\hat{K}}^{2}}\right) ^{\frac{1%
}{2}}\left( \hat{K}+\frac{\sigma _{\hat{K}}^{2}F\left( \hat{X},K_{\hat{X}%
}\right) }{f^{2}\left( \hat{X}\right) }\right) \right) \hat{\Psi}^{\left(
1\right) }\left( \hat{K},\hat{X}\right) \\
&\equiv &\hat{\Psi}_{\lambda ,C}^{\left( 0\right) }\left( \hat{K},\hat{X}%
\right) \hat{\Psi}^{\left( 1\right) }\left( \hat{K},\hat{X}\right)
\end{eqnarray*}%
and we look for $\hat{\Psi}^{\left( 1\right) }$ of the form:%
\begin{equation}
\hat{\Psi}^{\left( 1\right) }=\exp \left( \sigma _{X}^{2}h\left( K,X\right)
\right)  \label{cps}
\end{equation}%
Introducing the postulated form in (\ref{fdr}) we are lead to:

\begin{equation*}
\frac{\sigma _{X}^{2}}{2}\nabla _{\hat{X}}^{2}\left( \hat{\Psi}^{\left(
1\right) }\hat{\Psi}_{\lambda ,C}^{\left( 0\right) }\right) +\left( \frac{%
\sigma _{\hat{K}}^{2}}{2}\nabla _{\hat{K}}^{2}\hat{\Psi}^{\left( 1\right)
}\right) \hat{\Psi}_{\lambda ,C}^{\left( 0\right) }+\left( \nabla _{\hat{K}}%
\hat{\Psi}^{\left( 1\right) }\right) \left( \sigma _{\hat{K}}^{2}\nabla _{%
\hat{K}}\hat{\Psi}_{\lambda ,C}^{\left( 0\right) }-\hat{K}f\left( \hat{X}%
\right) \hat{\Psi}_{\lambda ,C}^{\left( 0\right) }\right) =0
\end{equation*}%
Written in terms of $h\left( \hat{K},\hat{X}\right) $, this equation becomes
at the first order in $\sigma _{X}^{2}$: 
\begin{equation}
\frac{\nabla _{\hat{X}}^{2}\hat{\Psi}_{\lambda ,C}^{\left( 0\right) }}{\hat{%
\Psi}_{\lambda ,C}^{\left( 0\right) }}+\sigma _{\hat{K}}^{2}\nabla _{\hat{K}%
}^{2}h\left( \hat{K},\hat{X}\right) +2\left( \nabla _{\hat{K}}h\left( \hat{K}%
,\hat{X}\right) \right) \left( \sigma _{\hat{K}}^{2}\frac{\nabla _{\hat{K}}%
\hat{\Psi}_{\lambda ,C}^{\left( 0\right) }}{\hat{\Psi}_{\lambda ,C}^{\left(
0\right) }}-\hat{K}f\left( \hat{X}\right) \right) =0  \label{qnh}
\end{equation}%
The solution of\ (\ref{qnh}) is of the type:

\begin{equation*}
\nabla _{\hat{K}}\left( h\left( K,X\right) \right) =C\left( \hat{K},X\right)
\exp \left( -2\int \left( \frac{\nabla _{\hat{K}}\hat{\Psi}_{\lambda
,C}^{\left( 0\right) }}{\hat{\Psi}_{\lambda ,C}^{\left( 0\right) }}-\frac{%
\hat{K}f\left( \hat{X}\right) }{\sigma _{\hat{K}}^{2}}\right) d\hat{K}%
\right) =C\left( \hat{K},X\right) \exp \left( -\left( 2\ln \hat{\Psi}%
_{\lambda ,C}^{\left( 0\right) }-\frac{\hat{K}^{2}}{\sigma _{\hat{K}}^{2}}%
f\left( \hat{X}\right) \right) \right)
\end{equation*}%
where $C\left( X\right) $\ satifies: 
\begin{equation*}
C^{\prime }\left( \hat{K},X\right) =-\frac{\nabla _{\hat{X}}^{2}\hat{\Psi}%
_{\lambda ,C}^{\left( 0\right) }}{\hat{\Psi}_{\lambda ,C}^{\left( 0\right)
}\sigma _{\hat{K}}^{2}}\exp \left( 2\ln \hat{\Psi}_{\lambda ,C}^{\left(
0\right) }-\frac{\hat{K}^{2}f\left( \hat{X}\right) }{\sigma _{\hat{K}}^{2}}%
\right) =-\frac{\nabla _{\hat{X}}^{2}\hat{\Psi}_{\lambda ,C}^{\left(
0\right) }}{\hat{\Psi}_{\lambda ,C}^{\left( 0\right) }}\left( \hat{\Psi}%
_{\lambda ,C}^{\left( 0\right) }\right) ^{2}\exp \left( -\frac{\hat{K}%
^{2}f\left( \hat{X}\right) }{\sigma _{\hat{K}}^{2}}\right)
\end{equation*}%
and the solution of (\ref{qnh}) is:%
\begin{equation*}
\nabla _{\hat{K}}\left( h\left( K,X\right) \right) =\exp \left( -\left( 2\ln 
\hat{\Psi}_{\lambda ,C}^{\left( 0\right) }-\frac{\hat{K}^{2}}{\sigma _{\hat{K%
}}^{2}}f\left( \hat{X}\right) \right) \right) \left( C-\int \frac{\nabla _{%
\hat{X}}^{2}\hat{\Psi}_{\lambda ,C}^{\left( 0\right) }}{\hat{\Psi}_{\lambda
,C}^{\left( 0\right) }\sigma _{\hat{K}}^{2}}\left( \hat{\Psi}_{\lambda
,C}^{\left( 0\right) }\right) ^{2}\exp \left( -\frac{\hat{K}^{2}}{\sigma _{%
\hat{K}}^{2}}f\left( \hat{X}\right) \right) d\hat{K}\right)
\end{equation*}%
letting $C=0$, we obtain:%
\begin{equation}
\nabla _{\hat{K}}\left( h\left( K,X\right) \right) =-\frac{1}{\sigma _{\hat{K%
}}^{2}\left( \hat{\Psi}_{\lambda ,C}^{\left( 0\right) }\right) ^{2}}\exp
\left( \frac{\hat{K}^{2}}{\sigma _{\hat{K}}^{2}}f\left( \hat{X}\right)
\right) \left( \int \frac{\nabla _{\hat{X}}^{2}\hat{\Psi}_{\lambda
,C}^{\left( 0\right) }}{\hat{\Psi}_{\lambda ,C}^{\left( 0\right) }}\left( 
\hat{\Psi}_{\lambda ,C}^{\left( 0\right) }\right) ^{2}\exp \left( -\frac{%
\hat{K}^{2}}{\sigma _{\hat{K}}^{2}}f\left( \hat{X}\right) \right) d\hat{K}%
\right)  \label{drg}
\end{equation}%
To compute $h\left( K,X\right) $, we have to estimate $\frac{\nabla _{\hat{X}%
}^{2}\hat{\Psi}_{\lambda ,C}^{\left( 0\right) }}{\hat{\Psi}_{\lambda
,C}^{\left( 0\right) }}$ in (\ref{drg}). To do so, we write, for $%
\varepsilon <<1$, i.e. $\left\vert f\left( \hat{X}\right) \right\vert >>1$:

\begin{eqnarray*}
&&\exp \left( \frac{\hat{K}^{2}}{\sigma _{\hat{K}}^{2}}f\left( \hat{X}%
\right) \right) D_{p\left( \hat{X},\hat{\lambda}\right) }\left( \left( \frac{%
\left\vert f\left( \hat{X}\right) \right\vert }{\sigma _{\hat{K}}^{2}}%
\right) ^{\frac{1}{2}}\left( \hat{K}+\frac{\sigma _{\hat{K}}^{2}F\left( \hat{%
X},K_{\hat{X}}\right) }{f^{2}\left( \hat{X}\right) }\right) \right) \\
&\simeq &\exp \left( \frac{\hat{K}^{2}}{\sigma _{\hat{K}}^{2}}f\left( \hat{X}%
\right) -\frac{\left( \hat{K}+\frac{\sigma _{\hat{K}}^{2}F\left( \hat{X},K_{%
\hat{X}}\right) }{f^{2}\left( \hat{X}\right) }\right) ^{2}\left\vert f\left( 
\hat{X}\right) \right\vert }{4\sigma _{\hat{K}}^{2}}\right) \left( \left( 
\frac{\left\vert f\left( \hat{X}\right) \right\vert }{\sigma _{\hat{K}}^{2}}%
\right) ^{\frac{1}{2}}\left( \hat{K}+\frac{\sigma _{\hat{K}}^{2}F\left( \hat{%
X},K_{\hat{X}}\right) }{f^{2}\left( \hat{X}\right) }\right) \right)
^{p\left( \hat{X},\hat{\lambda}\right) } \\
&=&\exp \left( \frac{\hat{K}^{2}}{\sigma _{\hat{K}}^{2}}f\left( \hat{X}%
\right) -\frac{\left( \hat{K}+\frac{\sigma _{\hat{K}}^{2}F\left( \hat{X},K_{%
\hat{X}}\right) }{f^{2}\left( \hat{X}\right) }\right) ^{2}\left\vert f\left( 
\hat{X}\right) \right\vert }{4\sigma _{\hat{K}}^{2}}\right) \\
&&\times \exp \left( \left( p\left( \hat{X},\hat{\lambda}\right) \right) \ln
\left( \left( \frac{\left\vert f\left( \hat{X}\right) \right\vert }{\sigma _{%
\hat{K}}^{2}}\right) ^{\frac{1}{2}}\left( \hat{K}+\frac{\sigma _{\hat{K}%
}^{2}F\left( \hat{X},K_{\hat{X}}\right) }{f^{2}\left( \hat{X}\right) }%
\right) \right) \right)
\end{eqnarray*}%
which allows to compute the successives derivatives of $\hat{\Psi}$. We
find, for $f>0$:%
\begin{eqnarray}
\frac{\nabla _{\hat{X}}^{2}\hat{\Psi}_{\lambda ,C}^{\left( 0\right) }}{\hat{%
\Psi}_{\lambda ,C}^{\left( 0\right) }} &\simeq &\left( \frac{-f^{\prime
}\sigma _{\hat{X}}^{2}\hat{\lambda}-g^{2}f^{\prime }+2fgg^{\prime }}{\sigma
_{\hat{X}}^{2}f^{2}}\ln \left( \left( \hat{K}+\frac{\sigma _{\hat{K}%
}^{2}F\left( \hat{X},K_{\hat{X}}\right) }{f^{2}\left( \hat{X}\right) }%
\right) \left( \frac{f\left( \hat{X}\right) }{\sigma _{\hat{K}}^{2}}\right)
^{\frac{1}{2}}\right) \right.  \label{dvt} \\
&&+\frac{1}{2}\left( \frac{\left( g\left( \hat{X}\right) \right) ^{2}+\sigma
_{\hat{X}}^{2}\left( f\left( \hat{X}\right) +\nabla _{\hat{X}}g\left( \hat{X}%
,K_{\hat{X}}\right) -\frac{\sigma _{\hat{K}}^{2}F^{2}\left( \hat{X},K_{\hat{X%
}}\right) }{2f^{2}\left( \hat{X}\right) }+\hat{\lambda}\right) }{\sigma _{%
\hat{X}}^{2}\sqrt{f^{2}\left( \hat{X}\right) }}+\frac{1}{2}\right) \frac{%
f^{\prime }}{f}  \notag \\
&&\left. +\frac{\hat{K}^{2}-\left( \frac{\hat{K}+\frac{\sigma _{\hat{K}%
}^{2}F\left( \hat{X},K_{\hat{X}}\right) }{f^{2}\left( \hat{X}\right) }}{2}%
\right) ^{2}}{\sigma _{\hat{K}}^{2}}f^{\prime }\right) ^{2}  \notag \\
&\simeq &\left( \frac{\left( 4\hat{K}^{2}-\left( \hat{K}+\frac{\sigma _{\hat{%
K}}^{2}F\left( \hat{X},K_{\hat{X}}\right) }{f^{2}\left( \hat{X}\right) }%
\right) ^{2}\right) f^{\prime }\left( X\right) }{4\sigma _{\hat{K}}^{2}}%
\right) ^{2}  \notag
\end{eqnarray}%
The same approximation is valid for $f<0$ and we find for this case:%
\begin{equation*}
\frac{\nabla _{\hat{X}}^{2}\hat{\Psi}_{\lambda ,C}^{\left( 0\right) }}{\hat{%
\Psi}_{\lambda ,C}^{\left( 0\right) }}\simeq \left( \frac{\left( 4\hat{K}%
^{2}+\left( \hat{K}+\frac{\sigma _{\hat{K}}^{2}F\left( \hat{X},K_{\hat{X}%
}\right) }{f^{2}\left( \hat{X}\right) }\right) ^{2}\right) f^{\prime }\left(
X\right) }{4\sigma _{\hat{K}}^{2}}\right) ^{2}
\end{equation*}%
Then, introducing $\mp $ to account for the sign of $-f$, (\ref{drg})
becomes:%
\begin{eqnarray}
&\nabla _{\hat{K}}\left( h\left( K,X\right) \right) =&-\frac{1}{\sigma _{%
\hat{K}}^{2}\left( \hat{\Psi}_{\lambda ,C}^{\left( 0\right) }\right) ^{2}}%
\exp \left( \frac{\hat{K}^{2}}{\sigma _{\hat{K}}^{2}}f\left( \hat{X}\right)
\right) \int \left( \frac{\left( 4\hat{K}^{2}\mp \left( \hat{K}+\frac{\sigma
_{\hat{K}}^{2}F\left( \hat{X},K_{\hat{X}}\right) }{f^{2}\left( \hat{X}%
\right) }\right) ^{2}\right) f^{\prime }\left( X\right) }{4\sigma _{\hat{K}%
}^{2}}\right) ^{2} \\
&&\times \left( \hat{\Psi}_{\lambda ,C}^{\left( 0\right) }\right) ^{2}\exp
\left( -\frac{\hat{K}^{2}}{\sigma _{\hat{K}}^{2}}f\left( \hat{X}\right)
\right) d\hat{K}  \notag \\
&\simeq &-\frac{1}{\sigma _{\hat{K}}^{2}\left( \hat{\Psi}_{\lambda
,C}^{\left( 0\right) }\right) ^{2}}\exp \left( \frac{\hat{K}^{2}}{\sigma _{%
\hat{K}}^{2}}f\left( \hat{X}\right) \right) \int \left( \frac{\left( 4\hat{K}%
^{2}\mp \left( \hat{K}+\frac{\sigma _{\hat{K}}^{2}F\left( \hat{X},K_{\hat{X}%
}\right) }{f^{2}\left( \hat{X}\right) }\right) ^{2}\right) f^{\prime }\left(
X\right) }{4\sigma _{\hat{K}}^{2}}\right) ^{2}  \notag \\
&&\times \exp \left( \frac{\hat{K}^{2}f\left( \hat{X}\right) -\frac{1}{2}%
\left( \hat{K}+\frac{\sigma _{\hat{K}}^{2}F\left( \hat{X},K_{\hat{X}}\right) 
}{f^{2}\left( \hat{X}\right) }\right) ^{2}\left\vert f\left( \hat{X}\right)
\right\vert }{\sigma _{\hat{K}}^{2}}\right) d\hat{K} \\
&\simeq &-\frac{1}{\sigma _{\hat{K}}^{2}\left( \hat{\Psi}_{\lambda
,C}^{\left( 0\right) }\right) ^{2}}\exp \left( \frac{\hat{K}^{2}}{\sigma _{%
\hat{K}}^{2}}f\left( \hat{X}\right) \right) \int \left( \sigma _{\hat{K}}^{2}%
\frac{\left( \hat{K}^{2}\mp \frac{1}{4}\left( \hat{K}+\frac{\sigma _{\hat{K}%
}^{2}F\left( \hat{X},K_{\hat{X}}\right) }{f^{2}\left( \hat{X}\right) }%
\right) ^{2}\right) f^{\prime }\left( X\right) }{\left( 2\hat{K}f\left( \hat{%
X}\right) -\left( \hat{K}+\frac{\sigma _{\hat{K}}^{2}F\left( \hat{X},K_{\hat{%
X}}\right) }{f^{2}\left( \hat{X}\right) }\right) \left\vert f\left( \hat{X}%
\right) \right\vert \right) ^{2}}\right) ^{2}  \notag \\
&&\times \partial _{\hat{K}}^{4}\exp \left( \frac{\hat{K}^{2}f\left( \hat{X}%
\right) -\frac{1}{2}\left( \hat{K}+\frac{\sigma _{\hat{K}}^{2}F\left( \hat{X}%
,K_{\hat{X}}\right) }{f^{2}\left( \hat{X}\right) }\right) ^{2}\left\vert
f\left( \hat{X}\right) \right\vert }{\sigma _{\hat{K}}^{2}}\right) d\hat{K}
\end{eqnarray}%
Assuming $\frac{\sigma _{\hat{K}}^{2}F\left( \hat{X},K_{\hat{X}}\right) }{%
f^{2}\left( \hat{X}\right) }<<1$, we have ultimately:%
\begin{eqnarray*}
\nabla _{\hat{K}}\left( h\left( K,X\right) \right) &\simeq &-\frac{1}{\sigma
_{\hat{K}}^{2}\left( \hat{\Psi}_{\lambda ,C}^{\left( 0\right) }\right) ^{2}}%
\exp \left( \frac{\hat{K}^{2}}{\sigma _{\hat{K}}^{2}}f\left( \hat{X}\right)
\right) \left( \sigma _{\hat{K}}^{2}\frac{\left( \hat{K}^{2}\mp \frac{1}{4}%
\left( \hat{K}+\frac{\sigma _{\hat{K}}^{2}F\left( \hat{X},K_{\hat{X}}\right) 
}{f^{2}\left( \hat{X}\right) }\right) ^{2}\right) f^{\prime }\left( X\right) 
}{2\hat{K}f\left( \hat{X}\right) -\left( \hat{K}+\frac{\sigma _{\hat{K}%
}^{2}F\left( \hat{X},K_{\hat{X}}\right) }{f^{2}\left( \hat{X}\right) }%
\right) \left\vert f\left( \hat{X}\right) \right\vert }\right) ^{2} \\
&&\times \partial _{\hat{K}}^{3}\exp \left( \frac{\hat{K}^{2}f\left( \hat{X}%
\right) -\frac{1}{2}\left( \hat{K}+\frac{\sigma _{\hat{K}}^{2}F\left( \hat{X}%
,K_{\hat{X}}\right) }{f^{2}\left( \hat{X}\right) }\right) ^{2}\left\vert
f\left( \hat{X}\right) \right\vert }{\sigma _{\hat{K}}^{2}}\right) \\
&=&-\frac{\left( \frac{\left( \hat{K}^{2}-\frac{1}{4}\left( \hat{K}+\frac{%
\sigma _{\hat{K}}^{2}F\left( \hat{X},K_{\hat{X}}\right) }{f^{2}\left( \hat{X}%
\right) }\right) ^{2}\right) f^{\prime }\left( X\right) }{\sigma _{\hat{K}%
}^{2}}\right) ^{2}}{2\hat{K}f\left( \hat{X}\right) -\left( \hat{K}+\frac{%
\sigma _{\hat{K}}^{2}F\left( \hat{X},K_{\hat{X}}\right) }{f^{2}\left( \hat{X}%
\right) }\right) \left\vert f\left( \hat{X}\right) \right\vert } \\
&=&-\frac{\left( \left( \hat{K}^{2}\mp \frac{1}{4}\left( \hat{K}+\frac{%
\sigma _{\hat{K}}^{2}F\left( \hat{X},K_{\hat{X}}\right) }{f^{2}\left( \hat{X}%
\right) }\right) ^{2}\right) f^{\prime }\left( X\right) \right) ^{2}}{\left(
\sigma _{\hat{K}}^{2}\right) ^{2}\left( 2\hat{K}f\left( \hat{X}\right)
-\left( \hat{K}+\frac{\sigma _{\hat{K}}^{2}F\left( \hat{X},K_{\hat{X}%
}\right) }{f^{2}\left( \hat{X}\right) }\right) \left\vert f\left( \hat{X}%
\right) \right\vert \right) }
\end{eqnarray*}

Replacing in first approximation $\hat{K}$ by $\frac{\left\Vert \Psi \left( 
\hat{X}\right) \right\Vert ^{2}\hat{K}_{\hat{X}}}{\hat{N}\left( \hat{X}%
\right) }$ in (\ref{dvt}), and using (\ref{drg}) and (\ref{cps}) leads to: 
\begin{equation*}
\hat{\Psi}^{\left( 1\right) }\left( \hat{X}\right) =\sqrt{C}\exp \left(
-\int \frac{\left( \left( \hat{K}^{2}\mp \frac{1}{4}\left( \hat{K}+\frac{%
\sigma _{\hat{K}}^{2}F\left( \hat{X},K_{\hat{X}}\right) }{f^{2}\left( \hat{X}%
\right) }\right) ^{2}\right) f^{\prime }\left( X\right) \right) ^{2}}{\left(
\sigma _{\hat{K}}^{2}\right) ^{2}\left( 2\hat{K}f\left( \hat{X}\right)
-\left( \hat{K}+\frac{\sigma _{\hat{K}}^{2}F\left( \hat{X},K_{\hat{X}%
}\right) }{f^{2}\left( \hat{X}\right) }\right) \left\vert f\left( \hat{X}%
\right) \right\vert \right) }d\hat{K}\right)
\end{equation*}%
with $C$ a constant to be computed using the normalization condition.

To find $\Psi ^{\dag }$, we need also $\hat{\Psi}^{\left( 1\right) \dag }$.
Writing:%
\begin{equation*}
\hat{\Psi}^{\left( 1\right) \dag }=\exp \left( \sigma _{X}^{2}g\left(
K,X\right) \right)
\end{equation*}%
with a function $g\left( K,X\right) $ that satisfies:%
\begin{equation*}
\frac{\nabla _{\hat{X}}^{2}\hat{\Psi}_{\lambda ,C}^{\left( 0\right) \dag }}{%
\hat{\Psi}_{\lambda ,C}^{\left( 0\right) }}+\sigma _{\hat{K}}^{2}\nabla _{%
\hat{K}}^{2}g\left( \hat{K},\hat{X}\right) +2\left( \nabla _{\hat{K}}g\left( 
\hat{K},\hat{X}\right) \right) \left( \sigma _{\hat{K}}^{2}\frac{\nabla _{%
\hat{K}}\hat{\Psi}_{\lambda ,C}^{\left( 0\right) \dag }}{\hat{\Psi}_{\lambda
,C}^{\left( 0\right) }}+\hat{K}f\left( \hat{X}\right) \right) =0
\end{equation*}%
with:%
\begin{equation*}
\hat{\Psi}_{\lambda ,C}^{\left( 0\right) \dag }=\exp \left( -\frac{\hat{K}%
^{2}}{\sigma _{\hat{K}}^{2}}f\left( \hat{X}\right) \right) D_{p\left( \hat{X}%
,\hat{\lambda}\right) }\left( \left( \frac{\left\vert f\left( \hat{X}\right)
\right\vert }{\sigma _{\hat{K}}^{2}}\right) ^{\frac{1}{2}}\left( \hat{K}+%
\frac{\sigma _{\hat{K}}^{2}F\left( \hat{X},K_{\hat{X}}\right) }{f^{2}\left( 
\hat{X}\right) }\right) \right)
\end{equation*}%
we find:%
\begin{equation*}
\nabla _{\hat{K}}\left( g\left( K,X\right) \right) =-\frac{\nabla _{\hat{X}%
}^{2}\hat{\Psi}_{\lambda ,C}^{\left( 0\right) \dag }}{\hat{\Psi}_{\lambda
,C}^{\left( 0\right) }}\exp \left( -\frac{\hat{K}^{2}}{\sigma _{\hat{K}}^{2}}%
f\left( \hat{X}\right) \right) \left( \int \frac{\nabla _{\hat{X}}^{2}\hat{%
\Psi}_{\lambda ,C}^{\left( 0\right) \dag }}{\hat{\Psi}_{\lambda ,C}^{\left(
0\right) \dag }}\left( \hat{\Psi}_{\lambda ,C}^{\left( 0\right) \dag
}\right) ^{2}\exp \left( \frac{\hat{K}^{2}}{\sigma _{\hat{K}}^{2}}f\left( 
\hat{X}\right) \right) d\hat{K}\right)
\end{equation*}%
and:

\begin{equation*}
\hat{\Psi}^{\left( 1\right) \dag }\left( \hat{X}\right) =\sqrt{C}\exp \left(
\int \frac{\left( \left( \hat{K}^{2}\pm \frac{1}{4}\left( \hat{K}+\frac{%
\sigma _{\hat{K}}^{2}F\left( \hat{X},K_{\hat{X}}\right) }{f^{2}\left( \hat{X}%
\right) }\right) ^{2}\right) f^{\prime }\left( X\right) \right) ^{2}}{\sigma
_{\hat{K}}^{2}\left( 2\hat{K}f\left( \hat{X}\right) +\left( \hat{K}+\frac{%
\sigma _{\hat{K}}^{2}F\left( \hat{X},K_{\hat{X}}\right) }{f^{2}\left( \hat{X}%
\right) }\right) \left\vert f\left( \hat{X}\right) \right\vert \right) }d%
\hat{K}\right)
\end{equation*}%
where $\pm $ accounts for the sign of $f$.\ 

Ultimately, coming back to the initial definition of the fields we obtain
for $\hat{\Psi}_{\lambda ,C}\left( \hat{K},\hat{X}\right) $ and $\hat{\Psi}%
_{\lambda ,C}^{\dag }\left( \hat{K},\hat{X}\right) $:%
\begin{eqnarray*}
\hat{\Psi}_{\lambda ,C}\left( \hat{K},\hat{X}\right) &=&\sqrt{C}\exp \left(
-\sigma _{X}^{2}\int \frac{\left( \left( \hat{K}^{2}\mp \frac{1}{4}\left( 
\hat{K}+\frac{\sigma _{\hat{K}}^{2}F\left( \hat{X},K_{\hat{X}}\right) }{%
f^{2}\left( \hat{X}\right) }\right) ^{2}\right) f^{\prime }\left( X\right)
\right) ^{2}}{\left( \sigma _{\hat{K}}^{2}\right) ^{2}\left( 2\hat{K}f\left( 
\hat{X}\right) -\left( \hat{K}+\frac{\sigma _{\hat{K}}^{2}F\left( \hat{X},K_{%
\hat{X}}\right) }{f^{2}\left( \hat{X}\right) }\right) \left\vert f\left( 
\hat{X}\right) \right\vert \right) }d\hat{K}\right) \\
&&\times \exp \left( \frac{1}{\sigma _{\hat{X}}^{2}}\int g\left( \hat{X}%
\right) d\hat{X}+\frac{\hat{K}^{2}}{\sigma _{\hat{K}}^{2}}f\left( \hat{X}%
\right) \right) D_{p\left( \hat{X},\hat{\lambda}\right) }\left( \hat{K}%
\left( \frac{\left\vert f\left( \hat{X}\right) \right\vert }{\sigma _{\hat{K}%
}^{2}}\right) ^{\frac{1}{2}}\right) \\
\hat{\Psi}_{\lambda ,C}^{\dag }\left( \hat{K},\hat{X}\right) &=&\sqrt{C}\exp
\left( \sigma _{X}^{2}\int \frac{\left( \left( \hat{K}^{2}\pm \frac{1}{4}%
\left( \hat{K}+\frac{\sigma _{\hat{K}}^{2}F\left( \hat{X},K_{\hat{X}}\right) 
}{f^{2}\left( \hat{X}\right) }\right) ^{2}\right) f^{\prime }\left( X\right)
\right) ^{2}}{\left( \sigma _{\hat{K}}^{2}\right) ^{2}\left( 2\hat{K}f\left( 
\hat{X}\right) +\left( \hat{K}+\frac{\sigma _{\hat{K}}^{2}F\left( \hat{X},K_{%
\hat{X}}\right) }{f^{2}\left( \hat{X}\right) }\right) \left\vert f\left( 
\hat{X}\right) \right\vert \right) }d\hat{K}\right) \\
&&\times \exp \left( -\left( \frac{1}{\sigma _{\hat{X}}^{2}}\int g\left( 
\hat{X}\right) d\hat{X}+\frac{\hat{K}^{2}}{\sigma _{\hat{K}}^{2}}f\left( 
\hat{X}\right) \right) \right) D_{p\left( \hat{X},\hat{\lambda}\right)
}\left( \hat{K}\left( \frac{\left\vert f\left( \hat{X}\right) \right\vert }{%
\sigma _{\hat{K}}^{2}}\right) ^{\frac{1}{2}}\right)
\end{eqnarray*}

\paragraph{A3.1.3.3 Computation of $\left\Vert \hat{\Psi}\left( \hat{K},\hat{%
X}\right) \right\Vert ^{2}$}

As a consequence of the previsous result, we can compute $\left\Vert \hat{%
\Psi}_{\lambda ,C}\left( \hat{K},\hat{X}\right) \right\Vert ^{2}$. We start
with $\hat{\Psi}^{\left( 1\right) \dag }\hat{\Psi}^{\left( 1\right) }$. We
have: 
\begin{eqnarray*}
\hat{\Psi}^{\left( 1\right) \dag }\hat{\Psi}^{\left( 1\right) } &=&C\exp
\left( -\sigma _{X}^{2}\int \left( \frac{\left( \left( \hat{K}^{2}\mp \frac{1%
}{4}\left( \hat{K}+\frac{\sigma _{\hat{K}}^{2}F\left( \hat{X},K_{\hat{X}%
}\right) }{f^{2}\left( \hat{X}\right) }\right) ^{2}\right) f^{\prime }\left(
X\right) \right) ^{2}}{\left( \sigma _{\hat{K}}^{2}\right) ^{2}\left( 2\hat{K%
}f\left( \hat{X}\right) -\left( \hat{K}+\frac{\sigma _{\hat{K}}^{2}F\left( 
\hat{X},K_{\hat{X}}\right) }{f^{2}\left( \hat{X}\right) }\right) \left\vert
f\left( \hat{X}\right) \right\vert \right) }\right. \right. \\
&&\left. \left. -\frac{\left( \left( \hat{K}^{2}\pm \frac{1}{4}\left( \hat{K}%
+\frac{\sigma _{\hat{K}}^{2}F\left( \hat{X},K_{\hat{X}}\right) }{f^{2}\left( 
\hat{X}\right) }\right) ^{2}\right) f^{\prime }\left( X\right) \right) ^{2}}{%
\left( \sigma _{\hat{K}}^{2}\right) ^{2}\left( 2\hat{K}f\left( \hat{X}%
\right) +\left( \hat{K}+\frac{\sigma _{\hat{K}}^{2}F\left( \hat{X},K_{\hat{X}%
}\right) }{f^{2}\left( \hat{X}\right) }\right) \left\vert f\left( \hat{X}%
\right) \right\vert \right) }d\hat{K}\right) \right) \\
&=&C\exp \left( -\sigma _{X}^{2}\int \left( \frac{\left( \left( \hat{K}^{2}-%
\frac{1}{4}\left( \hat{K}+\frac{\sigma _{\hat{K}}^{2}F\left( \hat{X},K_{\hat{%
X}}\right) }{f^{2}\left( \hat{X}\right) }\right) ^{2}\right) f^{\prime
}\left( X\right) \right) ^{2}}{\sigma _{\hat{K}}^{2}\left( 2\hat{K}%
\left\vert f\left( \hat{X}\right) \right\vert -\left( \hat{K}+\frac{\sigma _{%
\hat{K}}^{2}F\left( \hat{X},K_{\hat{X}}\right) }{f^{2}\left( \hat{X}\right) }%
\right) \left\vert f\left( \hat{X}\right) \right\vert \right) }\right.
\right. \\
&&-\left. \left. -\frac{\left( \left( \hat{K}^{2}+\frac{1}{4}\left( \hat{K}+%
\frac{\sigma _{\hat{K}}^{2}F\left( \hat{X},K_{\hat{X}}\right) }{f^{2}\left( 
\hat{X}\right) }\right) ^{2}\right) f^{\prime }\left( X\right) \right) ^{2}}{%
\left( \sigma _{\hat{K}}^{2}\right) ^{2}\left( 2\hat{K}\left\vert f\left( 
\hat{X}\right) \right\vert +\left( \hat{K}+\frac{\sigma _{\hat{K}%
}^{2}F\left( \hat{X},K_{\hat{X}}\right) }{f^{2}\left( \hat{X}\right) }%
\right) \left\vert f\left( \hat{X}\right) \right\vert \right) }d\hat{K}%
\right) \right)
\end{eqnarray*}%
And for $\frac{\sigma _{\hat{K}}^{2}F\left( \hat{X},K_{\hat{X}}\right) }{%
f^{2}\left( \hat{X}\right) }<<1$:%
\begin{eqnarray*}
\hat{\Psi}^{\left( 1\right) \dag }\hat{\Psi}^{\left( 1\right) } &\simeq
&C\exp \left( -\sigma _{X}^{2}\int \left( \frac{\left( \frac{3}{4}\hat{K}%
^{2}f^{\prime }\left( X\right) \right) ^{2}}{\left( \sigma _{\hat{K}%
}^{2}\right) ^{2}\hat{K}\left\vert f\left( \hat{X}\right) \right\vert }-%
\frac{\left( \frac{5}{4}\hat{K}^{2}f^{\prime }\left( X\right) \right) ^{2}}{%
3\left( \sigma _{\hat{K}}^{2}\right) ^{2}\hat{K}f\left( \hat{X}\right) }%
\right) d\hat{K}\right) \\
&=&C\exp \left( -\frac{\sigma _{X}^{2}\hat{K}^{4}\left( f^{\prime }\left(
X\right) \right) ^{2}}{96\left( \sigma _{\hat{K}}^{2}\right) ^{2}\left\vert
f\left( \hat{X}\right) \right\vert }\right)
\end{eqnarray*}%
Gathering the previous results, we obtain the norm of $\left\Vert \hat{\Psi}%
_{\lambda ,C}\left( \hat{K},\hat{X}\right) \right\Vert ^{2}$:%
\begin{equation}
\left\Vert \hat{\Psi}_{\lambda ,C}\left( \hat{K},\hat{X}\right) \right\Vert
^{2}\simeq C\exp \left( -\frac{\sigma _{X}^{2}\hat{K}^{4}\left( f^{\prime
}\left( X\right) \right) ^{2}}{96\left( \sigma _{\hat{K}}^{2}\right)
^{2}\left\vert f\left( \hat{X}\right) \right\vert }\right) D_{p\left( \hat{X}%
,\hat{\lambda}\right) }^{2}\left( \left( \frac{\left\vert f\left( \hat{X}%
\right) \right\vert }{\sigma _{\hat{K}}^{2}}\right) ^{\frac{1}{2}}\left( 
\hat{K}+\frac{\sigma _{\hat{K}}^{2}F\left( \hat{X},K_{\hat{X}}\right) }{%
f^{2}\left( \hat{X}\right) }\right) \right)  \label{spc}
\end{equation}

with:%
\begin{eqnarray}
f\left( \hat{X},K_{\hat{X}}\right) &=&\left( r\left( K_{\hat{X}},\hat{X}%
\right) -\gamma \left\Vert \Psi \left( \hat{X}\right) \right\Vert
^{2}+F_{1}\left( \frac{R\left( K_{\hat{X}},\hat{X}\right) }{\int R\left(
K_{X^{\prime }}^{\prime },X^{\prime }\right) \left\Vert \Psi \left(
X^{\prime }\right) \right\Vert ^{2}dX^{\prime }}\right) \right)  \label{ftf}
\\
g\left( \hat{X},K_{\hat{X}}\right) &=&\left( \frac{\nabla _{\hat{X}%
}F_{0}\left( R\left( K_{\hat{X}},\hat{X}\right) \right) }{\left\Vert \nabla
_{\hat{X}}R\left( K_{\hat{X}},\hat{X}\right) \right\Vert }+\nu \nabla _{\hat{%
X}}F_{1}\left( \frac{R\left( K_{\hat{X}},\hat{X}\right) }{\int R\left(
K_{X^{\prime }}^{\prime },X^{\prime }\right) \left\Vert \Psi \left(
X^{\prime }\right) \right\Vert ^{2}dX^{\prime }}\right) \right)  \label{gft}
\end{eqnarray}%
The solutions are parametrized by $C$ and $\hat{\lambda}$ and $\hat{K}_{\hat{%
X}}$. Using the constraint $\left\Vert \hat{\Psi}\left( \hat{K},\hat{X}%
\right) \right\Vert ^{2}=\hat{N}$ will reduce the solutions to a
one-parameter set of solutions. The computation of the average capital over
this set will lead to the defining equation for $\hat{K}_{\hat{X}}$.

Replacing in first approximation $\hat{K}$ by its average $\frac{\left\Vert
\Psi \left( \hat{X}\right) \right\Vert ^{2}\hat{K}_{\hat{X}}}{\hat{N}\left( 
\hat{X}\right) }$ in the first term yields:%
\begin{equation}
\left\Vert \hat{\Psi}_{\lambda ,C}\left( \hat{K},\hat{X}\right) \right\Vert
^{2}\simeq C\exp \left( -\frac{\sigma _{X}^{2}\left( \frac{\left\Vert \Psi
\left( \hat{X}\right) \right\Vert ^{2}\hat{K}_{\hat{X}}}{\hat{N}\left( \hat{X%
}\right) }\right) ^{4}\left( f^{\prime }\left( X\right) \right) ^{2}}{%
96\left( \sigma _{\hat{K}}^{2}\right) ^{2}\left\vert f\left( \hat{X}\right)
\right\vert }\right) D_{p\left( \hat{X},\hat{\lambda}\right) }^{2}\left(
\left( \frac{\left\vert f\left( \hat{X}\right) \right\vert }{\sigma _{\hat{K}%
}^{2}}\right) ^{\frac{1}{2}}\left( \hat{K}+\frac{\sigma _{\hat{K}%
}^{2}F\left( \hat{X},K_{\hat{X}}\right) }{f^{2}\left( \hat{X}\right) }%
\right) \right)  \label{bgH}
\end{equation}

\subsubsection*{A3.1.4 Estimation of $\ S_{3}\left( \hat{\Psi}_{\hat{\protect%
\lambda}}\left( \hat{K},\hat{X}\right) \right) +S_{4}\left( \hat{\Psi}_{\hat{%
\protect\lambda}}\left( \hat{K},\hat{X}\right) \right) $}

For later purposes, we compute an estimation of $S_{3}\left( \hat{\Psi}_{%
\hat{\lambda}}\left( \hat{K},\hat{X}\right) \right) +S_{4}\left( \hat{\Psi}_{%
\hat{\lambda}}\left( \hat{K},\hat{X}\right) \right) $ for any background
field $\hat{\Psi}_{\hat{\lambda}}\left( \hat{K},\hat{X}\right) $. We
multiply (\ref{hqn})by $\hat{\Psi}_{\hat{\lambda}}^{\dagger }\left( \hat{K},%
\hat{X}\right) $ on the left and integrate the equation over $\hat{K}$ and $%
\hat{X}$. It yields:%
\begin{equation*}
0=S_{3}\left( \hat{\Psi}_{\hat{\lambda}}\left( \hat{K},\hat{X}\right)
\right) +S_{4}\left( \hat{\Psi}_{\hat{\lambda}}\left( \hat{K},\hat{X}\right)
\right) -\hat{\lambda}\int \left\Vert \hat{\Psi}_{\hat{\lambda}}\left( \hat{K%
},\hat{X}\right) \right\Vert ^{2}d\hat{K}d\hat{X}-\int F\left( \hat{X},K_{%
\hat{X}}\right) \hat{K}\left\Vert \hat{\Psi}_{\hat{\lambda}}\left( \hat{K},%
\hat{X}\right) \right\Vert ^{2}d\hat{K}d\hat{X}
\end{equation*}

Using the constraint about the number of investors:%
\begin{equation*}
\int \left\Vert \hat{\Psi}_{\hat{\lambda}}\left( \hat{K},\hat{X}\right)
\right\Vert ^{2}d\hat{K}=\hat{N}
\end{equation*}%
we find:%
\begin{equation*}
S_{3}\left( \hat{\Psi}_{\hat{\lambda}}\left( \hat{K},\hat{X}\right) \right)
+S_{4}\left( \hat{\Psi}_{\hat{\lambda}}\left( \hat{K},\hat{X}\right) \right)
=\hat{\lambda}\hat{N}+\int F\left( \hat{X},K_{\hat{X}}\right) \hat{K}%
\left\Vert \hat{\Psi}_{\hat{\lambda}}\left( \hat{K},\hat{X}\right)
\right\Vert ^{2}d\hat{K}d\hat{X}
\end{equation*}%
Moreover, equation (\ref{Fct}) implies\footnote{%
All averages in the next formula are computed in state $\hat{\Psi}_{\hat{%
\lambda}}\left( \hat{K},\hat{X}\right) $.
\par
{}}:%
\begin{eqnarray}
&&\int F\left( \hat{X},K_{\hat{X}}\right) \hat{K}\left\Vert \hat{\Psi}_{\hat{%
\lambda}}\left( \hat{K},\hat{X}\right) \right\Vert ^{2}d\hat{K}d\hat{X} \\
&=&\int K_{\hat{X}}\nabla _{K_{\hat{X}}}\left( \frac{\left( g\left( \hat{X}%
,K_{\hat{X}}\right) \right) ^{2}}{2\sigma _{\hat{X}}^{2}}+\frac{1}{2}\nabla
_{\hat{X}}g\left( \hat{X},K_{\hat{X}}\right) +f\left( \hat{X},K_{\hat{X}%
}\right) \right) \left\Vert \hat{\Psi}\left( \hat{X}\right) \right\Vert ^{2}d%
\hat{X}  \notag \\
&&+\int K_{\hat{X}}\frac{\nabla _{K_{\hat{X}}}f^{2}\left( \hat{X},K_{\hat{X}%
}\right) }{\sigma _{\hat{K}}^{2}}\left\langle \hat{K}^{2}\right\rangle _{%
\hat{X}}d\hat{X}  \notag
\end{eqnarray}%
In our applications the involved functions are roughly power functions in $%
K_{\hat{X}}$, and as a consequence, the integral $\int F\left( \hat{X},K_{%
\hat{X}}\right) \hat{K}\left\Vert \hat{\Psi}_{\hat{\lambda}}\left( \hat{K},%
\hat{X}\right) \right\Vert ^{2}d\hat{K}d\hat{X}$ is of order:%
\begin{equation}
\int \left( \frac{\left( g\left( \hat{X},K_{\hat{X}}\right) \right) ^{2}}{%
2\sigma _{\hat{X}}^{2}}+\frac{1}{2}\nabla _{\hat{X}}g\left( \hat{X},K_{\hat{X%
}}\right) +f\left( \hat{X},K_{\hat{X}}\right) \right) \left\Vert \hat{\Psi}%
\left( \hat{X}\right) \right\Vert ^{2}d\hat{X}+\int \frac{f^{2}\left( \hat{X}%
,K_{\hat{X}}\right) }{\sigma _{\hat{K}}^{2}}\left\langle \hat{K}%
^{2}\right\rangle _{\hat{X}}d\hat{X}
\end{equation}%
Since $\left\langle \hat{K}^{2}\right\rangle _{\hat{X}}\simeq K_{\hat{X}}^{2}%
\frac{\left\Vert \Psi \left( \hat{X}\right) \right\Vert ^{2}}{\left\Vert 
\hat{\Psi}\left( \hat{X}\right) \right\Vert ^{2}}$, the second term in (\ref%
{ntrM}) is negligible if we assume $\frac{\left\Vert \Psi \left( \hat{X}%
\right) \right\Vert ^{2}}{\left\Vert \hat{\Psi}\left( \hat{X}\right)
\right\Vert ^{2}}<<1$, i.e. the number of firms is smaller than the number
of investors. As a consequence, (\ref{ntrM}) reduces to:%
\begin{eqnarray*}
\int \left( \frac{\left( g\left( \hat{X},K_{\hat{X}}\right) \right) ^{2}}{%
2\sigma _{\hat{X}}^{2}}+\frac{1}{2}\nabla _{\hat{X}}g\left( \hat{X},K_{\hat{X%
}}\right) +f\left( \hat{X},K_{\hat{X}}\right) \right) \left\Vert \hat{\Psi}%
\left( \hat{X}\right) \right\Vert ^{2}d\hat{X} &\lesssim &\int M\left\Vert 
\hat{\Psi}\left( \hat{X}\right) \right\Vert ^{2}d\hat{X} \\
&=&M\hat{N}
\end{eqnarray*}%
where $M$ is the lowest bound for $\left\vert \hat{\lambda}\right\vert $,
computed below in (\ref{mdf}) and (\ref{mdF}). Our previous estimation
relies on $\frac{\sigma _{\hat{K}}^{2}F^{2}\left( \hat{X},K_{\hat{X}}\right) 
}{2f^{2}\left( \hat{X}\right) }<<1$,which is true for $f^{2}\left( \hat{X}%
\right) >>1$. As a consequence:%
\begin{equation}
S_{3}\left( \hat{\Psi}_{\hat{\lambda}}\left( \hat{K},\hat{X}\right) \right)
+S_{4}\left( \hat{\Psi}_{\hat{\lambda}}\left( \hat{K},\hat{X}\right) \right)
=\left( \hat{\lambda}+M\right) \hat{N}=-\left( \left\vert \hat{\lambda}%
\right\vert -M\right) \hat{N}  \label{stT}
\end{equation}

\subsubsection*{A3.1.4 \textbf{Identification of} $K_{\hat{X}}$ and $N\left( 
\hat{X}\right) $:}

\paragraph*{A3.1.4.1 Formula depending on $\hat{\protect\lambda}$ and $C$}

In this paragraph, we compute the average capital $K_{\hat{X}}$ and the
number of investors $\hat{N}\left( \hat{X}\right) $ at $\hat{X}$ that are
defined by using (\ref{Kx}): 
\begin{eqnarray}
K_{\hat{X}}\left\Vert \Psi \left( \hat{X}\right) \right\Vert ^{2}
&=&\int_{0}^{\infty }\hat{K}C\exp \left( -\frac{\sigma _{X}^{2}u\left( \hat{X%
},\hat{K}_{\hat{X}}\right) }{\left( \sigma _{\hat{K}}^{2}\right) ^{2}}\right)
\label{nmk} \\
&&\times D_{p\left( \hat{X},\hat{\lambda}\right) }^{2}\left( \left( \frac{%
\left\vert f\left( \hat{X}\right) \right\vert }{\sigma _{\hat{K}}^{2}}%
\right) ^{\frac{1}{2}}\left( \hat{K}+\frac{\sigma _{\hat{K}}^{2}F\left( \hat{%
X},K_{\hat{X}}\right) }{f^{2}\left( \hat{X}\right) }\right) \right) d\hat{K}
\notag
\end{eqnarray}%
and:%
\begin{eqnarray*}
N\left( \hat{X}\right) &=&C\int_{0}^{\infty }\exp \left( -\frac{\sigma
_{X}^{2}u\left( \hat{X},\hat{K}_{\hat{X}}\right) }{\left( \sigma _{\hat{K}%
}^{2}\right) ^{2}}\right) \\
&&\times D_{p\left( \hat{X},\hat{\lambda}\right) }^{2}\left( \left( \frac{%
\left\vert f\left( \hat{X}\right) \right\vert }{\sigma _{\hat{K}}^{2}}%
\right) ^{\frac{1}{2}}\left( \hat{K}+\frac{\sigma _{\hat{K}}^{2}F\left( \hat{%
X},K_{\hat{X}}\right) }{f^{2}\left( \hat{X}\right) }\right) \right) d\hat{K}
\end{eqnarray*}%
with:%
\begin{equation}
u\left( \hat{X},\hat{K}_{\hat{X}}\right) =\frac{\left( \frac{\left\Vert \Psi
\left( \hat{X}\right) \right\Vert ^{2}\hat{K}_{\hat{X}}}{\hat{N}\left( \hat{X%
}\right) }\right) ^{4}\left( f^{\prime }\left( X\right) \right) ^{2}}{%
96\left\vert f\left( \hat{X}\right) \right\vert }  \label{dnl}
\end{equation}%
Note that in these formulas, $K_{\hat{X}}$ and $N\left( \hat{X}\right) $
depend implicitely of $\hat{\lambda}$ since they have been computed in the
state defined by the background field $\hat{\Psi}_{\lambda ,C}\left( \hat{K},%
\hat{X}\right) $. In the sequel, for the sake of simplicity, $\hat{\Psi}%
_{\lambda ,C}\left( \hat{K},\hat{X}\right) $, the indices $\lambda $ and $C$
may be omitted.

We will also need $\frac{K_{\hat{X}}\left\Vert \Psi \left( \hat{X}\right)
\right\Vert ^{2}}{\hat{N}\left( \hat{X}\right) }$ that arises in (\ref{dnl}%
): 
\begin{equation*}
\frac{K_{\hat{X}}\left\Vert \Psi \left( \hat{X}\right) \right\Vert ^{2}}{%
\hat{N}\left( \hat{X}\right) }=\frac{\int_{0}^{\infty }\hat{K}D_{p\left( 
\hat{X},\hat{\lambda}\right) }^{2}\left( \left( \frac{\left\vert f\left( 
\hat{X}\right) \right\vert }{\sigma _{\hat{K}}^{2}}\right) ^{\frac{1}{2}%
}\left( \hat{K}+\frac{\sigma _{\hat{K}}^{2}F\left( \hat{X},K_{\hat{X}%
}\right) }{f^{2}\left( \hat{X}\right) }\right) \right) d\hat{K}}{\int_{\frac{%
\sigma _{\hat{K}}^{2}F\left( \hat{X},K_{\hat{X}}\right) }{f^{2}\left( \hat{X}%
\right) }}^{\infty }\hat{K}D_{p\left( \hat{X},\hat{\lambda}\right)
}^{2}\left( \left( \frac{\left\vert f\left( \hat{X}\right) \right\vert }{%
\sigma _{\hat{K}}^{2}}\right) ^{\frac{1}{2}}\hat{K}\right) d\hat{K}}
\end{equation*}%
By a change of variable $\hat{K}+\frac{\sigma _{\hat{K}}^{2}F\left( \hat{X}%
,K_{\hat{X}}\right) }{f^{2}\left( \hat{X}\right) }\rightarrow \hat{K}$ we
can also write: 
\begin{equation*}
K_{\hat{X}}\left\Vert \Psi \left( \hat{X}\right) \right\Vert ^{2}\simeq
C\exp \left( -\frac{\sigma _{X}^{2}u\left( \hat{X},\hat{K}_{\hat{X}}\right) 
}{\sigma _{\hat{K}}^{2}}\right) \int_{\frac{\sigma _{\hat{K}}^{2}F\left( 
\hat{X},K_{\hat{X}}\right) }{f^{2}\left( \hat{X}\right) }}^{\infty }\hat{K}%
D_{p\left( \hat{X},\hat{\lambda}\right) }^{2}\left( \left( \frac{\left\vert
f\left( \hat{X}\right) \right\vert }{\sigma _{\hat{K}}^{2}}\right) ^{\frac{1%
}{2}}\hat{K}\right) d\hat{K}
\end{equation*}%
\begin{equation*}
\hat{N}\left( \hat{X}\right) \simeq C\exp \left( -\frac{\sigma
_{X}^{2}u\left( \hat{X},\hat{K}_{\hat{X}}\right) }{16\sigma _{\hat{K}}^{2}}%
\right) \int_{\frac{\sigma _{\hat{K}}^{2}F\left( \hat{X},K_{\hat{X}}\right) 
}{f^{2}\left( \hat{X}\right) }}^{\infty }D_{p\left( \hat{X},\hat{\lambda}%
\right) }^{2}\left( \left( \frac{\left\vert f\left( \hat{X}\right)
\right\vert }{\sigma _{\hat{K}}^{2}}\right) ^{\frac{1}{2}}\hat{K}\right) d%
\hat{K}
\end{equation*}%
and by a zeroth order expansion around $0$ of $\hat{K}D_{p\left( \hat{X},%
\hat{\lambda}\right) }^{2}$ and $D_{p\left( \hat{X},\hat{\lambda}\right)
}^{2}$ we have:%
\begin{equation}
K_{\hat{X}}\left\Vert \Psi \left( \hat{X}\right) \right\Vert ^{2}\simeq
C\exp \left( -\frac{\sigma _{X}^{2}u\left( \hat{X},\hat{K}_{\hat{X}}\right) 
}{16\sigma _{\hat{K}}^{2}}\right) \int_{0}^{\infty }\hat{K}D_{p\left( \hat{X}%
,\hat{\lambda}\right) }^{2}\left( \left( \frac{\left\vert f\left( \hat{X}%
\right) \right\vert }{\sigma _{\hat{K}}^{2}}\right) ^{\frac{1}{2}}\hat{K}%
\right) d\hat{K}  \label{mnk}
\end{equation}%
\begin{equation}
N\left( \hat{X}\right) \simeq C\exp \left( -\frac{\sigma _{X}^{2}u\left( 
\hat{X},\hat{K}_{\hat{X}}\right) }{16\sigma _{\hat{K}}^{2}}\right) \left(
\int_{0}^{\infty }D_{p\left( \hat{X},\hat{\lambda}\right) }^{2}\left( \left( 
\frac{\left\vert f\left( \hat{X}\right) \right\vert }{\sigma _{\hat{K}}^{2}}%
\right) ^{\frac{1}{2}}\hat{K}\right) d\hat{K}-\frac{\left( \frac{\left\vert
f\left( \hat{X}\right) \right\vert }{\sigma _{\hat{K}}^{2}}\right) ^{-\frac{1%
}{2}}2^{\frac{p\left( \hat{X},\hat{\lambda}\right) }{2}}\sqrt{\pi }}{\Gamma
\left( \frac{1-p\left( \hat{X},\hat{\lambda}\right) }{2}\right) }\frac{%
\sigma _{\hat{K}}^{2}F\left( \hat{X},K_{\hat{X}}\right) }{f^{2}\left( \hat{X}%
\right) }\right)  \label{nrp}
\end{equation}%
To compute $N\left( \hat{X}\right) $ we use that the function $D$ satisfies: 
\begin{equation*}
\int D_{p}^{2}=\frac{\sqrt{\pi }}{2^{\frac{3}{2}}}\frac{\func{Psi}\left( 
\frac{1}{2}-\frac{p}{2}\right) -\func{Psi}\left( -\frac{p}{2}\right) }{%
\Gamma \left( -p\right) }
\end{equation*}%
The computation of the norm implies a second change of variable $\hat{K}%
\rightarrow \hat{K}\left( \frac{\left\vert f\left( \hat{X}\right)
\right\vert }{\sigma _{\hat{K}}^{2}}\right) ^{\frac{1}{2}}$ and we obtain
for (\ref{nrp}):%
\begin{eqnarray}
&&\hat{N}\left( \hat{X}\right) =\int \left\Vert \hat{\Psi}_{\lambda
,C}\left( \hat{K},\hat{X}\right) \right\Vert ^{2}d\hat{K}  \label{fmn} \\
&=&C\exp \left( -\frac{\sigma _{X}^{2}u\left( \hat{X},\hat{K}_{\hat{X}%
}\right) }{16\sigma _{\hat{K}}^{2}}\right) \left( \int D_{p\left( \hat{X},%
\hat{\lambda}\right) }^{2}\left( \hat{K}\left( f^{2}\left( \hat{X}\right)
\right) ^{\frac{1}{4}}\right) dK-\frac{\left( \frac{\left\vert f\left( \hat{X%
}\right) \right\vert }{\sigma _{\hat{K}}^{2}}\right) ^{-\frac{1}{2}}2^{\frac{%
p\left( \hat{X},\hat{\lambda}\right) }{2}}\sqrt{\pi }}{\Gamma \left( \frac{%
1-p\left( \hat{X},\hat{\lambda}\right) }{2}\right) }\frac{\sigma _{\hat{K}%
}^{2}F\left( \hat{X},K_{\hat{X}}\right) }{f^{2}\left( \hat{X}\right) }\right)
\notag \\
&=&C\exp \left( -\frac{\sigma _{X}^{2}u\left( \hat{X},\hat{K}_{\hat{X}%
}\right) }{16\sigma _{\hat{K}}^{2}}\right) \left( \frac{\left\vert f\left( 
\hat{X}\right) \right\vert }{\sigma _{\hat{K}}^{2}}\right) ^{-\frac{1}{2}} 
\notag \\
&&\times \left( \frac{\sqrt{\pi }}{2^{\frac{3}{2}}}\frac{\func{Psi}\left( 
\frac{1-p\left( \hat{X},\hat{\lambda}\right) }{2}\right) -\func{Psi}\left( -%
\frac{p\left( \hat{X},\hat{\lambda}\right) }{2}\right) }{\Gamma \left(
-p\left( \hat{X},\hat{\lambda}\right) \right) }-\frac{2^{\frac{p\left( \hat{X%
},\hat{\lambda}\right) }{2}}\sqrt{\pi }}{\Gamma \left( \frac{1-p\left( \hat{X%
},\hat{\lambda}\right) }{2}\right) }\frac{\sigma _{\hat{K}}^{2}F\left( \hat{X%
},K_{\hat{X}}\right) }{f^{2}\left( \hat{X}\right) }\right)  \notag
\end{eqnarray}%
Expression (\ref{mnk}) is computed using that:%
\begin{equation*}
\int_{0}^{\infty }zD_{p}^{2}\left( z\right) dz=\int_{0}^{\infty
}D_{p+1}\left( z\right) D_{p}\left( z\right) dz+p\int_{0}^{\infty
}D_{p-1}\left( z\right) D_{p}\left( z\right) dz
\end{equation*}%
\begin{equation*}
\int_{0}^{\infty }zD_{p}^{2}\left( z\right) dz=\frac{\Gamma \left( -\frac{p+1%
}{2}\right) \Gamma \left( \frac{1-p}{2}\right) -\Gamma \left( -\frac{p}{2}%
\right) \Gamma \left( -\frac{p}{2}\right) }{2^{p+2}\Gamma \left( -p-1\right)
\Gamma \left( -p\right) }+p\frac{\Gamma \left( -\frac{p}{2}\right) \Gamma
\left( \frac{2-p}{2}\right) -\Gamma \left( -\frac{p-1}{2}\right) \Gamma
\left( -\frac{p-1}{2}\right) }{2^{p+1}\Gamma \left( -p\right) \Gamma \left(
-p+1\right) }
\end{equation*}%
and:%
\begin{equation*}
\int \hat{K}D_{p\left( \hat{X},\hat{\lambda}\right) }^{2}\left( \hat{K}%
\left( f^{2}\left( \hat{X}\right) \right) ^{\frac{1}{4}}\right) =\left(
f\left( \hat{X}\right) \right) ^{-1}\int uD_{p\left( \hat{X},\hat{\lambda}%
\right) }^{2}\left( u\right)
\end{equation*}%
We obtain:%
\begin{eqnarray}
K_{\hat{X}}\left\Vert \Psi \left( \hat{X}\right) \right\Vert ^{2} &\simeq
&\exp \left( -\frac{\sigma _{X}^{2}u\left( \hat{X},\hat{K}_{\hat{X}}\right) 
}{16\left( \sigma _{\hat{K}}^{2}\right) ^{2}}\right) \left( \frac{\left\vert
f\left( \hat{X}\right) \right\vert }{\sigma _{\hat{K}}^{2}}\right) ^{-1}C
\label{knm} \\
&&\times \left( \frac{\Gamma \left( -\frac{p+1}{2}\right) \Gamma \left( 
\frac{1-p}{2}\right) -\Gamma \left( -\frac{p}{2}\right) \Gamma \left( \frac{%
-p}{2}\right) }{2^{p+2}\Gamma \left( -p-1\right) \Gamma \left( -p\right) }+p%
\frac{\Gamma \left( -\frac{p}{2}\right) \Gamma \left( \frac{2-p}{2}\right)
-\Gamma \left( -\frac{p-1}{2}\right) \Gamma \left( -\frac{p-1}{2}\right) }{%
2^{p+1}\Gamma \left( -p\right) \Gamma \left( -p+1\right) }\right)  \notag
\end{eqnarray}%
where:%
\begin{equation}
p=p\left( \hat{X},\hat{\lambda}\right)  \label{pft}
\end{equation}%
Ultimately we can compute $\frac{K_{\hat{X}}\left\Vert \Psi \left( \hat{X}%
\right) \right\Vert ^{2}}{\hat{N}\left( \hat{X}\right) }$:%
\begin{eqnarray*}
\frac{K_{\hat{X}}\left\Vert \Psi \left( \hat{X}\right) \right\Vert ^{2}}{%
\hat{N}\left( \hat{X}\right) } &\simeq &\left( \frac{\left\vert f\left( \hat{%
X}\right) \right\vert }{\sigma _{\hat{K}}^{2}}\right) ^{-\frac{1}{2}}\frac{%
\frac{\Gamma \left( -\frac{p+1}{2}\right) \Gamma \left( \frac{1-p}{2}\right)
-\Gamma \left( -\frac{p}{2}\right) \Gamma \left( \frac{-p}{2}\right) }{%
2^{p+2}\Gamma \left( -p-1\right) \Gamma \left( -p\right) }+p\frac{\Gamma
\left( -\frac{p}{2}\right) \Gamma \left( \frac{2-p}{2}\right) -\Gamma \left(
-\frac{p-1}{2}\right) \Gamma \left( -\frac{p-1}{2}\right) }{2^{p+1}\Gamma
\left( -p\right) \Gamma \left( -p+1\right) }}{\frac{\sqrt{\pi }}{2^{\frac{3}{%
2}}}\frac{\func{Psi}\left( \frac{1-p}{2}\right) -\func{Psi}\left( -\frac{p}{2%
}\right) }{\Gamma \left( -p\right) }} \\
&\equiv &\left( \frac{\left\vert f\left( \hat{X}\right) \right\vert }{\sigma
_{\hat{K}}^{2}}\right) ^{-\frac{1}{2}}h\left( p\right) \\
&\simeq &\left( \frac{\left\vert f\left( \hat{X}\right) \right\vert }{\sigma
_{\hat{K}}^{2}}\right) ^{-\frac{1}{2}}\sqrt{p+\frac{1}{2}}
\end{eqnarray*}%
so that:%
\begin{equation}
\exp \left( -\frac{\sigma _{X}^{2}u\left( \hat{X},\hat{K}_{\hat{X}}\right) }{%
\left( \sigma _{\hat{K}}^{2}\right) ^{2}}\right) \simeq \exp \left( -\frac{%
\sigma _{X}^{2}\left( p+\frac{1}{2}\right) ^{2}\left( f^{\prime }\left(
X\right) \right) ^{2}}{96\left\vert f\left( \hat{X}\right) \right\vert ^{3}}%
\right)  \label{fxp}
\end{equation}%
We end this section by finding asymptotic form for $\hat{N}\left( \hat{X}%
\right) $ and $K_{\hat{X}}\left\Vert \Psi \left( \hat{X}\right) \right\Vert
^{2}$

For $\varepsilon <<1$ an asymptotic form yields that:%
\begin{equation}
D_{p\left( \hat{X},\hat{\lambda}\right) }\left( \hat{K}\left( f^{2}\left( 
\hat{X}\right) \right) ^{\frac{1}{4}}\right) \simeq \exp \left( -\frac{\hat{K%
}^{2}\left\vert f\left( \hat{X}\right) \right\vert }{4\sigma _{\hat{K}}^{2}}%
\right) \left( \hat{K}\left( \frac{\left\vert f\left( \hat{X}\right)
\right\vert }{\sigma _{\hat{K}}^{2}}\right) ^{\frac{1}{2}}\right) ^{p\left( 
\hat{X},\hat{\lambda}\right) }  \label{pmn}
\end{equation}%
and we obtain:%
\begin{eqnarray*}
\hat{N}\left( \hat{X}\right) &=&C\exp \left( -\frac{\sigma _{X}^{2}u\left( 
\hat{X},\hat{K}_{\hat{X}}\right) }{\sigma _{\hat{K}}^{2}}\right) \\
&&\times \int_{0}^{\infty }\exp \left( -\frac{\left( \hat{K}+\frac{\sigma _{%
\hat{K}}^{2}F\left( \hat{X},K_{\hat{X}}\right) }{f^{2}\left( \hat{X}\right) }%
\right) ^{2}\left\vert f\left( \hat{X}\right) \right\vert }{2\sigma _{\hat{K}%
}^{2}}\right) \left( \left( \hat{K}+\frac{\sigma _{\hat{K}}^{2}F\left( \hat{X%
},K_{\hat{X}}\right) }{f^{2}\left( \hat{X}\right) }\right) \left( \frac{%
\left\vert f\left( \hat{X}\right) \right\vert }{\sigma _{\hat{K}}^{2}}%
\right) ^{\frac{1}{2}}\right) ^{2p\left( \hat{X},\hat{\lambda}\right) }d\hat{%
K}
\end{eqnarray*}%
A change of variable $w=\frac{\left( \hat{K}+\frac{\sigma _{\hat{K}%
}^{2}F\left( \hat{X},K_{\hat{X}}\right) }{f^{2}\left( \hat{X}\right) }%
\right) ^{2}\left\vert f\left( \hat{X}\right) \right\vert }{2\sigma _{\hat{K}%
}^{2}}$ leads to:%
\begin{equation}
\hat{N}\left( \hat{X}\right) \simeq C\exp \left( -\frac{\sigma
_{X}^{2}u\left( \hat{X},\hat{K}_{\hat{X}}\right) }{\sigma _{\hat{K}}^{2}}%
\right) \left( \frac{\left\vert f\left( \hat{X}\right) \right\vert }{\sigma
_{\hat{K}}^{2}}\right) ^{-\frac{1}{2}}\left( 2^{p\left( \hat{X},\hat{\lambda}%
\right) -\frac{1}{2}}\Gamma \left( p\left( \hat{X},\hat{\lambda}\right) +%
\frac{1}{2}\right) -\frac{2^{\frac{p\left( \hat{X},\hat{\lambda}\right) }{2}}%
\sqrt{\pi }}{\Gamma \left( \frac{1-p\left( \hat{X},\hat{\lambda}\right) }{2}%
\right) }\frac{\sigma _{\hat{K}}^{2}F\left( \hat{X},K_{\hat{X}}\right) }{%
f^{2}\left( \hat{X}\right) }\right)  \label{smn}
\end{equation}

By the same token we can use the asymptotic form (\ref{pmn}) to find $K_{%
\hat{X}}$: 
\begin{eqnarray*}
K_{\hat{X}}\left\Vert \Psi \left( \hat{X}\right) \right\Vert ^{2} &\simeq
&C\exp \left( -\frac{\sigma _{X}^{2}u\left( \hat{X},\hat{K}_{\hat{X}}\right) 
}{\sigma _{\hat{K}}^{2}}\right) \int \hat{K}\exp \left( -\frac{\hat{K}%
^{2}\left\vert f\left( \hat{X}\right) \right\vert }{2\sigma _{\hat{K}}^{2}}%
\right) \left( \hat{K}\left( \frac{\left\vert f\left( \hat{X}\right)
\right\vert }{\sigma _{\hat{K}}^{2}}\right) ^{\frac{1}{2}}\right) ^{2p\left( 
\hat{X},\hat{\lambda}\right) }d\hat{K} \\
&=&\frac{\sigma _{\hat{K}}^{2}C\exp \left( -\frac{\sigma _{X}^{2}u\left( 
\hat{X},\hat{K}_{\hat{X}}\right) }{\sigma _{\hat{K}}^{2}}\right) }{%
\left\vert f\left( \hat{X}\right) \right\vert }\int y\exp \left( -\frac{y^{2}%
}{2}\right) y^{2p\left( \hat{X},\hat{\lambda}\right) }
\end{eqnarray*}%
We set $y=\sqrt{2w}$ and we obtain:%
\begin{eqnarray*}
K_{\hat{X}}\left\Vert \Psi \left( \hat{X}\right) \right\Vert ^{2} &\simeq
&C\exp \left( -\frac{\sigma _{X}^{2}u\left( \hat{X},\hat{K}_{\hat{X}}\right) 
}{\sigma _{\hat{K}}^{2}}\right) 2^{p\left( \hat{X},\hat{\lambda}\right) }%
\frac{\sigma _{\hat{K}}^{2}}{\left\vert f\left( \hat{X}\right) \right\vert }%
\int \exp \left( -w\right) w^{p\left( \hat{X},\hat{\lambda}\right) }dw \\
&=&C\exp \left( -\frac{\sigma _{X}^{2}u\left( \hat{X},\hat{K}_{\hat{X}%
}\right) }{\sigma _{\hat{K}}^{2}}\right) 2^{p\left( \hat{X},\hat{\lambda}%
\right) }\frac{\sigma _{\hat{K}}^{2}}{\left\vert f\left( \hat{X}\right)
\right\vert }\Gamma \left( p\left( \hat{X},\hat{\lambda}\right) +1\right)
\end{eqnarray*}

\paragraph*{A3.1.4.2 \textbf{Computation of} $C$ as a function of $\hat{%
\protect\lambda}$:}

Ultimately, we need to determine \ the value of the Lagrange multiplier $%
\hat{\lambda}$ and of the associated value of $C$. We do so by integrating (%
\ref{spc}) and the result is constrained to be $\hat{N}$, the total number
of agents:%
\begin{equation*}
\hat{N}=\int \left\Vert \hat{\Psi}_{\lambda ,C}\left( \hat{K},\hat{X}\right)
\right\Vert ^{2}d\hat{K}d\hat{X}=\int \hat{N}\left( \hat{X}\right) d\hat{X}
\end{equation*}%
Using (\ref{fmn}) and (\ref{fxp}), we have:%
\begin{eqnarray}
\hat{N} &=&\int \hat{N}\left( \hat{X}\right) \simeq \int C\exp \left( -\frac{%
\sigma _{X}^{2}u\left( \hat{X},\hat{K}_{\hat{X}}\right) }{\sigma _{\hat{K}%
}^{2}}\right) \left( \frac{\left\vert f\left( \hat{X}\right) \right\vert }{%
\sigma _{\hat{K}}^{2}}\right) ^{-\frac{1}{2}}  \label{nrb} \\
&&\times \left( \frac{\sqrt{\pi }}{2^{\frac{3}{2}}}\frac{\func{Psi}\left( 
\frac{1-p\left( \hat{X},\hat{\lambda}\right) }{2}\right) -\func{Psi}\left( -%
\frac{p\left( \hat{X},\hat{\lambda}\right) }{2}\right) }{\Gamma \left(
-p\left( \hat{X},\hat{\lambda}\right) \right) }-\frac{2^{\frac{p\left( \hat{X%
},\hat{\lambda}\right) }{2}}\sqrt{\pi }}{\Gamma \left( \frac{1-p\left( \hat{X%
},\hat{\lambda}\right) }{2}\right) }\frac{\sigma _{\hat{K}}^{2}F\left( \hat{X%
},K_{\hat{X}}\right) }{f^{2}\left( \hat{X}\right) }\right) d\hat{X}  \notag
\\
&\simeq &\int C\exp \left( -\frac{\sigma _{X}^{2}\sigma _{\hat{K}}^{2}\left(
p+\frac{1}{2}\right) ^{2}\left( f^{\prime }\left( X\right) \right) ^{2}}{%
96\left\vert f\left( \hat{X}\right) \right\vert ^{3}}\right) \left( \frac{%
\left\vert f\left( \hat{X}\right) \right\vert }{\sigma _{\hat{K}}^{2}}%
\right) ^{-\frac{1}{2}}\frac{\sqrt{\pi }}{2^{\frac{3}{2}}}\frac{\func{Psi}%
\left( \frac{1-p\left( \hat{X},\hat{\lambda}\right) }{2}\right) -\func{Psi}%
\left( -\frac{p\left( \hat{X},\hat{\lambda}\right) }{2}\right) }{\Gamma
\left( -p\left( \hat{X},\hat{\lambda}\right) \right) }d\hat{X}  \notag
\end{eqnarray}%
with $f$ and $g$ given by (\ref{ftf}) and (\ref{gft}). We thus obtain $C$ as
a function of $\hat{\lambda}$. For $f\left( \hat{X}\right) $\ slowly varying
around its average we can replace $\left\vert f\left( \hat{X}\right)
\right\vert $ and $f^{\prime }\left( X\right) $ by $\left\langle \left\vert
f\left( \hat{X}\right) \right\vert \right\rangle $ and $\left\langle
f^{\prime }\left( X\right) \right\rangle $, where the bracket $\left\langle
A\left( \hat{X}\right) \right\rangle $ represents the average of the
quantity $A\left( \hat{X}\right) $ over the sectors space. Given that the
integrated function is of order $\Gamma \left( p\right) $, we can replace
the integral by the maximal values of the integrand. As a consequence, we
have:%
\begin{equation}
C\left( \bar{p}\left( \hat{\lambda}\right) \right) \simeq \frac{\exp \left( -%
\frac{\sigma _{X}^{2}\sigma _{\hat{K}}^{2}\left( \frac{\left( \bar{p}\left( 
\hat{\lambda}\right) +\frac{1}{2}\right) f^{\prime }\left( X_{0}\right) }{%
f\left( \hat{X}_{0}\right) }\right) ^{2}}{96\left\vert f\left( \hat{X}%
_{0}\right) \right\vert }\right) \hat{N}\Gamma \left( -\bar{p}\left( \hat{%
\lambda}\right) \right) }{\left( \frac{\left\langle \left\vert f\left( \hat{X%
}\right) \right\vert \right\rangle }{\sigma _{\hat{K}}^{2}}\right) ^{-\frac{1%
}{2}}V_{r}\left( \func{Psi}\left( -\frac{\bar{p}\left( \hat{\lambda}\right)
-1}{2}\right) -\func{Psi}\left( -\frac{\bar{p}\left( \hat{\lambda}\right) }{2%
}\right) \right) }  \label{cld}
\end{equation}%
where:%
\begin{equation}
\bar{p}\left( \hat{\lambda}\right) =\left( -\frac{\frac{\left( g\left( \hat{X%
}_{0}\right) \right) ^{2}}{\sigma _{\hat{X}}^{2}}+\left( f\left( \hat{X}%
_{0}\right) +\frac{1}{2}\left\vert f\left( \hat{X}_{0}\right) \right\vert
+\nabla _{\hat{X}}g\left( \hat{X}_{0},K_{\hat{X}_{0}}\right) -\frac{\sigma _{%
\hat{K}}^{2}F^{2}\left( \hat{X}_{0},K_{\hat{X}_{0}}\right) }{2f^{2}\left( 
\hat{X}_{0}\right) }+\hat{\lambda}\right) }{\left\vert f\left( \hat{X}%
_{0}\right) \right\vert }\right)  \label{rpb}
\end{equation}%
and:%
\begin{equation}
\hat{X}_{0}=\arg \min_{\hat{X}}\left( \frac{\sigma _{X}^{2}\sigma _{\hat{K}%
}^{2}\left( \frac{\left( p\left( \hat{\lambda}\right) +\frac{1}{2}\right)
f^{\prime }\left( X\right) }{f\left( \hat{X}\right) }\right) ^{2}}{%
96\left\vert f\left( \hat{X}\right) \right\vert }\right)  \label{dfr}
\end{equation}%
and $V_{r}$ is the volume of the reduced space where the maximum is reached
defined by:%
\begin{equation*}
V_{r}=\sum_{\hat{X}/p\left( \hat{X},\hat{\lambda}\right) =\bar{p}\left( \hat{%
\lambda}\right) }\frac{1}{\left\vert \frac{\hat{N}^{\prime \prime }\left( 
\hat{X}\right) }{C}\right\vert }
\end{equation*}%
We thus can replace $C$ by $C\left( \hat{\lambda}\right) $ and we are left
with an infinite number of solutions of (\ref{knq}) parametrized by $\hat{%
\lambda}$ and given by (\ref{spc}). We write $\left\Vert \hat{\Psi}_{\hat{%
\lambda}}\left( \hat{K},\hat{X}\right) \right\Vert ^{2}$ the solution for $%
\hat{\lambda}$.

\paragraph*{A3.1.4.2 Identification equation for $K_{\hat{X}}$}

To each state $\left\Vert \hat{\Psi}_{\hat{\lambda}}\left( \hat{K},\hat{X}%
\right) \right\Vert ^{2}$, we can associate an average level of $K_{\hat{X},%
\hat{\lambda}}$ satisfying (\ref{knm}) rewritten as a function of $\hat{%
\lambda}$. Using (\ref{fxp}) we find:%
\begin{eqnarray}
K_{\hat{X},\hat{\lambda}}\left\Vert \hat{\Psi}_{\hat{\lambda}}\left( \hat{X}%
\right) \right\Vert ^{2} &=&\hat{K}_{\hat{X},\hat{\lambda}}  \label{knl} \\
&=&\exp \left( -\frac{\sigma _{X}^{2}\sigma _{\hat{K}}^{2}\left( p+\frac{1}{2%
}\right) ^{2}\left( f^{\prime }\left( X\right) \right) ^{2}}{96\left\vert
f\left( \hat{X}\right) \right\vert ^{3}}\right) \left( \frac{\left\vert
f\left( \hat{X}\right) \right\vert }{\sigma _{\hat{K}}^{2}}\right) ^{-1} 
\notag \\
&&\times C\left( \bar{p}\left( \hat{\lambda}\right) \right) \left( \frac{%
\Gamma \left( -\frac{p+1}{2}\right) \Gamma \left( \frac{1-p}{2}\right)
-\Gamma \left( -\frac{p}{2}\right) \Gamma \left( \frac{-p}{2}\right) }{%
2^{p+2}\Gamma \left( -p-1\right) \Gamma \left( -p\right) }+p\frac{\Gamma
\left( -\frac{p}{2}\right) \Gamma \left( \frac{2-p}{2}\right) -\Gamma \left(
-\frac{p-1}{2}\right) \Gamma \left( -\frac{p-1}{2}\right) }{2^{p+1}\Gamma
\left( -p\right) \Gamma \left( -p+1\right) }\right)  \notag
\end{eqnarray}

where:%
\begin{equation}
p\left( \hat{X},\hat{\lambda}\right) =-\frac{\left( g\left( \hat{X}\right)
\right) ^{2}+\sigma _{\hat{X}}^{2}\left( f\left( \hat{X}\right) +\nabla _{%
\hat{X}}g\left( \hat{X},K_{\hat{X}}\right) -\frac{\sigma _{\hat{K}%
}^{2}F^{2}\left( \hat{X},K_{\hat{X}}\right) }{2f^{2}\left( \hat{X}\right) }+%
\hat{\lambda}\right) }{\sigma _{\hat{X}}^{2}\sqrt{f^{2}\left( \hat{X}\right) 
}}-\frac{1}{2}
\end{equation}

As explained in the core of the paper, to compute $K_{\hat{X}}$ we have to
average (\ref{knl}) over $\hat{\lambda}$ with the weight $\exp \left(
-\left( S_{3}+S_{4}\right) \right) $. Given equation (\ref{fdr}), a solution
(\ref{spc}) for a given $\hat{\lambda}$ and taking into account the
constraint $\left\Vert \hat{\Psi}\left( \hat{K},\hat{X}\right) \right\Vert
^{2}=\hat{N}$, has the associated normalized weight (see (\ref{stT})):

\begin{equation*}
w\left( \left\vert \hat{\lambda}\right\vert \right) =\frac{\exp \left(
-\left( \left\vert \hat{\lambda}\right\vert -M\right) \hat{N}\right) }{%
\int_{\left\vert \hat{\lambda}\right\vert >M}\exp \left( -\left( \left\vert 
\hat{\lambda}\right\vert -M\right) \hat{N}\right) d\left\vert \hat{\lambda}%
\right\vert }
\end{equation*}%
with $M$ is the lower bound for $\left\vert \hat{\lambda}\right\vert $.

This lower bound is found by considering (\ref{nqk}) and adding the term
proportional to $\frac{\sigma _{\hat{X}}^{2}}{2}$:%
\begin{equation}
\frac{\sigma _{\hat{X}}^{2}}{2}\nabla _{\hat{X}}^{2}\hat{\Psi}+\nabla
_{y}^{2}\hat{\Psi}-\left( \sqrt{f^{2}\left( \hat{X}\right) }\frac{y^{2}}{4}+%
\frac{\left( g\left( \hat{X}\right) \right) ^{2}}{\sigma _{\hat{X}}^{2}}%
+\left( f\left( \hat{X}\right) +\nabla _{\hat{X}}g\left( \hat{X},K_{\hat{X}%
}\right) -\frac{\sigma _{\hat{K}}^{2}F^{2}\left( \hat{X},K_{\hat{X}}\right) 
}{2f^{2}\left( \hat{X}\right) }+\hat{\lambda}\right) \right) \Psi
\label{gmk}
\end{equation}%
multiplying (\ref{gmk}) by $\hat{\Psi}^{\dag }$ and integrating. It yields: 
\begin{eqnarray}
0 &=&-\frac{\sigma _{\hat{X}}^{2}}{2}\int \left( \nabla _{\hat{X}}\hat{\Psi}%
^{\dag }\right) \left( \nabla _{\hat{X}}\hat{\Psi}\right)  \label{kmg} \\
&&-\frac{1}{2}\int \sqrt{f^{2}\left( \hat{X}\right) }\left( \left( \nabla
_{y}\hat{\Psi}^{\dag }\right) \left( \nabla _{y}\hat{\Psi}\right) +\hat{\Psi}%
^{\dag }\frac{y^{2}}{4}\hat{\Psi}\right) +\int \hat{\Psi}_{y=0}^{\dag
}\left( \nabla _{y}\hat{\Psi}\right) _{y=0}  \notag \\
&&-\int \hat{\Psi}^{\dag }\left( \sqrt{f^{2}\left( \hat{X}\right) }\frac{%
y^{2}}{4}+\frac{\left( g\left( \hat{X}\right) \right) ^{2}}{\sigma _{\hat{X}%
}^{2}}+\left( f\left( \hat{X}\right) +\nabla _{\hat{X}}g\left( \hat{X},K_{%
\hat{X}}\right) -\frac{\sigma _{\hat{K}}^{2}F^{2}\left( \hat{X},K_{\hat{X}%
}\right) }{2f^{2}\left( \hat{X}\right) }+\hat{\lambda}\right) \right) \Psi 
\notag
\end{eqnarray}%
The first part of the right hand side in (\ref{kmg}):%
\begin{equation}
-\frac{\sigma _{\hat{X}}^{2}}{2}\int \left( \nabla _{\hat{X}}\hat{\Psi}%
^{\dag }\right) \left( \nabla _{\hat{X}}\hat{\Psi}\right) -\int \sqrt{%
f^{2}\left( \hat{X}\right) }\left( \frac{1}{2}\left( \nabla _{y}\hat{\Psi}%
^{\dag }\right) \left( \nabla _{y}\hat{\Psi}\right) +\hat{\Psi}^{\dag }\frac{%
y^{2}}{4}\hat{\Psi}\right)  \label{rhd}
\end{equation}%
includes the hamiltonian of a sum of harmonic oscillators, and thus (\ref%
{rhd}) is lower than $-\frac{\int \hat{\Psi}^{\dag }\sqrt{f^{2}\left( \hat{X}%
\right) }\hat{\Psi}}{2}$. As a consequence, we have the inequality for all $%
\hat{X}$:%
\begin{eqnarray*}
&&\hat{\Psi}_{y=0}^{\dag }\left( \nabla _{y}\hat{\Psi}\right) _{y=0}+\int 
\hat{\Psi}^{\dag }\left( \left\vert \hat{\lambda}\right\vert -\frac{\left(
g\left( \hat{X}\right) \right) ^{2}}{\sigma _{\hat{X}}^{2}}-\left( f\left( 
\hat{X}\right) +\nabla _{\hat{X}}g\left( \hat{X},K_{\hat{X}}\right) -\frac{%
\sigma _{\hat{K}}^{2}F^{2}\left( \hat{X},K_{\hat{X}}\right) }{2f^{2}\left( 
\hat{X}\right) }\right) \right) \Psi d\hat{K} \\
&>&\frac{\int \hat{\Psi}^{\dag }\sqrt{f^{2}\left( \hat{X}\right) }\hat{\Psi}d%
\hat{K}}{2}
\end{eqnarray*}%
Since: 
\begin{equation*}
\left\vert \hat{\lambda}\right\vert \int \left\vert \Psi \right\vert ^{2}d%
\hat{K}=\left\vert \hat{\lambda}\right\vert \hat{N}\left( \hat{X}\right)
\end{equation*}%
and $\hat{\Psi}_{y=0}^{\dag }\left( \nabla _{y}\hat{\Psi}\right) _{y=0}$ is
of\ order $1<<\hat{N}\left( \hat{X}\right) $ since it is integrated over $%
\hat{X}$ only. As a consequence, the condition reduces to:%
\begin{equation*}
\left\vert \hat{\lambda}\right\vert \hat{N}\left( \hat{X}\right) >\int \hat{%
\Psi}^{\dag }\left( \frac{\left( g\left( \hat{X}\right) \right) ^{2}}{\sigma
_{\hat{X}}^{2}}+f\left( \hat{X}\right) +\frac{1}{2}\sqrt{f^{2}\left( \hat{X}%
\right) }+\nabla _{\hat{X}}g\left( \hat{X},K_{\hat{X}}\right) -\frac{\sigma
_{\hat{K}}^{2}F^{2}\left( \hat{X},K_{\hat{X}}\right) }{2f^{2}\left( \hat{X}%
\right) }\right) \Psi d\hat{K}
\end{equation*}%
that is:%
\begin{equation*}
\left\vert \hat{\lambda}\right\vert >\frac{\left( g\left( \hat{X}\right)
\right) ^{2}}{\sigma _{\hat{X}}^{2}}+f\left( \hat{X}\right) +\frac{1}{2}%
\sqrt{f^{2}\left( \hat{X}\right) }+\nabla _{\hat{X}}g\left( \hat{X},K_{\hat{X%
}}\right) -\frac{\sigma _{\hat{K}}^{2}F^{2}\left( \hat{X},K_{\hat{X}}\right) 
}{2f^{2}\left( \hat{X}\right) }
\end{equation*}%
for each $\hat{X}$, and we have:%
\begin{equation}
M=\max_{\hat{X}}\left( \frac{\left( g\left( \hat{X}\right) \right) ^{2}}{%
\sigma _{\hat{X}}^{2}}+f\left( \hat{X}\right) +\frac{1}{2}\sqrt{f^{2}\left( 
\hat{X}\right) }+\nabla _{\hat{X}}g\left( \hat{X},K_{\hat{X}}\right) -\frac{%
\sigma _{\hat{K}}^{2}F^{2}\left( \hat{X},K_{\hat{X}}\right) }{2f^{2}\left( 
\hat{X}\right) }\right)  \label{mdf}
\end{equation}%
Note that in general, for $\varepsilon <<1$, $f\left( \hat{X}\right) >>1$
and: 
\begin{equation*}
\frac{\sigma _{\hat{K}}^{2}F^{2}\left( \hat{X},K_{\hat{X}}\right) }{%
2f^{2}\left( \hat{X}\right) }<<\frac{\left( g\left( \hat{X}\right) \right)
^{2}}{\sigma _{\hat{X}}^{2}}+f\left( \hat{X}\right) +\frac{1}{2}\sqrt{%
f^{2}\left( \hat{X}\right) }+\nabla _{\hat{X}}g\left( \hat{X},K_{\hat{X}%
}\right)
\end{equation*}%
so that:%
\begin{equation}
M\simeq \max_{\hat{X}}\left( \frac{\left( g\left( \hat{X}\right) \right) ^{2}%
}{\sigma _{\hat{X}}^{2}}+f\left( \hat{X}\right) +\frac{1}{2}\sqrt{%
f^{2}\left( \hat{X}\right) }+\nabla _{\hat{X}}g\left( \hat{X},K_{\hat{X}%
}\right) \right)  \label{mdF}
\end{equation}%
Having found $M$, this yields: 
\begin{equation}
w\left( \left\vert \hat{\lambda}\right\vert \right) =\hat{N}\exp \left(
-\left( \left\vert \hat{\lambda}\right\vert -M\right) \hat{N}\right)
\label{thw}
\end{equation}%
As a consequence, averaging equation (\ref{knl}) yields:%
\begin{equation*}
K_{\hat{X}}=\int K_{\hat{X},\hat{\lambda}}\hat{N}\exp \left( -\left(
\left\vert \hat{\lambda}\right\vert -M\right) \hat{N}\right) d\hat{\lambda}
\end{equation*}%
\begin{eqnarray}
K_{\hat{X}}\left\Vert \Psi \left( \hat{X}\right) \right\Vert ^{2} &=&\int
C\left( \hat{\lambda}\right) w\left( \left\vert \hat{\lambda}\right\vert
\right) \exp \left( -\frac{\sigma _{X}^{2}\sigma _{\hat{K}}^{2}\left( p+%
\frac{1}{2}\right) ^{2}\left( f^{\prime }\left( X\right) \right) ^{2}}{%
96\left\vert f\left( \hat{X}\right) \right\vert ^{3}}\right) \left( \frac{%
\left\vert f\left( \hat{X}\right) \right\vert }{\sigma _{\hat{K}}^{2}}%
\right) ^{-1}  \label{kvn} \\
&&\times \left( \frac{\Gamma \left( -\frac{p+1}{2}\right) \Gamma \left( 
\frac{1-p}{2}\right) -\Gamma \left( -\frac{p}{2}\right) \Gamma \left( \frac{%
-p}{2}\right) }{2^{p+2}\Gamma \left( -p-1\right) \Gamma \left( -p\right) }+p%
\frac{\Gamma \left( -\frac{p}{2}\right) \Gamma \left( \frac{2-p}{2}\right)
-\Gamma \left( -\frac{p-1}{2}\right) \Gamma \left( -\frac{p-1}{2}\right) }{%
2^{p+1}\Gamma \left( -p\right) \Gamma \left( -p+1\right) }\right) d\hat{%
\lambda}  \notag
\end{eqnarray}%
with $C\left( \bar{p}\left( \hat{\lambda}\right) \right) $ given by (\ref%
{cld}). \ Given (\ref{thw}), the average value of $\left\vert \hat{\lambda}%
\right\vert $ is $M+\frac{1}{\hat{N}}$ and have:%
\begin{eqnarray}
K_{\hat{X}}\left\Vert \Psi \left( \hat{X}\right) \right\Vert ^{2}\left\vert
f\left( \hat{X}\right) \right\vert &=&C\left( \bar{p}\left( -\left( M-\frac{1%
}{\hat{N}}\right) \right) \right) \sigma _{\hat{K}}^{2}  \label{nkv} \\
&&\times \left( \frac{\Gamma \left( -\frac{p+1}{2}\right) \Gamma \left( 
\frac{1-p}{2}\right) -\Gamma \left( -\frac{p}{2}\right) \Gamma \left( \frac{%
-p}{2}\right) }{2^{p+2}\Gamma \left( -p-1\right) \Gamma \left( -p\right) }+p%
\frac{\Gamma \left( -\frac{p}{2}\right) \Gamma \left( \frac{2-p}{2}\right)
-\Gamma \left( -\frac{p-1}{2}\right) \Gamma \left( -\frac{p-1}{2}\right) }{%
2^{p+1}\Gamma \left( -p\right) \Gamma \left( -p+1\right) }\right)  \notag
\end{eqnarray}

with:%
\begin{equation*}
p=-\frac{\left( g\left( \hat{X}\right) \right) ^{2}+\sigma _{\hat{X}%
}^{2}\left( f\left( \hat{X}\right) +\nabla _{\hat{X}}g\left( \hat{X},K_{\hat{%
X}}\right) -\frac{\sigma _{\hat{K}}^{2}F^{2}\left( \hat{X},K_{\hat{X}%
}\right) }{2f^{2}\left( \hat{X}\right) }-\left( M-\frac{1}{\hat{N}}\right)
\right) }{\sigma _{\hat{X}}^{2}\sqrt{f^{2}\left( \hat{X}\right) }}-\frac{1}{2%
}
\end{equation*}%
We can consider that $\frac{1}{\hat{N}}<<1$ so that $C\left( \bar{p}\left(
-\left( M-\frac{1}{\hat{N}}\right) \right) \right) \simeq C\left( \bar{p}%
\left( -M\right) \right) $. It amounts to consider $\left\vert \hat{\lambda}%
\right\vert =M$. We will also write $\bar{p}\left( -M\right) =\bar{p}$ and
given (\ref{rbp}) we have:%
\begin{equation}
\bar{p}=\left( \frac{M-\frac{\left( g\left( \hat{X}_{0}\right) \right) ^{2}}{%
\sigma _{\hat{X}}^{2}}+\left( f\left( \hat{X}_{0}\right) +\frac{1}{2}%
\left\vert f\left( \hat{X}_{0}\right) \right\vert +\nabla _{\hat{X}}g\left( 
\hat{X}_{0},K_{\hat{X}_{0}}\right) -\frac{\sigma _{\hat{K}}^{2}F^{2}\left( 
\hat{X}_{0},K_{\hat{X}_{0}}\right) }{2f^{2}\left( \hat{X}_{0}\right) }%
\right) }{\left\vert f\left( \hat{X}_{0}\right) \right\vert }\right)
\label{pbr}
\end{equation}%
and:%
\begin{equation}
p=\frac{M-\left( \frac{\left( g\left( \hat{X}\right) \right) ^{2}}{\sigma _{%
\hat{X}}^{2}}+\left( f\left( \hat{X}\right) +\frac{\sqrt{f^{2}\left( \hat{X}%
\right) }}{2}+\nabla _{\hat{X}}g\left( \hat{X},K_{\hat{X}}\right) -\frac{%
\sigma _{\hat{K}}^{2}F^{2}\left( \hat{X},K_{\hat{X}}\right) }{2f^{2}\left( 
\hat{X}\right) }\right) \right) }{\sqrt{f^{2}\left( \hat{X}\right) }}
\label{pdf}
\end{equation}%
Equation (\ref{cld}) rewrites:%
\begin{equation}
C\left( \bar{p}\right) \simeq \frac{\exp \left( -\frac{\sigma _{X}^{2}\sigma
_{\hat{K}}^{2}\left( \frac{\left( \bar{p}\left( \hat{\lambda}\right) +\frac{1%
}{2}\right) f^{\prime }\left( X_{0}\right) }{f\left( \hat{X}_{0}\right) }%
\right) ^{2}}{96\left\vert f\left( \hat{X}_{0}\right) \right\vert }\right) 
\hat{N}\Gamma \left( -\bar{p}\right) }{\left( \frac{\left\langle \left\vert
f\left( \hat{X}\right) \right\vert \right\rangle }{\sigma _{\hat{K}}^{2}}%
\right) ^{-\frac{1}{2}}V_{r}\left( \func{Psi}\left( -\frac{\bar{p}-1}{2}%
\right) -\func{Psi}\left( -\frac{\bar{p}}{2}\right) \right) }  \label{clk}
\end{equation}%
and (\ref{nkv}) reduces to:%
\begin{equation}
K_{\hat{X}}\left\Vert \Psi \left( \hat{X}\right) \right\Vert ^{2}\left\vert
f\left( \hat{X}\right) \right\vert =C\left( \bar{p}\right) \sigma _{\hat{K}%
}^{2}\hat{\Gamma}\left( p+\frac{1}{2}\right)  \label{Nkv}
\end{equation}%
with:%
\begin{eqnarray}
\hat{\Gamma}\left( p+\frac{1}{2}\right) &=&\exp \left( -\frac{\sigma
_{X}^{2}\sigma _{\hat{K}}^{2}\left( p+\frac{1}{2}\right) ^{2}\left(
f^{\prime }\left( X\right) \right) ^{2}}{96\left\vert f\left( \hat{X}\right)
\right\vert ^{3}}\right)  \label{gmh} \\
&&\times \left( \frac{\Gamma \left( -\frac{p+1}{2}\right) \Gamma \left( 
\frac{1-p}{2}\right) -\Gamma \left( -\frac{p}{2}\right) \Gamma \left( \frac{%
-p}{2}\right) }{2^{p+2}\Gamma \left( -p-1\right) \Gamma \left( -p\right) }+p%
\frac{\Gamma \left( -\frac{p}{2}\right) \Gamma \left( \frac{2-p}{2}\right)
-\Gamma \left( -\frac{p-1}{2}\right) \Gamma \left( -\frac{p-1}{2}\right) }{%
2^{p+1}\Gamma \left( -p\right) \Gamma \left( -p+1\right) }\right)  \notag
\end{eqnarray}%
We note that, asymptotically:%
\begin{equation}
\hat{\Gamma}\left( p+\frac{1}{2}\right) \sim _{\infty }\exp \left( -\frac{%
\sigma _{X}^{2}\sigma _{\hat{K}}^{2}\left( p+\frac{1}{2}\right) ^{2}\left(
f^{\prime }\left( X\right) \right) ^{2}}{96\left\vert f\left( \hat{X}\right)
\right\vert ^{3}}\right) \Gamma \left( p+\frac{3}{2}\right)
\end{equation}

\paragraph*{A3.1.4.3 \textbf{Replacing }$\left\Vert \Psi \left( X\right)
\right\Vert ^{2}$\textbf{\ in the} $K_{\hat{X}}$\protect\bigskip e\textbf{%
quation}}

We can isolate $K_{\hat{X}}$ in (\ref{nkv}) by using (\ref{psp}) and (\ref%
{spp}) to rewrite $\left\Vert \Psi \left( \hat{X}\right) \right\Vert ^{2}$:

Using (\ref{nbq}):%
\begin{eqnarray*}
D\left( \left\Vert \Psi \right\Vert ^{2}\right) &=&2\tau \left\Vert \Psi
\left( X\right) \right\Vert ^{2}+\frac{1}{2\sigma _{X}^{2}}\left( \nabla
_{X}R\left( X\right) \right) ^{2}H^{2}\left( \frac{\hat{K}_{X}}{\left\Vert
\Psi \left( X\right) \right\Vert ^{2}}\right) \left( 1-\frac{H^{\prime
}\left( \hat{K}_{X}\right) }{H\left( \hat{K}_{X}\right) }\frac{\hat{K}_{X}}{%
\left\Vert \Psi \left( X\right) \right\Vert ^{2}}\right) \\
&=&2\tau \left\Vert \Psi \left( X\right) \right\Vert ^{2}+\frac{1}{2\sigma
_{X}^{2}}\left( \nabla _{X}R\left( X\right) \right) ^{2}H^{2}\left(
K_{X}\right) \left( 1-\frac{H^{\prime }\left( K_{X}\right) }{H\left(
K_{X}\right) }K_{X}\right)
\end{eqnarray*}%
We rewrite $\left\Vert \Psi \left( X\right) \right\Vert ^{2}$ as a function
of $K_{X}$: 
\begin{equation}
\left\Vert \Psi \left( X\right) \right\Vert ^{2}=\frac{D\left( \left\Vert
\Psi \right\Vert ^{2}\right) -\frac{1}{2\sigma _{X}^{2}}\left( \nabla
_{X}R\left( X\right) \right) ^{2}H^{2}\left( K_{X}\right) \left( 1-\frac{%
H^{\prime }\left( K_{X}\right) }{H\left( K_{X}\right) }K_{X}\right) }{2\tau }%
\equiv D-\bar{H}\left( X,K_{X}\right)  \label{psrd}
\end{equation}%
Ultimately, the equation (\ref{Nkv}) for $K_{\hat{X}}$ can be rewritten: 
\begin{equation}
K_{\hat{X}}\left\vert f\left( \hat{X}\right) \right\vert =\frac{C\left( \bar{%
p}\right) \sigma _{\hat{K}}^{2}}{\left\Vert \Psi \left( X\right) \right\Vert
^{2}}\hat{\Gamma}\left( p+\frac{1}{2}\right) =\frac{C\left( \bar{p}\right)
\sigma _{\hat{K}}^{2}}{D-\bar{H}\left( X,K_{X}\right) }\hat{\Gamma}\left( p+%
\frac{1}{2}\right)  \label{qnk}
\end{equation}%
with $C\left( \bar{p}\right) $\ given by (\ref{clk}), $\hat{\Gamma}\left( p+%
\frac{1}{2}\right) $\ defined in (\ref{gmh}) and $p$ given by (\ref{pdf}).

\subsection*{A3.2 Approaches to solutions for $K_{\hat{X}}$}

We detail some computations of the three approaches detailed in the core of
the paper.

\subsubsection*{A3.2.1 First approach: Differential form of (\protect\ref%
{qtk})}

To understand the behavior of the solutions of (\ref{qtk}), we can write its
differential version. Assume a variation $\delta Y\left( \hat{X}\right) $
for any parameter of the system at point $\hat{X}$. This parameter $Y\left( 
\hat{X}\right) $ can be either $R\left( X\right) $, its gradient, or any
parameter arising in the definition of $f$ and $g$. This induces a variation 
$\delta K_{\hat{X}}$ for the average capital. The equation for $\delta K_{%
\hat{X}}$ is obtained by differentiation of (\ref{qtk}):

\begin{eqnarray}
\delta K_{\hat{X}} &=&\left( -\left( \frac{\frac{\partial f\left( \hat{X},K_{%
\hat{X}}\right) }{\partial K_{\hat{X}}}}{f\left( \hat{X},K_{\hat{X}}\right) }%
+\frac{\frac{\partial \left\Vert \Psi \left( \hat{X},K_{\hat{X}}\right)
\right\Vert ^{2}}{\partial K_{\hat{X}}}}{\left\Vert \Psi \left( \hat{X},K_{%
\hat{X}}\right) \right\Vert ^{2}}+l\left( \hat{X},K_{\hat{X}}\right) \right)
+k\left( p\right) \frac{\partial p}{\partial K_{\hat{X}}}\right) K_{\hat{X}%
}\delta K_{\hat{X}}  \label{dvr} \\
&&+\frac{\partial }{\partial Y\left( \hat{X}\right) }\left( \frac{\sigma _{%
\hat{K}}^{2}C\left( \bar{p}\right) 2\hat{\Gamma}\left( p+\frac{1}{2}\right) 
}{\left\vert f\left( \hat{X},K_{\hat{X}}\right) \right\vert \left\Vert \Psi
\left( \hat{X},K_{\hat{X}}\right) \right\Vert ^{2}}\right) \delta Y\left( 
\hat{X}\right)  \notag
\end{eqnarray}%
where we define:%
\begin{eqnarray*}
l\left( \hat{X},K_{\hat{X}}\right) &=&\frac{\sigma _{X}^{2}\sigma _{\hat{K}%
}^{2}\left( \nabla _{K_{\hat{X}}}\left( f^{\prime }\left( \hat{X}\right)
\right) ^{2}\left\vert f\left( \hat{X}\right) \right\vert -3\left( \nabla
_{K_{\hat{X}}}\left\vert f\left( \hat{X}\right) \right\vert \right) \left(
f^{\prime }\left( \hat{X}\right) \right) ^{2}\right) \left( p+\frac{1}{2}%
\right) ^{2}}{120\left\vert f\left( \hat{X}\right) \right\vert ^{4}} \\
&&+\frac{\partial p}{\partial K_{\hat{X}}}\frac{\sigma _{X}^{2}\sigma _{\hat{%
K}}^{2}\left( p+\frac{1}{2}\right) \left( f^{\prime }\left( X\right) \right)
^{2}}{48\left\vert f\left( \hat{X}\right) \right\vert ^{3}}
\end{eqnarray*}%
\begin{equation}
k\left( p\right) =\frac{\frac{d}{dp}\hat{\Gamma}\left( p+\frac{1}{2}\right) 
}{\hat{\Gamma}\left( p+\frac{1}{2}\right) }\sim _{\infty }\sqrt{\frac{p-%
\frac{1}{2}}{2}}-\frac{\sigma _{X}^{2}\sigma _{\hat{K}}^{2}\left( p+\frac{1}{%
2}\right) \left( f^{\prime }\left( X\right) \right) ^{2}}{48\left\vert
f\left( \hat{X}\right) \right\vert ^{3}}
\end{equation}%
and:%
\begin{equation*}
\frac{\partial p}{\partial K_{\hat{X}}}=\frac{\partial }{\partial K_{\hat{X}}%
}\frac{M-A\left( \hat{X},K_{\hat{X}}\right) }{\left\vert f\left( \hat{X},K_{%
\hat{X}}\right) \right\vert }=-\frac{\partial _{K_{\hat{X}}}\left\vert
f\left( \hat{X},K_{\hat{X}}\right) \right\vert p+\partial _{K_{\hat{X}%
}}A\left( \hat{X},K_{\hat{X}}\right) }{\left\vert f\left( \hat{X},K_{\hat{X}%
}\right) \right\vert }
\end{equation*}%
with:%
\begin{eqnarray*}
A\left( \hat{X},K_{\hat{X}}\right) &=&\frac{\left( g\left( \hat{X},K_{\hat{X}%
}\right) \right) ^{2}}{\sigma _{\hat{X}}^{2}}+\left( f\left( \hat{X},K_{\hat{%
X}}\right) +\frac{\left\vert f\left( \hat{X},K_{\hat{X}}\right) \right\vert 
}{2}+\nabla _{\hat{X}}g\left( \hat{X},K_{\hat{X}}\right) -\frac{\sigma _{%
\hat{K}}^{2}F^{2}\left( \hat{X},K_{\hat{X}}\right) }{2f^{2}\left( \hat{X},K_{%
\hat{X}}\right) }\right) \\
&\simeq &\frac{\left( g\left( \hat{X},K_{\hat{X}}\right) \right) ^{2}}{%
\sigma _{\hat{X}}^{2}}+f\left( \hat{X},K_{\hat{X}}\right) +\frac{\left\vert
f\left( \hat{X},K_{\hat{X}}\right) \right\vert }{2}+\nabla _{\hat{X}}g\left( 
\hat{X},K_{\hat{X}}\right)
\end{eqnarray*}

In an expanded form (\ref{dvr}) writes:

\begin{eqnarray*}
\delta K_{\hat{X}} &=&\left( k\left( p\right) \frac{\partial _{K_{\hat{X}%
}}\left( M-\left( \frac{\left( g\left( \hat{X},K_{\hat{X}}\right) \right)
^{2}}{\sigma _{\hat{X}}^{2}}+\nabla _{\hat{X}}g\left( \hat{X},K_{\hat{X}%
}\right) -\frac{\sigma _{\hat{K}}^{2}F^{2}\left( \hat{X},K_{\hat{X}}\right) 
}{2f^{2}\left( \hat{X}\right) }\right) \right) }{f\left( \hat{X},K_{\hat{X}%
}\right) }\right. \\
&&\left. -\left( \frac{\frac{\partial f\left( \hat{X},K_{\hat{X}}\right) }{%
\partial K_{\hat{X}}}\left( 1+\left( p+\mathcal{H}\left( f\left( \hat{X},K_{%
\hat{X}}\right) \right) +\frac{1}{2}\right) k\left( p\right) \right) }{%
f\left( \hat{X},K_{\hat{X}}\right) }+\frac{\frac{\partial \left\Vert \Psi
\left( \hat{X},K_{\hat{X}}\right) \right\Vert ^{2}}{\partial K_{\hat{X}}}}{%
\left\Vert \Psi \left( \hat{X},K_{\hat{X}}\right) \right\Vert ^{2}}+l\left( 
\hat{X},K_{\hat{X}}\right) \right) \right) K_{\hat{X}}\delta K_{\hat{X}} \\
&&+\frac{\partial }{\partial Y}\left( \frac{\sigma _{\hat{K}}^{2}C\left( 
\bar{p}\right) 2\hat{\Gamma}\left( p+\frac{1}{2}\right) }{\left\vert f\left( 
\hat{X},K_{\hat{X}}\right) \right\vert \left\Vert \Psi \left( \hat{X},K_{%
\hat{X}}\right) \right\Vert ^{2}}\right) \delta Y
\end{eqnarray*}%
with $\mathcal{H}$ the heaviside function. Moreover:%
\begin{eqnarray*}
&&\frac{\partial }{\partial Y}\left( \frac{\sigma _{\hat{K}}^{2}C\left( \bar{%
p}\right) 2\hat{\Gamma}\left( p+\frac{1}{2}\right) }{\left\vert f\left( \hat{%
X},K_{\hat{X}}\right) \right\vert \left\Vert \Psi \left( \hat{X},K_{\hat{X}%
}\right) \right\Vert ^{2}}\right) \delta Y \\
&=&\left( k\left( p\right) \frac{\partial _{Y}\left( M-\left( \frac{\left(
g\left( \hat{X},K_{\hat{X}}\right) \right) ^{2}}{\sigma _{\hat{X}}^{2}}%
+\nabla _{\hat{X}}g\left( \hat{X},K_{\hat{X}}\right) -\frac{\sigma _{\hat{K}%
}^{2}F^{2}\left( \hat{X},K_{\hat{X}}\right) }{2f^{2}\left( \hat{X}\right) }%
\right) \right) }{f\left( \hat{X},K_{\hat{X}}\right) }\right. \\
&&\left. -\left( \frac{\frac{\partial f\left( \hat{X},K_{\hat{X}}\right) }{%
\partial Y}\left( 1+\left( p+\mathcal{H}\left( f\left( \hat{X},K_{\hat{X}%
}\right) \right) +\frac{1}{2}\right) k\left( p\right) \right) }{f\left( \hat{%
X},K_{\hat{X}}\right) }+\frac{\frac{\partial \left\Vert \Psi \left( \hat{X}%
,K_{\hat{X}}\right) \right\Vert ^{2}}{\partial Y}}{\left\Vert \Psi \left( 
\hat{X},K_{\hat{X}}\right) \right\Vert ^{2}}+m_{Y}\left( \hat{X},K_{\hat{X}%
}\right) \right) \right) K_{\hat{X}}\delta Y
\end{eqnarray*}%
with:%
\begin{eqnarray*}
m_{Y}\left( \hat{X},K_{\hat{X}}\right) &=&\frac{\sigma _{X}^{2}\sigma _{\hat{%
K}}^{2}\left( \nabla _{Y}\left( f^{\prime }\left( \hat{X}\right) \right)
^{2}\left\vert f\left( \hat{X}\right) \right\vert -3\left( \nabla
_{Y}\left\vert f\left( \hat{X}\right) \right\vert \right) \left( f^{\prime
}\left( \hat{X}\right) \right) ^{2}\right) \left( p+\frac{1}{2}\right) ^{2}}{%
120\left\vert f\left( \hat{X}\right) \right\vert ^{4}} \\
&&+\nabla _{Y}p\frac{\sigma _{X}^{2}\sigma _{\hat{K}}^{2}\left( p+\frac{1}{2}%
\right) \left( f^{\prime }\left( X\right) \right) ^{2}}{48\left\vert f\left( 
\hat{X}\right) \right\vert ^{3}}
\end{eqnarray*}%
so that:%
\begin{eqnarray*}
\frac{\delta K_{\hat{X}}}{K_{\hat{X}}} &=&\left( k\left( p\right) \frac{%
\partial _{Y}\left( M-\left( \frac{\left( g\left( \hat{X},K_{\hat{X}}\right)
\right) ^{2}}{\sigma _{\hat{X}}^{2}}+\nabla _{\hat{X}}g\left( \hat{X},K_{%
\hat{X}}\right) -\frac{\sigma _{\hat{K}}^{2}F^{2}\left( \hat{X},K_{\hat{X}%
}\right) }{2f^{2}\left( \hat{X}\right) }\right) \right) }{f\left( \hat{X},K_{%
\hat{X}}\right) }\right. \\
&&\left. -\left( \frac{\frac{\partial f\left( \hat{X},K_{\hat{X}}\right) }{%
\partial Y}\left( 1+\left( p+\mathcal{H}\left( f\left( \hat{X},K_{\hat{X}%
}\right) \right) +\frac{1}{2}\right) k\left( p\right) \right) }{f\left( \hat{%
X},K_{\hat{X}}\right) }+\frac{\frac{\partial \left\Vert \Psi \left( \hat{X}%
,K_{\hat{X}}\right) \right\Vert ^{2}}{\partial Y}}{\left\Vert \Psi \left( 
\hat{X},K_{\hat{X}}\right) \right\Vert ^{2}}+m_{Y}\left( \hat{X},K_{\hat{X}%
}\right) \right) \right) \frac{\delta Y}{D}
\end{eqnarray*}%
with:%
\begin{eqnarray*}
D &=&1+\left( \left( \frac{\frac{\partial f\left( \hat{X},K_{\hat{X}}\right) 
}{\partial K_{\hat{X}}}\left( 1+\left( p+\mathcal{H}\left( f\left( \hat{X}%
,K_{\hat{X}}\right) \right) +\frac{1}{2}\right) k\left( p\right) \right) }{%
f\left( \hat{X},K_{\hat{X}}\right) }+\frac{\frac{\partial \left\Vert \Psi
\left( \hat{X},K_{\hat{X}}\right) \right\Vert ^{2}}{\partial K_{\hat{X}}}}{%
\left\Vert \Psi \left( \hat{X},K_{\hat{X}}\right) \right\Vert ^{2}}+l\left( 
\hat{X},K_{\hat{X}}\right) \right) \right. \\
&&\left. -k\left( p\right) \frac{\partial _{K_{\hat{X}}}\left( M-\left( 
\frac{\left( g\left( \hat{X},K_{\hat{X}}\right) \right) ^{2}}{\sigma _{\hat{X%
}}^{2}}+\nabla _{\hat{X}}g\left( \hat{X},K_{\hat{X}}\right) -\frac{\sigma _{%
\hat{K}}^{2}F^{2}\left( \hat{X},K_{\hat{X}}\right) }{2f^{2}\left( \hat{X}%
\right) }\right) \right) }{f\left( \hat{X},K_{\hat{X}}\right) }\right) K_{%
\hat{X}}
\end{eqnarray*}

\subsubsection*{A3.2.2 Second approach: Expansion around particular solutions%
}

As explained in the text, we choose to expand (\ref{Nkv}), or equivalently (%
\ref{qnk}), around solutions with $p=0$.

\paragraph*{A3.2.2.1 Particular solutions\protect\bigskip}

To find the solution with $p=0$, we maximize the function:%
\begin{equation*}
A\left( \hat{X}\right) =\frac{\left( g\left( \hat{X}\right) \right) ^{2}}{%
\sigma _{\hat{X}}^{2}}+f\left( \hat{X}\right) +\frac{1}{2}\sqrt{f^{2}\left( 
\hat{X}\right) }+\nabla _{\hat{X}}g\left( \hat{X},K_{\hat{X}}\right) -\frac{%
\sigma _{\hat{K}}^{2}F^{2}\left( \hat{X},K_{\hat{X}}\right) }{2f^{2}\left( 
\hat{X}\right) }
\end{equation*}%
We write:%
\begin{equation}
M=\max_{\hat{X}}A\left( \hat{X}\right)  \label{Mqn}
\end{equation}%
and denote by $\left( \hat{X}_{M},K_{\hat{X}_{M}}\right) $ the solutions $%
\hat{X}_{M}$ of (\ref{Mqn}) with $K_{\hat{X}_{M}}$ their associated value of
average capital per firm.

Given that $\hat{\Gamma}\left( \frac{1}{2}\right) =1$, (\ref{Nkv}) becomes
at points $\left( \hat{X}_{M},K_{\hat{X}_{M}}\right) $ and $p=0$:%
\begin{equation}
K_{\hat{X},M}\left\vert f\left( \hat{X}_{M},K_{\hat{X}_{M}}\right)
\right\vert \left\Vert \Psi \left( \hat{X}_{M},K_{\hat{X}_{M}}\right)
\right\Vert ^{2}\simeq \sigma _{\hat{K}}^{2}C\left( \bar{p}\right) \exp
\left( -\frac{\sigma _{X}^{2}\sigma _{\hat{K}}^{2}\left( f^{\prime }\left( 
\hat{X}_{M},K_{\hat{X}_{M}}\right) \right) ^{2}}{384\left\vert f\left( \hat{X%
}_{M},K_{\hat{X}_{M}}\right) \right\vert ^{3}}\right)  \label{qMK}
\end{equation}%
This equation has in general several solutions, depending on the assumptions
on $f\left( \hat{X}_{M},K_{\hat{X}_{M}}\right) $. These solutions are
discussed in the text. We also give the form of $K_{\hat{X}}$ for some
particular form of the parameter function $f\left( \hat{X},K_{\hat{X}%
}\right) $.

Note that once a solution $K_{\hat{X}}$ of (\ref{qnk}) is found, the value
of $C\left( \bar{p}\right) $ can be obtained by solving (\ref{pbr}) and
using (\ref{clk}).

The next paragraph computes the expansion of (\ref{Nkv}) around these
solutions with $p=0$. Remark that coming back to (\ref{Nkv}) and (\ref{qnk})
for general values of $p$ defined in (\ref{pdf}), the value of $C\left( \bar{%
p}\right) \sigma _{\hat{K}}^{2}$\ can be replaced by $K_{\hat{X}%
_{M}}\left\vert f\left( \hat{X}_{M},K_{\hat{X}_{M}}\right) \right\vert
\left\Vert \Psi \left( \hat{X}_{M},K_{\hat{X}_{M}}\right) \right\Vert ^{2}$
for any solution $\left( \hat{X}_{M},K_{\hat{X}_{M}}\right) $.

\paragraph*{A3.2.2.2 Expansion around particular solutions\protect\bigskip}

We can find approximate solutions to (\ref{Nkv}): 
\begin{equation}
K_{\hat{X}}\left\Vert \Psi \left( \hat{X}\right) \right\Vert ^{2}\left\vert
f\left( \hat{X}\right) \right\vert =C\left( \bar{p}\right) \sigma _{\hat{K}%
}^{2}\hat{\Gamma}\left( p+\frac{1}{2}\right)  \label{Nkvv}
\end{equation}%
with:%
\begin{eqnarray}
\hat{\Gamma}\left( p+\frac{1}{2}\right) &=&\exp \left( -\frac{\sigma
_{X}^{2}\sigma _{\hat{K}}^{2}\left( p+\frac{1}{2}\right) ^{2}\left(
f^{\prime }\left( X\right) \right) ^{2}}{96\left\vert f\left( \hat{X}\right)
\right\vert ^{3}}\right)  \label{gmhh} \\
&&\times \left( \frac{\Gamma \left( -\frac{p+1}{2}\right) \Gamma \left( 
\frac{1-p}{2}\right) -\Gamma \left( -\frac{p}{2}\right) \Gamma \left( \frac{%
-p}{2}\right) }{2^{p+2}\Gamma \left( -p-1\right) \Gamma \left( -p\right) }+p%
\frac{\Gamma \left( -\frac{p}{2}\right) \Gamma \left( \frac{2-p}{2}\right)
-\Gamma \left( -\frac{p-1}{2}\right) \Gamma \left( -\frac{p-1}{2}\right) }{%
2^{p+1}\Gamma \left( -p\right) \Gamma \left( -p+1\right) }\right)  \notag
\end{eqnarray}%
for general form of the functions $f\left( \hat{X}\right) $ and $g\left( 
\hat{X}\right) $ by expanding (\ref{Nkvv}), for each $\hat{X}$,around the
closest point $\hat{X}_{M}$ satisfying (\ref{Nkvv}) with $p=0$. We use that:%
\begin{eqnarray}
&&\left( \frac{\Gamma \left( -\frac{p+1}{2}\right) \Gamma \left( \frac{1-p}{2%
}\right) -\Gamma \left( -\frac{p}{2}\right) \Gamma \left( \frac{-p}{2}%
\right) }{2^{p+2}\Gamma \left( -p-1\right) \Gamma \left( -p\right) }+p\frac{%
\Gamma \left( -\frac{p}{2}\right) \Gamma \left( \frac{2-p}{2}\right) -\Gamma
\left( -\frac{p-1}{2}\right) \Gamma \left( -\frac{p-1}{2}\right) }{%
2^{p+1}\Gamma \left( -p\right) \Gamma \left( -p+1\right) }\right)
\label{dvp} \\
&=&1-p\left( \gamma _{0}+\ln 2-2\right) +o\left( p\right)  \notag
\end{eqnarray}%
with $\gamma _{0}$ the Euler-Mascheroni constant, as well as the following
relations:%
\begin{equation*}
\nabla _{K_{\hat{X}}}\left( -\frac{\sigma _{X}^{2}\sigma _{\hat{K}%
}^{2}h\left( p\right) \left( f^{\prime }\left( \hat{X}\right) \right) ^{2}}{%
96\left\vert f\left( \hat{X}\right) \right\vert ^{3}}\right) _{p=0}\simeq -%
\frac{\nabla _{K_{\hat{X}}}\left( f^{\prime }\left( \hat{X}\right) \right)
^{2}\left\vert f\left( \hat{X}\right) \right\vert -3\left( \nabla _{K_{\hat{X%
}}}\left\vert f\left( \hat{X}\right) \right\vert \right) \left( f^{\prime
}\left( \hat{X}\right) \right) ^{2}}{120\left\vert f\left( \hat{X}\right)
\right\vert ^{4}}
\end{equation*}%
and:%
\begin{eqnarray*}
&&\nabla _{\hat{X}}\left( -\frac{h\left( p\right) \left( f^{\prime }\left( 
\hat{X}\right) \right) ^{2}}{96\left\vert f\left( \hat{X}\right) \right\vert
^{3}}\right) _{p=0} \\
&\simeq &-\frac{2f^{\prime }\left( X\right) f^{\prime \prime }\left( \hat{X}%
\right) \left\vert f\left( \hat{X}\right) \right\vert -3\left( f^{\prime
}\left( \hat{X}\right) \right) ^{3}}{120\left\vert f\left( \hat{X}\right)
\right\vert ^{4}}=\frac{f^{\prime }\left( X\right) \left( 3\left( f^{\prime
}\left( \hat{X}\right) \right) ^{2}-2f^{\prime \prime }\left( X\right)
\left\vert f\left( \hat{X}\right) \right\vert \right) }{120\left\vert
f\left( \hat{X}\right) \right\vert ^{4}}
\end{eqnarray*}%
the expansion of (\ref{Nkvv}) at the lowest order, is:%
\begin{eqnarray*}
&&\left( 1+\frac{\frac{\partial f\left( \hat{X},K_{\hat{X}}\right) }{%
\partial K_{\hat{X}}}}{f\left( \hat{X},K_{\hat{X}}\right) }+\frac{\frac{%
\partial \left\Vert \Psi \left( \hat{X},K_{\hat{X}}\right) \right\Vert ^{2}}{%
\partial K_{\hat{X}}}}{\left\Vert \Psi \left( \hat{X},K_{\hat{X}}\right)
\right\Vert ^{2}}\right) _{K_{\hat{X},M}}\left( K_{\hat{X}}-K_{\hat{X}%
,M}\right) +\left( \frac{\frac{\partial f\left( \hat{X},K_{\hat{X}}\right) }{%
\partial \hat{X}}}{f\left( \hat{X},K_{\hat{X}}\right) }+\frac{\frac{\partial
\left\Vert \Psi \left( \hat{X},K_{\hat{X}}\right) \right\Vert ^{2}}{\partial 
\hat{X}}}{\left\Vert \Psi \left( \hat{X},K_{\hat{X}}\right) \right\Vert ^{2}}%
\right) _{K_{\hat{X},M}}\left( \hat{X}-\hat{X}_{M}\right) \\
&\simeq &-\left( \sigma _{X}^{2}\sigma _{\hat{K}}^{2}\frac{\nabla _{K_{\hat{X%
}}}\left( f^{\prime }\left( \hat{X}\right) \right) ^{2}\left\vert f\left( 
\hat{X}\right) \right\vert -3\left( \nabla _{K_{\hat{X}}}\left\vert f\left( 
\hat{X}\right) \right\vert \right) \left( f^{\prime }\left( \hat{X}\right)
\right) ^{2}}{120\left\vert f\left( \hat{X}\right) \right\vert ^{4}}\right)
_{K_{\hat{X},M}}\left( K_{\hat{X}}-K_{\hat{X},M}\right) \\
&&-\left( \sigma _{X}^{2}\sigma _{\hat{K}}^{2}\frac{2f^{\prime }\left(
X\right) f^{\prime \prime }\left( \hat{X}\right) \left\vert f\left( \hat{X}%
\right) \right\vert -3\left( f^{\prime }\left( \hat{X}\right) \right) ^{3}}{%
120\left\vert f\left( \hat{X}\right) \right\vert ^{4}}\right) _{K_{\hat{X}%
,M}}\left( \hat{X}-\hat{X}_{M}\right) \\
&&-b\frac{\partial _{K_{\hat{X}}}\left( \frac{\left( g\left( \hat{X},K_{\hat{%
X}}\right) \right) ^{2}}{\sigma _{\hat{X}}^{2}}+\left( f\left( \hat{X},K_{%
\hat{X}}\right) +\frac{\left\vert f\left( \hat{X},K_{\hat{X}}\right)
\right\vert }{2}+\nabla _{\hat{X}}g\left( \hat{X},K_{\hat{X}}\right) \right)
\right) \left( K_{\hat{X}}-K_{\hat{X},M}\right) }{\left\vert f\left( \hat{X}%
,K_{\hat{X}}\right) \right\vert } \\
&&-b\frac{\partial _{\hat{X}}\left( \frac{\left( g\left( \hat{X},K_{\hat{X}%
}\right) \right) ^{2}}{\sigma _{\hat{X}}^{2}}+\left( f\left( \hat{X},K_{\hat{%
X}}\right) +\frac{\left\vert f\left( \hat{X},K_{\hat{X}}\right) \right\vert 
}{2}+\nabla _{\hat{X}}g\left( \hat{X},K_{\hat{X}}\right) \right) \right)
\left( \hat{X}-\hat{X}_{M}\right) }{\left\vert f\left( \hat{X},K_{\hat{X}%
}\right) \right\vert }
\end{eqnarray*}%
Given the maximization (\ref{Mqn}), the two last terms in the right hand
side is equal to $0$.%
\begin{eqnarray}
\left( K_{\hat{X}}-K_{\hat{X},M}\right) &=&\frac{1}{D}\left( \sigma
_{X}^{2}\sigma _{\hat{K}}^{2}\frac{3\left( f^{\prime }\left( \hat{X}\right)
\right) ^{3}-2f^{\prime }\left( X\right) f^{\prime \prime }\left( \hat{X}%
\right) \left\vert f\left( \hat{X}\right) \right\vert }{120\left\vert
f\left( \hat{X}\right) \right\vert ^{4}}\right.  \label{prx} \\
&&\left. -\frac{\frac{\partial f\left( \hat{X},K_{\hat{X}}\right) }{\partial 
\hat{X}}}{f\left( \hat{X},K_{\hat{X}}\right) }-\frac{\frac{\partial
\left\Vert \Psi \left( \hat{X},K_{\hat{X}}\right) \right\Vert ^{2}}{\partial 
\hat{X}}}{\left\Vert \Psi \left( \hat{X},K_{\hat{X}}\right) \right\Vert ^{2}}%
\right) _{K_{\hat{X},M}}\left( \hat{X}-\hat{X}_{M}\right)  \notag \\
&&-\frac{1}{D}\frac{b}{2}\left( \hat{X}-\hat{X}_{M}\right) \nabla _{\hat{X}%
}^{2}\left( \frac{\left( g\left( \hat{X},K_{\hat{X}}\right) \right) ^{2}}{%
\sigma _{\hat{X}}^{2}}+\frac{3}{2}f\left( \hat{X},K_{\hat{X}}\right) +\nabla
_{\hat{X}}g\left( \hat{X},K_{\hat{X}}\right) \right) _{K_{\hat{X},M}}\left( 
\hat{X}-\hat{X}_{M}\right)  \notag
\end{eqnarray}%
with:%
\begin{equation*}
D=\left( 1+\frac{\frac{\partial f\left( \hat{X},K_{\hat{X}}\right) }{%
\partial K_{\hat{X}}}}{f\left( \hat{X},K_{\hat{X}}\right) }+\frac{\frac{%
\partial \left\Vert \Psi \left( \hat{X},K_{\hat{X}}\right) \right\Vert ^{2}}{%
\partial K_{\hat{X}}}}{\left\Vert \Psi \left( \hat{X},K_{\hat{X}}\right)
\right\Vert ^{2}}+\frac{\sigma _{X}^{2}\sigma _{\hat{K}}^{2}\left( \nabla
_{K_{\hat{X}}}\left( f^{\prime }\left( \hat{X}\right) \right) ^{2}\left\vert
f\left( \hat{X}\right) \right\vert -3\left( \nabla _{K_{\hat{X}}}\left\vert
f\left( \hat{X}\right) \right\vert \right) \left( f^{\prime }\left( \hat{X}%
\right) \right) ^{2}\right) }{120\left\vert f\left( \hat{X}\right)
\right\vert ^{4}}\right) _{K_{\hat{X}_{M}}}
\end{equation*}%
and $K_{\hat{X}_{M}}$\ solution of:%
\begin{eqnarray*}
&&K_{\hat{X},M}\left\vert f\left( \hat{X},K_{\hat{X},M}\right) \right\vert
\left\Vert \Psi \left( \hat{X},K_{\hat{X},M}\right) \right\Vert ^{2} \\
&\simeq &\sigma _{\hat{K}}^{2}\exp \left( -\frac{\sigma _{X}^{2}\sigma _{%
\hat{K}}^{2}\left( p+\frac{1}{2}\right) ^{2}\left( f^{\prime }\left(
X\right) \right) ^{2}}{96\left\vert f\left( \hat{X}\right) \right\vert ^{3}}%
\right) C\left( \bar{p}\right) \simeq \sigma _{\hat{K}}^{2}C\left( \bar{p}%
\right)
\end{eqnarray*}%
The maximization condition (\ref{Mqn}) cancels the contribution due to: 
\begin{equation*}
\frac{\left( g\left( \hat{X},K_{\hat{X}}\right) \right) ^{2}}{\sigma _{\hat{X%
}}^{2}}+\left( f\left( \hat{X},K_{\hat{X}}\right) +\frac{\left\vert f\left( 
\hat{X},K_{\hat{X}}\right) \right\vert }{2}+\nabla _{\hat{X}}g\left( \hat{X}%
,K_{\hat{X}}\right) \right)
\end{equation*}%
To find a contribution due to this term, we have to expand (\ref{Nkvv}) to
the second order. The second order contributions proportional to $\left( K_{%
\hat{X}}-K_{\hat{X},M}\right) ^{2}$ modifies slightly (\ref{prx}) and the
term $\left( K_{\hat{X}}-K_{\hat{X},M}\right) \left( \hat{X}-\hat{X}%
_{M}\right) $ shifts $D$ at the first order. Both modifications do not alter
the interpretation for (\ref{prx}). We can thus consider the sole term:%
\begin{equation*}
\frac{C\left( \bar{p}\right) \sigma _{\hat{K}}^{2}\hat{\Gamma}\left( p+\frac{%
1}{2}\right) }{\left\Vert \Psi \left( \hat{X}\right) \right\Vert
^{2}\left\vert f\left( \hat{X}\right) \right\vert }
\end{equation*}%
Due to (\ref{qnk}), for $H\left( K_{\hat{X}}\right) $ slowly varying, the
contribution due to the derivatives of $\left\Vert \Psi \left( \hat{X}%
\right) \right\Vert ^{2}$ can be neglected. Moreover the contribution due to
the derivative of $\left\vert f\left( \hat{X}\right) \right\vert $ are
negligible with respect to the first order terms. We can thus consider only
the second order contributions due to $\hat{\Gamma}\left( p+\frac{1}{2}%
\right) $. In the rhs of (\ref{gmhh}), the second term is dominant.
Moreover, we can check that in the second order expansion of (\ref{dvp}),
the term in $p^{2}$ can be neglected compared to $-p\left( \gamma _{0}+\ln
2-2\right) $. As a consequence, the relevant second order correction to (\ref%
{prx}) is :%
\begin{equation*}
b\left( \hat{X}-\hat{X}_{M}\right) \nabla _{\hat{X}}^{2}p\left( \hat{X}-\hat{%
X}_{M}\right) =b\left( \hat{X}-\hat{X}_{M}\right) \nabla _{\hat{X}%
}^{2}\left( \frac{M-\frac{\left( g\left( \hat{X},K_{\hat{X}}\right) \right)
^{2}}{\sigma _{\hat{X}}^{2}}+\frac{3}{2}f\left( \hat{X},K_{\hat{X}}\right)
+\nabla _{\hat{X}}g\left( \hat{X},K_{\hat{X}}\right) }{\left\vert f\left( 
\hat{X}\right) \right\vert }\right) \left( \hat{X}-\hat{X}_{M}\right)
\end{equation*}%
and the relevant contributions to (\ref{prx}) are:

\begin{eqnarray}
\left( K_{\hat{X}}-K_{\hat{X},M}\right) &=&\frac{1}{D}\left( \sigma
_{X}^{2}\sigma _{\hat{K}}^{2}\frac{3\left( f^{\prime }\left( \hat{X}\right)
\right) ^{3}-2f^{\prime }\left( X\right) f^{\prime \prime }\left( \hat{X}%
\right) \left\vert f\left( \hat{X}\right) \right\vert }{120\left\vert
f\left( \hat{X}\right) \right\vert ^{4}}\right.  \label{prZ} \\
&&\left. -\frac{\frac{\partial f\left( \hat{X},K_{\hat{X}}\right) }{\partial 
\hat{X}}}{f\left( \hat{X},K_{\hat{X}}\right) }-\frac{\frac{\partial
\left\Vert \Psi \left( \hat{X},K_{\hat{X}}\right) \right\Vert ^{2}}{\partial 
\hat{X}}}{\left\Vert \Psi \left( \hat{X},K_{\hat{X}}\right) \right\Vert ^{2}}%
\right) _{K_{\hat{X},M}}\left( \hat{X}-\hat{X}_{M}\right)  \notag \\
&&+\frac{1}{D}\frac{b}{2}\left( \hat{X}-\hat{X}_{M}\right) \nabla _{\hat{X}%
}^{2}\left( \frac{M-\frac{\left( g\left( \hat{X},K_{\hat{X}}\right) \right)
^{2}}{\sigma _{\hat{X}}^{2}}+\frac{3}{2}f\left( \hat{X},K_{\hat{X}}\right)
+\nabla _{\hat{X}}g\left( \hat{X},K_{\hat{X}}\right) }{\left\vert f\left( 
\hat{X}\right) \right\vert }\right) _{K_{\hat{X},M}}\left( \hat{X}-\hat{X}%
_{M}\right)  \notag
\end{eqnarray}

Interpretation of (\ref{prZ}) is given in the text.

\subsubsection*{A3.2.3 Third approach: Resolution for particular form for
the functions}

As stated in the text, we can find approximate solutions to (\ref{qnk}) by
choosing some forms for the parameters functions. The solutions are then
studied in some ranges for average capital per firm $K_{X}$: $K_{X}>>1$, $%
K_{X}>>>1$, $K_{X}<<1$\textbf{\ }and the intermediate range $\infty >K_{X}>1$
In the case $K_{X}>>>1$, the distinction between stable and unstable cases
has to be made.

\paragraph*{A3.2.3.1 Choice of functions $f$ and $g$}

We can find approximate solutions to (\ref{qnk}) if we give some particlar
standard forms to the functions involved in the system. We use the form
defined in the text. For $\left\Vert \Psi \left( X\right) \right\Vert ^{2}$:%
\begin{equation*}
\left\Vert \Psi \left( X\right) \right\Vert ^{2}\simeq \frac{D\left(
\left\Vert \Psi \right\Vert ^{2}\right) -\frac{F}{2\sigma _{X}^{2}}\left(
\nabla _{X}R\left( X\right) \right) ^{2}H\left( K_{X}\right) }{2\tau }
\end{equation*}%
we choose: 
\begin{equation}
H\left( K_{X}\right) =K_{X}^{\eta }  \label{prb}
\end{equation}%
so that we obtain:

\begin{equation}
\left\Vert \Psi \left( X\right) \right\Vert ^{2}\simeq \frac{D\left(
\left\Vert \Psi \right\Vert ^{2}\right) -\frac{F}{2\sigma _{X}^{2}}\left(
\nabla _{X}R\left( X\right) \right) ^{2}H\left( K_{X}\right) K_{X}^{\eta }}{%
2\tau }\equiv D-L\left( X\right) \left( \nabla _{X}R\left( X\right) \right)
^{2}K_{X}^{\eta }  \label{psr}
\end{equation}

\begin{equation}
f\left( \hat{X},\Psi ,\hat{\Psi}\right) =\frac{1}{\varepsilon }\left(
r\left( \hat{X}\right) K_{\hat{X}}^{\alpha -1}-\gamma \left\Vert \Psi \left( 
\hat{X}\right) \right\Vert ^{2}+b\arctan \left( \frac{K_{\hat{X}}^{\alpha
}R\left( \hat{X}\right) }{\left\langle K_{X}^{\alpha }\right\rangle
\left\langle R\left( X\right) \right\rangle }-1\right) \right)  \label{frt}
\end{equation}%
\begin{equation}
g\left( \hat{X},\Psi ,\hat{\Psi}\right) =\nabla _{\hat{X}}R\left( \hat{X}%
\right) \arctan \left( \frac{K_{\hat{X}}^{\alpha }}{\left\langle
K_{X}^{\alpha }\right\rangle }\right) +b\nabla _{\hat{X}}R\left( \hat{X}%
\right) \arctan \left( \frac{K_{\hat{X}}^{\alpha }R\left( \hat{X}\right) }{%
\left\langle K_{X}^{\alpha }\right\rangle \left\langle R\left( X\right)
\right\rangle }-1\right)  \label{grt}
\end{equation}%
and their approximation for $\frac{K_{\hat{X}}^{\alpha }R\left( \hat{X}%
\right) }{\left\langle K_{X}^{\alpha }\right\rangle \left\langle R\left(
X\right) \right\rangle }\simeq 1$ and $\eta =\alpha $:%
\begin{eqnarray*}
f\left( \hat{X},\Psi ,\hat{\Psi}\right) &=&\frac{1}{\varepsilon }\left(
\left( r\left( \hat{X}\right) +\frac{bR\left( \hat{X}\right) }{\left\langle
K_{\hat{X}}^{\alpha }\right\rangle \left\langle R\left( \hat{X}\right)
\right\rangle }+\gamma L\left( \hat{X}\right) \right) K_{\hat{X}}^{\alpha
}-\gamma D-b\right) \\
&\equiv &B_{1}\left( X\right) K_{\hat{X}}^{\alpha -1}+B_{2}\left( X\right)
K_{\hat{X}}^{\alpha }-C\left( X\right)
\end{eqnarray*}%
\begin{equation*}
g\left( \hat{X},\Psi ,\hat{\Psi}\right) \simeq K_{\hat{X}}^{\alpha }\nabla _{%
\hat{X}}R\left( \hat{X}\right) \left( 1+\frac{b}{\left\langle K_{\hat{X}%
}^{\alpha }\right\rangle \left\langle R\left( \hat{X}\right) \right\rangle }%
\right) \equiv \nabla _{\hat{X}}R\left( \hat{X}\right) A\left( \hat{X}%
\right) K_{\hat{X}}^{\alpha }
\end{equation*}%
\begin{equation*}
\nabla _{\hat{X}}g\left( \hat{X},\Psi ,\hat{\Psi}\right) \simeq \nabla _{%
\hat{X}}^{2}R\left( \hat{X}\right) \left( 1+\frac{b}{\left\langle K_{\hat{X}%
}^{\alpha }\right\rangle \left\langle R\left( \hat{X}\right) \right\rangle }%
\right) K_{\hat{X}}^{\alpha }\equiv \nabla _{\hat{X}}^{2}R\left( \hat{X}%
\right) A\left( \hat{X}\right) K_{\hat{X}}^{\alpha }
\end{equation*}

\paragraph*{A3.2.3.2 Solving (\protect\ref{qnk})}

Equation (\ref{qnk}) can be studied by considering four cases:

\subparagraph{Case 1 $K_{\hat{X}}>>1$}

In that case, we assume $K_{\hat{X}}$ relatively large, but such that the
approximation:%
\begin{equation}
\left\Vert \Psi \left( \hat{X}\right) \right\Vert ^{2}\simeq D  \label{bts}
\end{equation}%
is still valid.

Equations (\ref{frt}) and (\ref{grt}) imply that\ the function $f\left( \hat{%
X}\right) $ is independent of $K_{\hat{X}}$ and that $g\left( \hat{X}\right) 
$ is proportional to $\nabla _{\hat{X}}R\left( \hat{X}\right) $. Given (\ref%
{frt})\ the function $f\left( \hat{X}\right) $ can be rewritten:%
\begin{eqnarray*}
f\left( \hat{X}\right) &=&\frac{1}{\varepsilon }\left( r\left( \hat{X}%
\right) K_{\hat{X}}^{\alpha -1}-\gamma \left\Vert \Psi \left( \hat{X}\right)
\right\Vert ^{2}+b\arctan \left( \frac{K_{\hat{X}}^{\alpha }R\left( \hat{X}%
\right) }{\left\langle K_{X}^{\alpha }\right\rangle \left\langle R\left(
X\right) \right\rangle }-1\right) \right) \\
&\simeq &b\left( \frac{\pi }{2}-\frac{\left\langle K_{X}^{\alpha
}\right\rangle \left\langle R\left( X\right) \right\rangle }{K_{\hat{X}%
}^{\alpha }R\left( \hat{X}\right) }\right) -\gamma D \\
&\equiv &c-\frac{d}{K_{\hat{X}}^{\alpha }R\left( \hat{X}\right) }-\gamma D
\end{eqnarray*}%
As a consequence, the expression for $f^{\prime }\left( \hat{X}\right) $ is:%
\begin{equation}
f^{\prime }\left( \hat{X}\right) \simeq \frac{d\nabla _{\hat{X}}R\left( \hat{%
X}\right) }{K_{\hat{X}}^{\alpha }R^{2}\left( \hat{X}\right) }  \label{fpr}
\end{equation}%
Similarly,we can approximate (\ref{grt}) as:%
\begin{eqnarray}
g\left( \hat{X}\right) &\simeq &-\frac{\nabla _{\hat{X}}R\left( \hat{X}%
\right) f}{K_{\hat{X}}^{\alpha }R\left( \hat{X}\right) }  \label{prg} \\
\nabla _{\hat{X}}g\left( \hat{X}\right) &\simeq &-\frac{\nabla _{\hat{X}%
}^{2}R\left( \hat{X}\right) f}{K_{\hat{X}}^{\alpha }R\left( \hat{X}\right) }
\notag
\end{eqnarray}%
Given (\ref{bts}), equation (\ref{qnk}) is:%
\begin{equation}
K_{\hat{X}}D\left\vert f\left( \hat{X}\right) \right\vert =C\left( \bar{p}%
\right) \sigma _{\hat{K}}^{2}\exp \left( -\frac{\sigma _{X}^{2}\sigma _{\hat{%
K}}^{2}\left( p+\frac{1}{2}\right) ^{2}\left( f^{\prime }\left( X\right)
\right) ^{2}}{96\left\vert f\left( \hat{X}\right) \right\vert ^{3}}\right)
\Gamma \left( p+\frac{3}{2}\right)  \label{Lqg}
\end{equation}%
with: 
\begin{equation*}
p+\frac{1}{2}=\frac{M-\left( \frac{\left( g\left( \hat{X}\right) \right) ^{2}%
}{\sigma _{\hat{X}}^{2}}+\left( f\left( \hat{X}\right) +\nabla _{\hat{X}%
}g\left( \hat{X},K_{\hat{X}}\right) -\frac{\sigma _{\hat{K}}^{2}F^{2}\left( 
\hat{X},K_{\hat{X}}\right) }{2f^{2}\left( \hat{X}\right) }\right) \right) }{%
\sqrt{f^{2}\left( \hat{X}\right) }}
\end{equation*}%
Defining $V=\frac{1}{K_{\hat{X}}^{\alpha }}$, we can write (\ref{Lqg}) as an
equation for $V<<1$ by replacing all quantities in term of $V$ and then
perform a first order expansion. To do so, we first, we write (\ref{Lqg}) as:%
\begin{equation}
V-\frac{D\left\vert f\left( \hat{X}\right) \right\vert }{C\left( \bar{p}%
\right) \sigma _{\hat{K}}^{2}\exp \left( -\frac{\sigma _{X}^{2}\sigma _{\hat{%
K}}^{2}\left( p+\frac{1}{2}\right) ^{2}\left( f^{\prime }\left( X\right)
\right) ^{2}}{96\left\vert f\left( \hat{X}\right) \right\vert ^{3}}\right)
\Gamma \left( p+\frac{3}{2}\right) }=0  \label{lqG}
\end{equation}%
and then find an expansion in $V$ for $\Gamma \left( p+\frac{3}{2}\right) $.

The first order expansion in $V$ of $p+\frac{3}{2}$ is: 
\begin{eqnarray*}
p+\frac{3}{2} &=&\frac{M-\left( \frac{\left( g\left( \hat{X}\right) \right)
^{2}}{\sigma _{\hat{X}}^{2}}+\nabla _{\hat{X}}g\left( \hat{X},K_{\hat{X}%
}\right) -\frac{\sigma _{\hat{K}}^{2}F^{2}\left( \hat{X},K_{\hat{X}}\right) 
}{2f^{2}\left( \hat{X}\right) }\right) }{f\left( \hat{X}\right) } \\
&\simeq &\frac{M-\left( \frac{\left( g\left( \hat{X}\right) \right) ^{2}}{%
\sigma _{\hat{X}}^{2}}+\nabla _{\hat{X}}g\left( \hat{X},K_{\hat{X}}\right)
\right) }{c-\frac{d}{K_{\hat{X}}^{\alpha }R\left( \hat{X}\right) }} \\
&=&\frac{M-\left( \frac{\left( \nabla _{\hat{X}}R\left( \hat{X}\right)
\left( -\frac{fV}{R\left( \hat{X}\right) }\right) \right) ^{2}}{\sigma _{%
\hat{X}}^{2}}+\nabla _{\hat{X}}^{2}R\left( \hat{X}\right) \left( -\frac{fV}{%
R\left( \hat{X}\right) }\right) \right) }{c-\frac{dV}{R\left( \hat{X}\right) 
}} \\
&=&\frac{M}{c}+\frac{\nabla _{\hat{X}}^{2}R\left( \hat{X}\right) \frac{fV}{%
R\left( \hat{X}\right) }}{c}+\frac{M\frac{dV}{cR\left( \hat{X}\right) }}{c}
\end{eqnarray*}%
As a consequence, $\Gamma \left( p+\frac{3}{2}\right) $ arising in (\ref{Kgl}%
) is given by:%
\begin{eqnarray*}
\Gamma \left( \frac{M}{c}+\frac{\nabla _{\hat{X}}^{2}R\left( \hat{X}\right) 
\frac{fV}{R\left( \hat{X}\right) }}{c}+\frac{M\frac{dV}{cR\left( \hat{X}%
\right) }}{c}\right) &=&\Gamma \left( \frac{M}{c}\left( 1+\nabla _{\hat{X}%
}^{2}R\left( \hat{X}\right) \frac{fV}{MR\left( \hat{X}\right) }+\frac{dV}{%
cR\left( \hat{X}\right) }\right) \right) \\
&\simeq &\Gamma \left( \frac{M}{c}\right) +\frac{MV}{c}\left( \frac{\nabla _{%
\hat{X}}^{2}R\left( \hat{X}\right) f}{MR\left( \hat{X}\right) }+\frac{d}{%
cR\left( \hat{X}\right) }\right) \Gamma ^{\prime }\left( \frac{M}{c}\right)
\\
&=&\Gamma \left( \frac{M}{c}\right) \left( 1+\frac{MV}{c}\left( \frac{\nabla
_{\hat{X}}^{2}R\left( \hat{X}\right) f}{MR\left( \hat{X}\right) }+\frac{d}{%
cR\left( \hat{X}\right) }\right) \func{Psi}\left( \frac{M}{c}\right) \right)
\end{eqnarray*}%
Ultimately, using that at the first order:%
\begin{equation*}
\exp \left( -\frac{\sigma _{X}^{2}\sigma _{\hat{K}}^{2}\left( p+\frac{1}{2}%
\right) ^{2}\left( f^{\prime }\left( X\right) \right) ^{2}}{96\left\vert
f\left( \hat{X}\right) \right\vert ^{3}}\right) \simeq 1
\end{equation*}%
equation (\ref{lqG}) for $V$ becomes:%
\begin{equation*}
V-\frac{D\left\vert f\left( \hat{X}\right) \right\vert }{C\left( \bar{p}%
\right) \sigma _{\hat{K}}^{2}\Gamma \left( \frac{M}{c}\right) \left( 1+\frac{%
MV}{c}\left( \frac{\nabla _{\hat{X}}^{2}R\left( \hat{X}\right) f}{MR\left( 
\hat{X}\right) }+\frac{d}{cR\left( \hat{X}\right) }\right) \func{Psi}\left( 
\frac{M}{c}\right) \right) }=0
\end{equation*}%
that is:%
\begin{equation*}
V-\frac{D\left( c-\frac{dV}{R\left( \hat{X}\right) }-\gamma D\right) }{%
C\left( \bar{p}\right) \sigma _{\hat{K}}^{2}\Gamma \left( \frac{M}{c}\right)
\left( 1+\frac{MV}{c}\left( \frac{\nabla _{\hat{X}}^{2}R\left( \hat{X}%
\right) f}{MR\left( \hat{X}\right) }+\frac{d}{cR\left( \hat{X}\right) }%
\right) \func{Psi}\left( \frac{M}{c}\right) \right) }=0
\end{equation*}%
And a first order expansion yields:%
\begin{equation*}
V-\frac{D\left( c-\gamma D\right) }{C\left( \bar{p}\right) \sigma _{\hat{K}%
}^{2}\Gamma \left( \frac{M}{c}\right) }\left( 1-\frac{dV}{\left( c-\gamma
D\right) R\left( \hat{X}\right) }-MV\left( \frac{\nabla _{\hat{X}%
}^{2}R\left( \hat{X}\right) f}{MR\left( \hat{X}\right) }+\frac{d}{cR\left( 
\hat{X}\right) }\right) \func{Psi}\left( \frac{M}{c}\right) \right) =0
\end{equation*}%
with solution:%
\begin{equation*}
V=\left( K_{\hat{X}}^{\alpha }\right) ^{-1}=\frac{\frac{D\left( c-\gamma
D\right) }{C\left( \bar{p}\right) \sigma _{\hat{K}}^{2}\Gamma \left( \frac{M%
}{c}\right) }}{1+\frac{D\left( c-\gamma D\right) }{C\left( \bar{p}\right)
\sigma _{\hat{K}}^{2}\Gamma \left( \frac{M}{c}\right) }\left( \frac{d}{%
\left( c-\gamma D\right) R\left( \hat{X}\right) }+M\left( \frac{\nabla _{%
\hat{X}}^{2}R\left( \hat{X}\right) f}{MR\left( \hat{X}\right) }+\frac{d}{%
cR\left( \hat{X}\right) }\right) \func{Psi}\left( \frac{M}{c}\right) \right) 
}
\end{equation*}%
Coming back to $K_{\hat{X}}^{\alpha }$ we have:%
\begin{eqnarray}
K_{\hat{X}}^{\alpha } &=&\frac{C\left( \bar{p}\right) \sigma _{\hat{K}%
}^{2}\Gamma \left( \frac{M}{c}\right) }{D\left( c-\gamma D\right) }+\frac{d}{%
\left( c-\gamma D\right) R\left( \hat{X}\right) }+M\left( \frac{\nabla _{%
\hat{X}}^{2}R\left( \hat{X}\right) f}{MR\left( \hat{X}\right) }+\frac{d}{%
cR\left( \hat{X}\right) }\right) \func{Psi}\left( \frac{M}{c}\right)
\label{shk} \\
&=&\frac{C\left( \bar{p}\right) \sigma _{\hat{K}}^{2}\Gamma \left( \frac{M}{c%
}\right) }{D\left( c-\gamma D\right) }+\frac{d}{\left( c-\gamma D\right)
R\left( \hat{X}\right) }\left( 1+M\func{Psi}\left( \frac{M}{c}\right) \left(
1+\frac{\nabla _{\hat{X}}^{2}R\left( \hat{X}\right) f}{M\left( c-\gamma
D\right) }\right) \right)  \notag
\end{eqnarray}%
This solution satisfies the condition $K_{\hat{X}}>>1$ only if $\frac{%
C\left( \bar{p}\right) \sigma _{\hat{K}}^{2}\sqrt{\frac{M-c}{c}}}{Dc}>>1$:
formula (\ref{shk}) thus shows that the dependency of $K_{\hat{X}}^{\alpha }$
in the return \ depends on the sign of $1+M\func{Psi}\left( \frac{M}{c}%
\right) \left( 1+\frac{\nabla _{\hat{X}}^{2}R\left( \hat{X}\right) f}{%
M\left( c-\gamma D\right) }\right) $. If:%
\begin{equation*}
1+M\func{Psi}\left( \frac{M}{c}\right) \left( 1+\frac{\nabla _{\hat{X}%
}^{2}R\left( \hat{X}\right) f}{M\left( c-\gamma D\right) }\right) >0
\end{equation*}%
then $K_{\hat{X}}^{\alpha }$ decreases with $R\left( \hat{X}\right) $. As
stated in the text, this corresponds to an unstable equilibrium.

If:%
\begin{equation*}
1+M\func{Psi}\left( \frac{M}{c}\right) \left( 1+\frac{\nabla _{\hat{X}%
}^{2}R\left( \hat{X}\right) f}{M\left( c-\gamma D\right) }\right) <0
\end{equation*}%
a stable equilibrium is possible and $K_{\hat{X}}^{\alpha }$ is an
increasing function of $R\left( \hat{X}\right) $ and $f\left( \hat{X}\right) 
$. This corresponds to $\nabla _{\hat{X}}^{2}R\left( \hat{X}\right) <<0$,
which arises for instance for a maximum of $R\left( \hat{X}\right) $. In
such case an increase in $R\left( \hat{X}\right) $ allows for an increased
number $\left\Vert \Psi \left( \hat{X}\right) \right\Vert ^{2}$ of firms,
without reducing the average capital per firm.

\subparagraph{Case 2 $K_{\hat{X}}>>>1$ stable and unstable cases}

In that case, $K_{\hat{X}}>>>1$, and we assume in first approximation that
(discarding the factor $L\left( \hat{X}\right) $):%
\begin{equation}
\left\Vert \Psi \left( \hat{X}\right) \right\Vert ^{2}\simeq D-\left( \nabla
_{X}R\left( \hat{X}\right) \right) ^{2}K_{\hat{X}}^{\alpha }<<1  \label{nLT}
\end{equation}%
This corresponds to a very high level of capital. Consequently, equation (%
\ref{frt})\ implies that the function $f\left( \hat{X}\right) $ can be
rewritten:%
\begin{eqnarray*}
f\left( \hat{X}\right) &=&\frac{1}{\varepsilon }\left( r\left( \hat{X}%
\right) K_{\hat{X}}^{\alpha -1}-\gamma \left\Vert \Psi \left( \hat{X}\right)
\right\Vert ^{2}+b\arctan \left( \frac{K_{\hat{X}}^{\alpha }R\left( \hat{X}%
\right) }{\left\langle K_{X}^{\alpha }\right\rangle \left\langle R\left(
X\right) \right\rangle }-1\right) \right) \\
&\simeq &b\left( \frac{\pi }{2}-\frac{\left\langle K_{X}^{\alpha
}\right\rangle \left\langle R\left( X\right) \right\rangle }{K_{\hat{X}%
}^{\alpha }R\left( \hat{X}\right) }\right) \\
&\equiv &c-\frac{d}{K_{\hat{X}}^{\alpha }R\left( \hat{X}\right) }>0
\end{eqnarray*}%
As a consequence, the expressions for $f^{\prime }\left( \hat{X}\right) $ $%
g\left( \hat{X}\right) $ and $\nabla _{\hat{X}}g\left( \hat{X}\right) $ (\ref%
{fpr}) and (\ref{prg}) are still valid.

Two different cases arise in the resolution of (\ref{Nkv}).

First, we assume that $\left( \nabla _{\hat{X}}R\left( \hat{X}\right)
\right) ^{2}\neq 0$.

In this case, we will solve (\ref{Nkv}) by using (\ref{nLT}) to replace $K_{%
\hat{X}}\simeq \left( \frac{D}{\left( \nabla _{X}R\left( \hat{X}\right)
\right) ^{2}}\right) ^{\frac{1}{\alpha }}$. We also change the variable $%
\frac{D}{\left( \nabla _{X}R\left( \hat{X}\right) \right) ^{2}}\rightarrow D$
temporarily for the sake of simplicity.

Inequality (\ref{nLT}) along with $K_{\hat{X}}>>>1$ and (\ref{frt}) implies
that only the case $f>0$ has to be considered.

Note that using our results about stability, it is easy to check that in
that case, this solution is locally unstable. A very high level of capital
has the tendency to attract more investments.

Given our assumptions, equation (\ref{qtc}) becomes: 
\begin{equation}
\left( \nabla _{X}R\left( \hat{X}\right) \right) ^{2}D^{\frac{1}{\alpha }%
}\left( D-K_{\hat{X}}^{\alpha }\right) =C\left( \bar{p}\right) \sigma _{\hat{%
K}}^{2}\exp \left( -\frac{\sigma _{X}^{2}\sigma _{\hat{K}}^{2}\left( p+\frac{%
1}{2}\right) ^{2}\left( f^{\prime }\left( X\right) \right) ^{2}}{%
96\left\vert f\left( \hat{X}\right) \right\vert ^{3}}\right) \frac{\Gamma
\left( p+\frac{3}{2}\right) }{\left\vert f\left( \hat{X}\right) \right\vert }
\label{gdr}
\end{equation}%
or equivalently:%
\begin{equation}
K_{\hat{X}}^{\alpha }=D-\frac{C\left( \bar{p}\right) \sigma _{\hat{K}%
}^{2}\exp \left( -\frac{\sigma _{X}^{2}\sigma _{\hat{K}}^{2}\left( p+\frac{1%
}{2}\right) ^{2}\left( f^{\prime }\left( X\right) \right) ^{2}}{96\left\vert
f\left( \hat{X}\right) \right\vert ^{3}}\right) \Gamma \left( p+\frac{3}{2}%
\right) }{\left( \nabla _{X}R\left( \hat{X}\right) \right) ^{2}D^{\frac{1}{%
\alpha }}\left\vert f\left( \hat{X}\right) \right\vert }  \label{lgk}
\end{equation}

Then, defining $V=\frac{1}{K_{\hat{X}}^{\alpha }}$ as in the first case, we
can write (\ref{lgk}) as an equation for $V<<1$ by replacing all quantities
in term of $V$ and then perform a first order expansion.

First, we write (\ref{lgk}) as:%
\begin{equation}
V-\frac{1}{D-\frac{C\left( \bar{p}\right) \sigma _{\hat{K}}^{2}\exp \left( -%
\frac{\sigma _{X}^{2}\sigma _{\hat{K}}^{2}\left( p+\frac{1}{2}\right)
^{2}\left( f^{\prime }\left( X\right) \right) ^{2}}{96\left\vert f\left( 
\hat{X}\right) \right\vert ^{3}}\right) \Gamma \left( p+\frac{3}{2}\right) }{%
\left( \nabla _{X}R\left( \hat{X}\right) \right) ^{2}D^{\frac{1}{\alpha }%
}\left\vert f\left( \hat{X}\right) \right\vert }}=0  \label{Kgl}
\end{equation}%
As in the previous case, the first order expansion in $V$ of $\Gamma \left(
p+\frac{3}{2}\right) $ arising in (\ref{Kgl}) is given by:%
\begin{equation}
\Gamma \left( p+\frac{3}{2}\right) \simeq \Gamma \left( \frac{M}{c}\right) +%
\frac{MV}{c}\left( \frac{\nabla _{\hat{X}}^{2}R\left( \hat{X}\right) f}{%
MR\left( \hat{X}\right) }+\frac{d}{cR\left( \hat{X}\right) }\right) \Gamma
^{\prime }\left( \frac{M}{c}\right)  \label{pqr}
\end{equation}%
Moreover, at the first order:%
\begin{equation*}
\exp \left( -\frac{\sigma _{X}^{2}\sigma _{\hat{K}}^{2}\left( p+\frac{1}{2}%
\right) ^{2}\left( f^{\prime }\left( X\right) \right) ^{2}}{96\left\vert
f\left( \hat{X}\right) \right\vert ^{3}}\right) \simeq 1
\end{equation*}%
and (\ref{Kgl}) becomes:%
\begin{equation*}
V-\frac{\left( \nabla _{\hat{X}}R\left( \hat{X}\right) \right) ^{2}D^{\frac{1%
}{\alpha }}\left\vert f\left( \hat{X}\right) \right\vert }{\left( \nabla _{%
\hat{X}}R\left( \hat{X}\right) \right) ^{2}D^{1+\frac{1}{\alpha }}\left\vert
f\left( \hat{X}\right) \right\vert -C\left( \bar{p}\right) \sigma _{\hat{K}%
}^{2}\Gamma \left( p+\frac{3}{2}\right) }=0
\end{equation*}%
that is:%
\begin{equation}
V-\frac{\left( \nabla _{\hat{X}}R\left( \hat{X}\right) \right) ^{2}D^{\frac{1%
}{\alpha }}\left( c-\frac{dV}{R\left( \hat{X}\right) }\right) }{\left(
\nabla _{\hat{X}}R\left( \hat{X}\right) \right) ^{2}D^{1+\frac{1}{\alpha }%
}\left( c-\frac{dV}{R\left( \hat{X}\right) }\right) -C\left( \bar{p}\right)
\sigma _{\hat{K}}^{2}\Gamma \left( p+\frac{3}{2}\right) }=0  \label{frd}
\end{equation}%
Using (\ref{pqr}) the first order expansion of the dominator in (\ref{frd})
is:%
\begin{eqnarray*}
&&\left( \nabla _{\hat{X}}R\left( \hat{X}\right) \right) ^{2}D^{1+\frac{1}{%
\alpha }}\left( c-\frac{dV}{R\left( \hat{X}\right) }\right) -C\left( \bar{p}%
\right) \sigma _{\hat{K}}^{2}\Gamma \left( p+\frac{3}{2}\right) \\
&=&\left( \nabla _{\hat{X}}R\left( \hat{X}\right) \right) ^{2}D^{1+\frac{1}{%
\alpha }}c-C\left( \bar{p}\right) \sigma _{\hat{K}}^{2}\Gamma \left( \frac{M%
}{c}\right) \\
&&-\left( \left( \nabla _{\hat{X}}R\left( \hat{X}\right) \right) ^{2}D^{1+%
\frac{1}{\alpha }}\frac{d}{R\left( \hat{X}\right) }+\frac{C\left( \bar{p}%
\right) \sigma _{\hat{K}}^{2}M}{c}\left( \frac{\nabla _{\hat{X}}^{2}R\left( 
\hat{X}\right) f}{MR\left( \hat{X}\right) }+\frac{d}{cR\left( \hat{X}\right) 
}\right) \Gamma ^{\prime }\left( \frac{M}{c}\right) \right) V
\end{eqnarray*}%
so that (\ref{frd}) writes:%
\begin{eqnarray}
&&\frac{\left( \nabla _{\hat{X}}R\left( \hat{X}\right) \right) ^{2}D^{\frac{1%
}{\alpha }}c}{\left( \nabla _{\hat{X}}R\left( \hat{X}\right) \right)
^{2}D^{1+\frac{1}{\alpha }}c-C\left( \bar{p}\right) \sigma _{\hat{K}%
}^{2}\Gamma \left( \frac{M}{c}\right) }  \label{frD} \\
&=&\left( 1-\frac{\left( \nabla _{\hat{X}}R\left( \hat{X}\right) \right)
^{2}D^{\frac{1}{\alpha }}c\left( \left( \nabla _{\hat{X}}R\left( \hat{X}%
\right) \right) ^{2}D^{1+\frac{1}{\alpha }}\frac{d}{R\left( \hat{X}\right) }+%
\frac{C\left( \bar{p}\right) \sigma _{\hat{K}}^{2}M}{c}\left( \frac{\nabla _{%
\hat{X}}^{2}R\left( \hat{X}\right) f}{MR\left( \hat{X}\right) }+\frac{d}{%
cR\left( \hat{X}\right) }\right) \Gamma ^{\prime }\left( \frac{M}{c}\right)
\right) }{\left( \left( \nabla _{\hat{X}}R\left( \hat{X}\right) \right)
^{2}D^{1+\frac{1}{\alpha }}c-C\left( \bar{p}\right) \sigma _{\hat{K}%
}^{2}\Gamma \left( \frac{M}{c}\right) \right) ^{2}}\right) V  \notag \\
&&+\frac{\left( \nabla _{\hat{X}}R\left( \hat{X}\right) \right) ^{2}D^{\frac{%
1}{\alpha }}\frac{d}{R\left( \hat{X}\right) }}{\left( \nabla _{\hat{X}%
}R\left( \hat{X}\right) \right) ^{2}D^{1+\frac{1}{\alpha }}c-C\left( \bar{p}%
\right) \sigma _{\hat{K}}^{2}\Gamma \left( \frac{M}{c}\right) }V  \notag
\end{eqnarray}%
Equation (\ref{frD}) can be solved for $V$\ with solution:

\begin{equation*}
\frac{1}{V}=D-\frac{C\left( \bar{p}\right) \sigma _{\hat{K}}^{2}\Gamma
\left( \frac{M}{c}\right) }{\left( \nabla _{\hat{X}}R\left( \hat{X}\right)
\right) ^{2}D^{\frac{1}{\alpha }}c}+\frac{d}{cR\left( \hat{X}\right) }\left(
1-\frac{\left( 1+\frac{C\left( \bar{p}\right) \sigma _{\hat{K}}^{2}M\Gamma
\left( \frac{M}{c}\right) }{c\left( \nabla _{\hat{X}}R\left( \hat{X}\right)
\right) ^{2}D^{1+\frac{1}{\alpha }}}\left( \frac{\nabla _{\hat{X}%
}^{2}R\left( \hat{X}\right) f}{Md}+\frac{1}{c}\right) \func{Psi}\left( \frac{%
M}{c}\right) \right) }{\left( 1-\frac{C\left( \bar{p}\right) \sigma _{\hat{K}%
}^{2}\Gamma \left( \frac{M}{c}\right) }{\left( \nabla _{\hat{X}}R\left( \hat{%
X}\right) \right) ^{2}D^{1+\frac{1}{\alpha }}c}\right) }\right)
\end{equation*}%
Ultimatly, restoring the variable:%
\begin{equation*}
D\rightarrow \frac{D}{\left( \nabla _{\hat{X}}R\left( \hat{X}\right) \right)
^{2}}
\end{equation*}%
we obtain the solution $K_{\hat{X}}^{\alpha }=\frac{1}{V}$:%
\begin{eqnarray}
K_{\hat{X}}^{\alpha } &=&\frac{D}{\left( \nabla _{\hat{X}}R\left( \hat{X}%
\right) \right) ^{2}}-\frac{C\left( \bar{p}\right) \sigma _{\hat{K}%
}^{2}\Gamma \left( \frac{M}{c}\right) }{\left( \nabla _{\hat{X}}R\left( \hat{%
X}\right) \right) ^{2\left( 1-\frac{1}{\alpha }\right) }D^{\frac{1}{\alpha }%
}c}  \label{mGK} \\
&&+\frac{d}{cR\left( \hat{X}\right) }\left( 1-\frac{\left( 1+\frac{C\left( 
\bar{p}\right) \left( \nabla _{\hat{X}}R\left( \hat{X}\right) \right) ^{%
\frac{2}{\alpha }}\sigma _{\hat{K}}^{2}}{cD^{1+\frac{1}{\alpha }}}\left( 
\frac{M}{c}+\frac{\nabla _{\hat{X}}^{2}R\left( \hat{X}\right) f}{d}\right)
\Gamma ^{\prime }\left( \frac{M}{c}\right) \right) }{\left( 1-\frac{\left(
\nabla _{\hat{X}}R\left( \hat{X}\right) \right) ^{\frac{2}{\alpha }}C\left( 
\bar{p}\right) \sigma _{\hat{K}}^{2}}{cD^{1+\frac{1}{\alpha }}}\Gamma \left( 
\frac{M}{c}\right) \right) }\right)  \notag
\end{eqnarray}%
As stated in the text, this is increasing in $c$, i.e. in $f\left( \hat{X}%
\right) $ and in $R\left( \hat{X}\right) $. This corresponds to a stable
level of capital.

In a second case we consider that $\left( \nabla _{\hat{X}}R\left( \hat{X}%
\right) \right) ^{2}\rightarrow 0$ and formula (\ref{mGK}) is not valid
anymore. Coming back to (\ref{dfg}) leads rather to replace:%
\begin{equation*}
\left( \nabla _{X}R\left( X\right) \right) ^{2}\rightarrow \left( \nabla
_{X}R\left( X\right) \right) ^{2}+\sigma _{X}^{2}\frac{\nabla
_{X}^{2}R\left( K_{X},X\right) }{H\left( K_{X}\right) }=\sigma _{X}^{2}\frac{%
\nabla _{X}^{2}R\left( K_{X},X\right) }{H\left( K_{X}\right) }
\end{equation*}

If $\nabla _{X}^{2}R\left( K_{X},X\right) <0$, (\ref{gdr}) is replaced by:%
\begin{equation*}
K_{\hat{X}}^{\alpha }\left( D+\sigma _{X}^{2}\left\vert \nabla
_{X}^{2}R\left( K_{X},X\right) \right\vert K_{\hat{X}}^{\frac{\alpha }{2}%
}\right) =C\left( \bar{p}\right) \sigma _{\hat{K}}^{2}\frac{\Gamma \left( p+%
\frac{3}{2}\right) }{\left\vert f\left( \hat{X}\right) \right\vert }
\end{equation*}%
with:%
\begin{equation*}
p+\frac{3}{2}\simeq \frac{M-\nabla _{\hat{X}}g\left( \hat{X},K_{\hat{X}%
}\right) }{f\left( \hat{X}\right) }
\end{equation*}%
and the equation for $K_{X}$ writes:%
\begin{equation*}
\sigma _{X}^{2}\left\vert \nabla _{X}^{2}R\left( K_{X},X\right) \right\vert
K_{\hat{X}}^{\frac{3}{2}\alpha }=\frac{C\left( \bar{p}\right) \sigma _{\hat{K%
}}^{2}\Gamma \left( \frac{M-\nabla _{\hat{X}}g\left( \hat{X},K_{\hat{X}%
}\right) }{f\left( \hat{X}\right) }\right) }{\left\vert f\left( \hat{X}%
\right) \right\vert }
\end{equation*}%
Since, given our assumptions $f\left( \hat{X}\right) \rightarrow c$ we find:%
\begin{equation}
K_{\hat{X}}=\left( \frac{C\left( \bar{p}\right) \sigma _{\hat{K}}^{2}}{%
\left\vert \nabla _{X}^{2}R\left( K_{X},X\right) \right\vert c}\Gamma \left( 
\frac{M-\nabla _{\hat{X}}g\left( \hat{X},K_{\hat{X}}\right) }{c}\right)
\right) ^{\frac{2}{3\alpha }}  \label{drsn}
\end{equation}%
Note that given (\ref{drsn}), an equilibrium in the range $K_{\hat{X}}>>>1$
is only possible for $c<<1$ Otherwise, there is no equilibrium for a maximum
of $R\left( K_{X},X\right) $. This equilibrium value of $K_{\hat{X}}$
decreases with $c$, which corresponds to an unstable equilibrium.

On the other hand, if $\nabla _{X}R\left( X\right) =0$, expression (\ref{nLT}%
) becomes:%
\begin{equation*}
\left\Vert \Psi \left( \hat{X}\right) \right\Vert ^{2}\simeq D-\sigma
_{X}^{2}\frac{\nabla _{X}^{2}R\left( K_{X},X\right) }{H\left( K_{X}\right) }%
K_{\hat{X}}^{\alpha }=D-\sigma _{X}^{2}\nabla _{X}^{2}R\left( K_{X},X\right)
K_{\hat{X}}^{\frac{\alpha }{2}}
\end{equation*}%
and thus, if $\nabla _{X}^{2}R\left( K_{X},X\right) >0$:%
\begin{equation}
K_{\hat{X}}^{\alpha }\simeq \left( \frac{D}{\sigma _{X}^{2}\nabla
_{X}^{2}R\left( K_{X},X\right) }\right) ^{2}  \label{drsp}
\end{equation}%
However, this solution with $K_{X}>>1$ corresponds to points such that $%
\nabla _{X}^{2}R\left( K_{X},X\right) >0$ and $\nabla _{X}R\left( X\right)
=0 $. Then, these points are minima of $R\left( X\right) $. This equilibrium
may exist only if the level of capital (\ref{drsp}) is high enough to
compensate the weakness of the purely position dependent part of expected
return and match the condition:%
\begin{equation*}
\frac{K_{\hat{X}}^{\alpha }R\left( \hat{X}\right) }{\left\langle
K_{X}^{\alpha }\right\rangle \left\langle R\left( X\right) \right\rangle }%
-1>0
\end{equation*}

\subparagraph{Case $K_{\hat{X}}<<1$:}

This case is developped in the text.

\subparagraph{Intermediate case $\infty >K_{\hat{X}}>1$:}

We start with asymptotic form of (\ref{Nkv}):%
\begin{equation}
K_{\hat{X}}\left\Vert \Psi \left( \hat{X}\right) \right\Vert ^{2}\left\vert
f\left( \hat{X}\right) \right\vert =C\left( \bar{p}\right) \sigma _{\hat{K}%
}^{2}\exp \left( -\frac{\sigma _{X}^{2}\sigma _{\hat{K}}^{2}\left( p+\frac{1%
}{2}\right) ^{2}\left( f^{\prime }\left( X\right) \right) ^{2}}{96\left\vert
f\left( \hat{X}\right) \right\vert ^{3}}\right) \Gamma \left( p+\frac{3}{2}%
\right)  \label{smP}
\end{equation}%
$\allowbreak \allowbreak $Up to a constant that can be absorbed in the
definition of $C\left( \bar{p}\right) $, we have:%
\begin{equation*}
\Gamma \left( p+\frac{3}{2}\right) \sim _{\infty }\sqrt{p+\frac{1}{2}}\exp
\left( \left( p+\frac{1}{2}\right) \left( \ln \left( p+\frac{1}{2}\right)
-1\right) \right)
\end{equation*}%
and (\ref{smP}) can be rewritten as:%
\begin{equation}
K_{\hat{X}}\left\Vert \Psi \left( \hat{X}\right) \right\Vert ^{2}\left\vert
f\left( \hat{X}\right) \right\vert =C\left( \bar{p}\right) \sigma _{\hat{K}%
}^{2}\sqrt{p+\frac{1}{2}}\exp \left( -\frac{\sigma _{X}^{2}\sigma _{\hat{K}%
}^{2}\left( p+\frac{1}{2}\right) ^{2}\left( f^{\prime }\left( X\right)
\right) ^{2}}{96\left\vert f\left( \hat{X}\right) \right\vert ^{3}}+\left( p+%
\frac{1}{2}\right) \left( \ln \left( p+\frac{1}{2}\right) -1\right) \right)
\label{sMP}
\end{equation}%
Since we are in an intermediate range for the parameters, we can replace, in
first approximation, $\ln \left( p+\frac{1}{2}\right) $ by its average over
this range: $\ln \left( \bar{p}+\frac{1}{2}\right) $. The exponential in (%
\ref{sMP}) thus becomes: 
\begin{equation*}
\exp \left( -\frac{\sigma _{X}^{2}\sigma _{\hat{K}}^{2}\left( p+\frac{1}{2}-%
\frac{48\left\vert f\left( \hat{X}\right) \right\vert ^{3}}{\sigma
_{X}^{2}\sigma _{\hat{K}}^{2}\left( f^{\prime }\left( X\right) \right) ^{2}}%
\left( \ln \left( \bar{p}+\frac{1}{2}\right) -1\right) \right) ^{2}\left(
f^{\prime }\left( X\right) \right) ^{2}}{96\left\vert f\left( \hat{X}\right)
\right\vert ^{3}}+\frac{24\left\vert f\left( \hat{X}\right) \right\vert ^{3}%
}{\sigma _{X}^{2}\sigma _{\hat{K}}^{2}\left( f^{\prime }\left( X\right)
\right) ^{2}}\left( \ln \left( \bar{p}+\frac{1}{2}\right) -1\right)
^{2}\right)
\end{equation*}%
and equation (\ref{sMP}) rewrites:

\begin{eqnarray}
&&K_{\hat{X}}\left\Vert \Psi \left( \hat{X}\right) \right\Vert
^{2}\left\vert f\left( \hat{X}\right) \right\vert \left( \frac{\sigma
_{X}^{2}\sigma _{\hat{K}}^{2}\left( f^{\prime }\left( X\right) \right)
^{2}\exp \left( -\frac{96\left\vert f\left( \hat{X}\right) \right\vert ^{3}}{%
\sigma _{X}^{2}\sigma _{\hat{K}}^{2}\left( f^{\prime }\left( X\right)
\right) ^{2}}\left( \ln \left( \bar{p}+\frac{1}{2}\right) -1\right)
^{2}\right) }{96\left\vert f\left( \hat{X}\right) \right\vert ^{3}}\right) ^{%
\frac{1}{4}}  \label{dvM} \\
&=&C\left( \bar{p}\right) \sigma _{\hat{K}}^{2}\exp \left( -\frac{\sigma
_{X}^{2}\sigma _{\hat{K}}^{2}\left( p+\frac{1}{2}-\frac{48\left\vert f\left( 
\hat{X}\right) \right\vert ^{3}}{\sigma _{X}^{2}\sigma _{\hat{K}}^{2}\left(
f^{\prime }\left( X\right) \right) ^{2}}\left( \ln \left( \bar{p}+\frac{1}{2}%
\right) -1\right) \right) ^{2}\left( f^{\prime }\left( X\right) \right) ^{2}%
}{96\left\vert f\left( \hat{X}\right) \right\vert ^{3}}\right) \sqrt{\left(
p+\frac{1}{2}\right) \sqrt{\frac{\sigma _{X}^{2}\sigma _{\hat{K}}^{2}\left(
f^{\prime }\left( X\right) \right) ^{2}}{96\left\vert f\left( \hat{X}\right)
\right\vert ^{3}}}}  \notag
\end{eqnarray}%
To solve (\ref{dvM}) for $K_{\hat{X}}$, we proceed in two steps.

We first introduce an intermediate variable $W$ and rewrite (\ref{dvM}) as
an equation for $K_{\hat{X}}$ and $W$. We set:%
\begin{equation}
\sqrt{\frac{\sigma _{X}^{2}\sigma _{\hat{K}}^{2}\left( f^{\prime }\left(
X\right) \right) ^{2}}{96\left\vert f\left( \hat{X}\right) \right\vert ^{3}}}%
\left( p+\frac{1}{2}-\frac{48\left\vert f\left( \hat{X}\right) \right\vert
^{3}}{\sigma _{X}^{2}\sigma _{\hat{K}}^{2}\left( f^{\prime }\left( X\right)
\right) ^{2}}\left( \ln \left( \bar{p}+\frac{1}{2}\right) -1\right) \right)
=W  \label{lpw}
\end{equation}%
and rewrite equation (\ref{dvM}) partly in terms of $W$:%
\begin{eqnarray}
&&K_{\hat{X}}\left\Vert \Psi \left( \hat{X}\right) \right\Vert ^{2}\left( 
\frac{\sigma _{X}^{2}\left( f^{\prime }\left( X\right) \right)
^{2}\left\vert f\left( \hat{X}\right) \right\vert \exp \left( -\frac{%
96\left\vert f\left( \hat{X}\right) \right\vert ^{3}}{\sigma _{X}^{2}\sigma
_{\hat{K}}^{2}\left( f^{\prime }\left( X\right) \right) ^{2}}\left( \ln
\left( \bar{p}+\frac{1}{2}\right) -1\right) ^{2}\right) }{96\left( \sigma _{%
\hat{K}}^{2}\right) ^{3}}\right) ^{\frac{1}{4}}  \label{mdg} \\
&=&C\left( \bar{p}\right) \exp \left( -W^{2}\right) \sqrt{W+2\sqrt{\frac{%
96\left\vert f\left( \hat{X}\right) \right\vert ^{3}}{\sigma _{X}^{2}\sigma
_{\hat{K}}^{2}\left( f^{\prime }\left( X\right) \right) ^{2}}}\left( \ln
\left( \bar{p}+\frac{1}{2}\right) -1\right) }  \notag
\end{eqnarray}%
\bigskip

Note that, as seen from (\ref{lpw}), $W$ is a function of $p$ and as such
can be seen as a parameter depending on the shape of the sectors space.

Equation (\ref{mdg}) both depends on $K_{\hat{X}}$ and $W$, and in a second
step, we use (\ref{lpw}) to write $K_{\hat{X}}$ as a function of $W$. To do
so, we use that in the intermediate case $\infty >K_{\hat{X}}>1$, we can
assume that: 
\begin{equation}
f\left( \hat{X}\right) =B_{1}\left( X\right) K_{\hat{X}}^{\alpha
-1}+B_{2}\left( X\right) K_{\hat{X}}^{\alpha }-C\left( \hat{X}\right) \simeq
B_{2}\left( X\right) K_{\hat{X}}^{\alpha }  \label{smT}
\end{equation}%
and that:%
\begin{equation}
\frac{M-\left( \frac{\left( \nabla _{\hat{X}}R\left( \hat{X}\right) A\left( 
\hat{X}\right) K_{\hat{X}}^{\alpha }\right) ^{2}}{\sigma _{\hat{X}}^{2}}%
+\nabla _{\hat{X}}^{2}R\left( \hat{X}\right) A\left( \hat{X}\right) K_{\hat{X%
}}^{\alpha }\right) }{B_{1}\left( X\right) K_{\hat{X}}^{\alpha
-1}+B_{2}\left( X\right) K_{\hat{X}}^{\alpha }-C\left( \hat{X}\right) }-%
\frac{3}{2}\simeq \frac{M-\left( \frac{\left( \nabla _{\hat{X}}R\left( \hat{X%
}\right) A\left( \hat{X}\right) K_{\hat{X}}^{\alpha }\right) ^{2}}{\sigma _{%
\hat{X}}^{2}}+\nabla _{\hat{X}}^{2}R\left( \hat{X}\right) A\left( \hat{X}%
\right) K_{\hat{X}}^{\alpha }\right) }{B_{2}\left( X\right) K_{\hat{X}%
}^{\alpha }}-\frac{3}{2}  \label{ssp}
\end{equation}%
Moreover, we can approximate $\left\Vert \Psi \left( \hat{X}\right)
\right\Vert ^{2}$:%
\begin{equation}
\left\Vert \Psi \left( \hat{X}\right) \right\Vert ^{2}\simeq D  \label{ssP}
\end{equation}

Our assumptions (\ref{smT}), (\ref{ssp}) and (\ref{ssP}) allow to rewrite
the relation (\ref{lpw}) between $K_{\hat{X}}^{\alpha }$ and $W$ as:%
\begin{equation*}
\sqrt{\frac{\sigma _{X}^{2}\sigma _{\hat{K}}^{2}\left( B_{2}^{\prime }\left(
X\right) \right) ^{2}}{96B_{2}^{3}\left( X\right) K_{\hat{X}}^{\alpha }}}%
\left( p+\frac{1}{2}-\frac{48\left\vert f\left( \hat{X}\right) \right\vert
^{3}}{\sigma _{X}^{2}\sigma _{\hat{K}}^{2}\left( f^{\prime }\left( X\right)
\right) ^{2}}\left( \ln \left( \bar{p}+\frac{1}{2}\right) -1\right) \right)
=W
\end{equation*}%
that is:%
\begin{eqnarray}
W &=&\sqrt{\frac{\sigma _{X}^{2}\sigma _{\hat{K}}^{2}\left( B_{2}^{\prime
}\left( X\right) \right) ^{2}}{96B_{2}^{5}\left( X\right) K_{\hat{X}%
}^{3\alpha }}}  \label{knw} \\
&&\times \left( M-\left( \frac{\left( \nabla _{\hat{X}}R\left( \hat{X}%
\right) \right) ^{2}A\left( \hat{X}\right) }{\sigma _{\hat{X}}^{2}}+\frac{%
48B_{2}^{4}\left( X\right) \left( \ln \left( \bar{p}+\frac{1}{2}\right)
-1\right) }{\sigma _{X}^{2}\sigma _{\hat{K}}^{2}\left( B_{2}^{\prime }\left(
X\right) \right) ^{2}}\right) K_{\hat{X}}^{2\alpha }-\left( \nabla _{\hat{X}%
}^{2}R\left( \hat{X}\right) +B_{2}\left( X\right) \right) K_{\hat{X}%
}^{\alpha }\right)  \notag
\end{eqnarray}

To solve this equation for $K_{\hat{X}}^{\alpha }$, we consider $M$ as the
dominant parameter and find an approximate solution of (\ref{knw}). At the
lowest order, we write:%
\begin{equation*}
\sqrt{\frac{\sigma _{X}^{2}\sigma _{\hat{K}}^{2}\left( B_{2}^{\prime }\left(
X\right) \right) ^{2}}{96B_{2}^{5}\left( X\right) K_{\hat{X}}^{3\alpha }}}M=W
\end{equation*}%
with solution:%
\begin{equation*}
K_{\hat{X}}^{\alpha }=\left( \frac{\sigma _{X}^{2}\sigma _{\hat{K}%
}^{2}\left( B_{2}^{\prime }\left( X\right) \right) ^{2}M^{2}}{%
96B_{2}^{5}\left( X\right) W^{2}}\right) ^{\frac{1}{3}}
\end{equation*}%
Considering corrections to this result, the solution to (\ref{knw}) is
decomposed as:%
\begin{equation}
K_{\hat{X}}^{\alpha }=\left( \frac{\sigma _{X}^{2}\sigma _{\hat{K}%
}^{2}\left( B_{2}^{\prime }\left( X\right) \right) ^{2}M^{2}}{%
96B_{2}^{5}\left( X\right) W^{2}}\right) ^{\frac{1}{3}}+\chi  \label{slK}
\end{equation}%
and using the following intermediate results:%
\begin{equation*}
K_{\hat{X}}^{2\alpha }=\left( \frac{\sigma _{X}^{2}\sigma _{\hat{K}%
}^{2}\left( B_{2}^{\prime }\left( X\right) \right) ^{2}M^{2}}{%
96B_{2}^{5}\left( X\right) W^{2}}\right) ^{\frac{2}{3}}\left( 1+2\chi \left( 
\frac{\sigma _{X}^{2}\sigma _{\hat{K}}^{2}\left( B_{2}^{\prime }\left(
X\right) \right) ^{2}M^{2}}{96B_{2}^{5}\left( X\right) W^{2}}\right) ^{-%
\frac{1}{3}}\right)
\end{equation*}%
\begin{equation*}
K_{\hat{X}}^{3\alpha }=\left( \frac{\sigma _{X}^{2}\sigma _{\hat{K}%
}^{2}\left( B_{2}^{\prime }\left( X\right) \right) ^{2}M^{2}}{%
96B_{2}^{5}\left( X\right) W^{2}}\right) \left( 1+3\chi \left( \frac{\sigma
_{X}^{2}\sigma _{\hat{K}}^{2}\left( B_{2}^{\prime }\left( X\right) \right)
^{2}M^{2}}{96B_{2}^{5}\left( X\right) W^{2}}\right) ^{-\frac{1}{3}}\right)
\end{equation*}%
we are lead to rewrite (\ref{knw}) as an equation for $\chi $ at first order:%
\begin{eqnarray*}
&&\chi \left( \frac{3}{2}\left( \frac{\sigma _{X}^{2}\sigma _{\hat{K}%
}^{2}\left( B_{2}^{\prime }\left( X\right) \right) ^{2}M^{2}}{%
96B_{2}^{5}\left( X\right) W^{2}}\right) ^{-\frac{1}{3}}W\right. \\
&&\left. +2\frac{W}{M}\left( \frac{\left( \nabla _{\hat{X}}R\left( \hat{X}%
\right) \right) ^{2}A\left( \hat{X}\right) }{\sigma _{\hat{X}}^{2}}+\frac{%
48B_{2}^{4}\left( X\right) \left( \ln \left( \bar{p}+\frac{1}{2}\right)
-1\right) }{\sigma _{X}^{2}\sigma _{\hat{K}}^{2}\left( B_{2}^{\prime }\left(
X\right) \right) ^{2}}\right) \left( \frac{\sigma _{X}^{2}\sigma _{\hat{K}%
}^{2}\left( B_{2}^{\prime }\left( X\right) \right) ^{2}M^{2}}{%
96B_{2}^{5}\left( X\right) W^{2}}\right) ^{\frac{1}{3}}+\frac{W}{M}\left(
\nabla _{\hat{X}}^{2}R\left( \hat{X}\right) +B_{2}\left( X\right) \right)
\right) \\
&=&-\frac{W}{M}\left( \frac{\left( \nabla _{\hat{X}}R\left( \hat{X}\right)
\right) ^{2}A\left( \hat{X}\right) }{\sigma _{\hat{X}}^{2}}+\frac{%
48B_{2}^{4}\left( X\right) \left( \ln \left( \bar{p}+\frac{1}{2}\right)
-1\right) }{\sigma _{X}^{2}\sigma _{\hat{K}}^{2}\left( B_{2}^{\prime }\left(
X\right) \right) ^{2}}\right) \left( \frac{\sigma _{X}^{2}\sigma _{\hat{K}%
}^{2}\left( B_{2}^{\prime }\left( X\right) \right) ^{2}M^{2}}{%
96B_{2}^{5}\left( X\right) W^{2}}\right) ^{\frac{2}{3}} \\
&&-\frac{W}{M}\left( \nabla _{\hat{X}}^{2}R\left( \hat{X}\right)
+B_{2}\left( X\right) \right) \left( \frac{\sigma _{X}^{2}\sigma _{\hat{K}%
}^{2}\left( B_{2}^{\prime }\left( X\right) \right) ^{2}M^{2}}{%
96B_{2}^{5}\left( X\right) W^{2}}\right) ^{\frac{1}{3}}
\end{eqnarray*}%
whose solution is:%
\begin{equation*}
\chi =-\frac{\left( \frac{\left( \nabla _{\hat{X}}R\left( \hat{X}\right)
\right) ^{2}A\left( \hat{X}\right) }{\sigma _{\hat{X}}^{2}}+\frac{%
48B_{2}^{4}\left( X\right) \left( \ln \left( \bar{p}+\frac{1}{2}\right)
-1\right) }{\sigma _{X}^{2}\sigma _{\hat{K}}^{2}\left( B_{2}^{\prime }\left(
X\right) \right) ^{2}}\right) \left( \frac{\sigma _{X}^{2}\sigma _{\hat{K}%
}^{2}\left( B_{2}^{\prime }\left( X\right) \right) ^{2}M^{2}}{%
96B_{2}^{5}\left( X\right) W^{2}}\right) +\left( \nabla _{\hat{X}%
}^{2}R\left( \hat{X}\right) +B_{2}\left( X\right) \right) \left( \frac{%
\sigma _{X}^{2}\sigma _{\hat{K}}^{2}\left( B_{2}^{\prime }\left( X\right)
\right) ^{2}M^{2}}{96B_{2}^{5}\left( X\right) W^{2}}\right) ^{\frac{2}{3}}}{%
\frac{3}{2}M+2\left( \frac{\left( \nabla _{\hat{X}}R\left( \hat{X}\right)
\right) ^{2}A\left( \hat{X}\right) }{\sigma _{\hat{X}}^{2}}+\frac{%
48B_{2}^{4}\left( X\right) \left( \ln \left( \bar{p}+\frac{1}{2}\right)
-1\right) }{\sigma _{X}^{2}\sigma _{\hat{K}}^{2}\left( B_{2}^{\prime }\left(
X\right) \right) ^{2}}\right) \left( \frac{\sigma _{X}^{2}\sigma _{\hat{K}%
}^{2}\left( B_{2}^{\prime }\left( X\right) \right) ^{2}M^{2}}{%
96B_{2}^{5}\left( X\right) W^{2}}\right) ^{\frac{2}{3}}+\left( \nabla _{\hat{%
X}}^{2}R\left( \hat{X}\right) +B_{2}\left( X\right) \right) \left( \frac{%
\sigma _{X}^{2}\sigma _{\hat{K}}^{2}\left( B_{2}^{\prime }\left( X\right)
\right) ^{2}M^{2}}{96B_{2}^{5}\left( X\right) W^{2}}\right) ^{\frac{1}{3}}}
\end{equation*}%
so that (\ref{slK}) yields $K_{\hat{X}}^{\alpha }$:%
\begin{eqnarray}
&&K_{\hat{X}}^{\alpha }-\left( \frac{\sigma _{X}^{2}\sigma _{\hat{K}%
}^{2}\left( B_{2}^{\prime }\left( X\right) \right) ^{2}M^{2}}{%
96B_{2}^{5}\left( X\right) W^{2}}\right) ^{\frac{1}{3}}  \label{kxf} \\
&=&-\frac{\left( \frac{\left( \nabla _{\hat{X}}R\left( \hat{X}\right)
\right) ^{2}A\left( \hat{X}\right) }{\sigma _{\hat{X}}^{2}}+\frac{%
48B_{2}^{4}\left( X\right) \left( \ln \left( \bar{p}+\frac{1}{2}\right)
-1\right) }{\sigma _{X}^{2}\sigma _{\hat{K}}^{2}\left( B_{2}^{\prime }\left(
X\right) \right) ^{2}}\right) \left( \frac{\sigma _{X}^{2}\sigma _{\hat{K}%
}^{2}\left( B_{2}^{\prime }\left( X\right) \right) ^{2}M^{2}}{%
96B_{2}^{5}\left( X\right) W^{2}}\right) +\left( \nabla _{\hat{X}%
}^{2}R\left( \hat{X}\right) +B_{2}\left( X\right) \right) \left( \frac{%
\sigma _{X}^{2}\sigma _{\hat{K}}^{2}\left( B_{2}^{\prime }\left( X\right)
\right) ^{2}M^{2}}{96B_{2}^{5}\left( X\right) W^{2}}\right) ^{\frac{2}{3}}}{%
\frac{3}{2}M+2\left( \frac{\left( \nabla _{\hat{X}}R\left( \hat{X}\right)
\right) ^{2}A\left( \hat{X}\right) }{\sigma _{\hat{X}}^{2}}+\frac{%
48B_{2}^{4}\left( X\right) \left( \ln \left( \bar{p}+\frac{1}{2}\right)
-1\right) }{\sigma _{X}^{2}\sigma _{\hat{K}}^{2}\left( B_{2}^{\prime }\left(
X\right) \right) ^{2}}\right) \left( \frac{\sigma _{X}^{2}\sigma _{\hat{K}%
}^{2}\left( B_{2}^{\prime }\left( X\right) \right) ^{2}M^{2}}{%
96B_{2}^{5}\left( X\right) W^{2}}\right) ^{\frac{2}{3}}+\left( \nabla _{\hat{%
X}}^{2}R\left( \hat{X}\right) +B_{2}\left( X\right) \right) \left( \frac{%
\sigma _{X}^{2}\sigma _{\hat{K}}^{2}\left( B_{2}^{\prime }\left( X\right)
\right) ^{2}M^{2}}{96B_{2}^{5}\left( X\right) W^{2}}\right) ^{\frac{1}{3}}} 
\notag \\
&&\frac{\left( \frac{\left( \nabla _{\hat{X}}R\left( \hat{X}\right) \right)
^{2}A\left( \hat{X}\right) }{\sigma _{\hat{X}}^{2}}+\frac{48B_{2}^{4}\left(
X\right) \left( \ln \left( \bar{p}+\frac{1}{2}\right) -1\right) }{\sigma
_{X}^{2}\sigma _{\hat{K}}^{2}\left( B_{2}^{\prime }\left( X\right) \right)
^{2}}\right) \left( \frac{\sigma _{X}^{2}\sigma _{\hat{K}}^{2}\left(
B_{2}^{\prime }\left( X\right) \right) ^{2}M^{2}}{96B_{2}^{5}\left( X\right)
W^{2}}\right) +\left( \nabla _{\hat{X}}^{2}R\left( \hat{X}\right)
+B_{2}\left( X\right) \right) \left( \frac{\sigma _{X}^{2}\sigma _{\hat{K}%
}^{2}\left( B_{2}^{\prime }\left( X\right) \right) ^{2}M^{2}}{%
96B_{2}^{5}\left( X\right) W^{2}}\right) ^{\frac{2}{3}}}{\frac{3}{2}%
M+2\left( \frac{\left( \nabla _{\hat{X}}R\left( \hat{X}\right) \right)
^{2}A\left( \hat{X}\right) }{\sigma _{\hat{X}}^{2}}+\frac{48B_{2}^{4}\left(
X\right) \left( \ln \left( \bar{p}+\frac{1}{2}\right) -1\right) }{\sigma
_{X}^{2}\sigma _{\hat{K}}^{2}\left( B_{2}^{\prime }\left( X\right) \right)
^{2}}\right) \left( \frac{\sigma _{X}^{2}\sigma _{\hat{K}}^{2}\left(
B_{2}^{\prime }\left( X\right) \right) ^{2}M^{2}}{96B_{2}^{5}\left( X\right)
W^{2}}\right) ^{\frac{2}{3}}+\left( \nabla _{\hat{X}}^{2}R\left( \hat{X}%
\right) +B_{2}\left( X\right) \right) \left( \frac{\sigma _{X}^{2}\sigma _{%
\hat{K}}^{2}\left( B_{2}^{\prime }\left( X\right) \right) ^{2}M^{2}}{%
96B_{2}^{5}\left( X\right) W^{2}}\right) ^{\frac{1}{3}}}  \notag
\end{eqnarray}%
In a third step, we can use equation (\ref{kxf}) to rewrite (\ref{mdg}) in
an approximate form. Actually, expression (\ref{kxf}) implies that in the
intermediate case, where $K_{\hat{X}}^{\alpha }$ is of finite magnitude, we
have $W^{2}\sim \sigma _{X}^{2}\sigma _{\hat{K}}^{2}M^{2}$ and:%
\begin{eqnarray*}
&&\exp \left( -W^{2}+\frac{24\left\vert B_{2}\left( X\right) \right\vert
^{3}K_{\hat{X}}^{\alpha }}{\sigma _{X}^{2}\sigma _{\hat{K}}^{2}\left(
B_{2}^{\prime }\left( X\right) \right) ^{2}}\left( \ln \left( \bar{p}+\frac{1%
}{2}\right) -1\right) ^{2}\right) \\
&\simeq &\exp \left( \frac{24\left\vert B_{2}\left( X\right) \right\vert
^{3}K_{\hat{X}}^{\alpha }}{\sigma _{X}^{2}\sigma _{\hat{K}}^{2}\left(
B_{2}^{\prime }\left( X\right) \right) ^{2}}\left( \ln \left( \bar{p}+\frac{1%
}{2}\right) -1\right) ^{2}\right)
\end{eqnarray*}%
Moreover using that:%
\begin{equation*}
W+2\sqrt{\frac{96\left\vert f\left( \hat{X}\right) \right\vert ^{3}}{\sigma
_{X}^{2}\sigma _{\hat{K}}^{2}\left( f^{\prime }\left( X\right) \right) ^{2}}}%
\left( \ln \left( \bar{p}+\frac{1}{2}\right) -1\right) \simeq 2\sqrt{\frac{%
96\left\vert f\left( \hat{X}\right) \right\vert ^{3}}{\sigma _{X}^{2}\sigma
_{\hat{K}}^{2}\left( f^{\prime }\left( X\right) \right) ^{2}}}\left( \ln
\left( \bar{p}+\frac{1}{2}\right) -1\right)
\end{equation*}%
and that ultimately the left hand side of equation (\ref{mdg}) writes at the
first order:

\begin{eqnarray*}
&&K_{\hat{X}}\left\Vert \Psi \left( \hat{X}\right) \right\Vert ^{2}\left( 
\frac{\sigma _{X}^{2}\left( f^{\prime }\left( X\right) \right)
^{2}\left\vert f\left( \hat{X}\right) \right\vert \exp \left( -\frac{%
96\left\vert f\left( \hat{X}\right) \right\vert ^{3}}{\sigma _{X}^{2}\sigma
_{\hat{K}}^{2}\left( f^{\prime }\left( X\right) \right) ^{2}}\left( \ln
\left( \bar{p}+\frac{1}{2}\right) -1\right) ^{2}\right) }{96\left( \sigma _{%
\hat{K}}^{2}\right) ^{3}}\right) ^{\frac{1}{4}} \\
&=&\left( \frac{\sigma _{X}^{2}\left( B_{2}^{\prime }\left( X\right) \right)
^{2}\left\vert B_{2}\left( X\right) \right\vert }{96\left( \sigma _{\hat{K}%
}^{2}\right) ^{3}}\right) ^{\frac{1}{4}}K_{\hat{X}}^{1+\frac{3\alpha }{4}%
}\left\Vert \Psi \left( \hat{X}\right) \right\Vert ^{2}\exp \left( -\frac{%
24\left\vert f\left( \hat{X}\right) \right\vert ^{3}}{\sigma _{X}^{2}\sigma
_{\hat{K}}^{2}\left( f^{\prime }\left( X\right) \right) ^{2}}\left( \ln
\left( \bar{p}+\frac{1}{2}\right) -1\right) ^{2}\right) \\
&\simeq &D\left( \frac{\sigma _{X}^{2}\left( B_{2}^{\prime }\left( X\right)
\right) ^{2}\left\vert B_{2}\left( X\right) \right\vert }{96\left( \sigma _{%
\hat{K}}^{2}\right) ^{3}}\right) ^{\frac{1}{4}}K_{\hat{X}}^{1+\frac{3\alpha 
}{4}}\exp \left( -\frac{24\left\vert B_{2}\left( X\right) \right\vert ^{3}K_{%
\hat{X}}^{\alpha }}{\sigma _{X}^{2}\sigma _{\hat{K}}^{2}\left( B_{2}^{\prime
}\left( X\right) \right) ^{2}}\left( \ln \left( \bar{p}+\frac{1}{2}\right)
-1\right) ^{2}\right)
\end{eqnarray*}%
equation (\ref{mdg}) writes:%
\begin{eqnarray*}
&&D\left( \frac{\sigma _{X}^{2}\left( B_{2}^{\prime }\left( X\right) \right)
^{2}\left\vert B_{2}\left( X\right) \right\vert }{96\left( \sigma _{\hat{K}%
}^{2}\right) ^{3}}\right) ^{\frac{1}{4}}K_{\hat{X}}^{1+\frac{3\alpha }{4}%
}\exp \left( -\frac{24\left\vert B_{2}\left( X\right) \right\vert ^{3}K_{%
\hat{X}}^{\alpha }}{\sigma _{X}^{2}\sigma _{\hat{K}}^{2}\left( B_{2}^{\prime
}\left( X\right) \right) ^{2}}\left( \ln \left( \bar{p}+\frac{1}{2}\right)
-1\right) ^{2}\right) \\
&=&C\left( \bar{p}\right) \sqrt{2\sqrt{\frac{96\left\vert B_{2}\left(
X\right) \right\vert ^{3}K_{\hat{X}}^{\alpha }}{\sigma _{X}^{2}\sigma _{\hat{%
K}}^{2}\left( B_{2}^{\prime }\left( X\right) \right) ^{2}}}\left( \ln \left( 
\bar{p}+\frac{1}{2}\right) -1\right) }
\end{eqnarray*}%
that is:

\begin{equation}
K_{\hat{X}}^{1+\frac{\alpha }{2}}\exp \left( -\frac{24\left\vert B_{2}\left(
X\right) \right\vert ^{3}K_{\hat{X}}^{\alpha }}{\sigma _{X}^{2}\sigma _{\hat{%
K}}^{2}\left( B_{2}^{\prime }\left( X\right) \right) ^{2}}\left( \ln \left( 
\bar{p}+\frac{1}{2}\right) -1\right) ^{2}\right) =\frac{8C\left( \bar{p}%
\right) }{D}\sqrt{\frac{3\sigma _{\hat{K}}^{2}\left\vert B_{2}\left(
X\right) \right\vert }{\sigma _{X}^{2}\left( B_{2}^{\prime }\left( X\right)
\right) ^{2}}\left( \ln \left( \bar{p}+\frac{1}{2}\right) -1\right) }
\label{qTN}
\end{equation}%
Equation (\ref{qTN}) has the form: 
\begin{equation*}
x^{d}\exp \left( -ax\right) =c
\end{equation*}%
$\allowbreak $with solution:%
\begin{equation*}
x=c^{\frac{1}{d}}\exp \left( -W_{0}\left( -\frac{a}{d}c^{\frac{1}{d}}\right)
\right)
\end{equation*}%
where $W_{0}$ is the Lambert $W$ function with parameter $0$. Applying this
result to our case with:%
\begin{eqnarray*}
d &=&\frac{1+\alpha }{2\alpha } \\
x &=&K_{\hat{X}}^{\alpha } \\
a &=&\frac{24\left\vert B_{2}\left( X\right) \right\vert ^{3}}{\sigma
_{X}^{2}\sigma _{\hat{K}}^{2}\left( B_{2}^{\prime }\left( X\right) \right)
^{2}}\left( \ln \left( \bar{p}+\frac{1}{2}\right) -1\right) ^{2} \\
c &=&\frac{8C\left( \bar{p}\right) }{D}\sqrt{\frac{3\sigma _{\hat{K}%
}^{2}\left\vert B_{2}\left( X\right) \right\vert }{\sigma _{X}^{2}\left(
B_{2}^{\prime }\left( X\right) \right) ^{2}}\left( \ln \left( \bar{p}+\frac{1%
}{2}\right) -1\right) }
\end{eqnarray*}%
we obtain:%
\begin{eqnarray*}
K_{\hat{X}}^{\alpha } &=&\left( \frac{8C\left( \bar{p}\right) }{D}\sqrt{%
\frac{3\sigma _{\hat{K}}^{2}\left\vert B_{2}\left( X\right) \right\vert }{%
\sigma _{X}^{2}\left( B_{2}^{\prime }\left( X\right) \right) ^{2}}\left( \ln
\left( \bar{p}+\frac{1}{2}\right) -1\right) }\right) ^{\frac{2\alpha }{%
1+\alpha }} \\
&&\times \exp \left( -W_{0}\left( -\frac{48\alpha }{1+\alpha }\left( \sqrt{%
\frac{3\sigma _{\hat{K}}^{2}}{\sigma _{X}^{2}}}\frac{8C\left( \bar{p}\right) 
}{D}\right) ^{\frac{2\alpha }{1+\alpha }}\frac{\left\vert B_{2}\left(
X\right) \right\vert ^{3+\frac{\alpha }{1+\alpha }}}{\sigma _{X}^{2}\sigma _{%
\hat{K}}^{2}\left( B_{2}^{\prime }\left( X\right) \right) ^{2+\frac{2\alpha 
}{1+\alpha }}}\left( \ln \left( \bar{p}+\frac{1}{2}\right) -1\right) ^{2+%
\frac{\alpha }{1+\alpha }}\right) \right)
\end{eqnarray*}%
As stated in the text, this is an increasing function of $B_{2}\left(
X\right) $. Moreover, the corrections to this formula, given in (\ref{kxf})
show that $K_{\hat{X}}^{\alpha }$ is a decreasing function of $\left( \nabla
_{\hat{X}}R\left( \hat{X}\right) \right) ^{2}$ and $\nabla _{\hat{X}%
}^{2}R\left( \hat{X}\right) $.

\section*{Appendix 4. \textbf{Dynamics for }$K_{\hat{X}}$}

\subsection*{A4.1 Variation of the defining equation for $K_{\hat{X}}$}

\subsubsection*{A4.1.1 Compact formulation}

As claimed in the text, we consider the dynamics for $K_{\hat{X}}$ generated
by modification of the parameters. To do so, we compute the variation of
equation (\ref{qnk}). We need the variations of the functions involved in (%
\ref{qnk}) with respect to two dynamical variables $K_{\hat{X}}$ and $%
R\left( X\right) $. Starting with (\ref{qnk}):

\begin{equation}
K_{\hat{X}}\left( D-L\left( \hat{X}\right) K_{\hat{X}}^{\eta }\right) =\frac{%
C\left( \bar{p}\right) \sigma _{\hat{K}}^{2}}{\left\vert f\left( \hat{X}%
\right) \right\vert }\hat{\Gamma}\left( p+\frac{1}{2}\right)  \label{dkn}
\end{equation}%
where:%
\begin{eqnarray}
\hat{\Gamma}\left( p+\frac{1}{2}\right) &=&\exp \left( -\frac{\sigma
_{X}^{2}\sigma _{\hat{K}}^{2}\left( p+\frac{1}{2}\right) ^{2}\left(
f^{\prime }\left( X\right) \right) ^{2}}{96\left\vert f\left( \hat{X}\right)
\right\vert ^{3}}\right) \\
&&\times \left( \frac{\Gamma \left( -\frac{p+1}{2}\right) \Gamma \left( 
\frac{1-p}{2}\right) -\Gamma \left( -\frac{p}{2}\right) \Gamma \left( \frac{%
-p}{2}\right) }{2^{p+2}\Gamma \left( -p-1\right) \Gamma \left( -p\right) }+p%
\frac{\Gamma \left( -\frac{p}{2}\right) \Gamma \left( \frac{2-p}{2}\right)
-\Gamma \left( -\frac{p-1}{2}\right) \Gamma \left( -\frac{p-1}{2}\right) }{%
2^{p+1}\Gamma \left( -p\right) \Gamma \left( -p+1\right) }\right)  \notag
\end{eqnarray}%
We first compute the variations of the right hand side and use that, in
first approximation:

\begin{equation}
\frac{d}{dp}\left( \ln \left( \frac{\Gamma \left( -\frac{p+1}{2}\right)
\Gamma \left( \frac{1-p}{2}\right) -\Gamma \left( -\frac{p}{2}\right) \Gamma
\left( \frac{-p}{2}\right) }{2^{p+2}\Gamma \left( -p-1\right) \Gamma \left(
-p\right) }+p\frac{\Gamma \left( -\frac{p}{2}\right) \Gamma \left( \frac{2-p%
}{2}\right) -\Gamma \left( -\frac{p-1}{2}\right) \Gamma \left( -\frac{p-1}{2}%
\right) }{2^{p+1}\Gamma \left( -p\right) \Gamma \left( -p+1\right) }\right)
\right) \simeq \ln \left( p+\frac{1}{2}\right)
\end{equation}%
and:%
\begin{equation*}
\frac{d}{dp}\left( -\frac{\sigma _{X}^{2}\left( p+\frac{1}{2}\right)
^{2}\left( f^{\prime }\left( X\right) \right) ^{2}}{96\left\vert f\left( 
\hat{X}\right) \right\vert ^{3}}\right) =-\frac{\sigma _{X}^{2}\left( p+%
\frac{1}{2}\right) \left( f^{\prime }\left( X\right) \right) ^{2}}{%
48\left\vert f\left( \hat{X}\right) \right\vert ^{3}}
\end{equation*}%
so that:%
\begin{equation}
\frac{\frac{d}{dp}\hat{\Gamma}\left( p+\frac{1}{2}\right) }{\hat{\Gamma}%
\left( p+\frac{1}{2}\right) }\simeq \ln \left( p+\frac{1}{2}\right) -\frac{%
\sigma _{X}^{2}\sigma _{\hat{K}}^{2}\left( p+\frac{1}{2}\right) \left(
f^{\prime }\left( X\right) \right) ^{2}}{48\left\vert f\left( \hat{X}\right)
\right\vert ^{3}}  \label{pdr}
\end{equation}%
Assuming that $C\left( \bar{p}\right) $ is constant, (\ref{pdr}) allows to
rewrite the variation of of equation (\ref{dkn}):%
\begin{eqnarray*}
\nabla _{\theta }\left( K_{\hat{X}}\left( D-L\left( \hat{X}\right) K_{\hat{X}%
}^{\eta }\right) \right) &=&K_{\hat{X}}\left( D-L\left( \hat{X}\right) K_{%
\hat{X}}^{\eta }\right) \\
&&\times \left( -\frac{\nabla _{\theta }\left\vert f\left( \hat{X}\right)
\right\vert }{\left\vert f\left( \hat{X}\right) \right\vert }+\left( \ln
\left( p+\frac{1}{2}\right) -\frac{\sigma _{X}^{2}\sigma _{\hat{K}%
}^{2}\left( p+\frac{1}{2}\right) \left( f^{\prime }\left( X\right) \right)
^{2}}{48\left\vert f\left( \hat{X}\right) \right\vert ^{3}}\right) \nabla
_{\theta }p\right) \\
&&+\frac{\sigma _{X}^{2}\sigma _{\hat{K}}^{2}\left( p+\frac{1}{2}\right)
^{2}\left( f^{\prime }\left( X\right) \right) ^{2}}{96\left\vert f\left( 
\hat{X}\right) \right\vert ^{3}}\left( \frac{\nabla _{\theta }\left\vert
f\left( \hat{X}\right) \right\vert }{\left\vert f\left( \hat{X}\right)
\right\vert }-\frac{\nabla _{\theta }\left( \frac{f^{\prime }\left( X\right) 
}{f\left( \hat{X}\right) }\right) ^{2}}{\left( \frac{f^{\prime }\left(
X\right) }{f\left( \hat{X}\right) }\right) ^{2}}\right)
\end{eqnarray*}%
and we deduce from this equation, that the dynamic version of equation (\ref%
{dkn}) is:%
\begin{eqnarray}
\frac{\nabla _{\theta }K_{\hat{X}}}{K_{\hat{X}}}-\frac{\nabla _{\theta
}\left( L\left( \hat{X}\right) K_{\hat{X}}^{\eta }\right) }{D-L\left( \hat{X}%
\right) K_{\hat{X}}^{\eta }} &=&\left( \frac{\sigma _{X}^{2}\sigma _{\hat{K}%
}^{2}\left( p+\frac{1}{2}\right) \left( f^{\prime }\left( X\right) \right)
^{2}}{48\left\vert f\left( \hat{X}\right) \right\vert ^{3}}-1\right) \frac{%
\nabla _{\theta }\left\vert f\left( \hat{X}\right) \right\vert }{\left\vert
f\left( \hat{X}\right) \right\vert }  \label{dqt} \\
&&+\left( \ln \left( p+\frac{1}{2}\right) -\frac{\sigma _{X}^{2}\sigma _{%
\hat{K}}^{2}\left( p+\frac{1}{2}\right) \left( f^{\prime }\left( X\right)
\right) ^{2}}{48\left\vert f\left( \hat{X}\right) \right\vert ^{3}}\right)
\nabla _{\theta }p  \notag \\
&&-\frac{\sigma _{X}^{2}\sigma _{\hat{K}}^{2}\left( p+\frac{1}{2}\right) ^{2}%
}{96\left\vert f\left( \hat{X}\right) \right\vert }\left( \nabla _{\theta
}\left( \frac{f^{\prime }\left( X\right) }{f\left( \hat{X}\right) }\right)
^{2}\right)  \notag
\end{eqnarray}

\subsubsection*{A4.1.2 Expanded form of (\protect\ref{dqt})}

To find the dynamic equation for $K_{\hat{X}}$ we expand each side of (\ref%
{dqt}).

The left hand side of (\ref{dqt}) can be developped as: 
\begin{eqnarray*}
&&\left( 1-\eta \frac{L\left( \hat{X}\right) K_{\hat{X}}^{\eta }}{D-L\left( 
\hat{X}\right) K_{\hat{X}}^{\eta }}\right) \frac{\nabla _{\theta }K_{\hat{X}}%
}{K_{\hat{X}}}-\frac{L\left( \hat{X}\right) K_{\hat{X}}^{\eta }}{D-L\left( 
\hat{X}\right) K_{\hat{X}}^{\eta }}\frac{\nabla _{\theta }L\left( \hat{X}%
\right) }{L\left( \hat{X}\right) } \\
&=&\left( 1-\eta \frac{L\left( \hat{X}\right) K_{\hat{X}}^{\eta }}{%
\left\Vert \Psi \left( \hat{X}\right) \right\Vert ^{2}}\right) \frac{\nabla
_{\theta }K_{\hat{X}}}{K_{\hat{X}}}-\frac{L\left( \hat{X}\right) K_{\hat{X}%
}^{\eta }}{\left\Vert \Psi \left( \hat{X}\right) \right\Vert ^{2}}\frac{%
\nabla _{\theta }L\left( \hat{X}\right) }{L\left( \hat{X}\right) } \\
&=&\left( 1-\eta \frac{L\left( \hat{X}\right) K_{\hat{X}}^{\eta }}{%
\left\Vert \Psi \left( \hat{X}\right) \right\Vert ^{2}}\right) \frac{\nabla
_{\theta }K_{\hat{X}}}{K_{\hat{X}}}-\frac{L\left( \hat{X}\right) K_{\hat{X}%
}^{\eta }}{\left\Vert \Psi \left( \hat{X}\right) \right\Vert ^{2}}\frac{%
\nabla _{\theta }\left( \left( \nabla _{X}R\left( X\right) \right)
^{2}+\sigma _{X}^{2}\frac{\nabla _{X}^{2}R\left( K_{X},X\right) }{H\left(
K_{X}\right) }\right) }{\left( \nabla _{X}R\left( X\right) \right)
^{2}+\sigma _{X}^{2}\frac{\nabla _{X}^{2}R\left( K_{X},X\right) }{H\left(
K_{X}\right) }} \\
&\simeq &\left( 1-\eta \frac{L\left( \hat{X}\right) K_{\hat{X}}^{\eta }}{%
\left\Vert \Psi \left( \hat{X}\right) \right\Vert ^{2}}\right) \frac{\nabla
_{\theta }K_{\hat{X}}}{K_{\hat{X}}}-2\frac{D-\left\Vert \Psi \left( \hat{X}%
\right) \right\Vert ^{2}}{\left\Vert \Psi \left( \hat{X}\right) \right\Vert
^{2}}\frac{\nabla _{\theta }\left( \nabla _{X}R\left( X\right) \right) }{%
\nabla _{X}R\left( X\right) }
\end{eqnarray*}%
and (\ref{dqt}) becomes:%
\begin{eqnarray}
&&\left( 1-\eta \frac{D-\left\Vert \Psi \left( \hat{X}\right) \right\Vert
^{2}}{\left\Vert \Psi \left( \hat{X}\right) \right\Vert ^{2}}\right) \frac{%
\nabla _{\theta }K_{\hat{X}}}{K_{\hat{X}}}-2\frac{D-\left\Vert \Psi \left( 
\hat{X}\right) \right\Vert ^{2}}{\left\Vert \Psi \left( \hat{X}\right)
\right\Vert ^{2}}\frac{\nabla _{X}\left( \nabla _{\theta }R\left( X\right)
\right) }{\nabla _{X}R\left( X\right) }  \label{tqd} \\
&=&\left( \frac{\sigma _{X}^{2}\sigma _{\hat{K}}^{2}\left( p+\frac{1}{2}%
\right) ^{2}\left( f^{\prime }\left( X\right) \right) ^{2}}{96\left\vert
f\left( \hat{X}\right) \right\vert ^{3}}-1\right) \frac{\nabla _{\theta
}\left\vert f\left( \hat{X}\right) \right\vert }{\left\vert f\left( \hat{X}%
\right) \right\vert }+\left( \ln \left( p+\frac{1}{2}\right) -\frac{\sigma
_{X}^{2}\sigma _{\hat{K}}^{2}\left( p+\frac{1}{2}\right) \left( f^{\prime
}\left( X\right) \right) ^{2}}{48\left\vert f\left( \hat{X}\right)
\right\vert ^{3}}\right) \nabla _{\theta }p  \notag \\
&&-\frac{\sigma _{X}^{2}\sigma _{\hat{K}}^{2}\left( p+\frac{1}{2}\right) ^{2}%
}{96\left\vert f\left( \hat{X}\right) \right\vert }\nabla _{\theta }\left( 
\frac{f^{\prime }\left( X\right) }{f\left( \hat{X}\right) }\right) ^{2} 
\notag
\end{eqnarray}%
To compute the right hand side of (\ref{tqd}). We use that:

\begin{equation*}
p=-\frac{M-\left( g\left( \hat{X}\right) \right) ^{2}+\sigma _{\hat{X}%
}^{2}\left( \nabla _{\hat{X}}g\left( \hat{X},K_{\hat{X}}\right) \right) }{%
\sigma _{\hat{X}}^{2}f\left( \hat{X}\right) }-\frac{3}{2}
\end{equation*}%
so that, the variation $\nabla _{\theta }p$ is given by: 
\begin{equation*}
\nabla _{\theta }p=-\frac{\nabla _{\theta }\left\vert f\left( \hat{X}\right)
\right\vert }{\left\vert f\left( \hat{X}\right) \right\vert }\left( p+\frac{3%
}{2}\right) -\left( 2\frac{g\left( \hat{X}\right) \nabla _{\theta }g\left( 
\hat{X}\right) }{\sigma _{\hat{X}}^{2}}+\nabla _{\theta }\nabla _{\hat{X}%
}g\left( \hat{X}\right) \right)
\end{equation*}%
To compute $\nabla _{\theta }p$ we have to use the form of the functions
defined in Appendix 2. We thus obtain: 
\begin{eqnarray*}
\frac{\nabla _{\theta }g\left( \hat{X}\right) }{g\left( \hat{X}\right) } &=&%
\frac{\nabla _{\theta }\nabla _{\hat{X}}R\left( \hat{X}\right) }{\nabla _{%
\hat{X}}R\left( \hat{X}\right) }+\alpha \frac{\nabla _{\theta }K_{\hat{X}}}{%
K_{\hat{X}}} \\
\frac{\nabla _{\theta }\nabla _{\hat{X}}g\left( \hat{X}\right) }{\nabla _{%
\hat{X}}g\left( \hat{X}\right) } &=&\frac{\nabla _{\theta }\nabla _{\hat{X}%
}^{2}R\left( \hat{X}\right) }{\nabla _{\hat{X}}^{2}R\left( \hat{X}\right) }%
+\alpha \frac{\nabla _{\theta }K_{\hat{X}}}{K_{\hat{X}}}
\end{eqnarray*}%
and as a consequence:%
\begin{equation*}
\nabla _{\theta }p=-\frac{\nabla _{\theta }\left\vert f\left( \hat{X}\right)
\right\vert }{\left\vert f\left( \hat{X}\right) \right\vert }\left( p+\frac{3%
}{2}\right) -\left( 2\frac{g^{2}\left( \hat{X}\right) \left( \frac{\nabla
_{\theta }\nabla _{\hat{X}}R\left( \hat{X}\right) }{\nabla _{\hat{X}}R\left( 
\hat{X}\right) }+\alpha \frac{\nabla _{\theta }K_{\hat{X}}}{K_{\hat{X}}}%
\right) }{\sigma _{\hat{X}}^{2}\left\vert f\left( \hat{X}\right) \right\vert 
}+\frac{\nabla _{\hat{X}}g\left( \hat{X}\right) }{\left\vert f\left( \hat{X}%
\right) \right\vert }\left( \frac{\nabla _{\theta }\nabla _{\hat{X}%
}^{2}R\left( \hat{X}\right) }{\nabla _{\hat{X}}^{2}R\left( \hat{X}\right) }%
+\alpha \frac{\nabla _{\theta }K_{\hat{X}}}{K_{\hat{X}}}\right) \right)
\end{equation*}%
Ultimately, the right hand side of (\ref{tqd}) is given by:%
\begin{eqnarray*}
&&\left( \frac{\sigma _{X}^{2}\sigma _{\hat{K}}^{2}\left( p+\frac{1}{2}%
\right) \left( f^{\prime }\left( X\right) \right) ^{2}}{48\left\vert f\left( 
\hat{X}\right) \right\vert ^{3}}-1\right) \frac{\nabla _{\theta }\left\vert
f\left( \hat{X}\right) \right\vert }{\left\vert f\left( \hat{X}\right)
\right\vert }+\left( \ln \left( p+\frac{1}{2}\right) -\frac{\sigma
_{X}^{2}\sigma _{\hat{K}}^{2}\left( p+\frac{1}{2}\right) \left( f^{\prime
}\left( X\right) \right) ^{2}}{48\left\vert f\left( \hat{X}\right)
\right\vert ^{3}}\right) \nabla _{\theta }p \\
&&-\frac{\sigma _{X}^{2}\sigma _{\hat{K}}^{2}\left( p+\frac{1}{2}\right) ^{2}%
}{96\left\vert f\left( \hat{X}\right) \right\vert }\nabla _{\theta }\left( 
\frac{f^{\prime }\left( X\right) }{f\left( \hat{X}\right) }\right) ^{2} \\
&=&-\frac{\nabla _{\theta }\left\vert f\left( \hat{X}\right) \right\vert }{%
\left\vert f\left( \hat{X}\right) \right\vert }\left( \left( 1-\frac{\sigma
_{X}^{2}\sigma _{\hat{K}}^{2}\left( p+\frac{1}{2}\right) ^{2}\left(
f^{\prime }\left( X\right) \right) ^{2}}{96\left\vert f\left( \hat{X}\right)
\right\vert ^{3}}\right) +\left( p+\frac{3}{2}\right) \left( \ln \left( p+%
\frac{1}{2}\right) -\frac{\sigma _{X}^{2}\sigma _{\hat{K}}^{2}\left( p+\frac{%
1}{2}\right) \left( f^{\prime }\left( X\right) \right) ^{2}}{48\left\vert
f\left( \hat{X}\right) \right\vert ^{3}}\right) \right) \\
&&-\left( 2\frac{g^{2}\left( \hat{X}\right) }{\sigma _{\hat{X}%
}^{2}\left\vert f\left( \hat{X}\right) \right\vert }\frac{\nabla _{\theta
}\nabla _{\hat{X}}R\left( \hat{X}\right) }{\nabla _{\hat{X}}R\left( \hat{X}%
\right) }+\frac{\nabla _{\hat{X}}g\left( \hat{X}\right) }{\left\vert f\left( 
\hat{X}\right) \right\vert }\frac{\nabla _{\theta }\nabla _{\hat{X}%
}^{2}R\left( \hat{X}\right) }{\nabla _{\hat{X}}^{2}R\left( \hat{X}\right) }+%
\frac{\alpha \left( 2\frac{g^{2}\left( \hat{X}\right) }{\sigma _{\hat{X}}^{2}%
}+\nabla _{\hat{X}}g\left( \hat{X}\right) \right) }{\left\vert f\left( \hat{X%
}\right) \right\vert }\frac{\nabla _{\theta }K_{\hat{X}}}{K_{\hat{X}}}\right)
\\
&&\times \left( \ln \left( p+\frac{1}{2}\right) -\frac{\sigma _{X}^{2}\sigma
_{\hat{K}}^{2}\left( p+\frac{1}{2}\right) \left( f^{\prime }\left( X\right)
\right) ^{2}}{48\left\vert f\left( \hat{X}\right) \right\vert ^{3}}\right) -%
\frac{\sigma _{X}^{2}\sigma _{\hat{K}}^{2}\left( p+\frac{1}{2}\right) ^{2}}{%
96\left\vert f\left( \hat{X}\right) \right\vert }\nabla _{\theta }\left( 
\frac{f^{\prime }\left( X\right) }{f\left( \hat{X}\right) }\right) ^{2}
\end{eqnarray*}%
so that the variational equation for $K_{\hat{X}}$ (\ref{tqd}) writes:%
\begin{eqnarray}
&&\left( 1-\eta \frac{D-\left\Vert \Psi \left( \hat{X}\right) \right\Vert
^{2}}{\left\Vert \Psi \left( \hat{X}\right) \right\Vert ^{2}}\right) \frac{%
\nabla _{\theta }K_{\hat{X}}}{K_{\hat{X}}}-2\frac{D-\left\Vert \Psi \left( 
\hat{X}\right) \right\Vert ^{2}}{\left\Vert \Psi \left( \hat{X}\right)
\right\Vert ^{2}}\frac{\nabla _{X}\left( \nabla _{\theta }R\left( X\right)
\right) }{\nabla _{X}R\left( X\right) }  \label{vrq} \\
&=&-C_{3}\left( p,\hat{X}\right) \frac{\nabla _{\theta }\left\vert f\left( 
\hat{X}\right) \right\vert }{\left\vert f\left( \hat{X}\right) \right\vert }%
-C_{1}\left( p,\hat{X}\right) \frac{\nabla _{\theta }\left( \frac{f^{\prime
}\left( X\right) }{f\left( \hat{X}\right) }\right) ^{2}}{\left( \frac{%
f^{\prime }\left( X\right) }{f\left( \hat{X}\right) }\right) ^{2}}  \notag \\
&&-C_{2}\left( p,\hat{X}\right) \left( 2\frac{g^{2}\left( \hat{X}\right) }{%
\sigma _{\hat{X}}^{2}\left\vert f\left( \hat{X}\right) \right\vert }\frac{%
\nabla _{\theta }\nabla _{\hat{X}}R\left( \hat{X}\right) }{\nabla _{\hat{X}%
}R\left( \hat{X}\right) }+\frac{\nabla _{\hat{X}}g\left( \hat{X}\right) }{%
\left\vert f\left( \hat{X}\right) \right\vert }\frac{\nabla _{\theta }\nabla
_{\hat{X}}^{2}R\left( \hat{X}\right) }{\nabla _{\hat{X}}^{2}R\left( \hat{X}%
\right) }+\frac{\alpha \left( 2\frac{g^{2}\left( \hat{X}\right) }{\sigma _{%
\hat{X}}^{2}}+\nabla _{\hat{X}}g\left( \hat{X}\right) \right) }{\left\vert
f\left( \hat{X}\right) \right\vert }\frac{\nabla _{\theta }K_{\hat{X}}}{K_{%
\hat{X}}}\right)  \notag
\end{eqnarray}%
with:%
\begin{eqnarray}
C_{1}\left( p,\hat{X}\right) &=&\frac{\sigma _{X}^{2}\sigma _{\hat{K}%
}^{2}\left( p+\frac{1}{2}\right) ^{2}\left( f^{\prime }\left( X\right)
\right) ^{2}}{96\left\vert f\left( \hat{X}\right) \right\vert ^{3}}
\label{Cqt} \\
C_{2}\left( p,\hat{X}\right) &=&\ln \left( p+\frac{1}{2}\right) -\frac{%
2C_{1}\left( p,\hat{X}\right) }{p+\frac{1}{2}}  \notag \\
C_{3}\left( p,\hat{X}\right) &=&1-C_{1}\left( p,\hat{X}\right) +\left( p+%
\frac{3}{2}\right) C_{2}\left( p,\hat{X}\right)  \notag
\end{eqnarray}%
These term can be reordered and the general dynamic equation for $K_{\hat{X}%
} $ is ultimately written as:%
\begin{eqnarray}
&&\left( 1-\eta \frac{D-\left\Vert \Psi \left( \hat{X}\right) \right\Vert
^{2}}{\left\Vert \Psi \left( \hat{X}\right) \right\Vert ^{2}}+\frac{\alpha
\left( 2\frac{g^{2}\left( \hat{X}\right) }{\sigma _{\hat{X}}^{2}}+\nabla _{%
\hat{X}}g\left( \hat{X}\right) \right) }{\left\vert f\left( \hat{X}\right)
\right\vert }C_{2}\left( p,\hat{X}\right) \right) \frac{\nabla _{\theta }K_{%
\hat{X}}}{K_{\hat{X}}}  \label{gnf} \\
&&+2\left( \frac{g^{2}\left( \hat{X}\right) C_{2}\left( p,\hat{X}\right) }{%
\sigma _{\hat{X}}^{2}\left\vert f\left( \hat{X}\right) \right\vert }-\frac{%
D-\left\Vert \Psi \left( \hat{X}\right) \right\Vert ^{2}}{\left\Vert \Psi
\left( \hat{X}\right) \right\Vert ^{2}}\right) \frac{\nabla _{\theta }\nabla
_{\hat{X}}R\left( \hat{X}\right) }{\nabla _{\hat{X}}R\left( \hat{X}\right) }+%
\frac{\nabla _{\hat{X}}g\left( \hat{X}\right) C_{2}\left( p,\hat{X}\right) }{%
\left\vert f\left( \hat{X}\right) \right\vert }\frac{\nabla _{\theta }\nabla
_{\hat{X}}^{2}R\left( \hat{X}\right) }{\nabla _{\hat{X}}^{2}R\left( \hat{X}%
\right) }  \notag \\
&=&-C_{3}\left( p,\hat{X}\right) \frac{\nabla _{\theta }\left\vert f\left( 
\hat{X}\right) \right\vert }{\left\vert f\left( \hat{X}\right) \right\vert }%
-C_{1}\left( p,\hat{X}\right) \frac{\nabla _{\theta }\left( \frac{f^{\prime
}\left( X\right) }{f\left( \hat{X}\right) }\right) ^{2}}{\left( \frac{%
f^{\prime }\left( X\right) }{f\left( \hat{X}\right) }\right) ^{2}}  \notag
\end{eqnarray}

\subsubsection*{A4.1.3 Dynamic equation for particular forms of $f\left( 
\hat{X},K_{\hat{X}}\right) $ and $\left\Vert \Psi \left( \hat{X}\right)
\right\Vert ^{2}$\protect\bigskip}

We can put equation (\ref{gnf}) in a specific form, by using the explicit
formula for $f\left( \hat{X},K_{\hat{X}}\right) $ and $\left\Vert \Psi
\left( \hat{X}\right) \right\Vert ^{2}$\ given in appendix 2. We have:

\begin{eqnarray*}
\frac{\nabla _{\theta }f\left( \hat{X},K_{\hat{X}}\right) }{f\left( \hat{X}%
,K_{\hat{X}}\right) } &\simeq &\frac{r\left( K_{\hat{X}},\hat{X}\right)
\left( \frac{\nabla _{\theta }r\left( K_{\hat{X}},\hat{X}\right) }{r\left(
K_{\hat{X}},\hat{X}\right) }+\left( \alpha -1\right) \frac{\nabla _{\theta
}K_{\hat{X}}}{K_{\hat{X}}}\right) }{f\left( \hat{X}\right) } \\
&&+\frac{\gamma \left( \eta L\left( \hat{X}\right) K_{\hat{X}}^{\eta }\frac{%
\nabla _{\theta }K_{\hat{X}}}{K_{\hat{X}}}+2L\left( \hat{X}\right) K_{\hat{X}%
}^{\eta }\frac{\nabla _{\theta }\left( \nabla _{X}R\left( X\right) \right) }{%
\nabla _{X}R\left( X\right) }\right) +F_{1}^{\prime }\left( R\left( K_{\hat{X%
}},\hat{X}\right) \right) \frac{\nabla _{\theta }R\left( K_{\hat{X}},\hat{X}%
\right) }{R\left( K_{\hat{X}},\hat{X}\right) }}{f\left( \hat{X}\right) } \\
&\simeq &\frac{r\left( K_{\hat{X}},\hat{X}\right) \left( \frac{\nabla
_{\theta }r\left( \hat{X}\right) }{r\left( K_{\hat{X}},\hat{X}\right) }%
+\left( \alpha -1\right) \frac{\nabla _{\theta }K_{\hat{X}}}{K_{\hat{X}}}%
\right) }{f\left( \hat{X}\right) } \\
&&+\frac{\gamma \left( \eta L\left( \hat{X}\right) K_{\hat{X}}^{\eta }\frac{%
\nabla _{\theta }K_{\hat{X}}}{K_{\hat{X}}}+2L\left( \hat{X}\right) K_{\hat{X}%
}^{\eta }\frac{\nabla _{\theta }\left( \nabla _{X}R\left( X\right) \right) }{%
\nabla _{X}R\left( X\right) }\right) +\varsigma F_{1}\left( R\left( K_{\hat{X%
}},\hat{X}\right) \right) \frac{\nabla _{\theta }R\left( K_{\hat{X}},\hat{X}%
\right) }{R\left( K_{\hat{X}},\hat{X}\right) }}{f\left( \hat{X}\right) } \\
&=&\frac{r\left( K_{\hat{X}},\hat{X}\right) \frac{\nabla _{\theta }r\left( 
\hat{X}\right) }{r\left( K_{\hat{X}},\hat{X}\right) }}{f\left( \hat{X}%
\right) } \\
&&+\frac{\left( \gamma \eta L\left( \hat{X}\right) K_{\hat{X}}^{\eta
}+\left( \alpha -1\right) \right) \frac{\nabla _{\theta }K_{\hat{X}}}{K_{%
\hat{X}}}+\varsigma F_{1}\left( R\left( K_{\hat{X}},\hat{X}\right) \right) 
\frac{\nabla _{\theta }R\left( K_{\hat{X}},\hat{X}\right) }{R\left( K_{\hat{X%
}},\hat{X}\right) }+2\gamma L\left( \hat{X}\right) K_{\hat{X}}^{\eta }\frac{%
\nabla _{\theta }\left( \nabla _{X}R\left( X\right) \right) }{\nabla
_{X}R\left( X\right) }}{f\left( \hat{X}\right) }
\end{eqnarray*}%
To compute $\nabla _{\theta }\ln \left( \frac{f^{\prime }\left( X\right) }{%
f\left( \hat{X}\right) }\right) ^{2}$ arising in (\ref{gnf}), we use that in
first approximation, for relatively large $K_{\hat{X}}$:%
\begin{equation*}
\left( \frac{f^{\prime }\left( X\right) }{f\left( \hat{X}\right) }\right)
^{2}\simeq \left( \frac{F_{1}^{\prime }\left( R\left( K_{\hat{X}},\hat{X}%
\right) \right) \nabla _{\hat{X}}R\left( K_{\hat{X}},\hat{X}\right) }{%
F_{1}\left( R\left( K_{\hat{X}},\hat{X}\right) \right) }\right) ^{2}\simeq
\left( \varsigma \nabla _{\hat{X}}R\left( K_{\hat{X}},\hat{X}\right) \right)
^{2}
\end{equation*}%
that can be considered int the sequel negligible at the first order.

As a consequence, for the chosen forms of the parameter functions, the
dynamics equation (\ref{gnf}) becomes ultimately:%
\begin{equation}
k\frac{\nabla _{\theta }K_{\hat{X}}}{K_{\hat{X}}}+l\frac{\nabla _{\theta
}R\left( \hat{X}\right) }{R\left( \hat{X}\right) }-2m\frac{\nabla _{\hat{X}%
}\nabla _{\theta }R\left( \hat{X}\right) }{\nabla _{\hat{X}}R\left( \hat{X}%
\right) }+n\frac{\nabla _{\hat{X}}^{2}\nabla _{\theta }R\left( \hat{X}%
\right) }{\nabla _{\hat{X}}^{2}R\left( \hat{X}\right) }=-C_{3}\left( p,\hat{X%
}\right) \frac{\nabla _{\theta }r\left( \hat{X}\right) }{f\left( \hat{X}%
\right) }  \label{cnd}
\end{equation}%
with:%
\begin{eqnarray}
k &=&1-\eta \left( 1-\frac{\gamma C_{3}\left( p,\hat{X}\right) }{\left\vert
f\left( \hat{X}\right) \right\vert }\right) \frac{D-\left\Vert \Psi \left( 
\hat{X}\right) \right\Vert ^{2}}{\left\Vert \Psi \left( \hat{X}\right)
\right\Vert ^{2}}  \label{Cfn} \\
&&+\frac{\alpha \left( 2\frac{g^{2}\left( \hat{X}\right) }{\sigma _{\hat{X}%
}^{2}}+\nabla _{\hat{X}}g\left( \hat{X}\right) \right) C_{2}\left( p,\hat{X}%
\right) -\left( 1-\alpha \right) C_{3}\left( p,\hat{X}\right) }{\left\vert
f\left( \hat{X}\right) \right\vert }  \notag \\
l &=&\frac{\varsigma F_{1}\left( R\left( K_{\hat{X}},\hat{X}\right) \right)
C_{3}\left( p,\hat{X}\right) }{f\left( \hat{X}\right) }  \notag \\
m &=&\left( 1-\frac{\gamma C_{3}\left( p,\hat{X}\right) }{f\left( \hat{X}%
\right) }\right) \frac{D-\left\Vert \Psi \left( \hat{X}\right) \right\Vert
^{2}}{\left\Vert \Psi \left( \hat{X}\right) \right\Vert ^{2}}-\frac{%
g^{2}\left( \hat{X}\right) C_{2}\left( p,\hat{X}\right) }{\sigma _{\hat{X}%
}^{2}}  \notag \\
n &=&\frac{\nabla _{\hat{X}}g\left( \hat{X}\right) C_{2}\left( p,\hat{X}%
\right) }{\left\vert f\left( \hat{X}\right) \right\vert }  \notag
\end{eqnarray}

\subsection*{A4.2 Full dynamical system}

To make the system self-consistent, we introduce also a dynamics for $R$.

We assume that $R$ depends on $K_{\hat{X}},\hat{X}$ and $\nabla _{\theta }K_{%
\hat{X}}$, that leads to write: $R\left( K_{\hat{X}},\hat{X},\nabla _{\theta
}K_{\hat{X}}\right) $. The variation is assumed to follow a diffusion
process:%
\begin{eqnarray*}
\nabla _{\theta }R\left( \theta ,\hat{X}\right) &=&\int_{\theta ^{\prime
}<\theta }G_{1}\left( \left( \theta ,\hat{X}\right) ,\left( \theta ^{\prime
},\hat{X}^{\prime }\right) \right) \nabla _{\theta ^{\prime }}R\left( \theta
^{\prime },\hat{X}^{\prime }\right) d\left( \theta ^{\prime },\hat{X}%
^{\prime }\right) \\
&&+\int_{\theta ^{\prime }<\theta }G_{2}\left( \left( \theta ,\hat{X}\right)
,\left( \theta ^{\prime },\hat{X}^{\prime }\right) \right) \nabla _{\theta
^{\prime }}K_{\hat{X}^{\prime }}d\left( \theta ^{\prime },\hat{X}^{\prime
}\right)
\end{eqnarray*}%
The first orders expansion of the right hand side leads to the following
form for $\nabla _{\theta }R\left( \theta ,\hat{X}\right) $:

\begin{eqnarray}
\nabla _{\theta }R\left( \theta ,\hat{X}\right) &=&\int \left( \hat{X}-\hat{X%
}^{\prime }\right) \left( G_{1}\left( \left( \theta ,\hat{X}\right) ,\left(
\theta ^{\prime },\hat{X}^{\prime }\right) \right) \nabla _{\hat{X}}\nabla
_{\theta }R\left( \theta ,\hat{X}\right) +G_{2}\left( \left( \theta ,\hat{X}%
\right) ,\left( \theta ^{\prime },\hat{X}^{\prime }\right) \right) \nabla _{%
\hat{X}}\nabla _{\theta }K_{\hat{X}}\right)  \label{dtr} \\
&&+\frac{1}{2}\int \left( \hat{X}-\hat{X}^{\prime }\right) ^{2}\left(
G_{1}\left( \left( \theta ,\hat{X}\right) ,\left( \theta ^{\prime },\hat{X}%
^{\prime }\right) \right) \nabla _{\hat{X}}^{2}\nabla _{\theta }R\left(
\theta ,\hat{X}\right) +G_{2}\left( \left( \theta ,\hat{X}\right) ,\left(
\theta ^{\prime },\hat{X}^{\prime }\right) \right) \nabla _{\hat{X}%
}^{2}\nabla _{\theta }K_{\hat{X}}\right)  \notag \\
&&+\int \left( \theta -\theta ^{\prime }\right) \left( G_{1}\left( \left(
\theta ,\hat{X}\right) ,\left( \theta ^{\prime },\hat{X}^{\prime }\right)
\right) \nabla _{\theta }\nabla _{\theta }R\left( \theta ,\hat{X}\right)
+G_{2}\left( \left( \theta ,\hat{X}\right) ,\left( \theta ^{\prime },\hat{X}%
^{\prime }\right) \right) \nabla _{\theta }\nabla _{\theta }K_{\hat{X}%
}\right)  \notag \\
&&+\frac{1}{2}\int \left( \theta -\theta ^{\prime }\right) ^{2}\left(
G_{1}\left( \left( \theta ,\hat{X}\right) ,\left( \theta ^{\prime },\hat{X}%
^{\prime }\right) \right) \nabla _{\theta }^{2}\nabla _{\theta }R\left(
\theta ,\hat{X}\right) +G_{2}\left( \left( \theta ,\hat{X}\right) ,\left(
\theta ^{\prime },\hat{X}^{\prime }\right) \right) \nabla _{\theta
}^{2}\nabla _{\theta }K_{\hat{X}}\right)  \notag \\
&&+...  \notag
\end{eqnarray}%
where the crossed derivatives have been discarded for the sake of
simplicity. We assume $G_{1}\left( \left( \theta ,\hat{X}\right) ,\left(
\theta ,\hat{X}\right) \right) =0$ to avoid auto-interaction.

Performing the integrals yields:

\begin{eqnarray}
\nabla _{\theta }R\left( \theta ,\hat{X}\right) &=&a_{0}\left( \hat{X}%
\right) \nabla _{\theta }K_{\hat{X}}+a\left( \hat{X}\right) \nabla _{\hat{X}%
}\nabla _{\theta }K_{\hat{X}}+b\left( \hat{X}\right) \nabla _{\hat{X}%
}^{2}\nabla _{\theta }K_{\hat{X}}  \label{dtR} \\
&&+c\left( \hat{X}\right) \nabla _{\theta }\left( \nabla _{\theta }K_{\hat{X}%
}\right) +d\left( \hat{X}\right) \nabla _{\theta }^{2}\left( \nabla _{\theta
}K_{\hat{X}}\right)  \notag \\
&&+e\left( \hat{X}\right) \nabla _{\hat{X}}\left( \nabla _{\theta }R\left(
\theta ,\hat{X}\right) \right) +f\left( \hat{X}\right) \nabla _{\hat{X}%
}^{2}\left( \nabla _{\theta }R\left( \theta ,\hat{X}\right) \right)  \notag
\\
&&+g\left( \hat{X}\right) \nabla _{\theta }\left( \nabla _{\theta }R\left(
\theta ,\hat{X}\right) \right) +h\left( \hat{X}\right) \nabla _{\theta
}^{2}\left( \nabla _{\theta }R\left( \theta ,\hat{X}\right) \right)  \notag
\\
&&+u\left( \hat{X}\right) \nabla _{\hat{X}}\nabla _{\theta }\left( \nabla
_{\theta }K_{\hat{X}}\right) +v\left( \hat{X}\right) \nabla _{\hat{X}}\nabla
_{\theta }\left( \nabla _{\theta }R\left( \theta ,\hat{X}\right) \right) 
\notag
\end{eqnarray}%
We ssume that the coefficients are slowly varying, since their are obtained
by averages.

Gathering the dynamics (\ref{cnd}) and (\ref{dtR})\ for $\nabla _{\theta }K_{%
\hat{X}}$ and $\nabla _{\theta }R\left( \theta ,\hat{X}\right) $ leads to a
matricial system: 
\begin{eqnarray}
&&0=\left( 
\begin{array}{cc}
\frac{k}{K_{\hat{X}}} & \frac{l}{R\left( \hat{X}\right) } \\ 
-a_{0}\left( \hat{X}\right) & 1%
\end{array}%
\right) \left( 
\begin{array}{c}
\nabla _{\theta }K_{\hat{X}} \\ 
\nabla _{\theta }R%
\end{array}%
\right)  \label{mtr} \\
&&-\left( 
\begin{array}{cc}
0 & \frac{2m}{\nabla _{\hat{X}}R\left( \hat{X}\right) }\nabla _{\hat{X}} \\ 
a\left( \hat{X}\right) \nabla _{\hat{X}}+c\left( \hat{X}\right) \nabla
_{\theta } & e\left( \hat{X}\right) \nabla _{\hat{X}}+g\left( \hat{X}\right)
\nabla _{\theta }%
\end{array}%
\right) \left( 
\begin{array}{c}
\nabla _{\theta }K_{\hat{X}} \\ 
\nabla _{\theta }R%
\end{array}%
\right)  \notag \\
&&-\left( 
\begin{array}{cc}
0 & -\frac{n}{\nabla _{\hat{X}}^{2}R\left( \hat{X}\right) }\nabla _{\hat{X}%
}^{2} \\ 
d\left( \hat{X}\right) \nabla _{\theta }^{2}+b\left( \hat{X}\right) \nabla _{%
\hat{X}}^{2}+u\left( \hat{X}\right) \nabla _{\hat{X}}\nabla _{\theta } & 
e\left( \hat{X}\right) \nabla _{\theta }^{2}+f\left( \hat{X}\right) \nabla _{%
\hat{X}}^{2}+v\nabla _{\hat{X}}\nabla _{\theta }%
\end{array}%
\right) \left( 
\begin{array}{c}
\nabla _{\theta }K_{\hat{X}} \\ 
\nabla _{\theta }R%
\end{array}%
\right)  \notag
\end{eqnarray}

\subsection*{A4.3 Oscillatory solutions\protect\bigskip}

We look for a solution of (\ref{drk}) of the form:%
\begin{equation*}
\left( 
\begin{array}{c}
\nabla _{\theta }K_{\hat{X}} \\ 
\nabla _{\theta }R\left( \hat{X}\right)%
\end{array}%
\right) =\exp \left( i\Omega \left( \hat{X}\right) \theta +iG\left( \hat{X}%
\right) \hat{X}\right) \left( 
\begin{array}{c}
\nabla _{\theta }K_{0} \\ 
\nabla _{\theta }R_{0}%
\end{array}%
\right)
\end{equation*}%
with $G\left( \hat{X}\right) $ and $\Omega \left( \hat{X}\right) $ slowly
varying. As a consequence, the system (\ref{mtr}) writes:%
\begin{equation}
\left( 
\begin{array}{cc}
\frac{k}{K_{\hat{X}}} & \frac{l}{R\left( \hat{X}\right) }-i\frac{2m}{\nabla
_{\hat{X}}R\left( \hat{X}\right) }G-\frac{n}{\nabla _{\hat{X}}^{2}R\left( 
\hat{X}\right) }G^{2} \\ 
\begin{array}{c}
-a_{0}\left( \hat{X}\right) -ia\left( \hat{X}\right) G-ic\left( \hat{X}%
\right) \Omega \\ 
+d\Omega ^{2}+bG^{2}+u\Omega G%
\end{array}
& 
\begin{array}{c}
1-ie\left( \hat{X}\right) G-ig\left( \hat{X}\right) \Omega +e\Omega ^{2} \\ 
+fG^{2}+u\Omega G%
\end{array}%
\end{array}%
\right) \left( 
\begin{array}{c}
\nabla _{\theta }K_{\hat{X}} \\ 
\nabla _{\theta }R%
\end{array}%
\right) =0  \label{drk}
\end{equation}%
By canceling the determinant of the system, we are led to the following
relation between $\Omega \left( \hat{X}\right) $ and $G\left( \hat{X}\right) 
$: 
\begin{eqnarray*}
0 &=&\frac{k}{K_{\hat{X}}}\left( 1-ieG-ig\Omega \right) +\left( \frac{l}{%
R\left( \hat{X}\right) }-i\frac{2m}{\nabla _{\hat{X}}R\left( \hat{X}\right) }%
G\right) \left( a_{0}+iaG+ic\Omega \right) \\
&&-\frac{l}{R\left( \hat{X}\right) }\left( d\Omega ^{2}+bG^{2}+u\Omega
G\right) +\frac{k}{K_{\hat{X}}}\left( e\Omega ^{2}+fG^{2}+v\Omega G\right)
\end{eqnarray*}%
In the sequel, we restrict to the first order terms, which yields the
expression for $\Omega $:%
\begin{eqnarray*}
\Omega &=&\frac{i}{\left( \frac{lc}{R\left( \hat{X}\right) }-i\frac{2mc}{%
\nabla _{\hat{X}}R\left( \hat{X}\right) }G\right) -\frac{kg}{K_{\hat{X}}}}%
\left( \frac{k}{K_{\hat{X}}}\left( 1-ieG\right) +\left( \frac{l}{R\left( 
\hat{X}\right) }-i\frac{2m}{\nabla _{\hat{X}}R\left( \hat{X}\right) }%
G\right) \left( a_{0}+iaG\right) \right) \\
&=&\frac{\left( \frac{lc}{R\left( \hat{X}\right) }-\frac{kg}{K_{\hat{X}}}%
\right) +i\frac{2mc}{\nabla _{\hat{X}}R\left( \hat{X}\right) }G}{\left( 
\frac{lc}{R\left( \hat{X}\right) }-\frac{kg}{K_{\hat{X}}}\right) ^{2}+\left( 
\frac{2mc}{\nabla _{\hat{X}}R\left( \hat{X}\right) }G\right) ^{2}} \\
&&\times \left( \left( \frac{ke}{K_{\hat{X}}}+\left( \frac{2ma_{0}}{\nabla _{%
\hat{X}}R\left( \hat{X}\right) }-\frac{la}{R\left( \hat{X}\right) }\right)
\right) G+i\left( \frac{k}{K_{\hat{X}}}+\frac{a_{0}l}{R\left( \hat{X}\right) 
}+\frac{2ma}{\nabla _{\hat{X}}R\left( \hat{X}\right) }G^{2}\right) \right)
\end{eqnarray*}%
Or equivalently:%
\begin{eqnarray*}
&&\Omega =\frac{\left( \frac{lc}{R\left( \hat{X}\right) }-\frac{kg}{K_{\hat{X%
}}}\right) \left( \frac{ke}{K_{\hat{X}}}+\left( \frac{2ma_{0}}{\nabla _{\hat{%
X}}R\left( \hat{X}\right) }-\frac{la}{R\left( \hat{X}\right) }\right)
\right) G-\frac{2mc}{\nabla _{\hat{X}}R\left( \hat{X}\right) }G\left( \frac{k%
}{K_{\hat{X}}}+\frac{a_{0}l}{R\left( \hat{X}\right) }+\frac{2ma}{\nabla _{%
\hat{X}}R\left( \hat{X}\right) }G^{2}\right) }{\left( \frac{lc}{R\left( \hat{%
X}\right) }-\frac{kg}{K_{\hat{X}}}\right) ^{2}+\left( \frac{2mc}{\nabla _{%
\hat{X}}R\left( \hat{X}\right) }G\right) ^{2}} \\
&&+i\frac{\left( \frac{lc}{R\left( \hat{X}\right) }-\frac{kg}{K_{\hat{X}}}%
\right) \left( \frac{k}{K_{\hat{X}}}+\frac{a_{0}l}{R\left( \hat{X}\right) }+%
\frac{2ma}{\nabla _{\hat{X}}R\left( \hat{X}\right) }G^{2}\right) +\frac{2mc}{%
\nabla _{\hat{X}}R\left( \hat{X}\right) }\left( \frac{ke}{K_{\hat{X}}}%
+\left( \frac{2ma_{0}}{\nabla _{\hat{X}}R\left( \hat{X}\right) }-\frac{la}{%
R\left( \hat{X}\right) }\right) \right) G^{2}}{\left( \frac{lc}{R\left( \hat{%
X}\right) }-\frac{kg}{K_{\hat{X}}}\right) ^{2}+\left( \frac{2mc}{\nabla _{%
\hat{X}}R\left( \hat{X}\right) }G\right) ^{2}}
\end{eqnarray*}%
We focus on the influence of time variations of $\nabla _{\theta }K_{\hat{X}%
} $ \ on $\nabla _{\theta }R$, and we can assume $g\simeq 0$ so that there
is no self influence of $\nabla _{\theta }R$ on itself: $\nabla _{\theta }R$
depends on the variations of $\nabla _{\theta }K_{\hat{X}}$ as well as the
neighboorhood sectors variations of $\nabla _{\theta }R$. Moreover, the
coefficients $e$ and $a$, being obtained by integration or first order
expansion, can be considered as nul.

As a consequence, the equation for $\Omega $ reduces to:%
\begin{equation*}
\Omega =\frac{\frac{lc}{R\left( \hat{X}\right) }\left( \frac{2ma_{0}}{\nabla
_{\hat{X}}R\left( \hat{X}\right) }\right) G-\frac{2mc}{\nabla _{\hat{X}%
}R\left( \hat{X}\right) }G\left( \frac{k}{K_{\hat{X}}}+\frac{a_{0}l}{R\left( 
\hat{X}\right) }\right) }{\left( \frac{lc}{R\left( \hat{X}\right) }\right)
^{2}+\left( \frac{2mc}{\nabla _{\hat{X}}R\left( \hat{X}\right) }G\right) ^{2}%
}+i\frac{\frac{lc}{R\left( \hat{X}\right) }\left( \frac{k}{K_{\hat{X}}}+%
\frac{a_{0}l}{R\left( \hat{X}\right) }\right) +\frac{2mc}{\nabla _{\hat{X}%
}R\left( \hat{X}\right) }\left( \frac{2ma_{0}}{\nabla _{\hat{X}}R\left( \hat{%
X}\right) }\right) G^{2}}{\left( \frac{lc}{R\left( \hat{X}\right) }\right)
^{2}+\left( \frac{2mc}{\nabla _{\hat{X}}R\left( \hat{X}\right) }G\right) ^{2}%
}
\end{equation*}

\subsection*{A4.4 Stability}

The system is stable and the dynamics is dampening if:%
\begin{equation}
\frac{lc}{R\left( \hat{X}\right) }\left( \frac{k}{K_{\hat{X}}}+\frac{a_{0}l}{%
R\left( \hat{X}\right) }\right) +\frac{4m^{2}ca_{0}}{\left( \nabla _{\hat{X}%
}R\left( \hat{X}\right) \right) ^{2}}G^{2}>0  \label{stB}
\end{equation}%
To study the sign of (\ref{stB}) we need to estimate the coefficient $k$.

\subsubsection*{A4.4.1 Estimation of the coefficients $k$, $l$ and $m$}

We can estimate $k$ and $l$ by computing the factors $C_{i}\left( p,\hat{X}%
\right) $, for $i=1,2,3$.

This is done by estimating $p+\frac{1}{2}$. We start with the asymptotic
form of $\hat{\Gamma}\left( p+\frac{1}{2}\right) $:

\begin{equation*}
\hat{\Gamma}\left( p+\frac{1}{2}\right) \simeq \sqrt{p+\frac{1}{2}}\exp
\left( -\frac{\sigma _{X}^{2}\left( p+\frac{1}{2}\right) ^{2}\left(
f^{\prime }\left( X\right) \right) ^{2}}{96\left\vert f\left( \hat{X}\right)
\right\vert ^{3}}\right)
\end{equation*}%
and rewriting the equation for $K_{\hat{X}}$ as: 
\begin{equation}
K_{\hat{X}}\left\Vert \Psi \left( \hat{X}\right) \right\Vert ^{2}\left\vert
f\left( \hat{X}\right) \right\vert \left( \frac{\sigma _{X}^{2}\left(
f^{\prime }\left( X\right) \right) ^{2}}{96\left\vert f\left( \hat{X}\right)
\right\vert ^{3}}\right) ^{\frac{1}{4}}=C\left( \bar{p}\right) \sigma _{\hat{%
K}}^{2}\exp \left( -\frac{\sigma _{X}^{2}\left( p+\frac{1}{2}\right)
^{2}\left( f^{\prime }\left( X\right) \right) ^{2}}{96\left\vert f\left( 
\hat{X}\right) \right\vert ^{3}}\right) \sqrt{\left( p+\frac{1}{2}\right) 
\sqrt{\frac{\sigma _{X}^{2}\left( f^{\prime }\left( X\right) \right) ^{2}}{%
96\left\vert f\left( \hat{X}\right) \right\vert ^{3}}}}  \label{vrK}
\end{equation}%
Then, using (\ref{Cqt}), we set:%
\begin{equation}
\left( p+\frac{1}{2}\right) \sqrt{\frac{\sigma _{X}^{2}\left( f^{\prime
}\left( X\right) \right) ^{2}}{96\left\vert f\left( \hat{X}\right)
\right\vert ^{3}}}=\sqrt{C_{1}\left( p,\hat{X}\right) }
\end{equation}%
Equation (\ref{vrK}) writes:%
\begin{equation}
\frac{K_{\hat{X}}\left\Vert \Psi \left( \hat{X}\right) \right\Vert
^{2}\left\vert f\left( \hat{X}\right) \right\vert \left( \frac{\sigma
_{X}^{2}\left( f^{\prime }\left( X\right) \right) ^{2}}{96\left\vert f\left( 
\hat{X}\right) \right\vert ^{3}}\right) ^{\frac{1}{4}}}{C\left( \bar{p}%
\right) \sigma _{\hat{K}}^{2}}=\exp \left( -C_{1}\left( p,\hat{X}\right)
\right) \left( C_{1}\left( p,\hat{X}\right) \right) ^{\frac{1}{4}}
\label{Cpq}
\end{equation}%
$\allowbreak $and the solution to (\ref{Cpq}) is:%
\begin{eqnarray}
C_{1}\left( p,\hat{X}\right) &=&\frac{\sigma _{X}^{2}\left( p+\frac{1}{2}%
\right) ^{2}\left( f^{\prime }\left( X\right) \right) ^{2}}{96\left\vert
f\left( \hat{X}\right) \right\vert ^{3}}  \label{Cpo} \\
&=&C_{0}\left( \hat{X},K_{\hat{X}}\right) \exp \left( -W\left(
k,-4C_{0}\left( \hat{X},K_{\hat{X}}\right) \right) \right)  \notag
\end{eqnarray}%
with:%
\begin{equation*}
C_{0}\left( \hat{X},K_{\hat{X}}\right) =\left( \frac{K_{\hat{X}}\left\Vert
\Psi \left( \hat{X}\right) \right\Vert ^{2}\left\vert f\left( \hat{X}\right)
\right\vert }{C\left( \bar{p}\right) \sigma _{\hat{K}}^{2}}\right) ^{4}\frac{%
\sigma _{X}^{2}\left( f^{\prime }\left( X\right) \right) ^{2}}{96\left\vert
f\left( \hat{X}\right) \right\vert ^{3}}
\end{equation*}%
and where $W\left( k,x\right) $ is the Lambert $W$ function. The parameter $%
k=0$ for the stable case with low $K_{\hat{X}}$ and $k=-1$ for the unstable
case with $K_{\hat{X}}$ large.

We can deduce $p+\frac{1}{2}$ from (\ref{Cpo}): 
\begin{equation}
p+\frac{1}{2}=\frac{\sqrt{C_{1}\left( p,\hat{X}\right) }}{\sqrt{\frac{\sigma
_{X}^{2}\left( f^{\prime }\left( X\right) \right) ^{2}}{96\left\vert f\left( 
\hat{X}\right) \right\vert ^{3}}}}  \label{Pon}
\end{equation}%
$\allowbreak $

and $2\frac{C_{1}\left( p,\hat{X}\right) }{p+\frac{1}{2}}$:%
\begin{equation*}
2\frac{C_{1}\left( p,\hat{X}\right) }{p+\frac{1}{2}}=\sqrt{C_{1}\left( p,%
\hat{X}\right) }\frac{\sigma _{X}^{2}\left( f^{\prime }\left( X\right)
\right) ^{2}}{48\left\vert f\left( \hat{X}\right) \right\vert ^{3}}
\end{equation*}%
From (\ref{Pon}) and (\ref{Cpd}) we deduce:%
\begin{eqnarray}
C_{2}\left( p,\hat{X}\right) &=&\ln \left( p+\frac{1}{2}\right) -\frac{%
2C_{1}\left( p,\hat{X}\right) }{p+\frac{1}{2}}  \label{CpD} \\
&=&\frac{1}{2}\ln \frac{C_{1}\left( p,\hat{X}\right) }{\frac{\sigma
_{X}^{2}\left( f^{\prime }\left( X\right) \right) ^{2}}{96\left\vert f\left( 
\hat{X}\right) \right\vert ^{3}}}-\sqrt{C_{1}\left( p,\hat{X}\right) }\frac{%
\sigma _{X}^{2}\left( f^{\prime }\left( X\right) \right) ^{2}}{48\left\vert
f\left( \hat{X}\right) \right\vert ^{3}}\simeq \frac{1}{2}\ln \frac{%
96\left\vert f\left( \hat{X}\right) \right\vert ^{3}}{\sigma _{X}^{2}\left(
f^{\prime }\left( X\right) \right) ^{2}}  \notag
\end{eqnarray}%
We can also compute:%
\begin{eqnarray*}
\left( p+\frac{3}{2}\right) \ln \left( p+\frac{1}{2}\right) &\simeq &\frac{%
48\left\vert f\left( \hat{X}\right) \right\vert ^{3}C_{1}\left( p,\hat{X}%
\right) }{\sigma _{X}^{2}\left( f^{\prime }\left( X\right) \right) ^{2}}\ln 
\frac{96\left\vert f\left( \hat{X}\right) \right\vert ^{3}}{\sigma
_{X}^{2}\left( f^{\prime }\left( X\right) \right) ^{2}} \\
&=&\frac{1}{2}\left( \frac{K_{\hat{X}}\left\Vert \Psi \left( \hat{X}\right)
\right\Vert ^{2}\left\vert f\left( \hat{X}\right) \right\vert }{C\left( \bar{%
p}\right) \sigma _{\hat{K}}^{2}}\right) ^{4}\ln \frac{96\left\vert f\left( 
\hat{X}\right) \right\vert ^{3}}{\sigma _{X}^{2}\left( f^{\prime }\left(
X\right) \right) ^{2}} \\
&&\times \exp \left( -W\left( k,-4C_{0}\left( \hat{X},K_{\hat{X}}\right)
\right) \right)
\end{eqnarray*}%
so that:%
\begin{eqnarray}
C_{3}\left( p,\hat{X}\right) &=&1-C_{1}\left( p,\hat{X}\right) +\left( p+%
\frac{3}{2}\right) C_{2}\left( p,\hat{X}\right)  \label{Cpt} \\
&=&1-C_{1}\left( p,\hat{X}\right) +\left( p+\frac{3}{2}\right) \left( \ln
\left( p+\frac{1}{2}\right) -\frac{2C_{1}\left( p,\hat{X}\right) }{p+\frac{1%
}{2}}\right)  \notag \\
&\simeq &1+\left( \frac{48\left\vert f\left( \hat{X}\right) \right\vert ^{3}%
}{\sigma _{X}^{2}\left( f^{\prime }\left( X\right) \right) ^{2}}\ln \frac{%
96\left\vert f\left( \hat{X}\right) \right\vert ^{3}}{\sigma _{X}^{2}\left(
f^{\prime }\left( X\right) \right) ^{2}}-1\right) C_{1}\left( p,\hat{X}%
\right)  \notag \\
&\simeq &1+\frac{1}{2}\left( \frac{K_{\hat{X}}\left\Vert \Psi \left( \hat{X}%
\right) \right\Vert ^{2}\left\vert f\left( \hat{X}\right) \right\vert }{%
C\left( \bar{p}\right) \sigma _{\hat{K}}^{2}}\right) ^{4}\exp \left(
-W\left( k,-4C_{0}\left( \hat{X},K_{\hat{X}}\right) \right) \right) \ln 
\frac{96\left\vert f\left( \hat{X}\right) \right\vert ^{3}}{\sigma
_{X}^{2}\left( f^{\prime }\left( X\right) \right) ^{2}}  \notag
\end{eqnarray}%
Given that our assumptions $\sigma _{X}^{2}<1$ and in most cases $\frac{%
96\left\vert f\left( \hat{X}\right) \right\vert ^{3}}{\sigma _{X}^{2}\left(
f^{\prime }\left( X\right) \right) ^{2}}>>1$, then $\frac{96\left\vert
f\left( \hat{X}\right) \right\vert ^{3}}{\sigma _{X}^{2}\left( f^{\prime
}\left( X\right) \right) ^{2}}>>1$ and $C_{3}\left( p,\hat{X}\right) >>1$.

These computations allow to estimate $k$ and $l$. We start with $k$. Given
that (see (\ref{Cfn})): 
\begin{eqnarray*}
k &=&1-\eta \left( 1-\frac{\gamma C_{3}\left( p,\hat{X}\right) }{\left\vert
f\left( \hat{X}\right) \right\vert }\right) \frac{D-\left\Vert \Psi \left( 
\hat{X}\right) \right\Vert ^{2}}{\left\Vert \Psi \left( \hat{X}\right)
\right\Vert ^{2}} \\
&&+\frac{\alpha \left( 2\frac{g^{2}\left( \hat{X}\right) }{\sigma _{\hat{X}%
}^{2}}+\nabla _{\hat{X}}g\left( \hat{X}\right) \right) C_{2}\left( p,\hat{X}%
\right) -\left( 1-\alpha \right) C_{3}\left( p,\hat{X}\right) }{\left\vert
f\left( \hat{X}\right) \right\vert } \\
l &=&\frac{\varsigma F_{1}\left( R\left( K_{\hat{X}},\hat{X}\right) \right)
C_{3}\left( p,\hat{X}\right) }{f\left( \hat{X}\right) } \\
m &=&\left( 1-\frac{\gamma C_{3}\left( p,\hat{X}\right) }{f\left( \hat{X}%
\right) }\right) \frac{D-\left\Vert \Psi \left( \hat{X}\right) \right\Vert
^{2}}{\left\Vert \Psi \left( \hat{X}\right) \right\Vert ^{2}}
\end{eqnarray*}%
the sign of $k$ and $l$\ depend on the magnitude of $K_{\hat{X}}$.

\paragraph*{A4.4.1.1 $K_{\hat{X}}>>1$}

For $K_{\hat{X}}>>1$, using (\ref{mgk}) and:%
\begin{equation*}
\left\Vert \Psi \left( \hat{X}\right) \right\Vert ^{2}=D-\left( \nabla
_{X}R\left( \hat{X}\right) \right) ^{2}K_{\hat{X}}^{\alpha }
\end{equation*}%
we have:%
\begin{equation*}
K_{\hat{X}}^{\alpha }\simeq \frac{D}{\left( \nabla _{\hat{X}}R\left( \hat{X}%
\right) \right) ^{2}}-\frac{C\left( \bar{p}\right) \sigma _{\hat{K}}^{2}%
\sqrt{\frac{M-c}{c}}}{\left( \nabla _{\hat{X}}R\left( \hat{X}\right) \right)
^{2\left( 1-\frac{1}{\alpha }\right) }D^{\frac{1}{\alpha }}c}
\end{equation*}%
and:%
\begin{equation*}
\frac{D-\left\Vert \Psi \left( \hat{X}\right) \right\Vert ^{2}}{\left\Vert
\Psi \left( \hat{X}\right) \right\Vert ^{2}}\simeq \frac{D^{1+\frac{1}{%
\alpha }}c}{C\left( \bar{p}\right) \sigma _{\hat{K}}^{2}\sqrt{\frac{M-c}{c}}%
\left( \nabla _{\hat{X}}R\left( \hat{X}\right) \right) ^{\frac{2}{\alpha }}}
\end{equation*}%
The constant $c$ has been defined in appendix 3, and satisfies $c<<1$. As a
consequence:%
\begin{eqnarray*}
k &\simeq &\eta \frac{\gamma C_{3}\left( p,\hat{X}\right) }{\left\vert
f\left( \hat{X}\right) \right\vert }\frac{D-\left\Vert \Psi \left( \hat{X}%
\right) \right\Vert ^{2}}{\left\Vert \Psi \left( \hat{X}\right) \right\Vert
^{2}}-\left( 1-\alpha \right) \frac{C_{3}\left( p,\hat{X}\right) }{%
\left\vert f\left( \hat{X}\right) \right\vert } \\
&\simeq &\left( \frac{\eta \gamma D^{1+\frac{1}{\alpha }}c}{C\left( \bar{p}%
\right) \sigma _{\hat{K}}^{2}\sqrt{\frac{M-c}{c}}\left( \nabla _{\hat{X}%
}R\left( \hat{X}\right) \right) ^{\frac{2}{\alpha }}}-\left( 1-\alpha
\right) \right) \frac{C_{3}\left( p,\hat{X}\right) }{\left\vert f\left( \hat{%
X}\right) \right\vert }
\end{eqnarray*}%
This may be negative or positive depending on the relative magnitude of $%
\frac{\eta \gamma D^{1+\frac{1}{\alpha }}c}{C\left( \bar{p}\right) \sigma _{%
\hat{K}}^{2}\sqrt{\frac{M-c}{c}}\left( \nabla _{\hat{X}}R\left( \hat{X}%
\right) \right) ^{\frac{2}{\alpha }}}$ and $\left( 1-\alpha \right) $. The
first case correspond to the stable equilibrium with large $K_{\hat{X}}$ and
the second case to the stable case with large $K_{\hat{X}}$ studied in
appendix 2.

\subparagraph{Unstable case}

This case corresponds to:%
\begin{equation*}
\frac{D^{1+\frac{1}{\alpha }}c}{C\left( \bar{p}\right) \sigma _{\hat{K}}^{2}%
\sqrt{\frac{M-c}{c}}\left( \nabla _{\hat{X}}R\left( \hat{X}\right) \right) ^{%
\frac{2}{\alpha }}}>>1
\end{equation*}%
Moreover, using (\ref{Cpt}) and the following estimation, we have:%
\begin{equation}
k\simeq \eta \frac{\gamma C_{3}\left( p,\hat{X}\right) }{\left\vert f\left( 
\hat{X}\right) \right\vert }\frac{\eta \gamma D^{1+\frac{1}{\alpha }}c}{%
C\left( \bar{p}\right) \sigma _{\hat{K}}^{2}\sqrt{\frac{M-c}{c}}\left(
\nabla _{\hat{X}}R\left( \hat{X}\right) \right) ^{\frac{2}{\alpha }}}>>1
\label{kfK}
\end{equation}%
We can also estimate $\left\vert \frac{k}{K_{\hat{X}}}\right\vert $. In this
case:%
\begin{equation}
\frac{k}{K_{\hat{X}}}\simeq \eta \frac{\gamma C_{3}\left( p,\hat{X}\right) }{%
\left\vert f\left( \hat{X}\right) \right\vert }\frac{\eta \gamma D^{\frac{1}{%
\alpha }}c}{C\left( \bar{p}\right) \sigma _{\hat{K}}^{2}\sqrt{\frac{M-c}{c}}%
\left( \nabla _{\hat{X}}R\left( \hat{X}\right) \right) ^{\frac{2}{\alpha }}}%
>>1  \label{stK}
\end{equation}%
We can estimate $l$ by the same token: 
\begin{equation*}
l=\frac{\varsigma F_{1}\left( R\left( K_{\hat{X}},\hat{X}\right) \right)
C_{3}\left( p,\hat{X}\right) }{f\left( \hat{X}\right) }>>1
\end{equation*}%
and using (\ref{stK}) we have:%
\begin{equation*}
\left\vert \frac{k}{K_{\hat{X}}}\right\vert >>l
\end{equation*}%
The coefficient $m$ is obtained by using that in this case:%
\begin{equation*}
m\simeq \left( 1-\frac{\gamma C_{3}\left( p,\hat{X}\right) }{f\left( \hat{X}%
\right) }\right) \frac{D-\left\Vert \Psi \left( \hat{X}\right) \right\Vert
^{2}}{\left\Vert \Psi \left( \hat{X}\right) \right\Vert ^{2}}\simeq -\frac{1%
}{\eta }k
\end{equation*}

\subparagraph{Stable case}

For the stable case we have:%
\begin{equation*}
\frac{\eta \gamma D^{1+\frac{1}{\alpha }}c}{C\left( \bar{p}\right) \sigma _{%
\hat{K}}^{2}\sqrt{\frac{M-c}{c}}\left( \nabla _{\hat{X}}R\left( \hat{X}%
\right) \right) ^{\frac{2}{\alpha }}}-\left( 1-\alpha \right) <0
\end{equation*}%
and we write:%
\begin{equation*}
k\simeq -\left( 1-\alpha \right) \frac{C_{3}\left( p,\hat{X}\right) }{%
\left\vert f\left( \hat{X}\right) \right\vert }<0
\end{equation*}%
We have:%
\begin{equation*}
\left\vert k\right\vert >>1
\end{equation*}%
and moreover:%
\begin{equation}
\left\vert \frac{k}{K_{\hat{X}}}\right\vert \simeq \left( 1-\alpha \right) 
\frac{C_{3}\left( p,\hat{X}\right) }{K_{\hat{X}}\left\vert f\left( \hat{X}%
\right) \right\vert }=\frac{1-\alpha }{\varsigma F_{1}\left( R\left( K_{\hat{%
X}},\hat{X}\right) \right) K_{\hat{X}}}l<<l  \label{nst}
\end{equation}%
The coefficient $m$ is obtained by using that in the stable case:%
\begin{equation*}
m\simeq -\frac{\gamma }{\varsigma F_{1}\left( R\left( K_{\hat{X}},\hat{X}%
\right) \right) }l
\end{equation*}

\paragraph*{A4.4.1.2 $K_{\hat{X}}<<1$}

On the other hand, for $K_{\hat{X}}\leqslant 1$, we have:%
\begin{equation}
\frac{D-\left\Vert \Psi \left( \hat{X}\right) \right\Vert ^{2}}{\left\Vert
\Psi \left( \hat{X}\right) \right\Vert ^{2}}<<1  \label{ngb}
\end{equation}%
so that:%
\begin{equation*}
k\simeq 1+\frac{\alpha \left( 2\frac{g^{2}\left( \hat{X}\right) }{\sigma _{%
\hat{X}}^{2}}+\nabla _{\hat{X}}g\left( \hat{X}\right) \right) C_{2}\left( p,%
\hat{X}\right) -\left( 1-\alpha \right) C_{3}\left( p,\hat{X}\right) }{%
\left\vert f\left( \hat{X}\right) \right\vert }
\end{equation*}%
Given (\ref{CpD}) and (\ref{Cpt}), this yields:%
\begin{equation}
k\simeq -\frac{\left( 1-\alpha \right) C_{3}\left( p,\hat{X}\right) }{%
\left\vert f\left( \hat{X}\right) \right\vert }<0  \label{sgK}
\end{equation}%
and, as in the previous case:%
\begin{eqnarray*}
\left\vert k\right\vert &>&>1 \\
l &>&>1
\end{eqnarray*}%
Moreover, given that $K_{\hat{X}}<<1$:%
\begin{equation}
\left\vert \frac{k}{K_{\hat{X}}}\right\vert >>1  \label{vlk}
\end{equation}%
and:%
\begin{equation}
\left\vert \frac{k}{K_{\hat{X}}}\right\vert >>l
\end{equation}%
Moreover, given (\ref{ngb}):%
\begin{equation*}
\left\vert m\right\vert =\left\vert 1-\frac{\gamma C_{3}\left( p,\hat{X}%
\right) }{f\left( \hat{X}\right) }\right\vert \frac{D-\left\Vert \Psi \left( 
\hat{X}\right) \right\Vert ^{2}}{\left\Vert \Psi \left( \hat{X}\right)
\right\Vert ^{2}}<<\left\vert \frac{\gamma C_{3}\left( p,\hat{X}\right) }{%
f\left( \hat{X}\right) }\right\vert
\end{equation*}%
and:%
\begin{equation*}
\left\vert m\right\vert <<l
\end{equation*}

\paragraph*{A4.4.1.3 Intermediate case\protect\bigskip}

In this case, we can consider that $\frac{D-\left\Vert \Psi \left( \hat{X}%
\right) \right\Vert ^{2}}{\left\Vert \Psi \left( \hat{X}\right) \right\Vert
^{2}}$ is of order $1$:%
\begin{equation}
\frac{D-\left\Vert \Psi \left( \hat{X}\right) \right\Vert ^{2}}{\left\Vert
\Psi \left( \hat{X}\right) \right\Vert ^{2}}=O\left( 1\right)  \label{ngl}
\end{equation}
Assuming that $\gamma <<1$ we have:%
\begin{eqnarray*}
k &\simeq &1+\frac{\alpha \left( 2\frac{g^{2}\left( \hat{X}\right) }{\sigma
_{\hat{X}}^{2}}+\nabla _{\hat{X}}g\left( \hat{X}\right) \right) C_{2}\left(
p,\hat{X}\right) -\left( 1-\alpha \right) C_{3}\left( p,\hat{X}\right) }{%
\left\vert f\left( \hat{X}\right) \right\vert } \\
&\simeq &1+\frac{\frac{\alpha }{2}\left( 2\frac{g^{2}\left( \hat{X}\right) }{%
\sigma _{\hat{X}}^{2}}+\nabla _{\hat{X}}g\left( \hat{X}\right) \right) -%
\frac{1-\alpha }{2}\left( \frac{K_{\hat{X}}\left\Vert \Psi \left( \hat{X}%
\right) \right\Vert ^{2}\left\vert f\left( \hat{X}\right) \right\vert }{%
C\left( \bar{p}\right) \sigma _{\hat{K}}^{2}}\right) ^{4}\exp \left(
-W\left( k,-4C_{0}\left( \hat{X},K_{\hat{X}}\right) \right) \right) }{%
\left\vert f\left( \hat{X}\right) \right\vert }\ln \frac{96\left\vert
f\left( \hat{X}\right) \right\vert ^{3}}{\sigma _{X}^{2}\left( f^{\prime
}\left( X\right) \right) ^{2}}
\end{eqnarray*}%
Given that the intermediate case is stable (see appendix 2), the relation
between $K_{\hat{X}}$ and $R\left( \hat{X}\right) $ is positive, we can
assume that $k<0$ and:%
\begin{eqnarray*}
k &\simeq &1+\frac{\alpha \left( 2\frac{g^{2}\left( \hat{X}\right) }{\sigma
_{\hat{X}}^{2}}+\nabla _{\hat{X}}g\left( \hat{X}\right) \right) C_{2}\left(
p,\hat{X}\right) -\left( 1-\alpha \right) C_{3}\left( p,\hat{X}\right) }{%
\left\vert f\left( \hat{X}\right) \right\vert } \\
&\simeq &-\frac{\frac{1-\alpha }{2}\left( \frac{K_{\hat{X}}\left\Vert \Psi
\left( \hat{X}\right) \right\Vert ^{2}\left\vert f\left( \hat{X}\right)
\right\vert }{C\left( \bar{p}\right) \sigma _{\hat{K}}^{2}}\right) ^{4}\exp
\left( -W\left( 0,-4C_{0}\left( \hat{X},K_{\hat{X}}\right) \right) \right) }{%
\left\vert f\left( \hat{X}\right) \right\vert }\ln \frac{96\left\vert
f\left( \hat{X}\right) \right\vert ^{3}}{\sigma _{X}^{2}\left( f^{\prime
}\left( X\right) \right) ^{2}}
\end{eqnarray*}%
and:%
\begin{eqnarray*}
l &=&\frac{\varsigma F_{1}\left( R\left( K_{\hat{X}},\hat{X}\right) \right)
C_{3}\left( p,\hat{X}\right) }{f\left( \hat{X}\right) }\simeq l \\
&=&\frac{\varsigma F_{1}\left( R\left( K_{\hat{X}},\hat{X}\right) \right)
\left( \frac{K_{\hat{X}}\left\Vert \Psi \left( \hat{X}\right) \right\Vert
^{2}\left\vert f\left( \hat{X}\right) \right\vert }{C\left( \bar{p}\right)
\sigma _{\hat{K}}^{2}}\right) ^{4}\exp \left( -W\left( 0,-4C_{0}\left( \hat{X%
},K_{\hat{X}}\right) \right) \right) }{f\left( \hat{X}\right) }\ln \frac{%
96\left\vert f\left( \hat{X}\right) \right\vert ^{3}}{\sigma _{X}^{2}\left(
f^{\prime }\left( X\right) \right) ^{2}}
\end{eqnarray*}%
Note that in this case:%
\begin{equation*}
k\simeq -\frac{1-\alpha }{\varsigma F_{1}\left( R\left( K_{\hat{X}},\hat{X}%
\right) \right) }l
\end{equation*}%
and, given (\ref{ngl}):%
\begin{equation*}
m\simeq -\gamma \frac{D-\left\Vert \Psi \left( \hat{X}\right) \right\Vert
^{2}}{\left\Vert \Psi \left( \hat{X}\right) \right\Vert ^{2}\varsigma
F_{1}\left( R\left( K_{\hat{X}},\hat{X}\right) \right) }l
\end{equation*}

\subsubsection*{A4.4.2 Stability conditions}

\paragraph*{A4.4.2.1 Case $K_{\hat{X}}>>1$}

\subparagraph{Stable case}

As shown above, $k<0$, $\left\vert k\right\vert >>1$, $l>>1$ and $\left\vert 
\frac{k}{K_{\hat{X}}}\right\vert <<l$. Coefficients $l$ and $m$ are of the
same order. Thus (\ref{stB}) becomes:%
\begin{equation*}
\frac{l^{2}a_{0}c}{\left( R\left( \hat{X}\right) \right) ^{2}}+\frac{%
4m^{2}ca_{0}}{\left( \nabla _{\hat{X}}R\left( \hat{X}\right) \right) ^{2}}%
G^{2}>0
\end{equation*}%
That is, for $c>0$ the oscillations are stable, whereas for $c<0$ they are
unstable.

\subparagraph{Unstable case}

In this case, $k>0$, $\left\vert k\right\vert >>1$, $l>>1$ and $\left\vert 
\frac{k}{K_{\hat{X}}}\right\vert >>l$. We have also $m\simeq -\frac{1}{\eta }%
k$ and (\ref{stB}) writes: 
\begin{equation}
\frac{cl}{R\left( \hat{X}\right) }\frac{k}{K_{\hat{X}}}+\frac{4k^{2}ca_{0}}{%
\eta ^{2}\left( \nabla _{\hat{X}}R\left( \hat{X}\right) \right) ^{2}}G^{2}>0
\end{equation}%
That is, for $c>0$ the oscillations are stable, whereas for $c<0$ they are
unstable.

\paragraph*{A4.4.2.1.2 Case $K_{\hat{X}}<<1$}

Equations (\ref{sgK}) and (\ref{vlk}) show that $k<0$, $\left\vert
k\right\vert >>1$, $l>>1$, $\left\vert m\right\vert <<l$ and $\left\vert 
\frac{k}{K_{\hat{X}}}\right\vert >>l$. Equation (\ref{stB}) thus writes: 
\begin{equation}
\frac{cl}{R\left( \hat{X}\right) }\frac{k}{K_{\hat{X}}}>0
\end{equation}%
That is, for $c>0$ the oscillations are unstable, whereas for $c<0$ they are
stable.

\paragraph*{A4.4.2.3 Intermediate case}

In this case, we have seen above that $k<0$:%
\begin{equation*}
k\simeq -\frac{1-\alpha }{\varsigma F_{1}\left( R\left( K_{\hat{X}},\hat{X}%
\right) \right) }l
\end{equation*}%
and:%
\begin{equation*}
m\simeq -\gamma \frac{D-\left\Vert \Psi \left( \hat{X}\right) \right\Vert
^{2}}{\left\Vert \Psi \left( \hat{X}\right) \right\Vert ^{2}\varsigma
F_{1}\left( R\left( K_{\hat{X}},\hat{X}\right) \right) }
\end{equation*}
As a consequence, equation (\ref{stB}) particularizes as:

\begin{equation*}
\frac{l^{2}c}{R\left( \hat{X}\right) }\left( \frac{a_{0}}{R\left( \hat{X}%
\right) }-\frac{1-\alpha }{\varsigma K_{\hat{X}}F_{1}\left( R\left( K_{\hat{X%
}},\hat{X}\right) \right) }\right) +4ca_{0}\left( \frac{\gamma \left(
D-\left\Vert \Psi \left( \hat{X}\right) \right\Vert ^{2}\right) }{\varsigma
\nabla _{\hat{X}}R\left( \hat{X}\right) F_{1}\left( R\left( K_{\hat{X}},\hat{%
X}\right) \right) \left\Vert \Psi \left( \hat{X}\right) \right\Vert ^{2}}%
\right) ^{2}G^{2}>0
\end{equation*}%
Given the definition of $a_{0}$ and the stability of the intermediate case,
we assume $a_{0}>0$. Thus, 2 possibilities arise.

\subparagraph{Coefficient $c>0$}

In this case, the oscillations are stable if: 
\begin{equation*}
\frac{a_{0}}{R\left( \hat{X}\right) }-\frac{1-\alpha }{\varsigma K_{\hat{X}%
}F_{1}\left( R\left( K_{\hat{X}},\hat{X}\right) \right) }>0
\end{equation*}
or if:%
\begin{equation*}
\frac{a_{0}}{R\left( \hat{X}\right) }-\frac{1-\alpha }{\varsigma K_{\hat{X}%
}F_{1}\left( R\left( K_{\hat{X}},\hat{X}\right) \right) }<0
\end{equation*}%
and:%
\begin{equation*}
G^{2}>\frac{l^{2}\left( \nabla _{\hat{X}}R\left( \hat{X}\right) \right) ^{2}%
}{4a_{0}R\left( \hat{X}\right) }\left( \frac{\varsigma \left( \nabla _{\hat{X%
}}R\left( \hat{X}\right) F_{1}\left( R\left( K_{\hat{X}},\hat{X}\right)
\right) \left\Vert \Psi \left( \hat{X}\right) \right\Vert ^{2}\right) }{%
\gamma \left( D-\left\Vert \Psi \left( \hat{X}\right) \right\Vert
^{2}\right) }\right) ^{2}\left\vert \frac{a_{0}}{R\left( \hat{X}\right) }-%
\frac{1-\alpha }{\varsigma F_{1}\left( R\left( K_{\hat{X}},\hat{X}\right)
\right) }\right\vert
\end{equation*}%
the oscillations are unstable.

\subparagraph{Coefficient $c<0$}

The oscillations are stable if:%
\begin{equation*}
\frac{a_{0}}{R\left( \hat{X}\right) }-\frac{1-\alpha }{\varsigma K_{\hat{X}%
}F_{1}\left( R\left( K_{\hat{X}},\hat{X}\right) \right) }<0
\end{equation*}%
and:%
\begin{equation*}
G^{2}<\frac{l^{2}\left( \nabla _{\hat{X}}R\left( \hat{X}\right) \right) ^{2}%
}{4a_{0}R\left( \hat{X}\right) }\left( \frac{\varsigma \left( \nabla _{\hat{X%
}}R\left( \hat{X}\right) F_{1}\left( R\left( K_{\hat{X}},\hat{X}\right)
\right) \left\Vert \Psi \left( \hat{X}\right) \right\Vert ^{2}\right) }{%
\gamma \left( D-\left\Vert \Psi \left( \hat{X}\right) \right\Vert
^{2}\right) }\right) ^{2}\left\vert \frac{a_{0}}{R\left( \hat{X}\right) }-%
\frac{1-\alpha }{\varsigma F_{1}\left( R\left( K_{\hat{X}},\hat{X}\right)
\right) }\right\vert
\end{equation*}

\section*{References}


\begin{description}
\item Abergel F, Chakraborti A, Muni Toke I and Patriarca M (2011a)
Econophysics review: I. Empirical facts, Quantitative Finance, Vol. 11, No.
7, 991-1012

\item Abergel F, Chakraborti A, Muni Toke I and Patriarca M (2011b)
Econophysics review: II. Agent-based models, Quantitative Finance, Vol. 11,
No. 7, 1013-1041

\item Bernanke,B., Gertler, M. and S. Gilchrist (1999), "The financial
accelerator in a quantitative business cycle framework", Chapter 21 in
Handbook of Macroeconomics, 1999, vol. 1, Part C, pp 1341-1393

\item Bensoussan A, Frehse J, Yam P (2018) Mean Field Games and Mean Field
Type Control Theory. Springer, New York

\item B\"{o}hm, V., Kikuchi, T., Vachadze, G.: Asset pricing and
productivity growth: the role of consumption scenarios. Comput. Econ. 32,
163--181 (2008)

\item Caggese A, Orive A P, The Interaction between Household and Firm
Dynamics and the Amplification of Financial Shocks. Barcelona GSE Working
Paper Series, Working Paper n%
${{}^o}$
866, 2015

\item Campello, M., Graham, J. and Harvey, C.R. (2010). "The Real Effects of
Financial Constraints: Evidence from a Financial Crisis," Journal of
Financial Economics, vol. 97(3), 470-487.

\item Gaffard JL and Napoletano M Editors (2012): Agent-based models and
economic policy. OFCE, Paris

\item Gomes DA, Nurbekyan L, Pimentel EA (2015) Economic Models and
Mean-Field Games Theory, Publica\c{c}\~{o}es Matem\'{a}ticas do IMPA, 30o Col%
\'{o}quio Brasileiro de Matem\'{a}tica, Rio de Janeiro

\item Gosselin P, Lotz A and Wambst M (2017) A Path Integral Approach to
Interacting Economic Systems with Multiple Heterogeneous Agents. IF\_PREPUB.
2017. hal-01549586v2

\item Gosselin P, Lotz A and Wambst M (2020) A Path Integral Approach to
Business Cycle Models with Large Number of Agents. Journal of Economic
Interaction and Coordination volume 15, pages 899--942

\item Gosselin P, Lotz A and Wambst M (2021) A statistical field approach to
capital accumulation. Journal of Economic Interaction and Coordination 16,
pages 817--908 (2021)

\item Grassetti, F., Mammana, C. \& Michetti, E. A dynamical model for real
economy and finance. Math Finan Econ (2022).
https://doi.org/10.1007/s11579-021-00311-3

\item Grosshans, D., Zeisberger, S.: All's well that ends well? on the
importance of how returns are achieved. J. Bank. Finance 87, 397--410 (2018)

\item Holmstrom, B., and Tirole, J. (1997). Financial intermediation,
loanable funds, and the
\end{description}

real sector. Quarterly Journal of Economics, 663-691.

\begin{description}
\item Jackson M (2010) Social and Economic Networks. Princeton University
Press, Princeton

\item Jermann, U.J. and Quadrini, V., (2012). "Macroeconomic Effects of
Financial Shocks," American Economic Review, Vol. 102, No. 1.

\item Khan, A., and Thomas, J. K. (2013). "Credit Shocks and Aggregate
Fluctuations in an Economy with Production Heterogeneity," Journal of
Political Economy, 121(6), 1055-1107.

\item Kaplan G, Violante L (2018) Microeconomic Heterogeneity and
Macroeconomic Shocks, Journal of Economic Perspectives, Vol. 32, No. 3,
167-194

\item Kleinert H (1989) Gauge fields in condensed matter Vol. I , Superflow
and vortex lines, Disorder Fields, Phase Transitions, Vol. II, Stresses and
defects, Differential Geometry, Crystal Melting. World Scientific, Singapore

\item Kleinert H (2009) Path Integrals in Quantum Mechanics, Statistics,
Polymer Physics, and Financial Markets 5th edition. World Scientific,
Singapore

\item Krugman P (1991) Increasing Returns and Economic Geography. Journal of
Political Economy, 99(3), 483-499

\item Lasry JM, Lions PL, Gu\'{e}ant O (2010a) Application of Mean Field
Games to Growth Theory \newline
https://hal.archives-ouvertes.fr/hal-00348376/document

\item Lasry JM, Lions PL, Gu\'{e}ant O (2010b) Mean Field Games and
Applications. Paris-Princeton lectures on Mathematical Finance, Springer.%
\textbf{\ }https://hal.archives-ouvertes.fr/hal-01393103

\item Lux T (2008) Applications of Statistical Physics in Finance and
Economics. Kiel Institute for the World Economy (IfW), Kiel

\item Lux T (2016) Applications of Statistical Physics Methods in Economics:
Current state and perspectives. Eur. Phys. J. Spec. Top. (2016) 225: 3255.
https://doi.org/10.1140/epjst/e2016-60101-x

\item Mandel A, Jaeger C, F\"{u}rst S, Lass W, Lincke D, Meissner F,
Pablo-Marti F, Wolf S (2010). Agent-based dynamics in disaggregated growth
models. Documents de travail du Centre d'Economie de la Sorbonne. Centre
d'Economie de la Sorbonne, Paris

\item Mandel A (2012) Agent-based dynamics in the general equilibrium model.
Complexity Economics 1, 105--121

\item Monacelli, T., Quadrini, V. and A. Trigari (2011). "Financial Markets
and Unemployment," NBER Working Papers 17389, National Bureau of Economic
Research.

\item Sims C A (2006) Rational inattention: Beyond the Linear Quadratic
Case, American Economic Review, vol. 96, no. 2, 158-163

\item Yang J (2018) Information theoretic approaches to economics, Journal
of Economic Survey, Vol. 32, No. 3, 940-960

\item Cochrane, J.H. (ed.): Financial Markets and the Real Economy,
International Library of Critical Writings in Financial Economics, vol. 18.
Edward Elgar (2006)
\end{description}

\end{document}